 \documentclass[twocolumn]{aastex631}

\newcommand{\kms}{km s$^{-1}$}

\newcommand{\vlsr}{$V_\mathrm{LSR}$}

\newcommand{\co}{$^{12}$CO}
\newcommand{\Tco}{$^{13}$CO}
\newcommand{\Tcoj}{$^{13}$CO~(1--0)}
\newcommand{\cEo}{C$^{18}$O}
\newcommand{\cEoj}{C$^{18}$O~(1--0)}
\newcommand{\nThp}{N$_2$H$^{+}$}
\newcommand{\nThpj}{N$_2$H$^{+}$~(1--0)}

\newcommand{\nhTd}{NH$_2$D}
\newcommand{\nhTdj}{ortho-NH$_2$D~(1$_{11}$--1$_{01}$)}
\newcommand{\cs}{CS}
\newcommand{\csj}{CS (2--1)}
\newcommand{\so}{SO}
\newcommand{\soj}{SO~(3$_2$--2$_1$)}
\newcommand{\hcop}{HCO$^+$}
\newcommand{\hcopj}{HCO$^+$~(1--0)}
\newcommand{\hTcop}{H$^{13}$CO$^+$}
\newcommand{\hTcopj}{H$^{13}$CO$^+$~(1--0)}

\shorttitle{Filaments and dense cores in Orion~B}
\shortauthors{Yoo et al.}
\graphicspath{{./}{figures/}}

\begin{document}

\title{TRAO Survey of Nearby Filamentary Molecular clouds, the Universal Nursery of Stars (TRAO-FUNS). 
\\ III. Filaments and dense cores in the NGC~2068 and NGC~2071 regions of Orion~B}

\correspondingauthor{Hyunju Yoo}
\email{hyunju527@gmail.com}

\author[0000-0002-8578-1728]{Hyunju Yoo}
\affiliation{Department of Astronomy and Space Science, Chungnam National University, 99 Daehak-ro, Yuseong-gu, \\ Daejeon 34134, Republic of Korea}
\affiliation{Korea Astronomy and Space Science Institute, 776 Daedeokdae-ro, Yuseong-gu, Daejeon 34055, Republic of Korea}


\author[0000-0002-3179-6334]{Chang Won Lee}
\affiliation{Korea Astronomy and Space Science Institute, 776 Daedeokdae-ro, Yuseong-gu, Daejeon 34055, Republic of Korea}
\affiliation{University of Science and Technology, Korea (UST), 217 Gajeong-ro, Yuseong-gu, Daejeon 34113, Republic of Korea}

\author[0000-0003-0014-1527]{Eun Jung Chung}
\affiliation{Department of Astronomy and Space Science, Chungnam National University, 99 Daehak-ro, Yuseong-gu, \\ Daejeon 34134, Republic of Korea}

\author[0000-0001-9333-5608]{Shinyoung Kim}
\affiliation{Korea Astronomy and Space Science Institute, 776 Daedeokdae-ro, Yuseong-gu, Daejeon 34055, Republic of Korea}

\author[0000-0002-2569-1253]{Mario Tafalla}
\affiliation{Observatorio Astronómico Nacional (IGN), Alfonso XII 3, E-28014 Madrid, Spain}

\author[0000-0003-1481-7911]{Paola Caselli}
\affiliation{Max-Planck-Institut für Extraterrestrische Physik, Gießenbachstrasse 1, 85748 Garching bei München, Germany}

\author[0000-0002-2885-1806]{Philip C. Myers}
\affiliation{Harvard-Smithsonian Center for Astrophysics, 60 Garden Street, Cambridge, MA 02138, USA}

\author[0000-0001-9597-7196]{Kyoung Hee Kim}
\affiliation{The Korea Astronomy Society, 776, Daedeok-daero, Yuseong-gu, Daejeon, Republic of Korea}

\author[0000-0002-5286-2564]{Tie Liu}
\affiliation{Key Laboratory for Research in Galaxies and Cosmology, Shanghai Astronomical Observatory, Chinese Academy of Sciences, \\ 80 Nandan Road, Shanghai 200030, People's Republic of China}

\author[0000-0003-4022-4132]{Woojin Kwon}
\affiliation{Department of Earth Science Education, Seoul National University, 1 Gwanak-ro, Gwanak-gu, Seoul 08826, Republic of Korea}
\affiliation{SNU Astronomy Research Center, Seoul National University, 1 Gwanak-ro, Gwanak-gu, Seoul 08826, Republic of Korea}

\author[0000-0002-6386-2906]{Archana Soam}
\affiliation{Indian Institute of Astrophysics, II Block, Koramangala, Bengaluru 560034, India}
\affiliation{SOFIA Science Center, Universities Space Research Association, NASA Ames Research Center, Moffett Field, CA 94035, USA}

\author[0000-0002-1229-0426]{Jongsoo Kim}
\affiliation{Korea Astronomy and Space Science Institute, 776 Daedeokdae-ro, Yuseong-gu, Daejeon 34055, Republic of Korea}







\begin{abstract}

We present the results of molecular line observations performed toward the NGC~2068 and NGC~2071 regions of the Orion~B cloud as the TRAO-FUNS project to study the roles of the filamentary structure in the formation of dense cores and stars in the clouds.
Gaussian decomposition for the \cEo\ spectra with multiple velocity components and application of a Friends-of-Friends algorithm for the decomposed components allowed us to identify a few tens of velocity coherent filaments. 
We also identified 48 dense cores from the observations of \nThp\ using a core finding tool, $\texttt{FellWalker}$. 
We made the virial analysis for these filaments and dense cores, finding that the filaments with \nThp\ dense core are thermally supercritical, and the filaments with larger ratio between the line mass and the thermal critical line mass
tend to have more dense cores. 
We investigated the contribution of the nonthermal motions in dense cores and filaments, showing the dense cores are mostly in transonic/subsonic motions while their natal filaments are mostly in supersonic motions. This may indicates that gas turbulent motions in the filaments have been dissipated at the core scale to form the dense cores there.
The filaments with (dynamically evolved) dense cores in infalling motions or with \nhTd\ bright (or chemically evolved) dense cores are all found to be gravitationally critical.
Therefore, the criticality of the filament is thought to provide a key condition for its fragmentation, the formation of dense cores, and their kinematical and chemical evolution.

\end{abstract}



\section{Introduction} \label{sec:intro}

The prevalence of morphologically elongated structures, so called filament, has been discovered over a wide range of spatial scales and is generally accepted as a key feature of star formation.  
According to observations at infrared/sub-millimeter wavelengths, the interstellar molecular clouds are found to be composed of a network of the filamentary structures \citep{andre10, andre14, arzoumanian11, arzoumanian19, hacar11, hacar13, hacar18, menshchikov10, molinari10}.
Especially, recent observational facilities (e.g., Herschel Space Observatory, and Atacama Large Millimeter/submillimeter Array) have been expanding the horizon for understanding star formation in early phase and gave opportunities to study the relation between diffuse filamentary structures and dense cores in depth \citep{andre10, hacar18}. 
Analytic approaches have advocated that the formation of filamentary structure in the diffuse medium is a spontaneous phenomenon 
\citep{padoan01, gomez14, wareing16}.
Although the importance of the filaments in star-forming regions is emphasized from many observations and numerical simulations in recent decades, the role of filamentary structures on the formation of dense cores and/or stars and the formation process of filaments themselves are still debated. The main difficulty to understand the structure and physical properties of filaments arises from the fact that we are looking at the complexity of the three-dimensional organization of molecular clouds along the line-of-sight. 
Even if we spatially resolve the small-scale structure of the molecular cloud with high angular resolution, we still need spectroscopic information to distinguish the multiple components with different velocities overlapped in the line-of-sight from  optically thin molecular line emission. Spatially and spectroscopically adequate observations are highly required for improving our understanding on the dense core and star formation in the filamentary clouds.

The Orion~B molecular cloud is the most suitable place to study both ongoing low- and high-mass star formation since it contains not only various evolutionary stages of young stellar objects 
but also massive stars. Furthermore, its relatively close distance 
($\sim$423~pc; \citealt{menten07, zucker19}) allows us to have less contamination from foreground materials than other massive star-forming clouds. 

The large-scale structures of Orion B have been revealed from molecular line observations. In the early studies, the bulk of the molecular gas with CO emission peaks has been shown. However, the velocity gradients and the detailed small-scale gas distribution within sub-structures have not been resolved due to low spatial and spectral resolution 
\citep{kutner77, maddalena86, bally91, wilson05}.
Molecular gas surveys with higher resolution have been carried out to investigate kinematics and distribution of condensed structures. 
\citet{lada91} identified clumpy substructures and provided a census of dense cores from unbiased CS (2--1) survey with 1.7~arcmin resolution. Dense cores have also been found in the dense gas tracer \hTcop\
\citep{aoyama01, ikeda09, walkersmith13} and the core properties were listed at the sub-mm regime 
\citep{johnstone01, motte01, mitchell01, kirk16}.
Especially, the survey observations in the CO isotopologues 
\citep{buckle10, walkersmith13, nishimura15} revealed that the Orion~B molecular cloud is composed of filamentary structures in the form of overdense gas. 

A large-scale molecular line survey has been carried out to map over 1.9 deg$^2$ around the NGC~2023 and NGC~2024 regions in the Orion~B molecular cloud with the IRAM-30m telescope in the context of the ORION-B project \citep{pety17}. Among their various molecular transitions observed at the 3mm wavelength band, \citet{orkisz19} used \cEoj\ integrated intensity map to search for a network of filaments and study the kinematical properties and statistical characteristics of the filaments in NGC 2023 and NGC 2024 region. 
We note that the region observed by \citet{pety17} is the southern part of the Orion~B complex (i.e. L1630N) and the target region for this paper is the northern part of the Orion~B, which includes two bright reflection nebulae NGC~2068 and NGC~2071 (see overview by \citealt{bally08, gibb08} and references therein). 

Using the extensive imaging from the Herschel Gould Belt Survey, \citet{schneider13} obtained a probability distribution function (PDF) from the column density map and presented that the PDF of Orion~B (covering both NGC~2023/24 and NGC~2068/71) shows a log-normal distribution with a power-law tail at high column density range. They explained that the existence of the power-law tail may be caused by global gravitational collapse of large covering areas like filaments and ridges. Based on the column density map at 18.2\arcsec~angular resolution, \citet{konyves20} identified filament networks in Orion~B and presented the distributions of dense cores.

As highlighted in \citet{hacar13}, however, the spectral information from optically thin molecular lines (e.g., \cEoj) is required to reveal multiple velocity components overlapped along the line of the sight. Even the filaments identified by Herschel continuum image in high spatial resolution can reveal detailed structure of the molecular cloud, they suffer from a lack of information on kinematic properties and thus can be in some cases multiple identities overlapped to the line-of-sight. 
Therefore, it would be highly requiring to present spectroscopic observations toward Orion B in part to explore complex gas kinematics in the three-dimensional position-position-velocity (PPV) space, and also to study evolutionary signatures in Orion~B. In this paper we present physical properties of filamentary structures identified by the \cEoj\ line and dense cores defined from \nThpj\ emission.

The paper is organized as follows. We summarize the previous studies toward the observing regions in Section~\ref{sec:intro} and describe the observation and data reduction in Section~\ref{sec:obs}. The results on identification and physical properties of filaments and dense cores are given in Section~\ref{sec:result}. We discuss on the kinematic and chemical properties of filaments and dense cores in Section~\ref{sec:disc} and give the summary of the study in Section~\ref{sec:sum}.

\section{Observation and Data Reduction} \label{sec:obs}
\subsection{TRAO-FUNS survey}

\begin{deluxetable*}{ccccccccc}
\tablenum{1}
\tablecaption{Information for TRAO observations  \label{tab:obs}}
\tablecolumns{9}
\tablewidth{0pt}
\tablehead{
\colhead{Molecular line} &
\colhead{Frequency} &
\colhead{$\theta_{\rm B}$} & 
\colhead{Area} & 
\colhead{Pixel size} & 
\colhead{$\Delta v$}  & 
\colhead{\textit{RMS}} &  
\colhead{\textit{RMS}$_{\rm m0}$} &
\colhead{\textit{n}$_{\rm crit}$} \\ 
\colhead{} & \colhead{(GHz)} &\colhead{(arcsec)} & 
\colhead{(degree$^2$)} & \colhead{(arcsec)} & \colhead{(\kms)} & 
\colhead{(K)} & 
\colhead{(K \kms)} & 
\colhead{(cm$^{-3}$)} 
}
\decimalcolnumbers
\startdata
\nhTdj &  85.9262780 & 57.0 & 0.106 & 22 & 0.1 & 0.049 & 0.031 & 6.5$\times10^4$\\
\hTcopj &  86.7542884 & 56.6 & 0.106 & 22 & 0.06 & 0.067 & 0.026 & 6.2$\times10^4$\\
\hcopj &  89.1885247 & 55.6 & 0.559 & 22 & 0.06 & 0.069 & 0.041 & 6.8$\times10^4$\\
\nThpj & 93.1737637 & 54.1 & 0.559 & 22 & 0.1 & 0.053 & 0.034 & 6.1$\times10^4$\\
\csj & 97.9809533 & 52.4 & 0.394 & 22 & 0.06 & 0.068 & 0.037 & 1.3$\times10^5$\\
\soj & 99.2998700 & 51.9 & 0.394 & 22 & 0.1 & 0.053 & 0.036 & 1.9$\times10^3$\\
\cEoj & 109.7821734 & 48.7 & 1.365 & 22 & 0.1 & 0.114 & 0.075 & 1.9$\times10^3$\\
\Tcoj & 110.2013543 & 48.6 & 1.365 & 22 & 0.1 & 0.109 & 0.064 & 3.5$\times10^5$\\
\enddata
\tablecomments{
Column 1: The molecular line transition.
Column 2: The rest frequencies are from \citet{lovas04} for \Tcoj, \citet{pickett98} for \soj, \citet{lee01} for \cEoj, \nThpj, and \csj. We refer to \citet{muller01} for \hcopj, \nhTdj, and \hTcopj.  
Column 3: The half power beam width (HPBW). 
Column 4: The total area of the observed region. 
Column 5: The pixel size in arcsec. 
Column 6: The velocity channel width. 
Column 7 and 8: The Noise levels of the data cubes and integrated intensity maps, respectively, in T$^{\star}_A$. 
Column 9: The critical densities are from \citet{wienen21} for \nhTdj, \citet{shirley15} for \hTcopj, \hcopj, \nThpj, and \csj, and \citet{chung21} for \soj, \cEoj, and \Tcoj. }
\end{deluxetable*}

The mapping observations toward the star-forming molecular cloud of Orion~B were carried out using the multibeam receiver SEcond QUabbin Observatory Imaging Array (SEQUOIA) of Taeduk Radio Astronomy Observatory (TRAO)\footnote{https://radio.kasi.re.kr/trao/}, which is a 14-m single-dish radio telescope located in South Korea. The receiver SEQUOIA-TRAO is equipped with 16 pixel MMIC (Monolithic Microwave Integrated Circuit) pre-amplifiers in a 4$\times$4 array and provides an observing frequency range from 85 to 115~GHz. The backend system, a FFT spectrometer, has 8192 channels with a spectral resolution of $\sim$15~kHz which corresponds to a velocity resolution of $\sim$0.04 \kms\ at 110~GHz and a full spectral bandwidth of $\sim$60~MHz. The narrow band second IF modules enable to observe two frequencies simultaneously within the 85--100 or 100--115~GHz bands. 
The main beam efficiencies of the telescope are 0.48 at 98 GHz, and 0.46 at 110 GHz, respectively \citep{jeong19}. 
The observations were made from December 2016 to October 2018 as part of the TRAO Key Science Program, TRAO Survey of Nearby Filamentary Molecular clouds, the Universal Nursery of Stars (TRAO-FUNS).

\begin{figure}[ht!]
\includegraphics[width=0.49\textwidth]{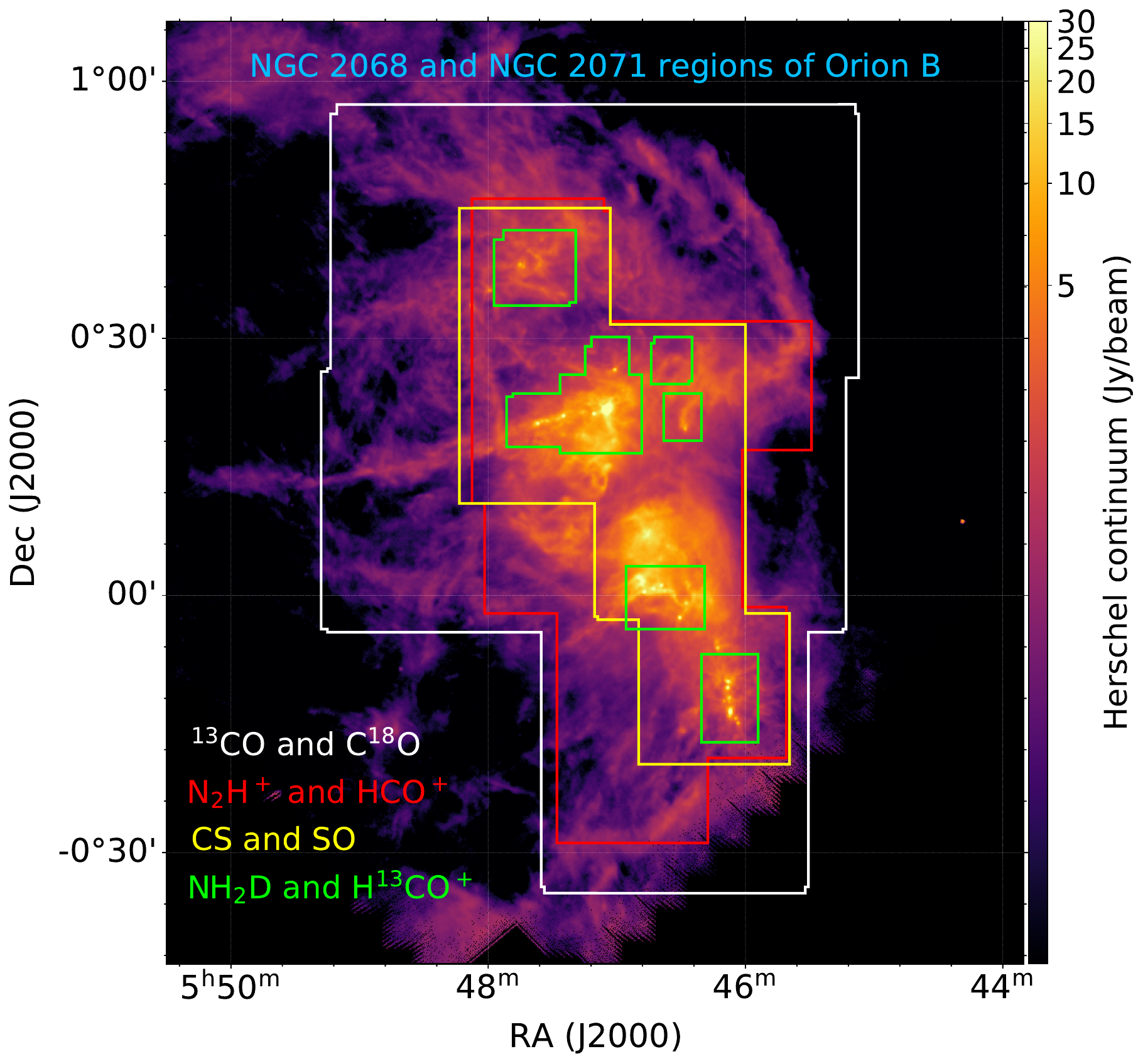}
\caption{The survey areas of Orion~B. The areas observed with various molecular lines are marked with white (\Tco~and \cEo), red (\nThp~and \hcop), yellow (\cs~and \so), and green (\nhTd~and \hTcop) boxes on the Herschel 250 $\mu m$ continuum image.
\label{fig:hsh}}
\end{figure}

In order to investigate the physical properties and the chemical characteristics of dense cores and surrounding filamentary structures in Orion~B including two reflection nebulae (NGC~2071 and NGC~2068), we selected eight molecular lines; \Tcoj, \cEoj, \nThpj, \nhTdj, \csj, \soj, \hcopj, and \hTcopj. 
\Tcoj\ and \cEoj\ lines were chosen to trace the distribution of molecular gas in the star-forming interstellar molecular cloud. Systematic observations of \csj\ and \hcopj\ lines were performed to examine the evidence of infall motion in dense cores. The \nThpj\ and \hTcopj\ lines can be used to determine the systemic velocities of the dense cores since they are usually optically thin. Furthermore, \soj\ and \nhTdj\ were selected to test the chemical state of the dense cores. The observational details including the rest frequency, the FWHM beam size, observing area, the pixel size, the channel width and the rms noise level for each transition are given in Table~\ref{tab:obs}. 

On-The-Fly (OTF) mapping observations were carried out and shown in the regions marked in Figure~\ref{fig:hsh} with the Herschel 250 $\mu m$ continuum \citep{andre10, HGBS} on the background, making the telescope system to continuously scan the mapping region along the scanning axis with a smooth and rapid speed. This observing mode benefits from the speedy observing capability of a large area with the minimum change in the setup of the observing system and improves the observing efficiency.  
We observed the most wide area in \Tco\ and \cEo\ lines ($\sim$1.37 deg$^2$, white box in Figure~\ref{fig:hsh}) to trace all the most extended faint structures. \nThp~and \hcop~lines ($\sim$0.56 deg$^2$, red box in Figure~\ref{fig:hsh}) and \cs~and \so~lines ($\sim$0.39 deg$^2$, yellow box in Figure~\ref{fig:hsh}) were observed toward the dense core areas including their nearby envelopes. Finally, the observations of \nhTd~and \hTcop~lines were focused toward the densest regions in the cloud ($\sim$0.11 deg$^2$, green boxes in Figure~\ref{fig:hsh}).

\begin{figure*}
  \centering 
  \includegraphics[width=\textwidth]{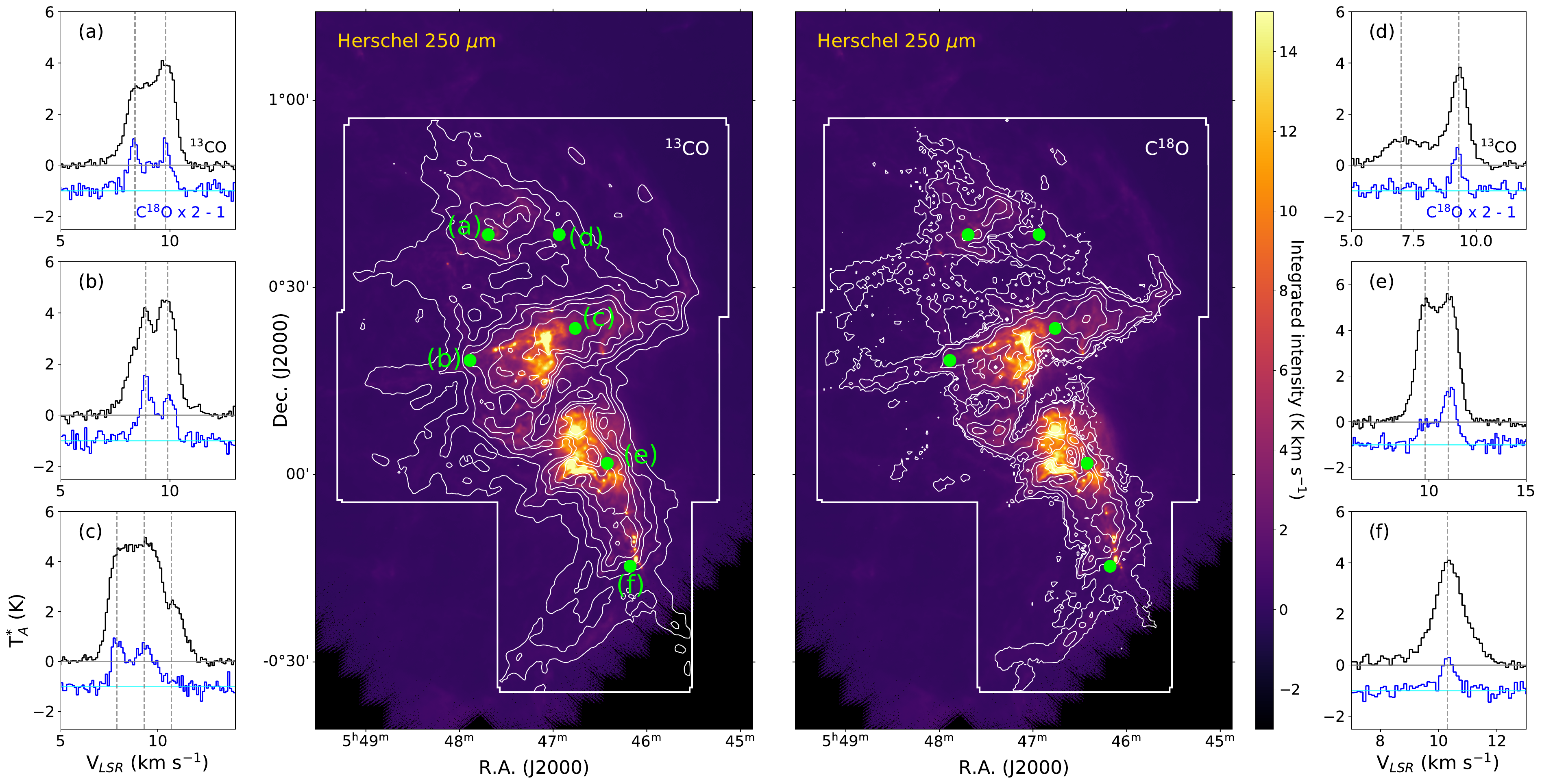}
  \caption{The color scale presents Herschel 250 $\mu m$ continuum image of the Orion~B region. Contours are from the integrated intensity maps of \Tco~(middle-left panel) and \cEo~(middle-right panel). The contour levels for the \Tco~integrated intensity map are 
  [1, 3, 5, 7, 9, 11] K \kms~and those for \cEo~integrated intensity map are 
  [0.25, 0.75, 1, 1.5, 2.2, 3.5, 6] K \kms. 
 Panels (a) to (f) are representative spectra of \Tco~(black line) and \cEo~(blue line) at the central positions marked with green dots in the middle-left panel and the middle-right panel. In the panel (a) to (f), the \cEo~spectrum is scaled up by a factor of 2 and shifted by --1 in intensity. The gray and cyan horizontal lines are baselines of the \Tco~and \cEo~spectra, respectively. 
\label{fig:hsh_sp}}
\end{figure*}

\subsection{Data reduction}
The raw data were obtained in the shape of a tile, an OTF scan individually observed, toward line emitting regions for each set of molecular lines (see Figure~\ref{fig:hsh}). After a first-order baseline correction using the OTFTOOL, the raw data were re-gridded to have a cell size of 22\arcsec\ and converted into the CLASS format. A noise-weighting is applied to have better signal-to-noise ratio. After the preprocessing, the baselines of the spectra were removed by an iterative second-order polynomial fitting. The spectra were resampled with a channel width of 0.06 \kms. The velocity range was cropped to have the brightest emission of \Tco~($\sim$9 \kms) at the line center and a full coverage of 60 \kms. Then, all the tiles were combined to make one averaged scan map for each transition. The observation was continuously performed until the rms level of each average tile is $\lesssim$ 0.1 K [T$^{*}_{A,rms}$] at 0.1 \kms\ channel width for \cEoj\ and \Tcoj, $\lesssim$ 0.07 K [T$^{*}_{A,rms}$] at 0.1 \kms\ channel width for \soj\ and \nhTdj, and $\lesssim$ 0.07 K [T$^{*}_{A,rms}$] at 0.06 \kms\ channel width for \nThpj, \hcopj, \hTcopj, and \csj. In the final step, spectra for four molecular lines (those of \nhTdj, \soj, \nThpj, \cEoj, and \Tcoj\ transition) were smoothed to have their channel width of 0.1 \kms. The channel widths of the \hTcopj, \hcopj, and \csj~molecular lines remained as 0.06 \kms\ (see Table~\ref{tab:obs}).

\section{Results} \label{sec:result}

\subsection{Filamentary structures}
Overall distribution of molecular emission in Orion B can be best traced with \Tco\ and \cEo\ lines simultaneously observed. Figure~\ref{fig:hsh_sp} indicates that the \Tco\ emission depicts the large-scale distribution of molecular gas in Orion~B (middle-left panel), while the \cEo\ emission traces dense filamentary structures (middle-right panel). 

Panels (a) to (f) in Figure~\ref{fig:hsh_sp} show various aspects of \Tco\ and \cEo\ line profiles. 
We find that the multiple velocity components are common in Orion~B region and double-peaked profile in \cEo\ spectrum is more clearly separable than \Tco\ spectrum. 
The distribution of both molecular transitions in the middle panels and the undetected emission peaks in panels (c) and (d) indicate that the \cEoj\ transition is tracing more compact region with higher density than the \Tcoj\ transition and the \Tco~emission is extended over a wider region, tracing the bulk motion of less dense molecular gas. 
Meanwhile, in panels (b) and (e), two peaks in the \Tco~emission have comparable intensities. On the other hand, the \cEo~emission peaks have different intensity levels. This may arise from optical depth effects or from different excitation.
Panel (f) shows that both the \Tco~and \cEo~emissions have the same velocity centroid at $\sim$10.3 \kms\, but the \Tco~emission has almost two times wider line width.
Based on the spectral information, the dips of \Tco~spectra in panel (a) to (e) are not self-absorption features, but due to multiple velocity components. 
For the purpose of this paper, the \cEoj\ line is used to resolve velocity components on the line-of-sight and identify velocity-coherent filamentary structures in Orion B in Section~\ref{sec:fil}.

\subsubsection{Identification of velocity-coherent filaments} \label{sec:fil}

\begin{figure*}
  \centering 
  \includegraphics[width=0.95\textwidth]{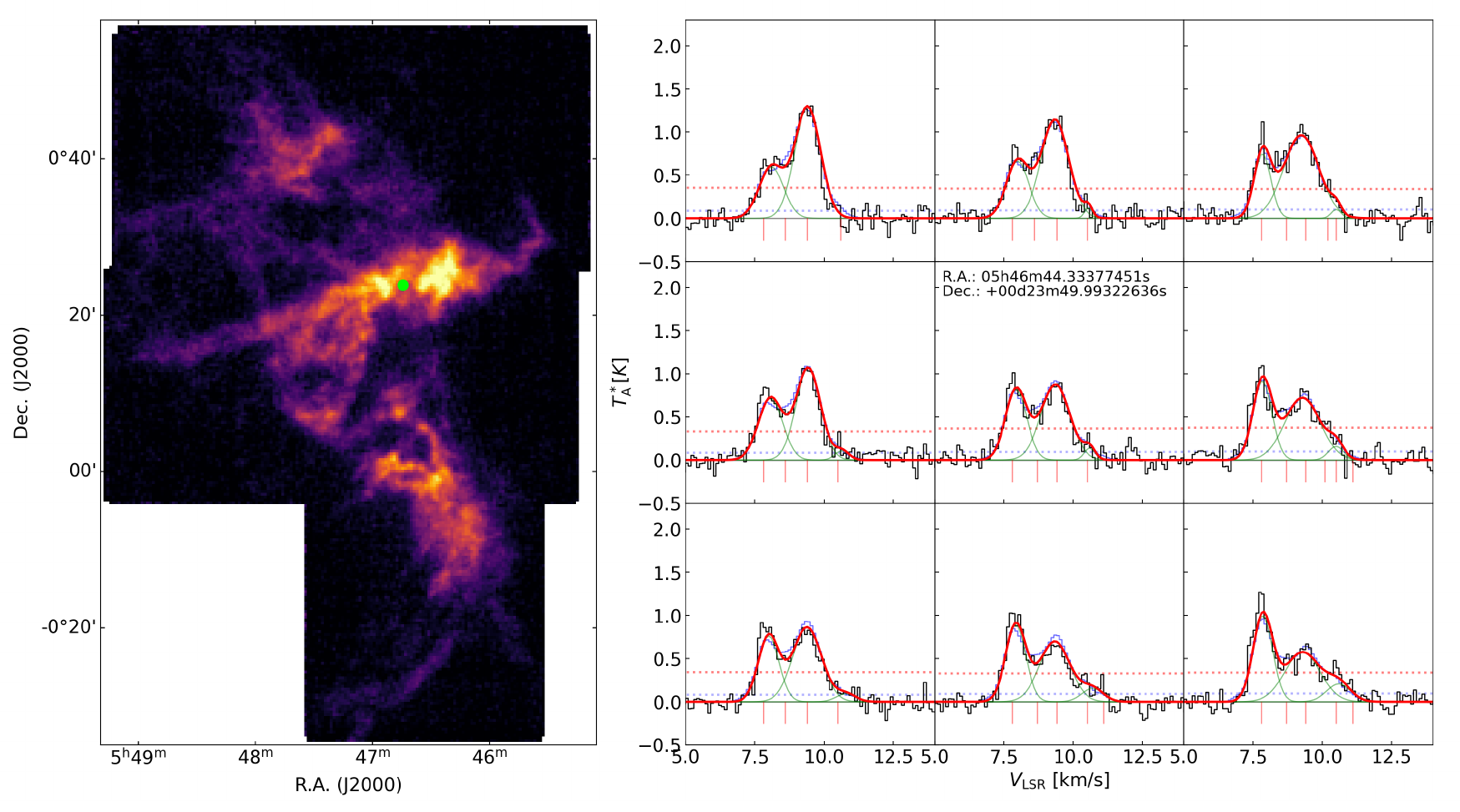}
  \caption{
  Left panel : Integrated intensity map of \cEoj\ of the Orion~B. The green dot is the position of the central pixel of the 3$\times$3 spectra in the right panel. 
  Right panel : Representative spectra are shown with results from the step-by-step decomposing procedure described in Section~\ref{sec:fil}. 
      Black solid line : Original spectrum at each pixel. 
      Blue solid line : Smoothed profile with the initial smoothing parameters ($\tt{Step 1}$). 
      Red vertical line : Initial guesses of velocity component found in $\tt{Step 1}$. 
      Green solid line : Decomposed Gaussian profiles determined in $\tt{Step 3}$ with modified initial guesses from $\tt{Step 2}$.
      Red thick solid line : The profiles combined by decomposed Gaussian components. 
      Red and blue horizontal dotted lines : 3$\sigma$ levels of original and smoothed profiles, respectively. 
\label{fig:gauss}}
\end{figure*}

In order to identify velocity-coherent structures in the Orion~B which are seen as overlapped to the line-of-sight, we decompose the multiple velocity components of the \cEoj\ spectrum at the each pixel 
by using the automated Gaussian decomposing code $\texttt{FUNStools.Decompose}$\footnote{https://github.com/radioshiny/funstools}, which is an optimally developed python class for TRAO-FUNS data. 
The brief step-by-step description of the procedure is as follows 
(see the details in Kim et al. in preparation):
	\begin{enumerate}
		\item $\tt{Step 1}$ : In the beginning  we do not know how many multiple velocity components exist and where their 
		                    centroid velocities are located in a single spectrum, and thus initial guesses for those are required. 
					        First, velocity smoothing process in a spectrum is performed to improve its signal-to-noise ratio, which would reduce the confusion from noise spikes.
					        Then spatial smoothing process is made using a two-dimensional Gaussian kernel to make spatial continuity and avoid local fluctuation for determination of the number of velocity components. 
					        From the smoothed spectrum, the peak velocity positions of the expected Gaussian components are determined by applying $\texttt{find\textunderscore peaks}$ algorithm in $\texttt{scipy.signal}$ package to find zero second derivatives of the spectrum. 
					        Based on the number of components and their velocity positions, 
					        initial multiple Gaussian fits are performed to the smoothed spectra. 
					        The final returns at $\tt{Step 1}$ are the initial guesses for each decomposed Gaussian component for the next step.
					        In this step, smoothing parameters should be selected depending on the characteristics 
					        (e.g., distribution and dynamic range of emissions) of the molecular cloud and the spectral resolution.  
					     
		\item $\tt{Step 2}$ : Using the initial guesses determined from $\tt{Step 1}$, a Gaussian fitting for the spectra is performed                     using these initial guesses to find the best-fit spectrum and returns the resulting Gaussian fitting      
		                    parameters for each velocity component of the spectrum. 
		
		\item $\tt{Step 3}$ : A final Gaussian fitting is performed to enhance the spatial continuity on the decomposed results. 
		                    The initial guesses for parameters are modified based on the best-fit parameters of the surrounding pixels obtained from $\tt{Step 2}$.
                            Then, a multiple Gaussian fit is performed again with the adjusted initial guesses.
				            Note that both velocity and spatial smoothing factors in this step are recommended to be smaller than the parameters 
				            in the $\tt{Step 1}$ of this tool. 
	\end{enumerate}

In order to improve the reliability of the $\tt{Step 1}$, the original spectrum was smoothed with 4 channels (0.4 \kms) in velocity after extracting the spectra above 3$\sigma$ level in their intensity. The spatial smoothing factor adopted in the step is adopted as a scale of 4 pixels (88\arcsec) which corresponds to the FWHM of a Gaussian kernel in the $\tt{Step 1}$. 
In case of the $\tt{Step 3}$, smoothing in smaller ranges, 2 channels (0.2 \kms) in velocity and 2 pixels (44\arcsec) in space, was performed to make a better fit to the original spectrum. 
The multiple Gaussian functions were used to fit the observed line profile with the initial guesses described in $\tt{Step 1}$. The optimization of the Gaussian fit to the observed line profile was performed using the Python routine $\tt{curve\_fit}$ in $\tt{scipy.optimize}$ package. The routine adopts the non-linear least-square fit based on the TRF (Trust Region Reflective) algorithm.
Figure~\ref{fig:gauss} presents an example of the Gaussian decomposing around the position marked with a green dot in the left panel. The green Gaussian profiles are the velocity components decomposed from $\texttt{FUNStools.Decompose}$ and the red spectra are the profiles combined by the decomposed components. 
Using the automated Gaussian decomposing code, 10,756 Gaussian components were identified with their peak intensities of S/N $>$ 3.

\begin{figure}[ht!]
\includegraphics[width=0.48\textwidth]{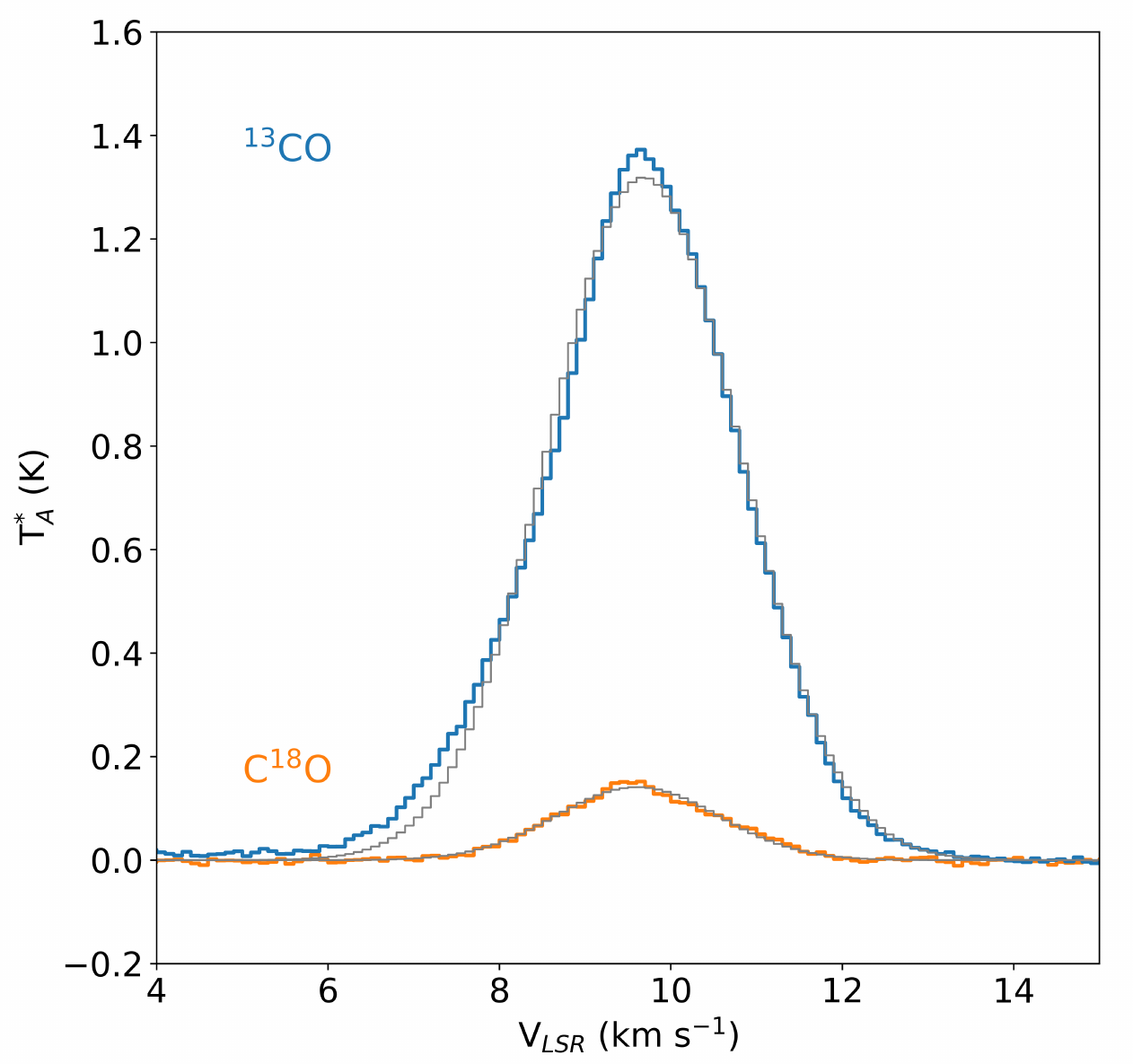}
\caption{The averaged J = 1--0 line profiles of \Tco~(blue) and \cEo~(orange) throughout whole line emitting region. Gray lines are to indicate the best-fit Gaussian profiles. 
\label{fig:avg_sp}}
\end{figure}

\begin{figure}[ht!]
\includegraphics[width=0.5\textwidth]{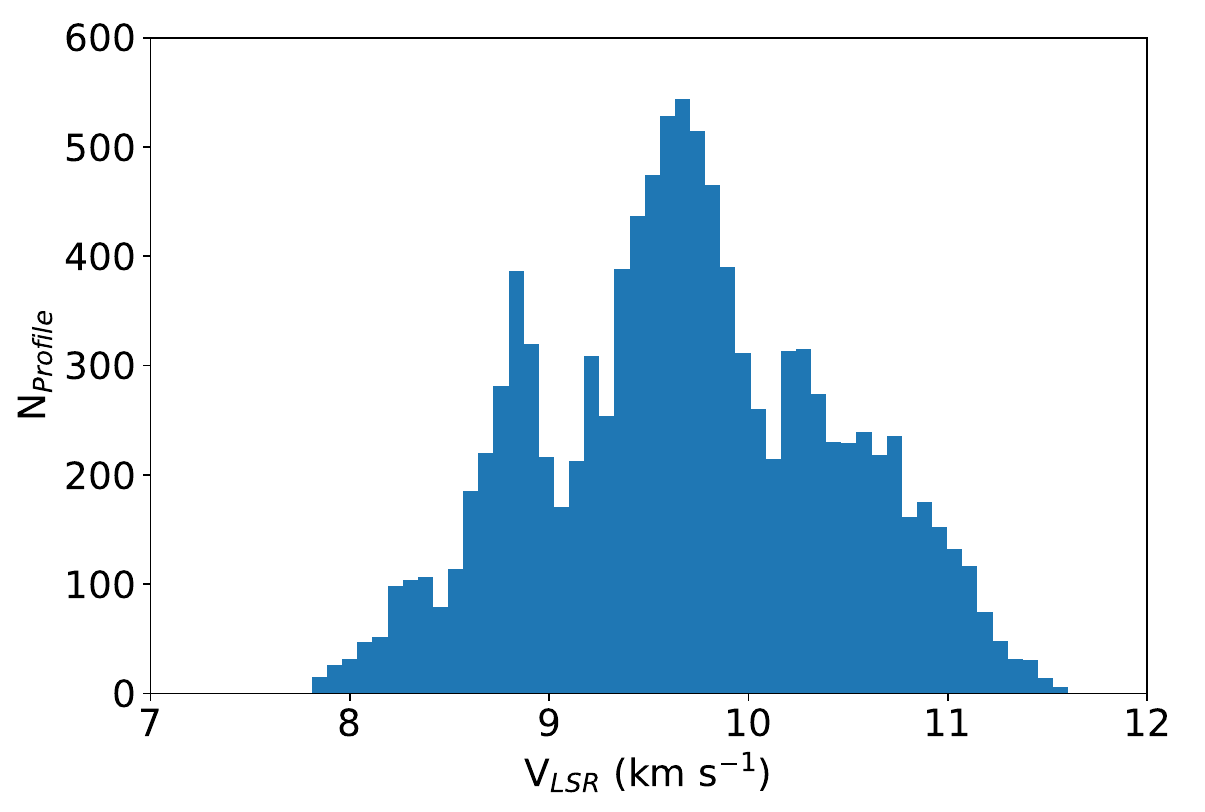}
\caption{The velocity distribution of \cEoj\ Gaussian-fit components decomposed by the Gaussian decomposition process (see Section~\ref{sec:fil})
\label{fig:histo}}
\end{figure}

Figure~\ref{fig:avg_sp} presents the averaged \Tco~and \cEo~spectra over the entire observing region.
After a careful decomposition of multiple Gaussians with line observations with 22\arcsec~grid size obtained by our TRAO-FUNS survey, we find that there are multiple velocity components with a wide range of velocity ($\sim$7.8 to 11.5 \kms) and present the histogram of decomposed velocity components in Figure~\ref{fig:histo}.

\begin{figure*}
\centering 
\includegraphics[width=0.87\textwidth]{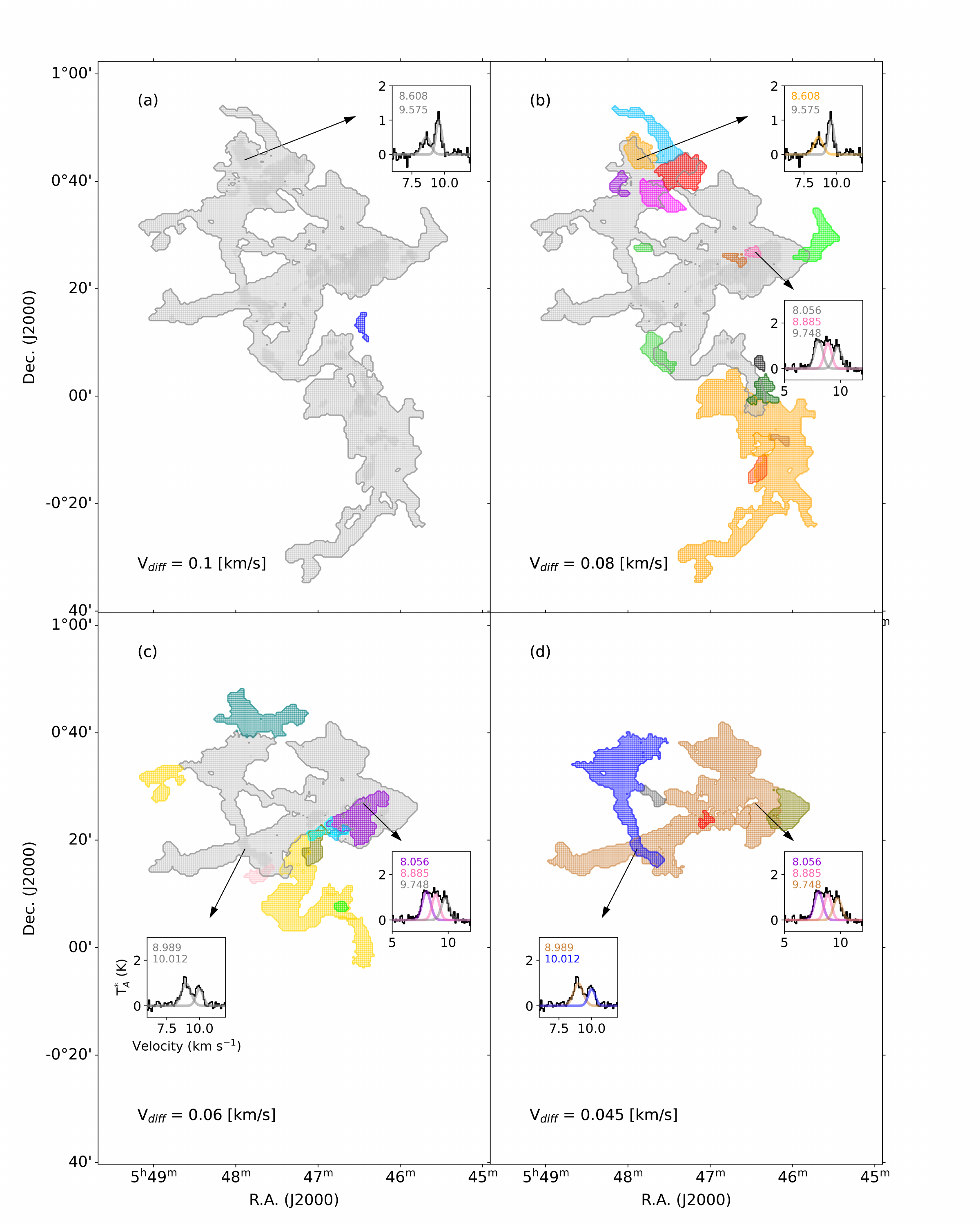}
\caption{Procedures identifying velocity-coherent filamentary structures with the FoF algorithm in the Orion~B region. The colored regions are the filamentary structures identified from FoF with the criteria for spatial separation ($\sqrt{2}$ pixel size) and velocity difference $\tt{V_{diff}}$ (from 0.1 to 0.045 \kms) written on the bottom of the panels. The gray region in each panel is  a continuously connected structure within the velocity criteria and the shades (darker regions) in the gray region are to indicate that there are the multiple velocity components in the line-of-sight in the velocity coherent structure. The insets in the panels are \cEoj\ spectra and their decomposed Gaussian components at a pixel point. The numbers in insets indicate the velocity centroids of the Gaussian components at the start point of the arrow. The colors of the velocity centriod numbers and Gaussian fit profiles  correspond to the colors of the filamentary structures. 
 \label{fig:hfof}}
\end{figure*}

\begin{figure*}
\centering 
\includegraphics[width=.93\textwidth]{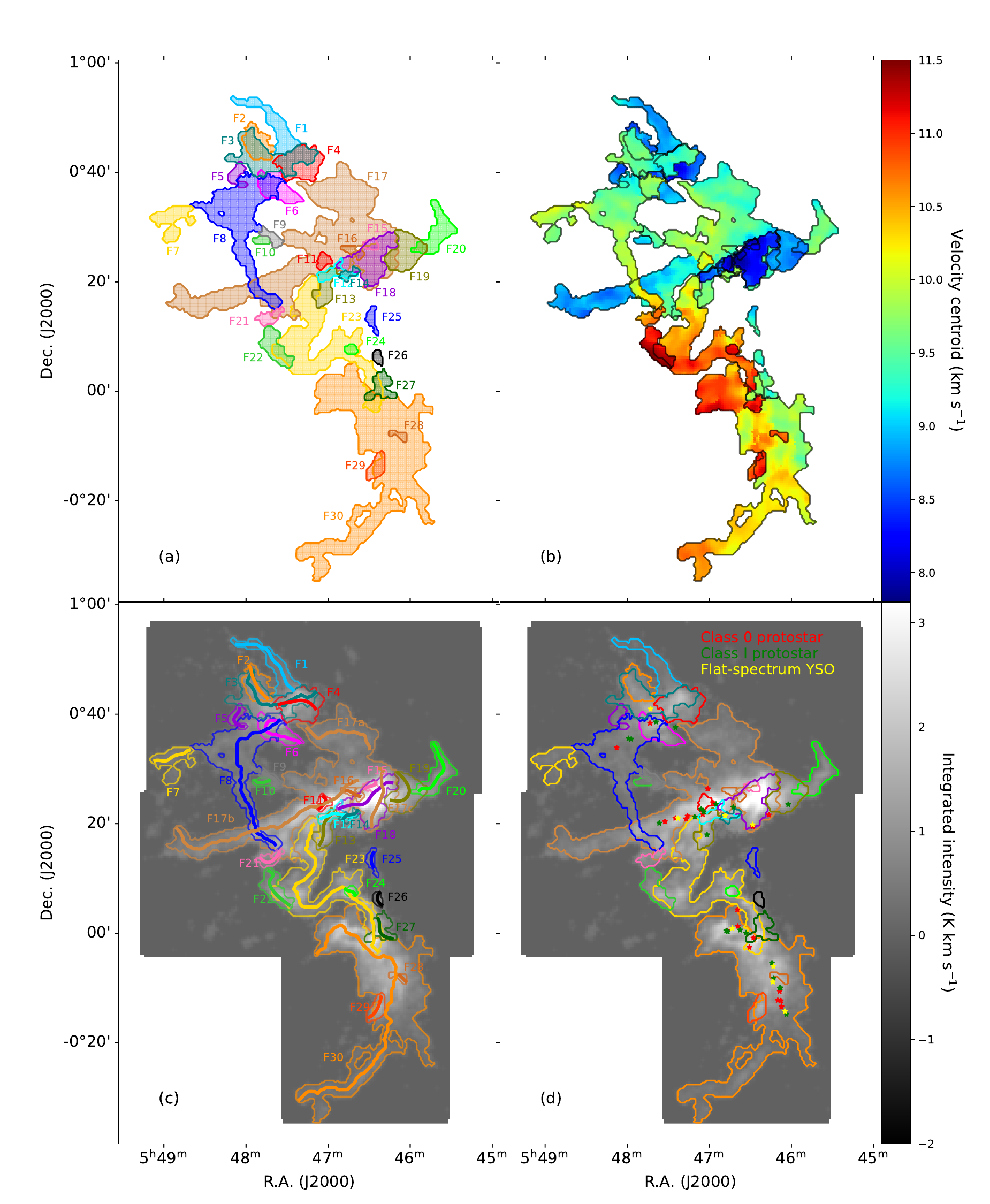}
\caption{Velocity-coherent filaments in Orion~B identified with the FoF algorithm. 
(a) Morphologies of the filaments. The filament names are labeled with the same colors of each filament. 
(b) The distribution of velocity centroids ranging from 7.8 to 11.5 \kms\ within identified filaments. 
(c) The spines of each filament are drawn with thick lines.
(d)  The distribution of young stellar objects identified and classified in \citet{furlan16}. Red asterisk symbol : Class 0 protostars. Green asterisk symbol : Class I protostars. Yellow asterisk symbol : Flat-spectrum YSOs. The colored contours delineate the filaments identified from this study. Gray tones in the two lower panels indicate the distribution of the integrated intensity of \cEoj\ line. 
\label{fig:fil}}
\end{figure*}

\begin{figure*}
  \centering 
  \includegraphics[width=0.97\textwidth]{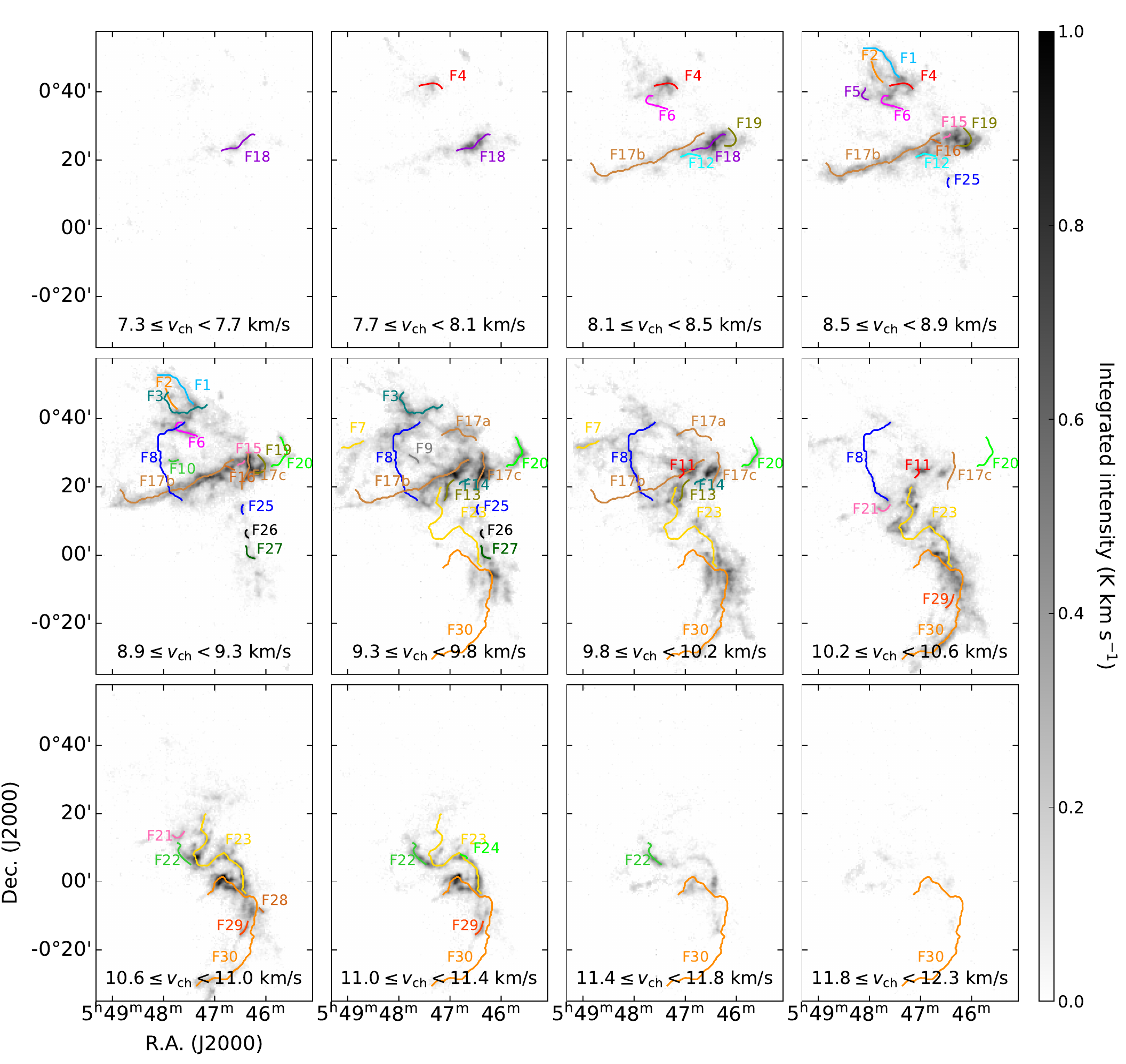}
  \caption{The velocity channel maps of \cEoj\ emission in relation with the identified filamentary structures.
  The velocity ranges of integration are indicated in the bottom of each panel. The filament spines covering the velocity channel range of each panel are overlaid with the same colors as the ones for the filaments in Figure~\ref{fig:fil}.
  \label{fig:chmap}}
\end{figure*}

With the decomposed velocity components, identification processes of the filaments using the Friends-of-Friends (FoF) algorithm were performed to eventually find velocity-coherent filamentary structures. 
The detailed processes of identifying the filaments are summarized as follows: 
First, all velocity components decomposed from the Gaussian fits are compared with a reference velocity component at the pixel with the brightest emission. If the velocity difference ($\tt{V_{diff}}$) between the component and the reference one is less than the spectral resolution $\sim$0.1 \kms~and the positional distance between two comparing pixels is $\leq$ $\sqrt{2}$ pixel size  (maximum distance between the reference pixel and one of neighboring 8 pixels) for their continuous positional connection, they become a part of a velocity coherent structure. Here, we set a number of pixels for a velocity coherent structure to be at least larger than 25, which corresponds to the number of pixels in 5 $\times$ $\theta_{\rm B}$ area (here, $\theta_{\rm B}$ is the half power beam width of \cEoj\ $\sim$ 49\arcsec). 
This process was iteratively done until there is no component left to be compared. The parameters can be empirically determined, and probably would be dependent on the properties of the molecular cloud. 

Now we explain how we identify the velocity-coherent filaments in the Orion~B region by using the FoF algorithm (Figure~\ref{fig:hfof}). In the starting run of the FoF, we set the difference between adjacent velocity centroids of the Gaussian components to be less than 0.1 \kms. 
Then, all the velocity components, except one small part of the cloud, are identified as one big velocity-coherent structure (see panel (a) in Figure~\ref{fig:hfof}). 
The shade parts in the gray region are meant to indicate that those regions have two or more velocity components at the single pixel (i.e. at the same line-of-sight direction), but are chosen as a part of one velocity-coherent structure. As the inset of Figure~\ref{fig:hfof}(a) shows, the spectrum shows obvious double peaked components. 
We simply assume that the \cEoj\ line is mostly optically thin and thus any asymmetric \cEoj\ line emission would be due to a sum of multiple Gaussian components. Hence every single filament identified here corresponds to a  cloud component that has a well-defined velocity at each pixel position.
This forces us to run the FoF algorithm again with different, in this case  smaller $\tt{V_{diff}}$ parameter to enable to distinguish these multiple velocity components.
Therefore, for the gray filamentary structure (with multiple velocity components) obtained in the previous step (see Figure~\ref{fig:hfof}(a)), we perform FoF again with $\tt{V_{diff}}$ of 0.08 \kms~, being able to identify more independent structures having single velocity component (colored regions in in Figure~\ref{fig:hfof}(b)). However, even in the second run of the FoF, we find several regions where multiple velocity components still exist even after identifying a velocity coherent structure. For example, 
in the middle region inset of panel (b) of Figure~\ref{fig:hfof}, the pink colored region is identified as a velocity coherent structure, but the gray-colored structure still contain two velocity components. 
Because of this problem, the third trial of FoF for these structures is performed with a parameter of $\tt{V_{diff}}$ = 0.06 \kms\ in panel (b). In this way, we perform the FoF until every identified filamentary structures contain a single velocity component only.  Panel (d) shows the results of our last trial of FoF performed with $\tt{V_{diff}}$ of 0.045 \kms~ after multiple tests with $\tt{V_{diff}}$ of 0.03, 0.035, 0.04, 0.045, and 0.05 \kms, indicating that the cloud regions previously marked by gray velocity profiles (insets in panel (b) and (c) in Figure~\ref{fig:hfof}) are finally well decomposed. 
Lastly, we check whether any Gaussian components with significant emission ($\sim$3$\sigma$) are missed during the FoF running by visual inspection and recover those in the final filamentary structure. 

\begin{figure}[ht!]
\includegraphics[width=0.49\textwidth]{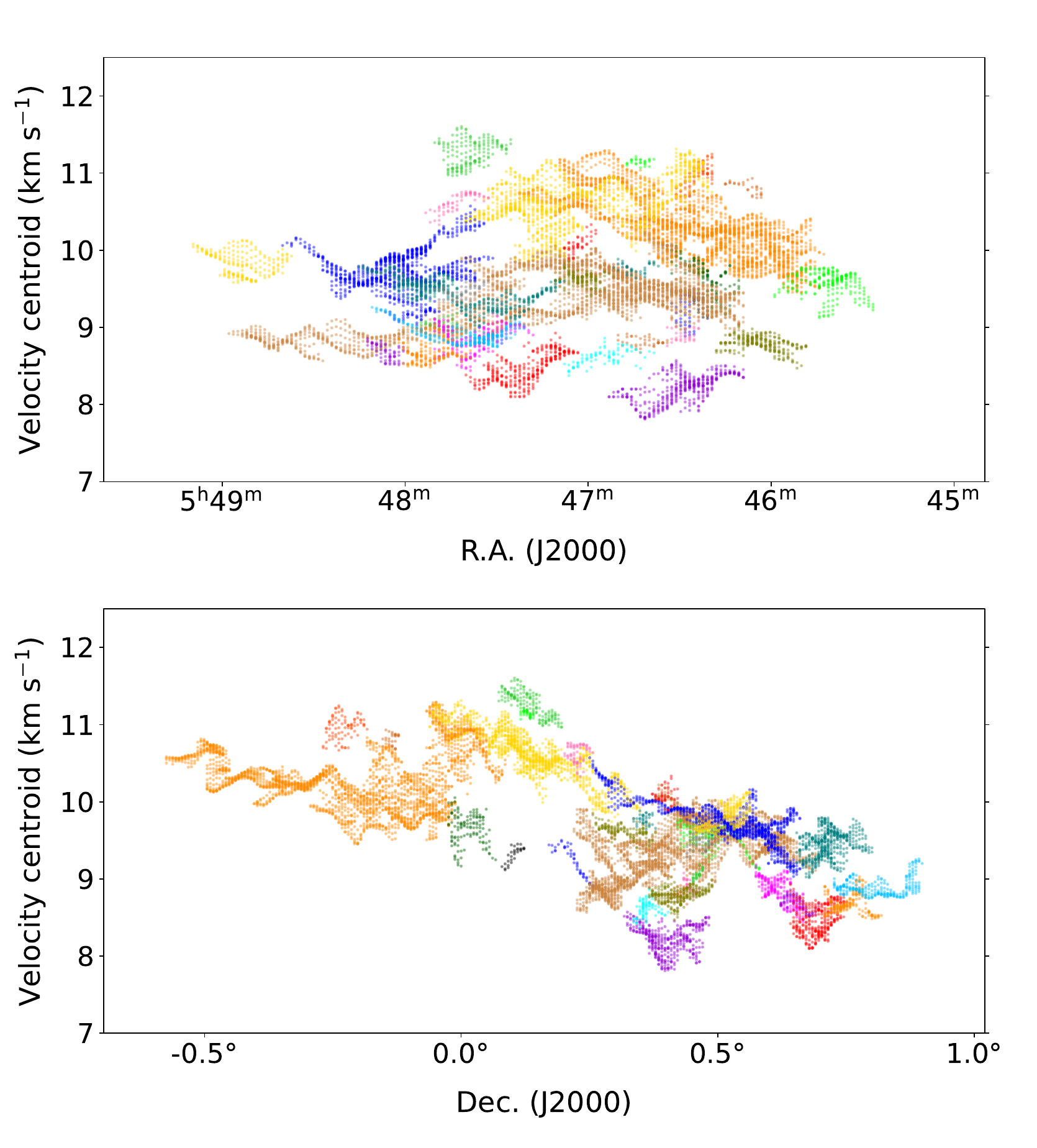}
\caption{The oscillatory features of the \cEoj\ velocity centroids of the filaments, along the R.A. (upper panel) and Dec. (lower panel) directions. The colors of the dots are identical to the colors of filaments in the upper-left panel of Figure~\ref{fig:fil}.
\label{fig:radec}}
\end{figure}

Through all these steps, a total of 30 velocity coherent structures were finally identified as filaments in the Orion~B region. Figure~\ref{fig:fil} presents morphologies, velocity structures, spine structures of the filaments, and the distribution of young stellar objects (YSOs) classified from \citet{furlan16} in the filaments. These will be main constituents for discussing the physical properties of the filaments in the coming sections.

\begin{figure}[ht!]
\centering
\includegraphics[width=0.49 \textwidth]{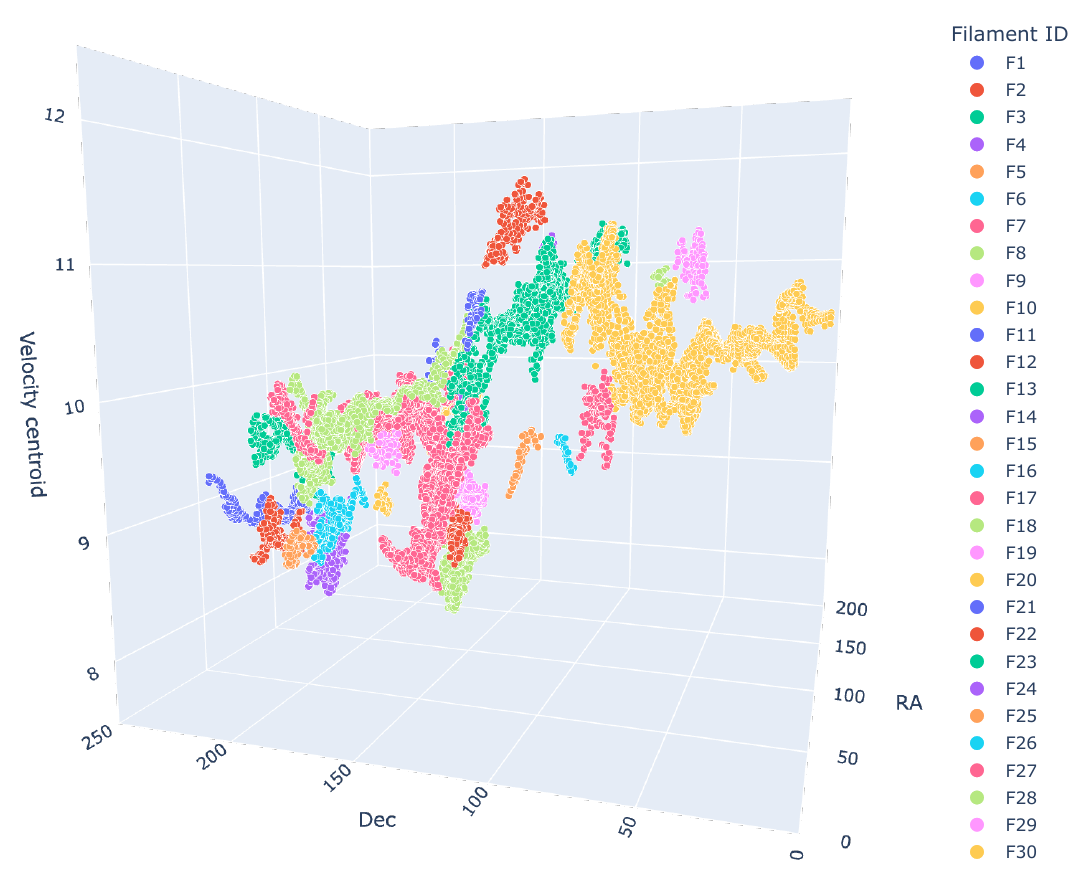}
\caption{The distribution of the \cEoj\ velocity centroids of 30 identified filaments in three-dimensional PPV space in the units of pixels (along R.A. and Dec. directions) and \kms\ (for velocity centroid). The colors of the dots are not identical to the colors of filaments in Figure~\ref{fig:fil}. Please refer to the colors and names in the legend at the viewing mode. 
Interactive viewing mode of this Figure is available in https://hyunjuyoo.github.io/Orion-Filaments/  
\label{fig:radec3d}}
\end{figure}

In order to display how all the filamentary structures are related with the velocity structure in detail, we provide the velocity channel maps of the \cEoj~emission with intervals of 0.4 \kms~from 7.3 to 12.3 \kms\ in Figure~\ref{fig:chmap}. 
The northern part of the molecular cloud is dominated by the lower velocity component and the middle and southern parts have bright emission at the higher velocity ranges. 
As Figure~\ref{fig:chmap}, Figure~\ref{fig:histo} and the upper-right panel of Figure~\ref{fig:fil} present, the filaments in Orion~B have velocity ranges from $\sim$7.8 to $\sim$11.5 \kms~with gradually increasing LSR velocities from North to South. 
For velocity centroids of the overlapped regions in Figure~\ref{fig:fil}(b), we present the detailed information for each filament in the Appendix~\ref{A1} and provide the 3-dimensional visualization in Figure~\ref{fig:radec3d}.
The ranges of the velocity centroids and the averaged line dispersions for each filament are tabulated in Table~\ref{tab:fil}.

The position-velocity (PV) plots in Figure~\ref{fig:radec} along R.A. and Dec. directions indicate that the \cEoj\ velocity centroid for individual filaments smoothly changes in velocity and position with interesting oscillatory patterns in a single structure. 
The PV-diagram drawn using the \co~(2--1) molecular line survey \citep{nishimura15} has also revealed such oscillating features in \vlsr~(see also L1630-N in the upper panel of Figure 3 in \citealt{Grosschedl21}). While the \co~(2--1) LSR velocity shows the averaged large-scale bulk motion of the cloud, our decomposed \cEo~(1--0) LSR velocity is able to reveal individual motions of substructures along the line of sight (Figure~\ref{fig:radec} and Figure~\ref{fig:radec3d}). We provide a link to the interactive viewing mode of velocity centroids in three-dimensional PPV space in Figure~\ref{fig:radec3d}. 
Further analysis and discussion on these velocity oscillations are presented in Section~\ref{sec:disc} (especially in Section~\ref{sec:f23}).

\begin{figure*}
  \centering 
  \includegraphics[width=1\textwidth]{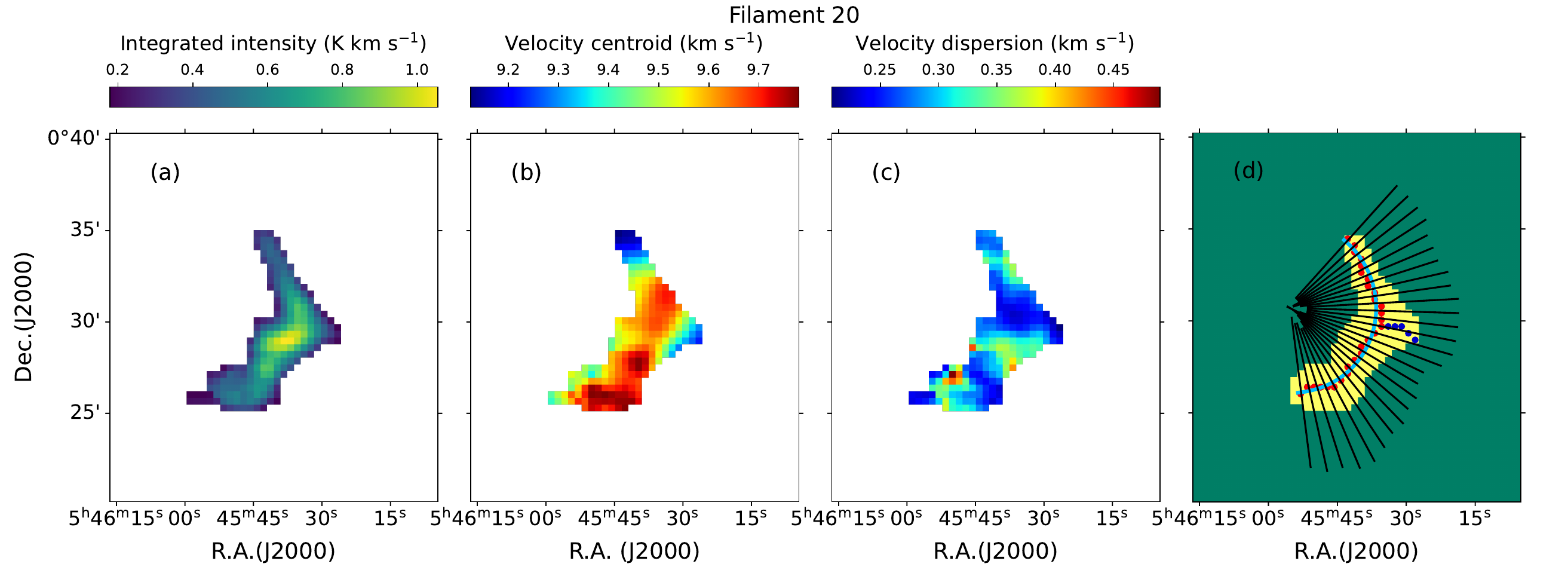}
  \caption{Defining main properties of the identified velocity-coherent filament. Here, F20 is given as an example. (a) The integrated intensity (for Gaussian area) image. (b) The velocity centroids distribution. (c) : The distribution of velocity dispersion. (d) Spine and branch, filament length, and radial cuts of the filament to produce the radial profiles. The path of the spine and branch are indicated with red points and blue points in panel (d). The red dots are the path of the spine, blue dots are for indicating an additional branch, and the curve (sky blue color) is the smoothed path of spine which is used for estimating the length of the filament. The black straight thin lines in panel (d) are drawn perpendicularly across the spine at the evenly sampled data points for drawing radial profiles in Figure~\ref{fig:radpro}.
   \label{fig:Fil23}}
\end{figure*}

\subsubsection{Lengths and widths of the filaments} \label{sec:lw}

\begin{figure}[ht!]
\includegraphics[width=0.45\textwidth]{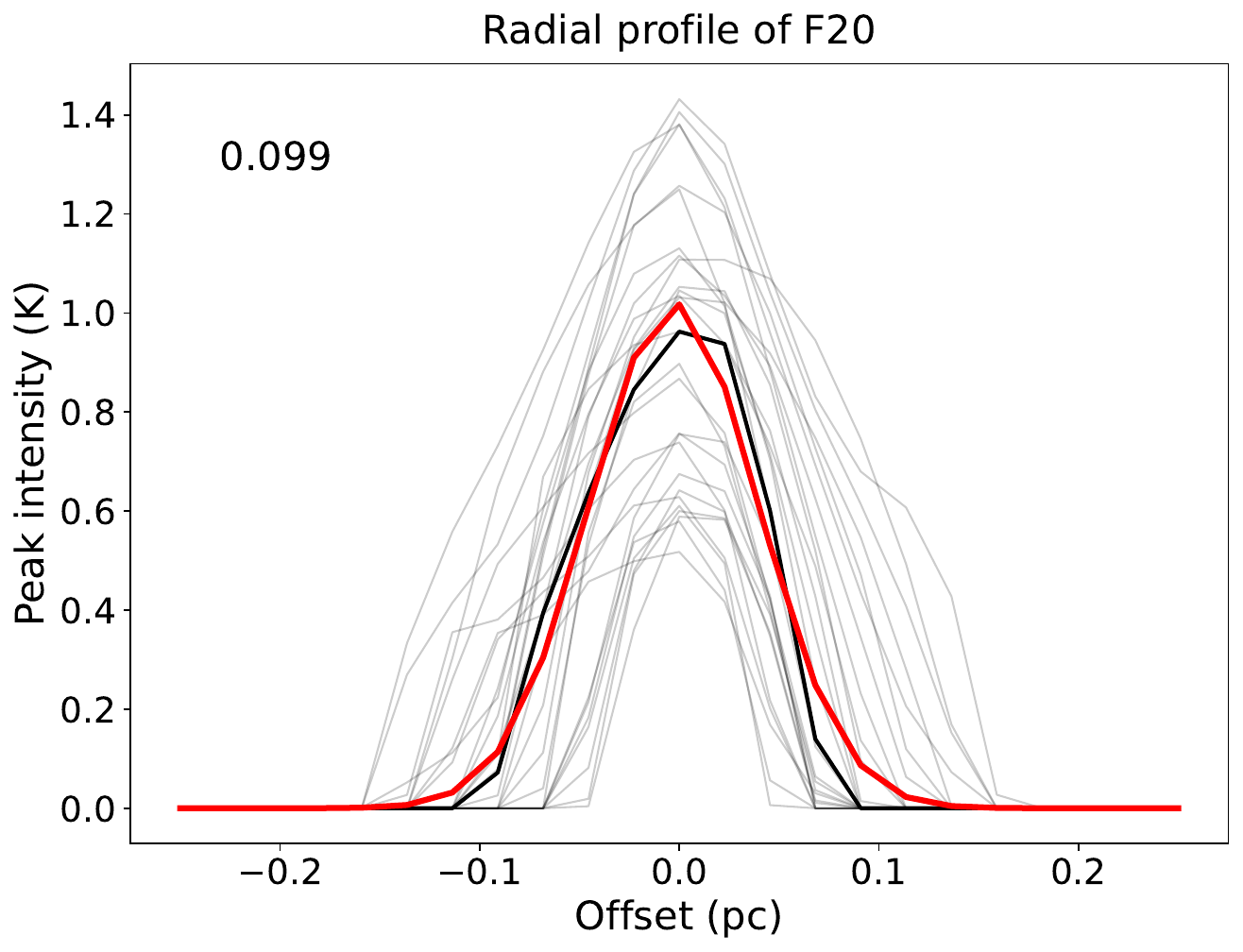}
\caption{The radial profiles of F20 drawn along the black line cuts in Figure~\ref{fig:Fil23}(d), which are evenly made with an interval of 25\arcsec. 
The gray lines are individual radial profiles and the black line is the average profile. 
The red line depicts the Gaussian-fitted profile. 
\label{fig:radpro}}
\end{figure}

\begin{figure}[ht!]
  \includegraphics[width=0.49\textwidth]{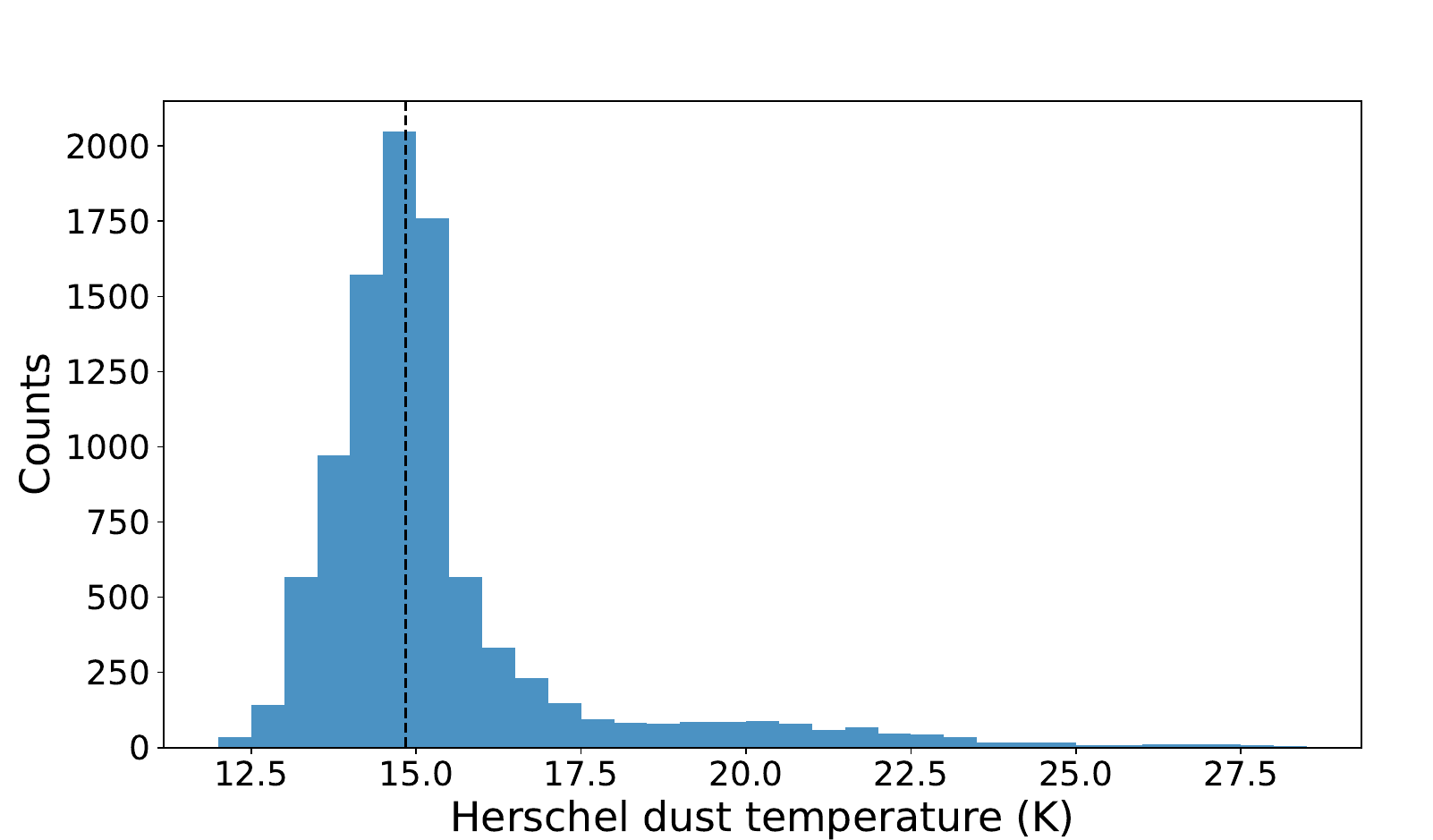}
  \caption{The dust temperatures obtained using the Herschel data toward the filamentary structures. The column densities for the filaments were calculated using these dust temperatures. The cyan vertical line indicates the median value (14.8~K) of Herschel dust temperature distribution toward the filaments. 
   \label{fig:hist}}
\end{figure}

Here we attempt to quantify the properties of the filaments. For this purpose we define the main structures of the filaments using a Python package $\textsc{filfinder}$ \citep{koch15} which extracts the main structure from the two-dimensional image via a Medial Axis Transform \citep{blum67}. First we produce the integrated intensity image for each filament from their resultant parameters (e.g., the amplitude, the velocity centroid, and the velocity dispersion) derived from the multiple Gaussian fits to the spectra for each filament (see panel (a)--(c) in  Figure~\ref{fig:Fil23}). The $\textsc{filfinder}$ algorithm provides not only the information for the spine of the filaments (see the lower-left panel in Figure~\ref{fig:fil}) but also the ones for the branches stretched out from the spine. 
We use the length of the spine for each filament as the filament's length $l$, and list them in Table~\ref{tab:fil}. The largest length of the filament is $\sim$6.5 pc and the median length for the entire sample of the filaments is 0.8~pc. 
This median length of the filaments is comparable to the case of B213-L1495 in Taurus (1.0$\pm$0.2 pc) but longer than the cases of Integral Shape Filament (ISF) in Orion (0.2$\pm$0.1 pc) or NGC~1333 in Perseus (0.4$\pm$0.2 pc) \citep{hacar18}.

We use the averaged radial profile to estimate the filament's width. Along each spine, we chose the sampling points on the spine with an interval of 25\arcsec~and made a cut across the spine to build the transversal profile. In Figure~\ref{fig:Fil23}(d), the cyan curve is the smoothed line of the spine of Filament 20 and the black lines indicate the transversal lines used to produce the radial profiles in Figure~\ref{fig:radpro}. The gray lines in Figure~\ref{fig:radpro} are individual radial profile obtained across the cuts in Figure~\ref{fig:Fil23}(d) and the black line is an average one of these radial profiles. Once the average profile is produced using all the transversal profiles, the 1D-Gaussian model fits the averaged profile to determine the FWHM as a width ($W$) of the filament. 
The median value of the filament widths is estimated to be 0.1~pc with a range of 0.05 to 0.26~pc. 
It should be noted that the TRAO beam $\sim$50\arcsec is comparable to the physical scale of the median width of the filaments ($\sim$0.1 pc) at the distance of Orion~B, and hence this measured width is likely affected by our observing beam size (see also \citealt{panopoulou22}).

The aspect ratio, which is defined as the ratio of the length $l$ and the width $W$ of the filament (\textit{AR} = $l/W$), is used to define how the filamentary structure is elongated (\citealt{andre14}). In our study, we define the \cEo~structures as filaments if their aspect ratios (\textit{AR}) are lager than 3. 
Note that F17 has quite large and complicate structure. Therefore, we divide it into three parts (F17a, F17b, and F17c) in this case for calculating their physical parameters. 
The lengths, widths and the aspect ratios AR for the entire velocity-coherent structures are given in Table~\ref{tab:fil}. 
The uncertainty on length comes mostly from the uncertainty on distance ($\sim$5\%; \citealt{zucker19}).

\subsubsection{Column densities, masses, and line masses of filaments}


The \cEo~column densities of the filaments were estimated using the following equation obtained under the assumption of local thermodynamic equilibrium (LTE)
(e.g., \citealt{garden91} and \citealt{pattle15}): 
\begin{eqnarray} \label{eq:N}
	N({\mathrm{C^{18}O}}) & = & \frac{3k_\mathrm{B}}{8\pi^{3}B\mu^{2}_D} \frac{e^{hBJ(J+1)/k_\mathrm{B}T_\mathrm{ex}}}{J+1} \nonumber \quad \\ 
	& & \times \frac{T_\mathrm{ex} + \frac{hB}{3k_\mathrm{B}}}{1 - e^{-h\nu / k_\mathrm{B} T_\mathrm{ex}}} \int{\tau} \textrm{d}v,
\end{eqnarray}
where \textit{k}$_\mathrm{B}$ is the Boltzmann constant, \textit{B} = 5.79384 $\times$ 10$^{10}$~s$^{-1}$ is the rotational constant, $\mu_D$ $\sim$0.112~D is the permanent dipole moment of the molecule, and \textit{J} is the lower rotational level of the \cEoj~transition. The excitation temperature \textit{T}$_\mathrm{ex}$ is assumed to be the same as the dust temperature \textit{T}$_\mathrm{dust}$ obtained from the Herschel dust continuum emission in Orion~B region \citep{schneider13, konyves20}. In the UV-dominated regions, the gas kinematic temperature and the dust temperature are difficult to derive. However, the \cEo~excitation temperature becomes closer to \textit{T}$_\mathrm{dust}$ in the dense cloud regions where \textit{T}$_\mathrm{dust}$ is lower than 20 K \citep{roueff20}. In our observed region, most \textit{T}$_\mathrm{dust}$ values are lower than 20~K (see Figure~\ref{fig:hist}), and thus, we use \textit{T}$_\mathrm{dust}$ as \textit{T}$_\mathrm{ex}$ in the column density calculation (see the averaged dust temperature used for calculation for each filament in Table~\ref{tab:fil}).

The integration of the optical depth, which is the rightmost term of the Equation~(\ref{eq:N}), can be expressed as below:
\begin{eqnarray} \label{eq:tau}
	\int\tau\mathrm{d}v & = & \frac{1}{J(T_\mathrm{ex}) - J(T_\mathrm{bg})} \int\frac{\tau(v)}{1 - e^{-\tau(v)}} T_\mathrm{mb} (v) \mathrm{d}v   \nonumber      \qquad \\
	& \approx &   \frac{1}{J(T_\mathrm{ex}) - J(T_\mathrm{bg})} \frac{\tau(v_0)}{1 - e^{-\tau(v_0)}} \int T_\mathrm{mb} (v) \mathrm{d}v,                  \qquad
\end{eqnarray}
where \textit{J(T)} = \textit{T$_0$}/(e\textit{$^{T_0/T}$} -- 1) is the source function (\textit{T$_0$} = \textit{h$\nu$}/\textit{k$_\mathrm{B}$}), \textit{T}$_\mathrm{bg}$ = 2.73 K is the cosmic microwave background temperature, \textit{T}$_\mathrm{mb} (v)$ is the observed main beam temperature at velocity $v$, and \textit{v}${_0}$ is the centroid velocity. Here, the ranges of integrated velocity is decided by each decomposed \cEo\ spectrum for individual filaments.

The optical depth at the central velocity $\tau(v_0)$ in Equation~(\ref{eq:tau}) is derived by the following relation, assuming that the \Tco~and \cEo~lines trace the similar densities and are excited with the same temperature:

\begin{equation}
	\frac{T\mathrm{_{max}^{C^{18}O}}} {T\mathrm{_{max}^{{}^{13}CO}}} \approx \frac{1 - e^{-\tau\mathrm{_{C^{18}O}}}} {1 - e^{-\tau\mathrm{_{{}^{13}CO}}}},
\end{equation}
where \textit{T}$\mathrm{_{max}^{C^{18}O}}$ and \textit{T}$\mathrm{_{max}^{{}^{13}CO}}$ are the intensities of the \cEoj~and \Tcoj\ transitions at the centroid velocity of the \cEo~line profiles, respectively. 
Here, we use an isotopic ratio [\Tco/\cEo] = 10 to calculate the optical depth, although the value [\Tco/\cEo] = 5.5 \citep{frerking82} is widely used. This is because the abundance ratio of \cEo\ in Orion molecular cloud is believed to be different from the terrestrial one \citep{nishimura15}. For example, the abundance ratios in NGC 2023 and NGC 2024 regions of Orion~B and the Orion A giant molecular cloud are reported to range from 8 to 50 with a mean value of 12.5 \citep{roueff20}, and from 5.7 to 33 with mean values of a factor of 2--3 larger than the terrestrial value, respectively, depending on the effect of the selective photodissociation of \cEo\ by the FUV radiation \citep{shimajiri14}. 
In this way, the \cEo\ column densities for the filaments were estimated and found to be in the wide range between 10$^{14}$ -- 10$^{16}$~cm$^{-2}$.

The masses of the velocity coherent filaments are estimated from the Herschel column density map \citep{andre10, konyves20}, which is convolved to the resolution of the TRAO beam ($\theta_{\rm B}$ = 48.7\arcsec\ for \cEoj). Since decomposed velocity components cannot be distinguished in the Herschel continuum map, we modify the Herschel masses based on the fraction of the intensity of \cEo\ emission associated with the specific structure used to calculate filament's mass.

The H$_2$ column densities of the velocity coherent filaments range from 2.8~$\times$~10$^{21}$~cm$^{-2}$ to 1.4~$\times$~10$^{22}$~cm$^{-2}$ with a median of 6.1~$\times$~10$^{21}$~cm$^{-2}$. The minimum and maximum masses of these filaments are 2.5 $M_{\odot}$ and 700 $M_{\odot}$ with a median of 35 $M_{\odot}$.  Line masses ($M_\mathrm{line}$), which are masses within the velocity-coherent structures divided by the lengths, are calculated to range between 14 $M_{\odot}$~pc$^{-1}$ and 175 $M_{\odot}$~pc$^{-1}$ (Table~\ref{tab:fil}).
The uncertainties of $M$ and $M_\mathrm{line}$ are estimated from the observational rms error and uncertainty on the length. The typical uncertainties of $M$ and $M_\mathrm{line}$ are $\sim$13\% and $\sim$12\%, respectively.

\subsubsection{Velocity dispersion}

We introduce four velocity dispersions in describing the kinematical information of the clouds and cores in Orion B such as the observed, thermal, nonthermal, and total velocity dispersions ($\sigma_\mathrm{obs}$, $\sigma_\mathrm{th}$, $\sigma_\mathrm{nth}$, and $\sigma_\mathrm{tot}$, respectively). 
The observed velocity dispersion ($\sigma_\mathrm{obs}$) can be given with the FWHM ($\Delta v_\mathrm{obs}$) of the line profile observed at the position of the peak emission, in a relation of $\Delta v_\mathrm{obs}$ = $\sqrt{8 \ln 2}$ $\sigma_\mathrm{obs}$. 
The thermal velocity dispersion is defined as $\sigma_\mathrm{th}$=$\sqrt{ k_\mathrm{B} T_\mathrm{dust}/m_\mathrm{obs}}$ for the observing molecule with a mass of $m_\mathrm{obs}$ and can be related with the nonthermal contribution ($\sigma_\mathrm{nth}$) in the turbulent clouds as follows \citep{myers91}:
\begin{eqnarray}
	\sigma_{nth} = \sqrt{ \sigma_\mathrm{obs}^2 - \sigma_\mathrm{th}^2 },     \quad\qquad \nonumber \\
	                    = \sqrt{ \frac{\Delta \textit{v}^2_\mathrm{obs}}{8\mathrm{ln}2}- \frac{k_\mathrm{B} T_\mathrm{dust}}{m_\mathrm{obs} } }.
\label{eq:signth}
\end{eqnarray}
In case of the \cEoj~line,  $m_\mathrm{obs}$ = $\mu_\mathrm{C^{18}O} m_\mathrm{H}$ is used, where $\mu_\mathrm{C^{18}O}$ = 30 is the molecular weight of \cEo~and $m_\mathrm{H}$ = 1.67 $\times$ 10$^{-24}$ g is the hydrogen mass.

The total velocity dispersion $\sigma_\mathrm{tot}$ of the mean molecules including both thermal and nonthermal effects is expressed as below:
\begin{equation}
\label{eq:sigtot}
	\sigma_\mathrm{tot} = \sqrt{\sigma_\mathrm{nth}^2 +\frac{k_\mathrm{B} T_\mathrm{dust}}{m_\mathrm{H}} \frac{1}{\mu_\mathrm{mean}} }, \quad
\end{equation}
where 
$\mu_\mathrm{mean}$ is the mean molecular weight per free particle (= 2.37) assuming the mean mass of gas with the cosmic composition of 71\% of hydrogen, 27\% of helium, and 2\% of metal \citep{kauffmann08}. 
The average total velocity dispersion $<$$\sigma_\mathrm{tot}$$>$ of the filaments are in the range of 0.31--0.54 \kms. 
The uncertainty on $\sigma_\mathrm{tot}$ is estimated from the uncertainty on the observed velocity dispersion (Table~\ref{tab:fil}).

\subsubsection{Critical line mass}
\label{sec:crit_Mline}
Assuming that the filamentary structure is a self-gravitating isothermal cylinder with an infinite length, the thermal critical line mass of the filament in equilibrium state under thermal and gravitational pressure, and the nonthermal critical line mass supported by turbulence can be estimated as below:
\begin{eqnarray}
	{M}_\mathrm{line}^\mathrm{th, crit}  & = & 2 c_\mathrm{s}^{2}/G,  \\		\quad \qquad		
	{M}_\mathrm{line}^\mathrm{nth, crit} & = & 2 \sigma^2_\mathrm{nth, C^{18}O}/G,		\quad \qquad
\end{eqnarray}
where $c_\mathrm{s}$ is the isothermal sound speed using the typical dust temperature inside the filament, $G$ is the gravitational constant, $\sigma_\mathrm{nth}$ is the average nonthermal velocity dispersion of filaments traced by \cEoj~emission using Equation~(\ref{eq:signth}). 
${M}_\mathrm{line}^\mathrm{th, crit}$ and ${M}_\mathrm{line}^\mathrm{nth, crit}$ are found in the ranges of 21--36 $M_{\odot}$ pc$^{-1}$ with a median value of 24 $M_{\odot}$ pc$^{-1}$ and 17--102 $M_{\odot}$ pc$^{-1}$ with a median value of 55 $M_{\odot}$ pc$^{-1}$, respectively. 

The stability of the filament can be examined by both thermal and nonthermal contributions in comparison with the virial mass of the filament below where the average total velocity dispersion plays an important role \citep{bertoldi92, ohashi16}: 
\begin{eqnarray}
	M_\mathrm{vir}^\mathrm{fil} = \frac{2~l}{G} \sigma_\mathrm{tot, C^{18}O}^2,	
\end{eqnarray}
where  $l$ is the filament length, and $\sigma_\mathrm{tot, C^{18}O}$ is the average total velocity dispersion from the \cEoj~emission calculated by using Equation~(\ref{eq:sigtot}) with \cEo~molecular weight ($\mu_\mathrm{obs}$= 30). 
The filament virial mass per unit length (=$M_\mathrm{vir}^\mathrm{fil}/l$) is referred to 
$M_\mathrm{line, vir}$ = $2 \sigma_\mathrm{tot, C^{18}O}^{2}/G$ \citep{fiege00, arzoumanian13}, which is  $M_\mathrm{line}^\mathrm{tot, crit}$ in here.
${M}_\mathrm{line}^\mathrm{tot, crit}$ is found in the range of 45--136 $M_{\odot}$ pc$^{-1}$ with a median value of 80 $M_{\odot}$ pc$^{-1}$.
The proportions of critical filaments among filaments with/without \nThp\ dense cores are 92\% and 65\%, respectively, indicating the criticality of filament is important condition on dense core formation (see details in Section~\ref{sec:stability_fil}). 
The uncertainties of thermal, nonthermal, and total critical line masses are estimated from the uncertainties on the thermal, nonthermal, and total velocity dispersions, respectively.

\subsection{Identification of dense cores and their virial parameters}\label{Sec:core}

\begin{figure}[ht!]
\centering 
\includegraphics[width=0.48\textwidth]{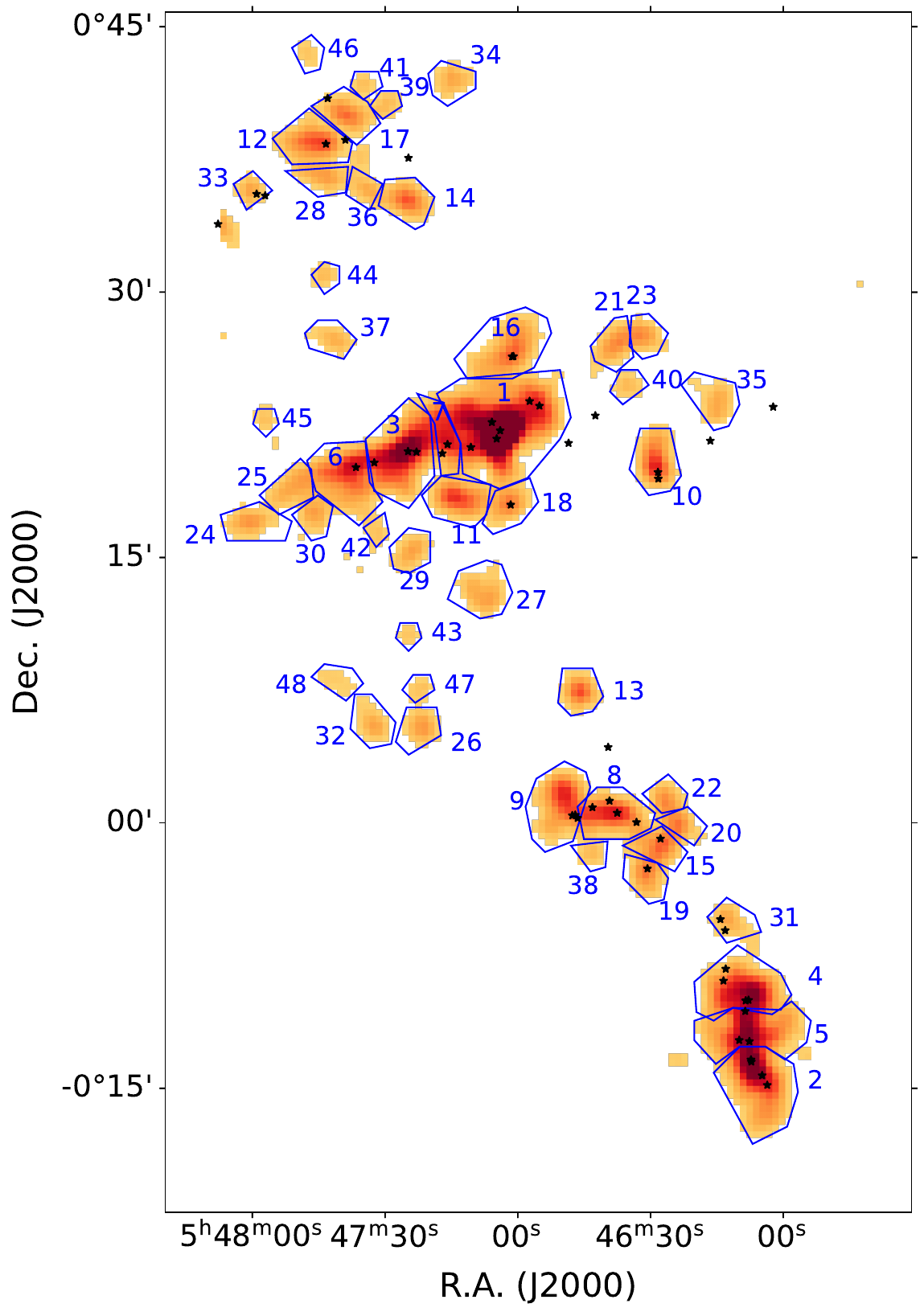}
\caption{The distribution of \nThp~dense cores identified by $\texttt{FellWalker}$ algorithm in Orion~B region (blue polygons). The color tones indicate the integrated intensity image of the \nThp. 
Asterisks are YSOs from \citet{furlan16} represented regardless of evolutionary stage.
The dense core ID is labeled in a decreasing sequence of peak brightness. 
\label{fig:core}}
\end{figure}

\nThp~is a common dense gas tracer which gets excited at relatively high densities ($\simeq$~10$^5$ cm$^{-3}$ for $J$ = 1--0), and thus, the bright \nThpj~emission can reveal the distribution and morphologies of dense cores. We use the $\texttt{FellWalker}$ algorithm \citep{berry15} in the $\texttt{CUPID}$ package \citep{berry07} in the $\texttt{Starlink}$ software to define the clumpy structures as dense cores in the \nThpj~integrated intensity map. The \nThpj~integrated intensity map of Orion~B presented in the background of Figure~\ref{fig:core} is produced including intensities of all the hyperfine components. 
We identified 48 \nThp~dense cores in the observed region (see polygons in Figure~\ref{fig:core}).  
To determine the LSR velocity and velocity dispersion for each dense core, we perform a hyperfine line-fitting using the $\texttt{curve\_fit}$ function in the $\texttt{SciPy}$ module \citep{virtanen20}. 
The physical information of the identified cores such as the peak intensity positions, and major and minor axes of the cores, their LSR velocities and line widths determined by the hyperfine fitting of the \nThpj~profile, their average \nThp~column densities and average H$_2$ column densities, their average dust temperatures, observed masses and virial masses, and virial parameters are listed in Table~\ref{tab:core}.

We derive the \nThp\ column densities of the dense cores using the \nThpj~emission, following the equation from \citet{caselli02} as below:
\begin{eqnarray} 
N(\mathrm{N_{2}H^+})  =  \frac{8\pi W_\mathrm{int}}{\lambda^{3}A}\frac{g_\mathrm{l}}{g_\mathrm{u}}\frac{1}{J_{\nu}(T_\mathrm{ex}) - J_{\nu}(T_\mathrm{bg})}      \nonumber \quad\qquad \\
                     \times   \frac{1}{1-\mathrm{exp}({-h\nu/k_\mathrm{B}T_\mathrm{ex}})}\frac{Q_\mathrm{rot}}{g_\mathrm{l}~\mathrm{exp}(-E_{l}/k_\mathrm{B}T_\mathrm{ex})}, \quad
\end{eqnarray}
where $W_\mathrm{int}$ is the integrated intensity of the \nThpj~emission, $\lambda$ and $\nu$ are the wavelength and frequency of the observed transition, respectively, $\textit{A}$ is the Einstein coefficient (= 3.628 $\times$ 10$^{-5}$ s$^{-1}$), $Q_\mathrm{rot}$ is the partition function, and E$_l$ is the energy of the lower level. 
The statistical weights of the lower level $g_\mathrm{l}$ and the upper level $g_\mathrm{u}$ are 1 and 3, respectively. 
Using our \nThpj~column density map derived from the \nThp\ integrated intensity where the signal-to-noise ratio exceeds 7 and Herschel H$_2$ column density, the \nThp\ fractional abundance [\nThp]/[H$_2$] = 1.6$\pm$0.6 $\times$ 10$^{-10}$ is obtained. 
In deriving the \nThp\ fractional abundance of the cores using their H$_2$ column densities, we considered that cores are having a single velocity component since the \nThp\ line profile is not resolved into multiple components. 
Then, we estimate the H$_2$ column densities \textit{N}($\mathrm{H_2}$) over the cores using the determined \nThp\ fractional abundance and observed \nThpj~column densities to calculate the core mass $M_\mathrm{core}$ by adding the H$_2$ column densities \textit{N}($\mathrm{H_2}$) over the cores. In these calculations we used the mean molecular weight per hydrogen molecule $\mu_\mathrm{H_2}$ of 2.8 \citep{kauffmann08}.
The \nThp~core masses range from 0.37 to 247 $M_{\odot}$ with a median value of 7 $M_{\odot}$ and the average uncertainty of core mass is 43\%. 
Note that the several bright \nThp~cores in this study are not fully resolved with the current resolution.

Here we examine whether the dense cores form stars or not, by identifying Young Stellar Objects (YSOs).
\citet{furlan16} have presented a catalogue of 330 YSOs using their 1.2--870 $\mu m$ spectral energy distributions from the Spitzer Orion Survey \citep{megeath12} and Herschel Orion Protostar Survey (HOPS). 
In this study, we use this catalogue to examine whether the dense cores are protostellar or starless. 
Among the HOPS sources, 22 Class 0 protostars, 21 Class I protostars, and 8 flat-spectrum YSOs are located in the observed cores of this study (see asterisks in Figure~\ref{fig:fil}(d) in red, green, and yellow colors, respectively). 
If a \nThp\ dense core has positional coincidence with any protostars, we define it as  a \nThp\ dense core with YSOs. If not, we define it as a starless dense core. 

In order to quantify the dynamical balance of the dense core, we determine the virial mass of the \nThp~cores using the equation \citep{maclaren88}:  
\begin{equation}
	M_\mathrm{vir}^\mathrm{core} = \left(\frac{5-2e}{3-e}\right) \frac{3~R_\mathrm{core}}{G} \sigma_\mathrm{tot, N_{2}H^+}^2,
\end{equation}
where $R_\mathrm{core}$ (=$\sqrt{ (R_\mathrm{maj} \times R_\mathrm{min})^{2} - \theta_{\rm B}^{2}}$) is the deconvolved radius of the core determined by beam deconvolution of the geometric mean radius of the major radius ($R_\mathrm{maj}$) and minor radius ($R_\mathrm{min}$) of core.
$\sigma_\mathrm{tot, N_{2}H^+}$ is the total velocity dispersion estimated from the Equation~(\ref{eq:sigtot}) using the averaged \nThpj~spectrum of the core
, $\theta_{\rm B}$ is the HPBW, and $e$ is a constant depending on the density profile, $\rho(r) \propto r^{-e}$. 
Assuming that the dense cores have isothermal density profile in the form of $\rho(r) \propto r^{-2}$, we obtain a correction factor (5 -- 2$\textit{e}$)/(3 -- $\textit{e}$) to 1 in this work. 
The virial parameter, the ratio of virial mass to core mass, 	$\alpha_\mathrm{vir}^\mathrm{core}$ = ${M}_\mathrm{vir}^\mathrm{core}/{M}_\mathrm{core}$, is measured to examine the dynamical state of \nThp~dense cores. 
Dense cores with YSOs show the virial parameter in the range of 0.05--1.5 and starless dense cores show the virial parameter in the range of 0.27--7.75 (see Section~\ref{sec:stability_core} for details). 
The average uncertainty in $\mathrm{M}_\mathrm{vir}^\mathrm{core}$ is 16\% and that in $\alpha_\mathrm{vir}^\mathrm{core}$ is 46\%. 
The information of the identified dense cores and their estimated physical quantities are listed in Table~\ref{tab:core}.

\section{Discussion}\label{sec:disc}

\subsection{Comparison of \cEo~filaments with Herschel continuum filaments}
\label{sec:hsc_fil}

\begin{figure*}
  \centering 
  \includegraphics[width=.7\textwidth]{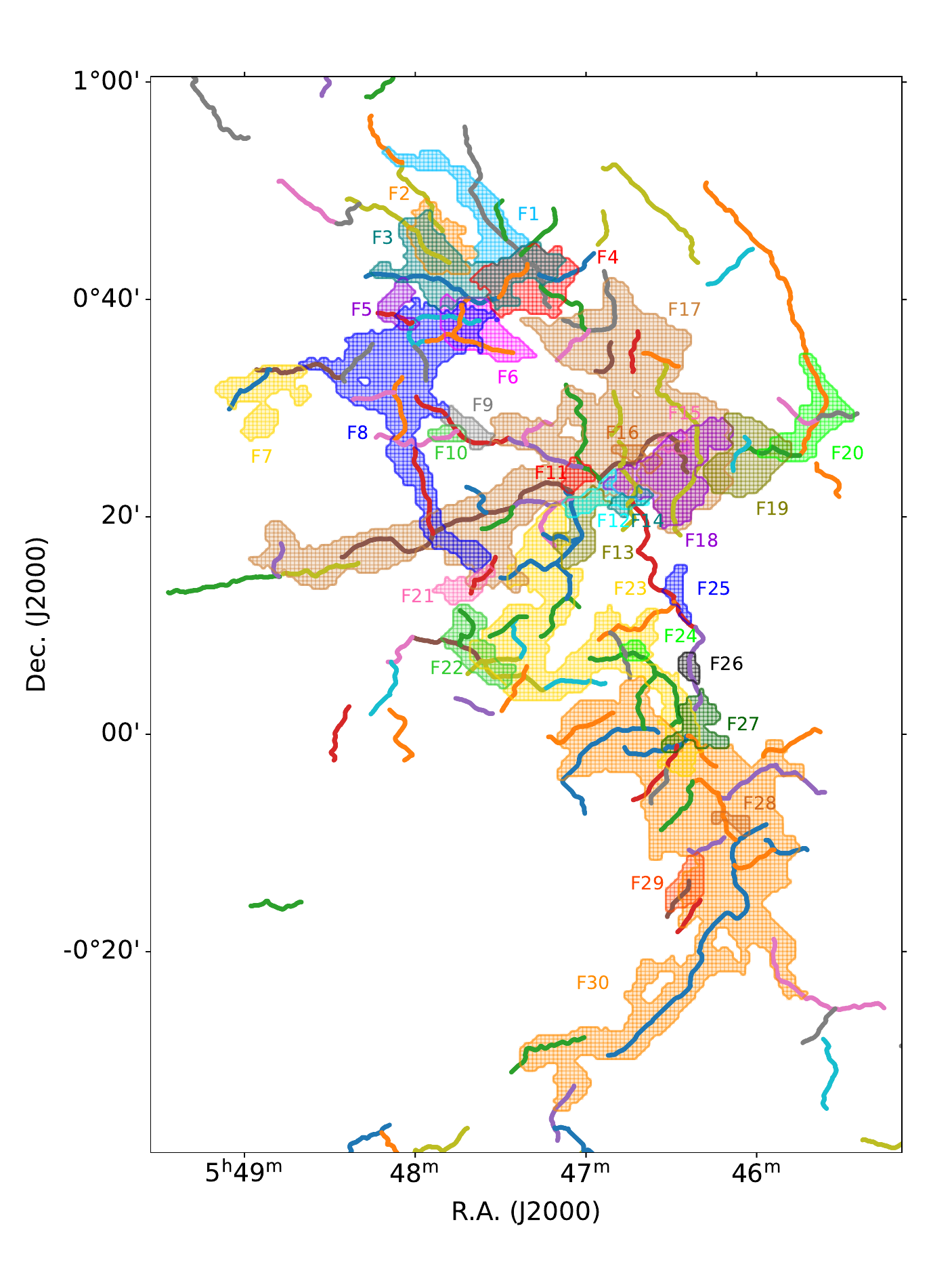}
  \caption{The velocity coherent filaments and Herschel continuum filaments. The velocity coherent filaments identified with the \cEoj~line emission in this study are presented by colored shades with their filament names. The Herschel continuum filaments  defined by \citet{arzoumanian19} using Herschel column density map are indicated with curved lines.  
 \label{fig:funs_hsc_fil}}
\end{figure*}

Using the continuum data obtained by the Herschel Gould Belt survey, \citet{arzoumanian19} and \citet{konyves20} identified filament networks toward the NGC~2068 and NGC~2071 regions. 
We compare the velocity-coherent filaments identified in this study with the Herschel continuum filaments in Figure~\ref{fig:funs_hsc_fil}. 
Although both the identification methods and the data are different, they are found to be coincident, especially in high density regions.

Figure~\ref{fig:funs_hsc_fil} shows that Herschel filaments exist not only in the regions where bright \cEo~filaments are discernible but also in the other regions  where \cEo~filaments are not well defined due to their faintness, mostly due to the better sensitivity in Herschel continuum data.
Some Herschel continuum filaments are identified in a single filament only through multiple velocity coherent filaments where evidently different velocity components exist
(e.g., olive line between F2 and F3, green line between F19 and F20, orange line between F23 and F25).
\citet{chung21} have compared the Herschel continuum filaments and \cEo~filaments in the IC~5146 molecular cloud, finding similar results. 
Those results place important emphasis on the need of velocity information for identifying velocity-coherent structures and studying their dynamical state. 

We note that our observations have a coarser spatial resolution than previous studies  of similar systems \citep{hacar13, tafalla15, hacar18}, so they may be missing some smaller-scale fibers like those seen by previous authors.  
Observation with higher spatial resolution will be helpful to understand on how large velocity-coherent filaments consist of smaller fiber structures and how they form with their relations.

\subsection{Gravitational stability in filaments}
\label{sec:stability_fil}

\begin{figure*}
  \centering 
  \includegraphics[width=\textwidth]{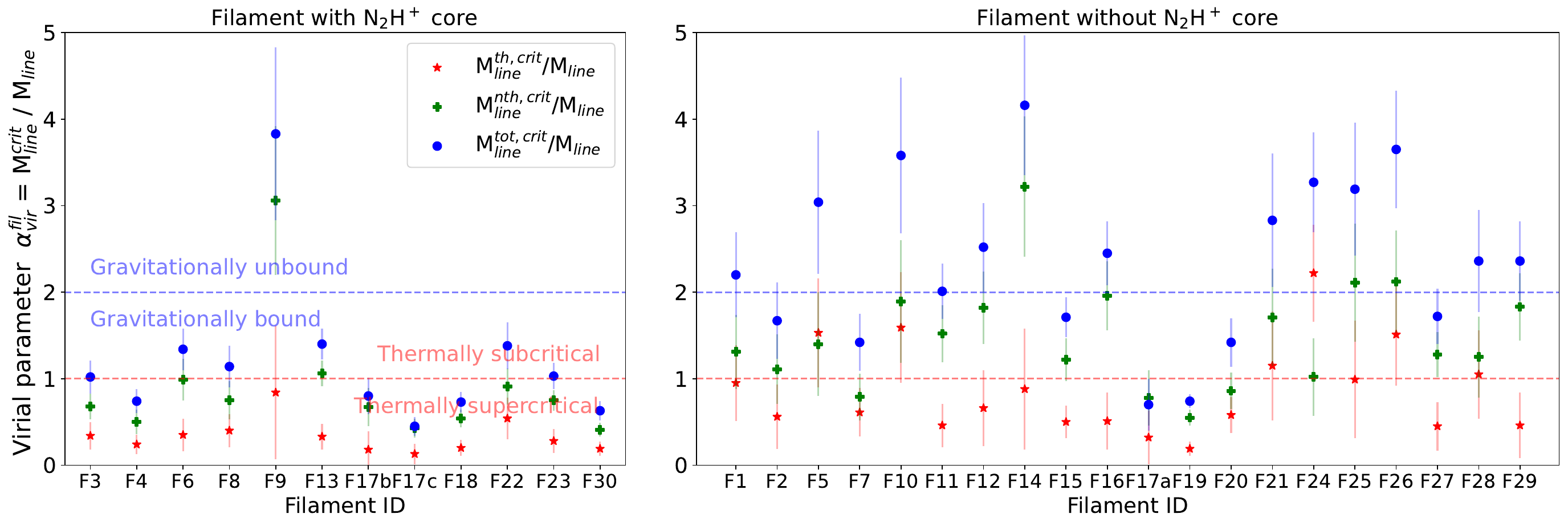}
  \caption{Virial parameters for the filaments with (left panel) and without (right panel) dense cores. 
The virial parameters for the filaments derived considering thermal support only, nonthermal support only, and both thermal plus nonthermal support are represented with red asterisk, green plus, and blue circle symbols, respectively. The horizontal lines indicate the filament's virial parameter (${M}_\mathrm{line}^\mathrm{crit}/{M}_\mathrm{line}$) of 2 (blue dashed line) and  unity (red dashed line). 
\label{fig:fil_vp}}
\end{figure*}

\begin{figure*}[ht!]
\centering 
\includegraphics[width=0.95\textwidth]{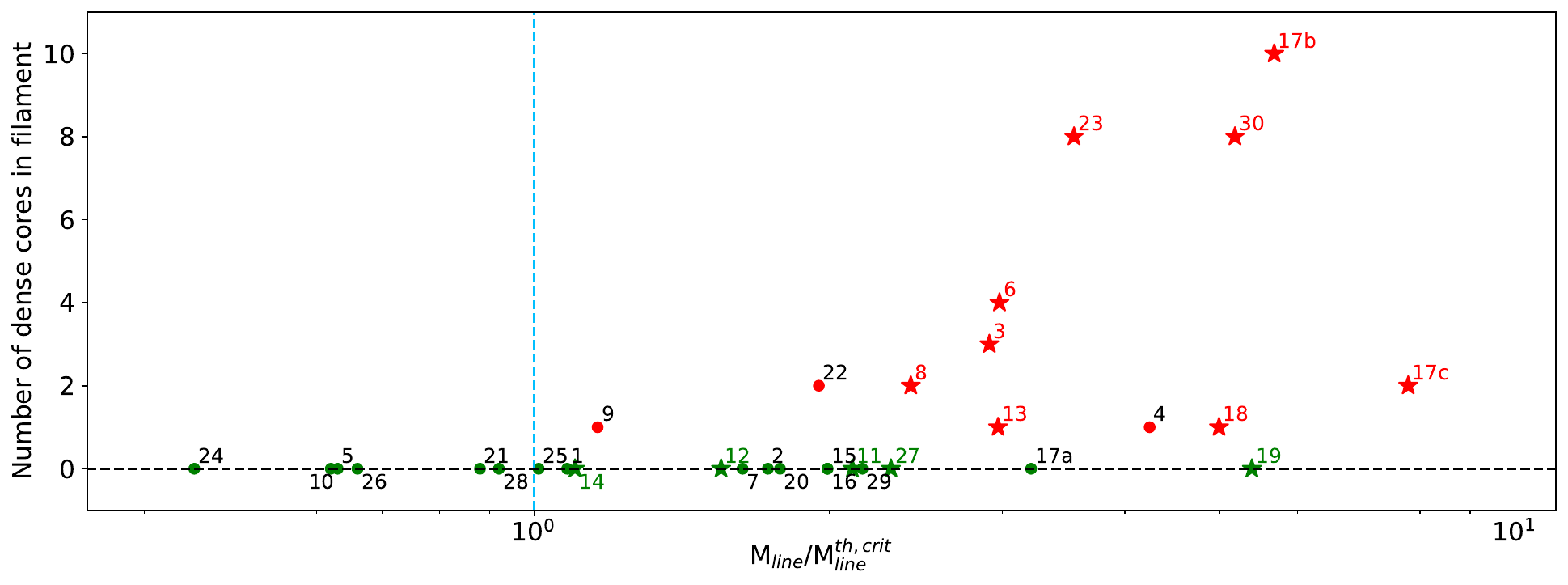}
\caption{Number of dense cores in each filament as a function of the ratio between ${M}_\mathrm{line}$ and ${M}_\mathrm{line}^\mathrm{th,crit}$. The cyan vertical line indicates the location where ${M}_\mathrm{line}$ equals to ${M}_\mathrm{line}^\mathrm{th,crit}$ and the black dashed line is to show the location where the filament does not contain any dense core. 
The circle and asterisk symbols are meant to indicate that the filaments do not contain any YSO and do contain YSOs, respectively.
Colors distinguish the filaments with dense core (red) from the filaments without dense core (green). 
The numbers indicate the filament's ID.  
\label{fig:corenum}}
\end{figure*}

As the filamentary structures in the clouds are regarded to play a key role in the formation of dense cores and stars in there, it would be highly important to examine whether our identified filaments are gravitationally stable or unstable. In this section we discuss this matter by looking into the criticalities of the filaments. 
Most of all, before determining the gravitational stability of filaments, we divide the filaments into two groups according to the existence of dense cores to explore how gravitational fragmentation occurs along the filament. 
For this, we determine which core is related to which filament using the velocity centroid of the filament in decomposed \cEoj~spectra and the \vlsr~derived from the hyperfine fitting of the \nThp~spectrum at the peak-intensity position in the dense core. 
Refer to the Figures in Appendix~\ref{A2} to find the positional and spectral coincidences of dense cores and filaments that lead to define two groups, \textit{the filaments with \nThp~dense cores} and \textit{the filaments without \nThp~dense cores (coreless filaments)}. 

In Figure~\ref{fig:fil_vp}, filaments having positional coincidences with \nThp~dense cores, which we call \textit{the filaments with \nThp~dense cores}, are on the left panel and \textit{the filaments without \nThp~dense cores} are on the right panel. 
The virial parameter of the filament $\alpha_\mathrm{fil}^\mathrm{vir}$ = $M_\mathrm{line}^\mathrm{crit}/M_\mathrm{line}$ is defined as $M_\mathrm{line}^\mathrm{tot, crit}/M_\mathrm{line}$ and presented in Figure~\ref{fig:fil_vp} with blue circles with additionally presented ${M}_\mathrm{line}^\mathrm{th, crit}/M_\mathrm{line}$ (red asterisks) and ${M}_\mathrm{line}^\mathrm{nth, crit}/M_\mathrm{line}$ (green plus symbols) depending on the contributing motion.
For the critical value of $M_\mathrm{line}^\mathrm{tot, crit}/M_\mathrm{line}$, we draw blue dashed horizontal line at $\alpha$ = 2, which is the lowest critical virial parameter proposed for non-magnetized clouds \citep{kauffmann13}. 
\citet{arzoumanian13} also used the same criteria for the virial parameter of filament.  
The estimated critical line masses are tabulated in Table~\ref{tab:fil}, and the given uncertainties of ${M}_\mathrm{line}^\mathrm{th, crit}/M_\mathrm{line}$, ${M}_\mathrm{line}^\mathrm{nth, crit}/M_\mathrm{line}$ and $M_\mathrm{line}^\mathrm{tot, crit}/M_\mathrm{line}$ are calculated from the propagation of uncertainties of observational velocity dispersion and line mass.

If the thermal critical line mass ${M}_\mathrm{line}^\mathrm{th,crit}$ is larger than the filament's line mass ${M}_\mathrm{line}$, the filament is thermally supported against the gravitational collapse (i.e. thermally subcritical). 
On the other hand, the ratio ${M}_\mathrm{line}^\mathrm{th,crit}/{M}_\mathrm{line}$ less than unity would indicate that the filamentary structure is under radial collapse or fragmentation (i.e. thermally supercritical).


In case of the filaments with \nThp~cores (left panel of Figure~\ref{fig:fil_vp}), virial parameters are less than 2, indicating the filamentary structures are gravitationally bound and have environments in which dense cores can be formed. 
The nonthermal velocity dispersions in units of sound speed ($\sigma_\mathrm{nth}/c_\mathrm{s}$) for each filament show that most of the filamentary structures are transonic/supersonic (see Section~\ref{sec:f23} and Appendix~\ref{A1}). 
The outlier F9 was identified as a filament containing a core, however its virial parameter is estimated to be large due to its large nonthermal contribution. 
The nonthermal velocity dispersion of F9 at the center of Core~37 is probably overestimated since the core spans two nearby filaments, F9 and F10 (see locations of F9 and F10 in Figures in Appendix~\ref{A2}). 
We adopt the critical value of $M_\mathrm{line}^\mathrm{th, crit}/M_\mathrm{line}$ $\sim$1 (red dashed horizontal line) to judge whether the filament is in thermally subcritical or supercritical state \citep{arzoumanian13}. 
As shown in the left panel of Figure~\ref{fig:fil_vp}, the ratio between ${M}_\mathrm{line}^\mathrm{th, crit}$ and ${M}_\mathrm{line}$ (red asterisks) presents that filaments with dense cores are all thermally supercritical as several previous studies have shown that dense cores form mostly within supecritical filaments \citep{andre10, Polychroni13, konyves15}.
On the other hand, coreless filaments (right panel of Figure~\ref{fig:fil_vp}) do not present consistent features. Some filaments (F2, F19, F20, and F27) are thermally supercritical and present virial parameters, for all the filaments but F9, less than 2 (gravitationally bound). Those filaments may have an environment potentially capable of forming dense cores. The other filaments (F5, F12, F14, F16, F21, F25, and F29) are gravitationally unbound and may disperse due to nonthermal contribution. Note that several filaments, F1, F7, and F17a, were not fully observed in \nThpj, and thus the existence of \nThp~cores could not be tested.

All filaments (except F5) have larger ${M}_\mathrm{line}^\mathrm{nth, crit}$ than ${M}_\mathrm{line}^\mathrm{th, crit}$ by a factor of 2--4, indicating that the effects of nonthermal motions are significant throughout entire molecular cloud. 
The Mach numbers (= $\sigma_\mathrm{nth}/c_\mathrm{s}$) along the filament's spine indicate that the filament has transonic or supersonic motions over those regions (see also Figures in Appendix~\ref{A1} for details on properties of each filament). 

In Taurus L1495/B213 molecular cloud, one group of their filaments (sterile fibers) do not contain cores, but the other group (fertile fibers) contained around three cores on average \citep{tafalla15}. According to their result, filaments either have no cores or have multiple cores, indicating that the distribution of cores between filaments may not be made randomly, but somehow systematically. This is consistent with the idea that if a filament is unstable (${M}_\mathrm{line}$ $>$ ${M}_\mathrm{line}^\mathrm{th, crit}$) for fragmentation, it will likely make several cores, while if it is stable (${M}_\mathrm{line}$ $<$ ${M}_\mathrm{line}^\mathrm{th, crit}$), it will make none. 
Figure~\ref{fig:corenum} shows the number of cores per filament. 
The number of dense core is zero for filaments in a stable state (${M}_\mathrm{line}/{M}_\mathrm{line}^\mathrm{crit}$ $<$ 1). 
For filaments in an unstable state (${M}_\mathrm{line}/{M}_\mathrm{line}^\mathrm{crit}$ $>$ 1), the larger ${M}_\mathrm{line}/{M}_\mathrm{line}^\mathrm{crit}$ tends to have the large number of dense cores in the filament. 
Hence, understanding the criticality of a filament can provide insights on the fragmentation of filaments and formation of dense cores/YSOs.
Note that the presence of YSOs in filaments with no dense core, F11, F12, F14, F19, and F27 (green star symbol in Figure~\ref{fig:corenum}) , are ambiguous to properly interpret, since some YSOs are located in the overlapped region between two filaments (e.g., F11, F12, and F14 are sharing YSOs with F17b, and F27 shares a YSO with F30).

\subsection{Virial state of the \nThp~dense cores}
\label{sec:stability_core}

\begin{figure*}
  \centering 
  \includegraphics[width=0.98\textwidth]{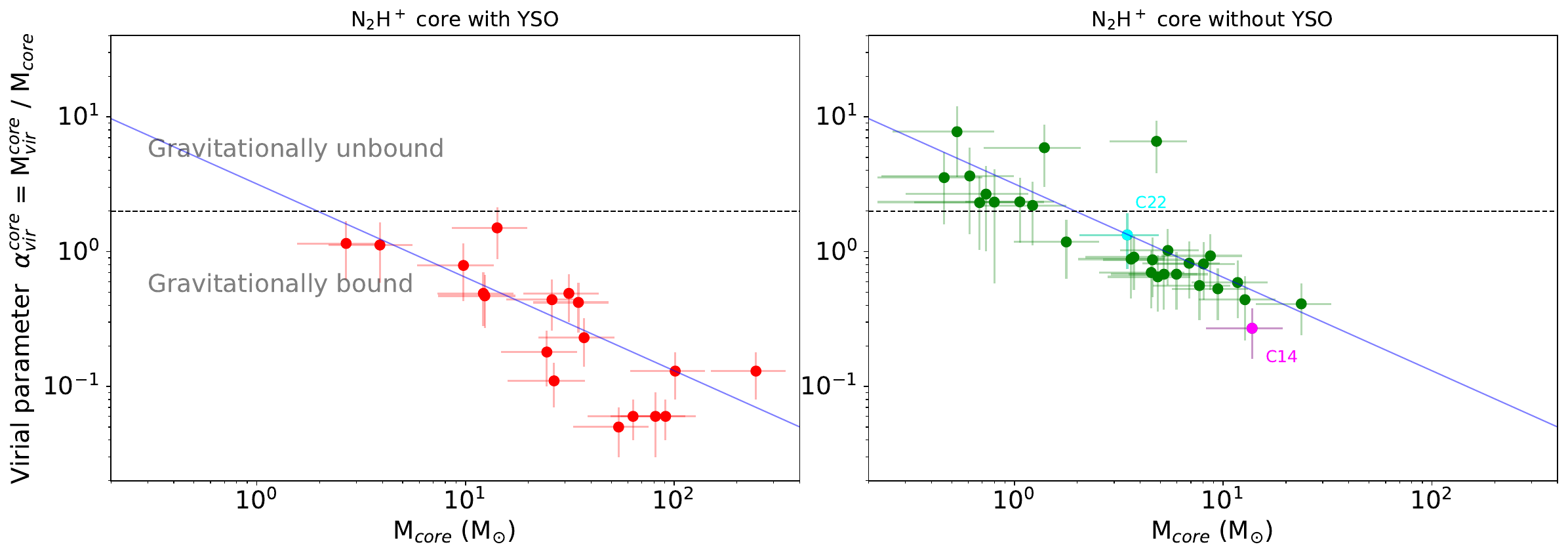}
  \caption{Virial parameters for the dense cores with (left panel) and without (right panel) young stellar objects. 
  The horizontal line marks the location of the critical value of the virial parameter, $\alpha_\mathrm{crit}$ = 2. The blue line indicates the best-fit of data points with a power-law index of \textminus0.7. Magenta and cyan circles are to indicate the starless dense cores C14 and C22 with NH$_2$D detection (see Section~\ref{sec:chem}).
\label{fig:core_vp}}
\end{figure*}

\begin{figure}[ht!]
\centering 
\includegraphics[width=0.48\textwidth]{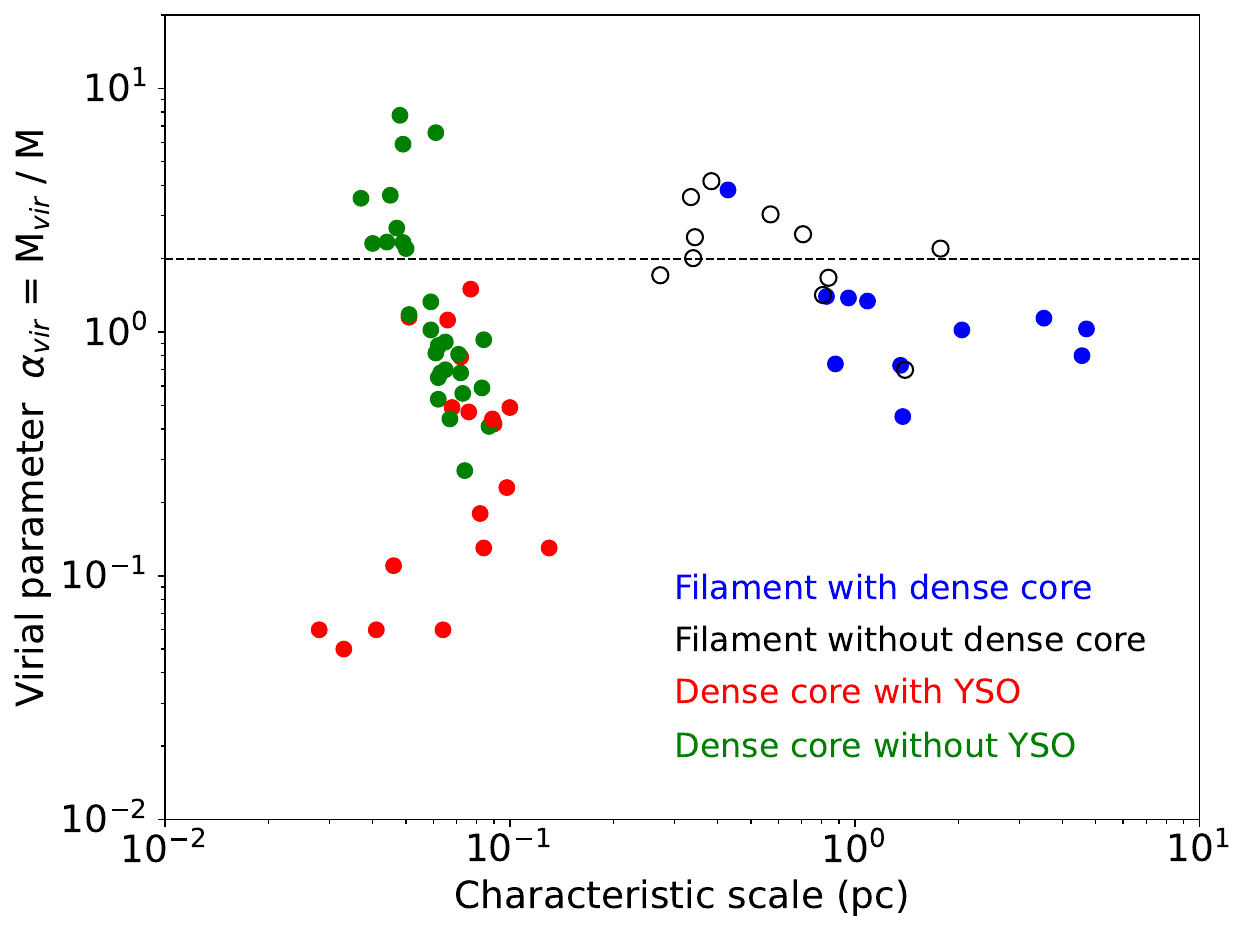}
\caption{Virial parameters for the dense cores and filaments  as a function of their characteristic scales. 
The characteristic scale is given as the deconvolved core radius \textit{R}$_\mathrm{core}$ for dense cores and the filament length \textit{L} for filaments. 
Red and green filled circles indicate dense cores with and without YSOs, respectively. Blue filled and Black open circles are to display filaments with and without dense cores, respectively. 
The horizontal line marks the virial parameter of 2. 
\label{fig:size_avir}}
\end{figure}


In molecular clouds that are gravitationally collapsing, gravity dominates the energy balance, and the kinetic energy and the absolute value of gravitational energy become comparable, making the system to be in the virial equilibrium. 
The equipartition between gravity and kinetic energy approaches at $\alpha_\mathrm{vir}^\mathrm{core}$ (=~$M_\mathrm{vir}^\mathrm{core}/M_\mathrm{core}$) = 2 for non-magnetic, uniformly density spheres \citep{Vazquez-Semadeni07, Ballesteros-Paredes11, Ramirez-Galeano22}. 

In case of  $\alpha_\mathrm{vir}^\mathrm{core}$ $<$ $\alpha_\mathrm{crit}$, the gravitational binding energy is more significant than the kinetic energy, and thus, 
the dense core is gravitationally unstable and can collapse. 
On the other hand, if $\alpha_\mathrm{vir}^\mathrm{core}$ $>$ $\alpha_\mathrm{crit}$, the kinetic energy is dominant over the gravitational energy, 
so that the dense core is gravitationally unbound and unlikely to form young stars.

Figure~\ref{fig:core_vp} presents the virial parameters of dense cores as a function of the core mass derived from \nThpj. 
The virial parameter is inversely proportional to the core mass, with a power-law index of \textminus0.7 for this region.
\nThp~cores with YSOs (left panel of Figure~\ref{fig:core_vp}) have $\alpha_\mathrm{vir}^\mathrm{core}$ of 0.05$-$1.5 (the average value of $\alpha_\mathrm{vir}^\mathrm{core}$ is 0.44). 
All cores with YSOs have $\alpha_\mathrm{vir}^\mathrm{core}$ lower than the critical value, indicating that they are likely gravitationally bound. 
On the other hand, starless \nThp~cores (right panel of Figure~\ref{fig:core_vp}) have $\alpha_\mathrm{vir}^\mathrm{core}$ of 0.27$-$7.75 with an average value of 1.85, indicating that some of them are gravitationally bound and others are not.
It is also noted that starless cores over 2 $M_{\odot}$ tend to be gravitationally unstable.

In a study by \citet{ohashi16}, dense cores in the Infrared Dark Cloud G14.225-0.506 and their parental filaments show a trend that their virial parameters decrease with decreasing scales from filament to clump to dense cores (see Figure 6 in \citealt{ohashi16}). 
Figure~\ref{fig:size_avir} shows the virial parameters for the characteristic scales for dense cores (the deconvolved radius $\textit{R}_\mathrm{core}$) and filaments (the filament lengths $\textit{L}$). 
In \citet{ohashi16}, virial parameters of prestellar cores and protostellar cores were indistinguishable. 
In our result, however, this is not the case. 
All the dense cores with YSOs have $\alpha_\mathrm{vir}^\mathrm{core}$ $<$ 2 while nearly half of the starless cores have $\alpha_\mathrm{vir}^\mathrm{core}$ $>$ 2. 
In the case of filaments, we divide them into filament with (blue dots) and without (black open circles) dense cores. 
The filaments in gravitationally bound state ($\alpha_\mathrm{vir}^\mathrm{fil}$ $<$ 2) tend to produce dense cores, and dense cores with the virial parameter less than its critical value tend to be in a more active star-forming environment. 
We also observed a decreasing trend with decreasing characteristic scale from filament to core when we take gravitationally bound samples (blue dots and red dots in Figure~\ref{fig:size_avir}).

However, we note that the virial parameter alone cannot strongly represent the environmental feature of star formation 
(note also the large uncertainty of $\alpha_\mathrm{vir}^\mathrm{core}$ $\sim$46\%). 
Further investigation on dynamical and chemical status  of the dense cores with molecular lines can provide supplementary factors for understanding the evolutionary status of the dense cores  (Section~\ref{sec:chem}). 
Furthermore, the \nThp~dense cores identified in this study are not fully resolved with the current angular resolution ($\theta_{\rm B}$ $\sim$54$''$).
Therefore, observation with higher angular resolution is needed to better examine core properties with virial analysis. 

\subsection{Kinematic information of filaments}\label{sec:kinematic}

\begin{figure}[ht!]
\centering 
\includegraphics[width=0.49\textwidth]{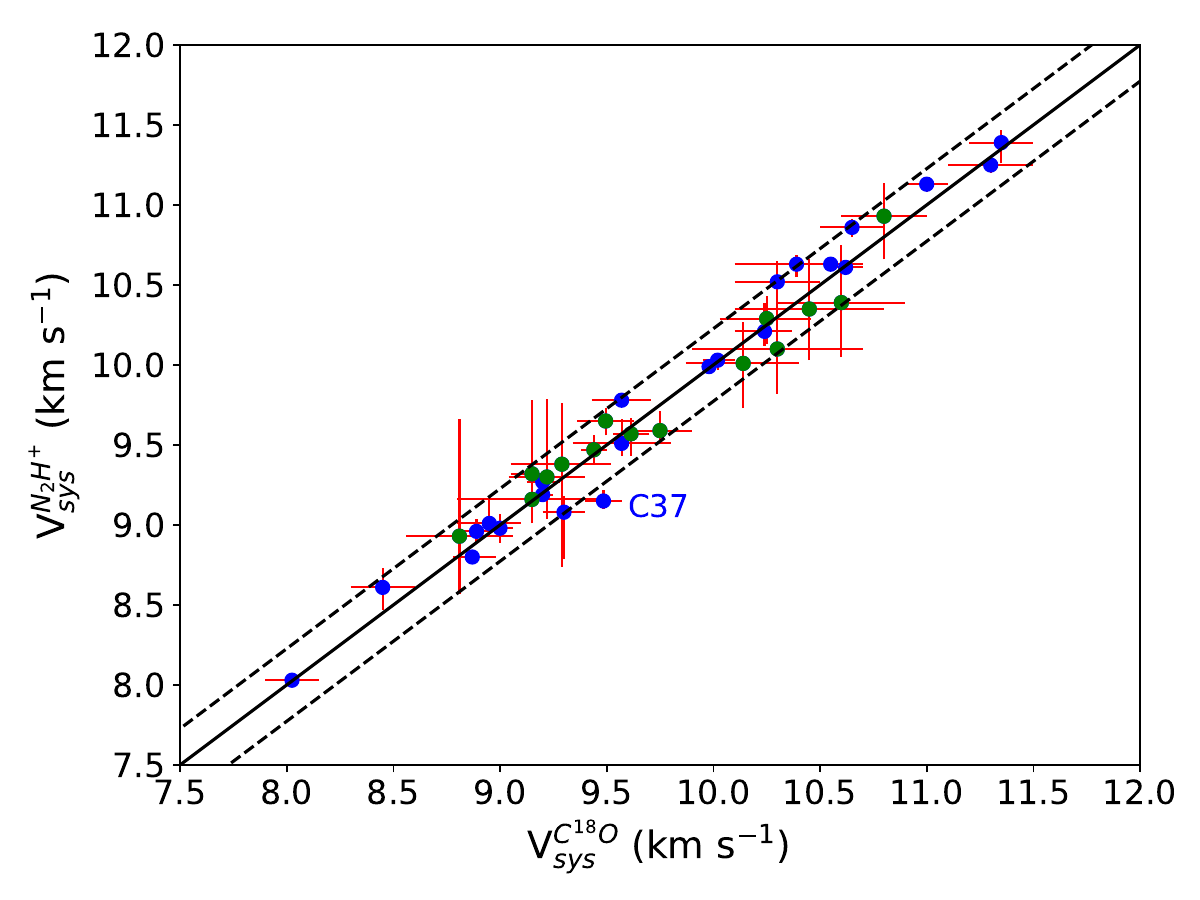}
\caption{The systemic velocities of dense cores measured by \nThpj~spectra in comparison with those of surrounding filaments obtained by \cEoj~spectra. 
Green and blue dots represent dense cores with and without YSOs. Red bars indicate 1$\sigma$ error bars for both data sets. A solid line indicates the result for the least squares fit for the data and dashed lines show the range from the fit result by the sound speed of 0.23 \kms, at the average dust temperature of 14.8 K. 
\label{fig:veldiff}}
\end{figure}

Discussions on the difference between the systemic velocities of filaments and dense cores have been made on several studies to investigate how the cores form in the filaments. 
\citep{kirk07, hacar11, chung19, chung21}. 
A core-to-envelope velocity difference less than the sound speed may imply that they share a similar kinematics from core to filament scale. 
Figure~\ref{fig:veldiff} shows that the systemic velocities of \cEoj~and \nThpj~have a tight correlation with difference less than 0.23 \kms~(sound speed at 14.8 K, the median value of dust temperature in Figure~\ref{fig:hist}). 
The systemic velocities of \cEo\ and \nThp\ spectra were obtained from either single Gaussian fits for each component  in \cEo\ or multiple Gaussian fits for the hyperfine components in  \nThp\ spectra, respectively. We independently give the initial guesses and conditions for finding best-fit results. 
The velocity coincidence between filaments and related dense cores indicates that dense cores and the surrounding filaments move together with a close kinematical relation between two objects. We are not able to find any significant difference between the two groups, the dense cores with (green dots) and without (blue dots) young stellar objects.

The outlier in Figure~\ref{fig:veldiff} (dense core C37) is located at the border between F9 and F10 (see also note in Section~\ref{sec:stability_fil}). 
F9 and F10 have slightly different velocity ranges within a sound speed (refer to column 2 in Table~\ref{tab:fil}). However, they are distinguishable by identification criteria in this study. Since the position of the emission peak of C37 belongs to the region of F9, we define that the dense core C37 is related with F9. The ambiguity may be solved in further studies with higher angular and spectral resolutions. 

As we mentioned in the last paragraph of Section~\ref{sec:fil}, the \vlsr~spans $\sim$4 \kms~through the entire cloud material and the average range of internal velocities of each filament is about 0.7 \kms\ (see Column 5 in Table~\ref{tab:fil}). 
In order to evaluate whether the velocity change comes from velocity dispersion due to internal structures or the large-scale velocity gradient along the filament, 
we trace the changes of intensity and velocity centroid along the spine of the filaments in the right-middle panel of Figure~\ref{fig:f23_civw} (see also Figures in Appendix~\ref{A1}). 
In the case of F23 in Figure~\ref{fig:f23_civw}, for example, there is a large-scale change of systemic velocity along the filament from south to north, with the approximate end-by-end velocity gradient of 0.26 \kms~pc$^{-1}$. Monotonic large-scale gradients are also seen in F8 (0.4 \kms~pc$^{-1}$) and F17b (0.2 \kms~pc$^{-1}$) with local velocity fluctuations at sub-pc scale 
(see also Figures in Appendix~\ref{A1} for the properties of the other filaments harboring dense cores). 
If the gas flow is converging to the core center, the velocity field would reverse at the density peak position of the core. 
Such pattern of velocity field is observed in surrounding filaments around several \nThp~cores 
(e.g., C39 and C41 in F3, C12 in F6, C14 in F6, C33 in F8, C40 in F18, C27 in F23, and C26 in F23). 

In most of filaments and their dense cores, the velocity dispersions are smaller in dense cores than in the surrounding filament clouds, while the velocity centroids of the filaments and cores are nearly identical. 
However, in case of F17b (Figure~\ref{fig:fA18}), the velocity centroids for massive dense cores (especially for C1, C3 and C7) spread over a remarkably larger range than those in the dense cores along other filaments. 
Moreover, the non-thermal velocity dispersions of these cores (C1, C3, and C7) are larger than that of F17b, contrary to the expectation that the turbulent motions of dense cores would be usually lower than their surrounding filaments. 
One of the most massive filaments, F30, also contains massive cores (C2, C4, and C5) which have a wide range of velocity dispersion. 
These distributions of centroid velocities and velocity dispersions of the cores can be explained as these cores are in a very complex location with a large number of YSOs and possible inflows along the filaments. 
The density and velocity structures given by F17, F18, and other several small filaments around the brightest dense core C1 appear to be very dynamic, almost like those shown in a hub-filament system \citep{myers09, peretto13}.
It is likely possible that these structures can be split into a number of individual overlapping filaments as a result of our choice of low velocity difference used in the last step of FoF algorithm in this study.



\subsection{A possible longitudinal fragmentation in a filament} \label{sec:f23}

\begin{figure*}
  \centering 
  \includegraphics[width=\textwidth]{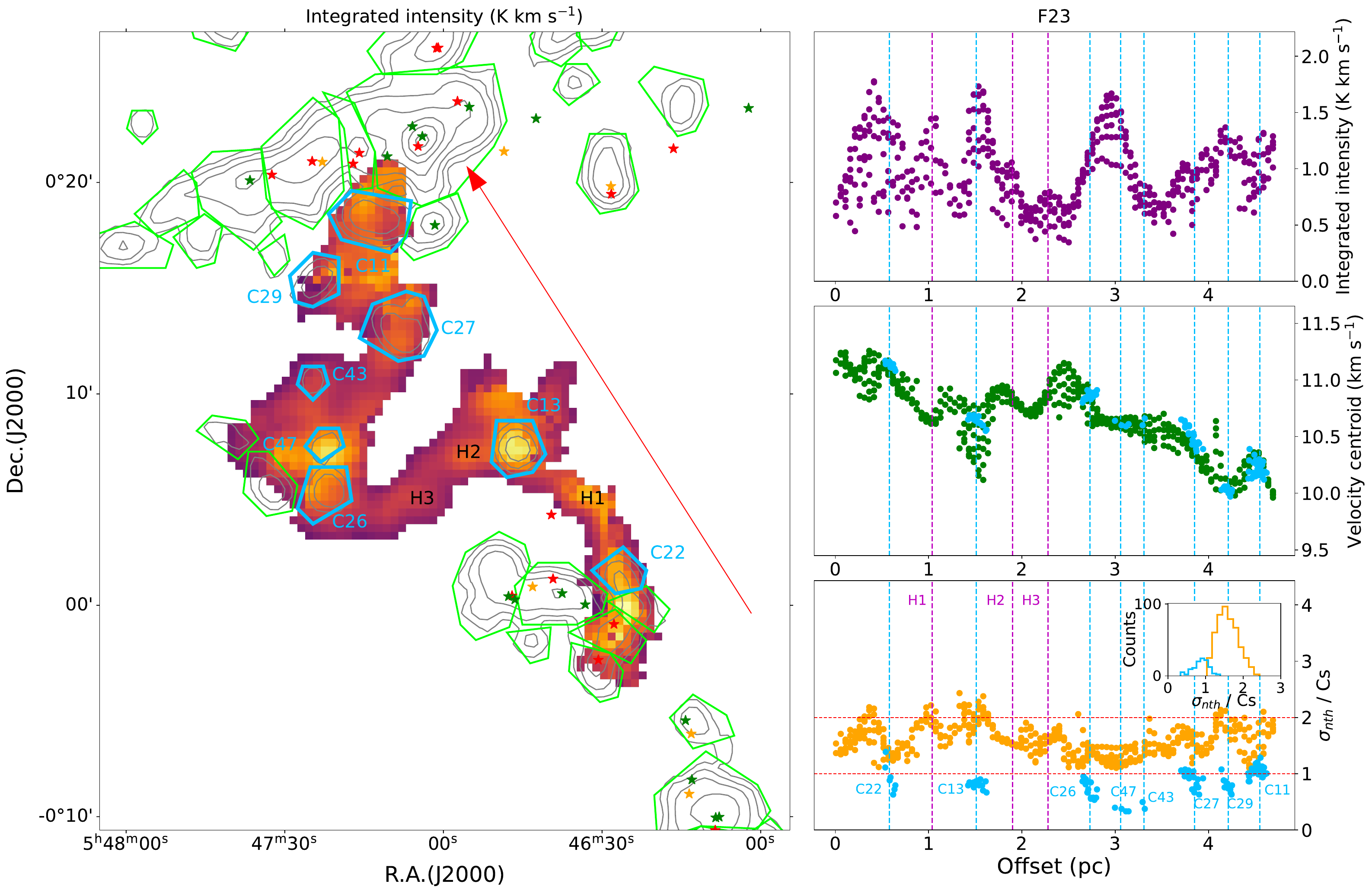}
  \caption{Left panel : The distribution of the YSOs and the dense cores in the \cEoj\ filament F23. The color tone is to indicate the image of \cEoj\ integrated intensity. The contours are the integrated intensity levels of \nThpj\ [0.08, 0.3, 0.8, 2.0, 4.0, 6.0, 8.0, 10.0] K \kms~in. The blue and green polygons are to display the dense cores seemingly associated with the filament in LSR velocities and the dense cores not associated with the filament, respectively. The asterisks indicate young stellar objects classified in \citet{furlan16}. Red : Class 0 protostars. Green : Class I protostars. Orange : Flat-spectrum YSOs. 
Right-upper panel : The variation of the integrated intensity along the filament spine. The red line displays the averaged intensity variation.
Right-middle panel : The variation of the velocity centroids along the filament spine from \cEoj\ (green dots) 
				and the dense cores from \nThpj\ (blue dots) associated with the filament. 
Right-lower panel : The variation of the nonthermal velocity dispersion divided by the sound speed along the spine of filament.
The orange and blue dots are estimated from \cEoj\ (for filaments) and \nThpj\ (for dense cores), respectively. 
The vertical dashed lines are representative positions of the dense cores embedded in the filament (the blue polygons in left-panel). Red horizontal lines indicate the locations for the Mach number (= $\sigma_\mathrm{nth}/c_\mathrm{s}$) of 1 and 2. 
The inset shows histograms of the Mach number distributions of filaments and cores with the same colors as the ones in the main plots. 
\label{fig:f23_civw}}
\end{figure*}

\begin{figure}[ht!]
\centering 
\includegraphics[width=0.47\textwidth]{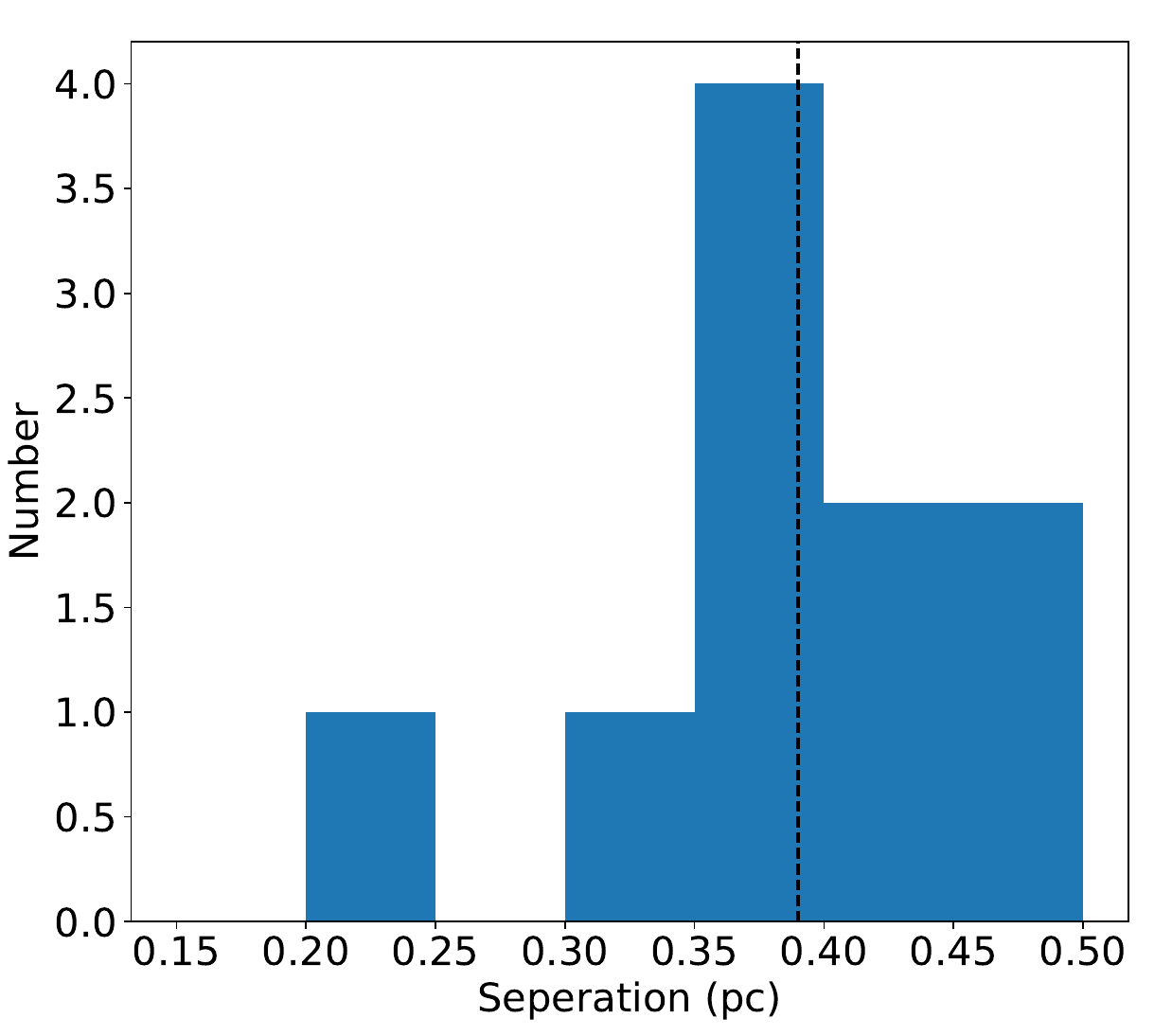}
\caption{The histogram of core-to-core separations in the filament F23 using offset positions of identified \nThp~dense cores and less dense, but possible core forming regions in future. The vertical line is to indicate the median value of the separation (0.39~pc).
\label{fig:F23_hist}}
\end{figure}

Left panel of Figure~\ref{fig:f23_civw} shows how the dense cores (polygons) are distributed in the filament F23 (color tone). 
Eight dense cores (C22, C13, C26, C47, C43, C27, C29, and C11 from south to north, marked with blue polygons) are located along the curved filament with similar \vlsr.
We also found the other dense cores off the filament in space or velocity dimension are drawn with green polygons. 
Note that the dense cores associated with the filament F23 do not contain any young stellar objects, implying that the dense cores in F23 are not yet fully evolved to form protostars. 

The right three panels of Figure~\ref{fig:f23_civw} show the distributions of integrated intensity of F23, velocity centroids, and Mach number, defined as $\sigma_\mathrm{nth}/c_\mathrm{s}$, along the spine of filament and dense cores located in the filament.
The red arrow in the left panel of Figure~\ref{fig:f23_civw} indicates the scanning direction along the spine of filament to trace the property changes. 
We present the locations of dense cores along F23 with blue dashed vertical lines on the three right panels. 
The right-upper panel of Figure~\ref{fig:f23_civw} shows the variation of \cEoj~integrated intensity distribution along the filament main axis. 
The positions of dense cores are generally well matched to the filament's intensity peaks along the spine. However, along the density fluctuation, there are three additional bumps at offset $\sim$1~pc, $\sim$1.9~pc and $\sim$2.3~pc (see purple vertical dashed lines) where there are no dense cores detected. So we believe that these are less dense fragments (H1, H2, and H3) where the \nThpj~emission is not strong enough to be detected, but possibly any dense core may form in future. 
Using the positions of both identified dense cores and possible core forming regions in future, we derive an average separation of 0.39~pc (vertical line in Figure~\ref{fig:F23_hist}) from the quasi-periodic locations of the fragments on the plane of the sky.

To explain the quasi-periodic inter-core separation, we bring an analytic theory on the instability that describes gravitational collapse of isothermal filaments. 
\citet{inutsuka92} proposed that a self-gravitating cylinder becomes unstable when the line mass is greater than its thermal critical line mass and then fragments with separation of four times the width of the cylinder. Observational evidence supporting the gravitational fragmentation scenario has been reported on various scales \citep{wang11, tafalla15, orkisz19, shimajiri19, zhang20}. 

The line mass of F23 (104 $M_{\odot}$) is 3.6 times larger than its thermal critical line mass (29 $M_{\odot}$). Sufficient line mass is supporting the underlying condition of the theory on longitudinal fragmentation. 
The width of F23 is estimated to be 0.13~pc and the average separation is 0.39~pc,  which is in a fairly good agreement with expectation from the theory (4 $\times$ width of F23 $\sim$0.52~pc).

However, several issues need to be addressed before reaching a definitive conclusion about the fragmentation process associated with the formation of dense cores in the filamentary clouds in Orion~B.

First of all, our measured separation is the one projected on the plane of the sky. Thus, this separation would be fairly consistent with the theoretical separation that can be resulted from the gravitational fragmentation in the isothermal filament, when the inclination angle between filament’s major axis and the plane of sky is close to 40 degree.
Since the filament is not perfectly straight, the geometrical effect  can also influence the oscillation pattern and the evolution of clumps \citep{gritschneder17}. 

We also note the theoretical expectation in \citet{inutsuka92} does not consider the effects of magnetic fields. A magnetic field perpendicular to the filament axis can suppress fragmentation and reduce the growth rate of instability \citep{hanawa17}. On the other hand, a magnetic field parallel to the filament axis does not suppress the instability of filament \citep{hanawa93}, and can even produce a core spacing comparable to the filament diameter, if it has a high magnetic field strength \citep{nakamura93}.
In fact the study of \citet{konyves20} may show the case. In that study the cores were identified in three times smaller scale than ours (see also Section~\ref{sec:hsc_fil}) for the same region of Orion B as our study and the median value of $\sim$0.14 pc Herschel core separation is equivalent to the typical width ($\sim$0.10) pc of Herschel filaments, implying that the core separation can be dependent with the tracing scale or some physical conditions that are not well known. 
Hence, further studies including information on magnetic fields (e.g., orientation, strength) are needed to understand the fragmentation of internal structure of the filament in the Orion~B. 

In order to study the dynamics of the dense cores and the surrounding material, we compare the velocity centroids of \nThp~and \cEo~emissions in the right-middle panel of Figure~\ref{fig:f23_civw}.
As the figure shows, the systemic velocities of \nThp~cores present a close correlation with those of the surrounding material of the filament. 

The Mach number along the filaments estimated with \cEo~spectra is distributed between 1 and 2, indicating that the motions in the filament are mostly transonic (see panel (i) of Figures in Appendix~\ref{A1}).    
On the other hand, the Mach number measured from \nThpj\ inside the dense cores associated with the filament is $\lesssim$ 1 indicating that the \nThp~dense cores are transonic/subsonic. 
The inset in the right-bottom panel of Figure~\ref{fig:f23_civw} presents the histograms for the distribution of the measurements of Mach numbers for both lines (orange for \cEo~and blue for \nThp).
Considering insignificant velocity shift between \cEo~and \nThp~lines and the lower Mach numbers in dense cores than those in surrounding filaments, 
turbulence dissipation may has been processed to form the dense cores in the filament.

\subsection{Cores in infalling motions}
\label{sec:infall}

\begin{figure*}
  \centering 
  \includegraphics[width=0.49\textwidth]{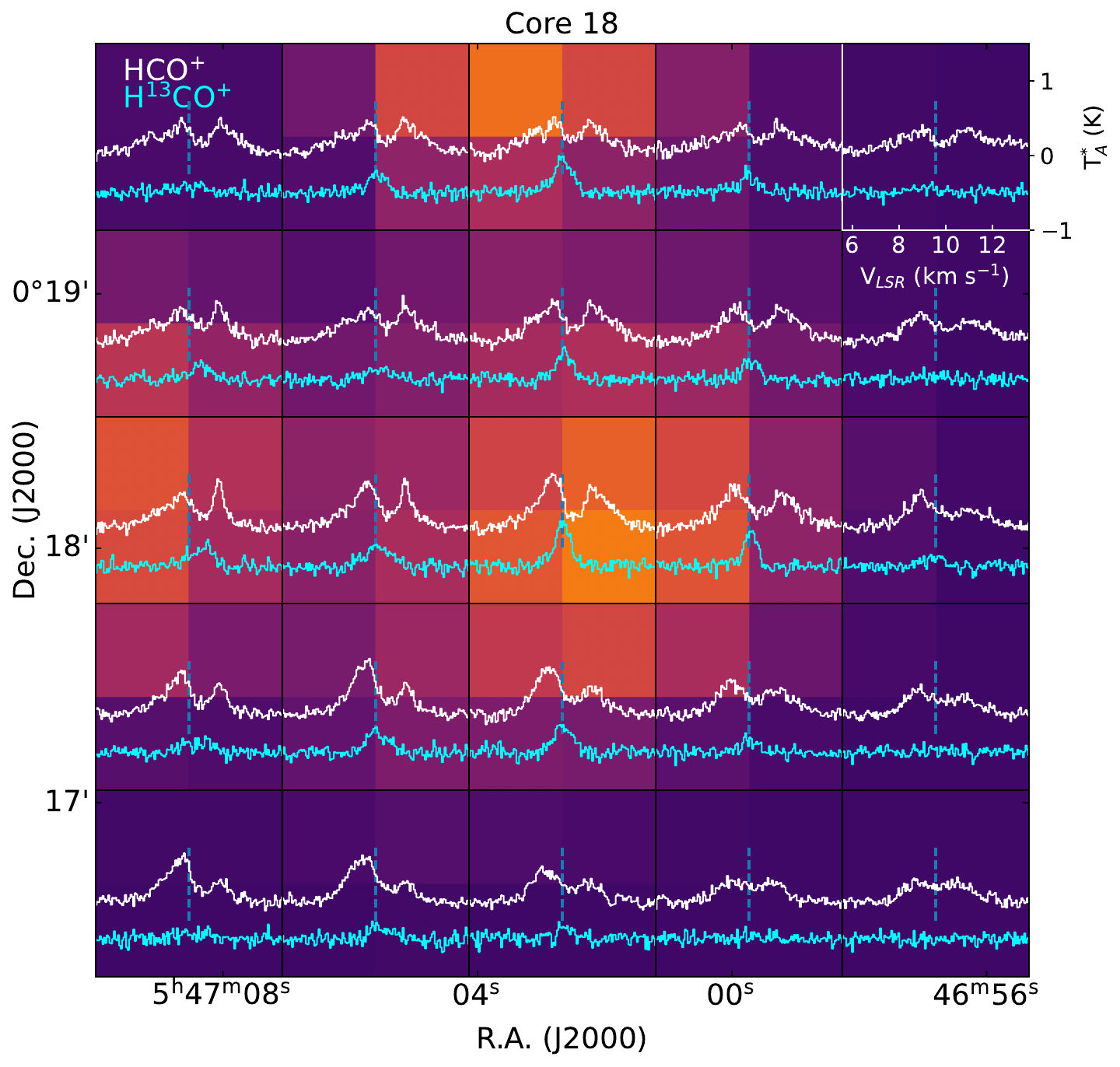}
  \includegraphics[width=0.49\textwidth]{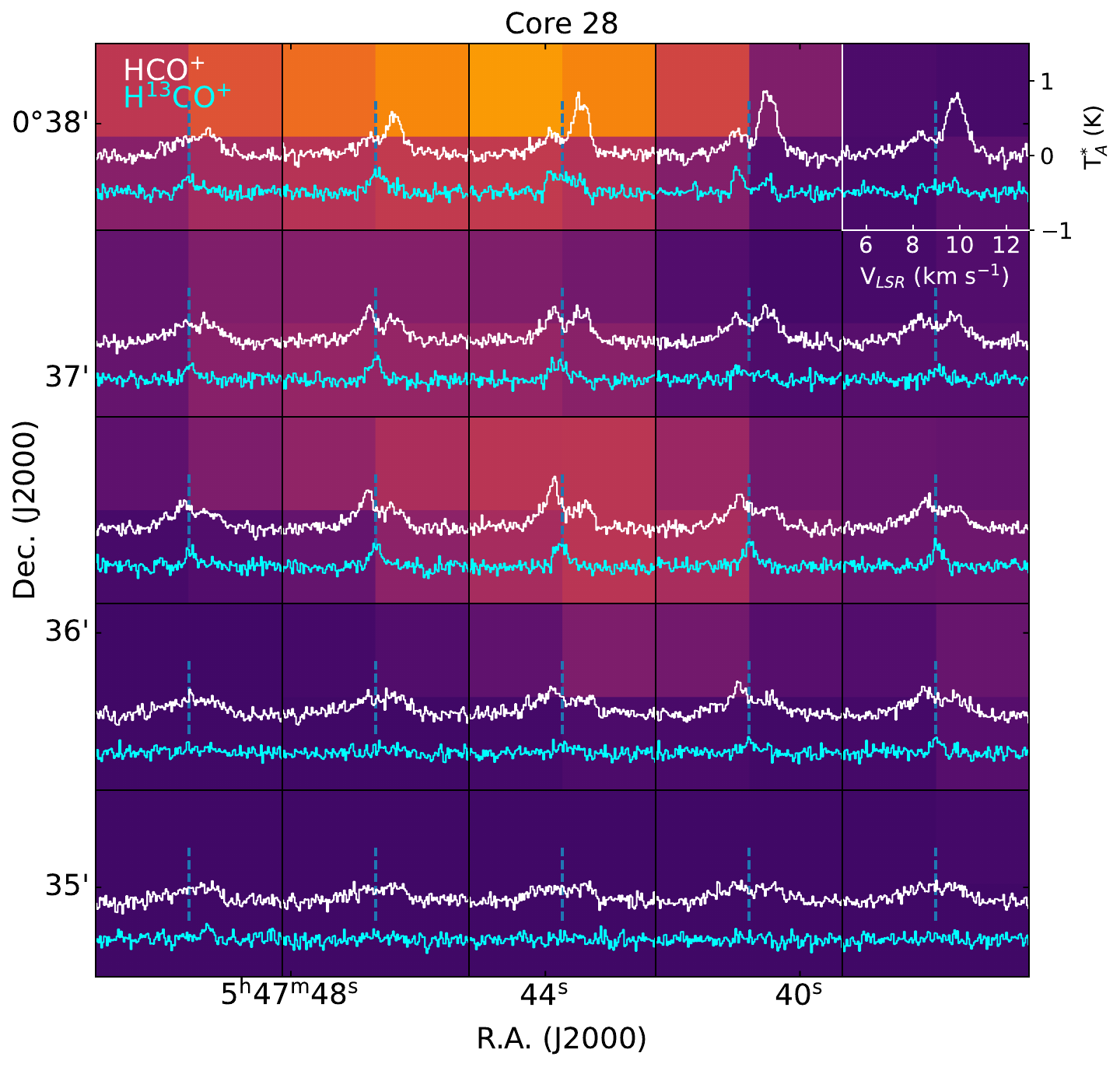}
  \caption{Infall asymmetries appeared in the \hcopj~(white) and \hTcopj~(cyan) line profiles toward C18 (central 3$\times$3 spectra in left panel) and C28 (central 3$\times$3 spectra in right panel). The color tones in the background are to indicate the \nThpj\ integrated intensity. The dashed blue vertical line at the center indicates the systemic velocity measured from the hyperfine fitting of \nThpj. Each spectrum is the average of the spectra for the 2$\times$2 pixels shown in the background of each spectrum. \hTcopj\ is shown with an offset of \textminus0.5 K in the y axis.
\label{fig:infall}}
\end{figure*}

\begin{figure}[ht!]
\centering 
\includegraphics[width=0.45\textwidth]{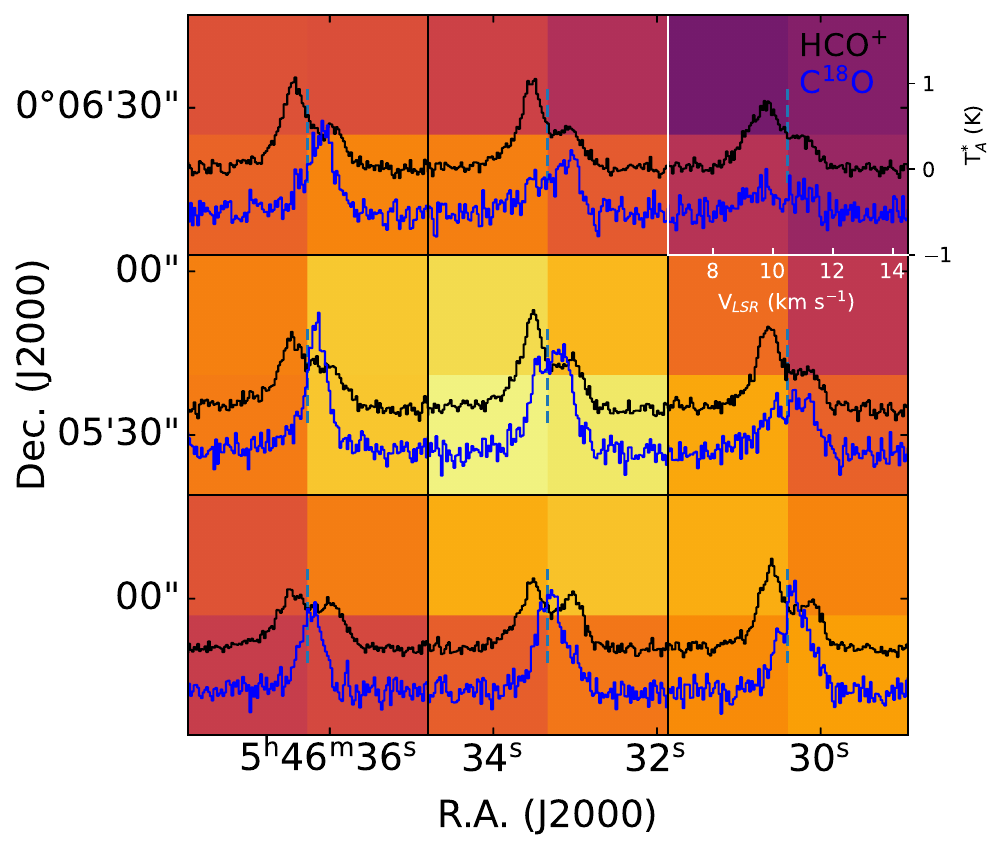}
\caption{Infall asymmetries shown \hcopj~(black) and \cEoj~(blue) line profiles at the region of H1 marked in Figure~\ref{fig:f23_civw}. 
Each line profile is the average of the spectra for the 2$\times$2 pixels shown in the background of each spectrum.
The background color scale indicates the \cEoj~integrated intensity. The baselines of \hcopj\ and \cEoj\ are set to be zero and  \textminus0.5 K, respectively.
\label{fig:H1}}
\end{figure}
One way to diagnose how much the dense cores are evolved is to see whether they are experiencing gas infalling motions which may play a role to enhance the evolution of the dense cores by helping to condense gaseous material in the cores.
Thus we attempted to search for ``blue asymmetries'' in the line profile of the optically thick molecular species \citep{evans03}. In order to make sure that the asymmetric double-peaked line profile is not coming from multiple velocity components along the line of sight, we need to compare the optically thick line profile with an optically thin line. 
Since \hcopj, which is often used as a tracer of infall motion in molecular clouds, is optically thick \citep{shimajiri17} and \hTcopj\ is optically thin in the observed region \citep{aoyama01}, we searched for infall signatures toward our dense cores using these two isotopologues of \hcop.

Figure~\ref{fig:infall} shows \hcopj\ and \hTcopj\ spectra of the two dense cores C18 and C28 over the \nThpj\ integrated intensity maps in color tones. The \hcop\ spectra around the bright core region (see central 3$\times$3 spectra) show double peaked profiles with higher blue peaks (so called blue profiles), indicating the core is undergoing contraction motions.
Cores C13, C14, C25, C26, C27, C29, and C36 also present such blue profiles in \hcopj\ lines (see Section~\ref{A3} in Appendix). 
Among them, C18 is the only protostellar dense core. Other eight dense cores are all starless dense cores. 
The filaments (F6, F13, F17, and F23), which contain nine dense cores presenting infall asymmetry in \hcopj\ line profile, are all thermally critical in virial analysis (see Figure~\ref{fig:fil_vp} and Figure~\ref{fig:corenum}).

Figure~\ref{fig:H1} shows an interesting case where infalling gas motions are seen in \hcopj\ toward a fragment at H1 region of filament F23, which seems not yet condensed enough to be detected in the \nThpj~line.
Probably the gravitationally critical status of its parent filament F23 may be triggering the condensing process toward a fragment to produce a future prestellar core. 
We may be watching for a moment of very early condensing process before the stage of a prestellar core. 
Indeed, there are several Herschel robust prestellar cores identified on the high resolution continuum map in \citet{konyves20}  (see also Section~\ref{sec:chem}).
In the cases of H2 and H3, \hcopj\ spectra do not present any significant infall signatures and there are no Herschel prestellar core. 
We think that these may be even younger fragments than H1 and their destines can not be determined at this moment.

\subsection{Chemical differentiation in the dense cores}
\label{sec:chem}
Chemical differentiation between the dense gas tracers can be also an important factor to examine how the dense cores are evolved.
In this section, we discuss how our molecular tracers show different or similar distributions in the dense cores and what the distributions would imply.

In dense cores with relatively low temperature ($T$ $\sim$10--20 K), C-bearing molecules (e.g., CO) are easily depleted 
and the abundance of H$_{3}^{+}$ can be enhanced because of less abundant CO as the main destroyer of H$_{3}^{+}$. 
The reaction of H$_{3}^{+}$ with HD can produce gas-phase progenitors of the deuterated species 
(H$_{2}$D$^{+}$, D$_{2}$H$^{+}$, and D$_3^+$) and lead to NH$_2$D~molecule formation from the reaction of the deuterated ions 
with NH$_3$ and the dissociative recombination of NH$_3$D$^+$ \citep{lepp87, pagani92, caselli03, sipila15}. 
Moreover, the ammonia molecule is less susceptible to freeze-out onto dust grains than C- or O-bearing molecules. 
Deuterated ammonia (NH$_2$D) molecules are especially good in tracing dynamically evolved starless cores (or pre-stellar cores; see e.g., \citealt{crapsi05, harju17}) although it has been recently found that \nhTd\ (as well as ammonia and other volatiles) sigificantly freezes out in the central few thousands au of prestellar cores \citep{caselli22, pineda22}.
Therefore, NH$_2$D can be used to trace highly evolved starless cores in cold molecular clouds (the critical density of \nhTdj\ transition is 6.52 $\times$ 10$^4$ cm$^{-3}$ at $\sim$10 K; \citealt{wienen21}).

On the other hand, sulfur monoxide (SO) molecule is proposed to be the most sensitive indicator of the evolutionary status of dense cores and can be used as a tracer of very young cores \citep{tafalla06}. 
\citet{frau12} showed that dense cores abundant in oxo-sulfurated (e.g., SO, SO$_2$) species are in a chemical transitional stage between the less dense cloud and the more evolved dense core. 
However, note that the molecule \so\ is well detected in shocks in star forming regions (see, e.g. \citealt{sakai14}). 

\begin{figure*}
  \centering 
  \includegraphics[width=.72\textwidth]{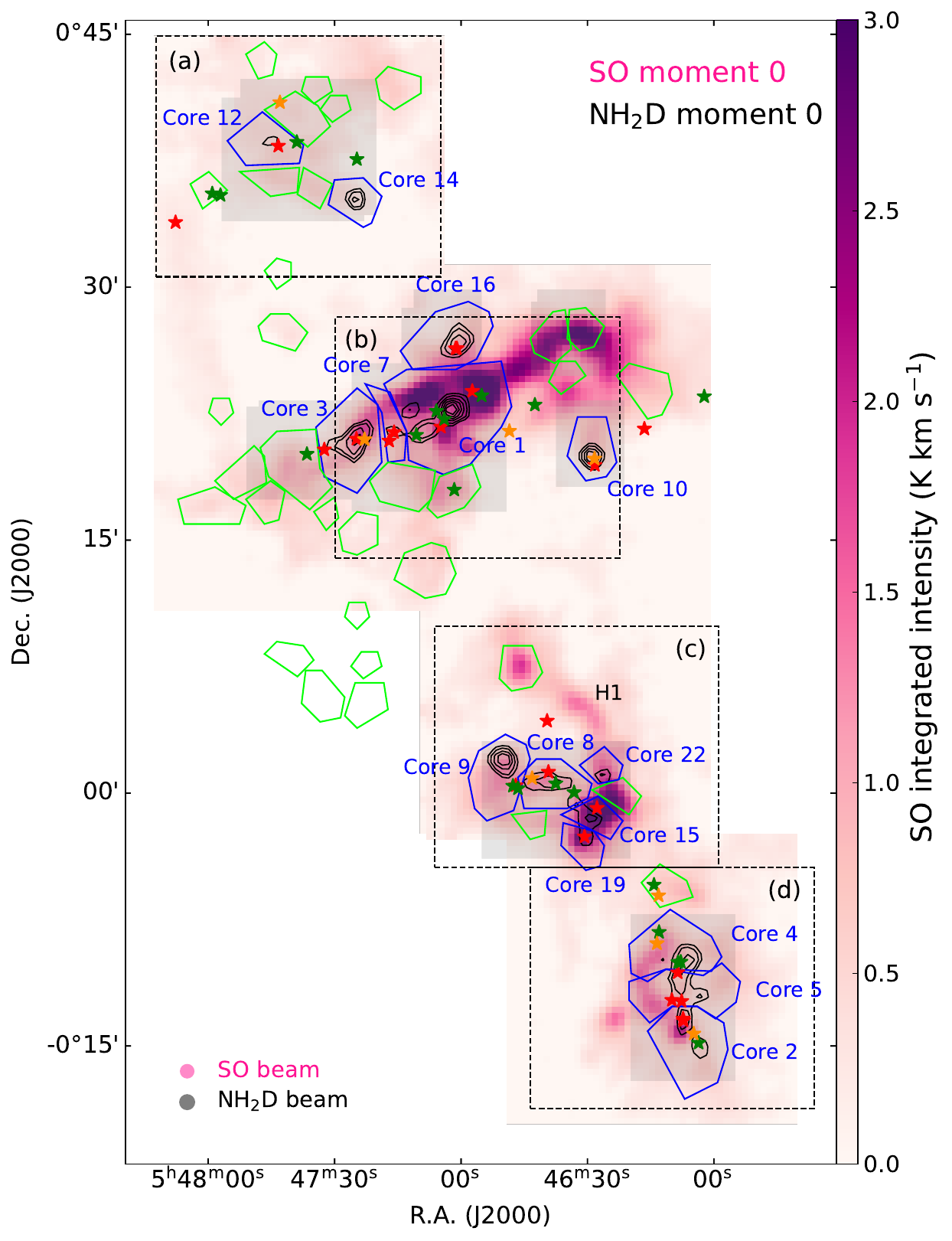}
  \caption{The distributions of SO and NH$_2$D in \nThp~dense cores.  The integrated intensities of \soj\ and \nhTdj\ are shown in magenta color tones and black contours with levels of 0.031 K \kms~$\times$ [3, 5, 8, 11, 14, 17]), respectively. 
  \nThp~dense cores are drawn in blue or green depending whether if \nhTd\ is detected or not detected. 
  Small gray shaded areas indicate the \nhTdj\ observed regions.
  The YSOs from \citet{furlan16} are shown with asterisk symbols (red for Class 0 protostars, green for Class I protostars, and orange for Flat-spectrum YSOs).  
  Core names are labeled only for \nhTd\ detected dense cores. 
  Boxes drawn in dashed line with labels (a)--(d) are the regions for close-up views in Figure~\ref{fig:h2_so_nh2d}.
\label{fig:so_nh2d}}
\end{figure*}

\begin{figure*}
  \centering 
  \gridline{\fig{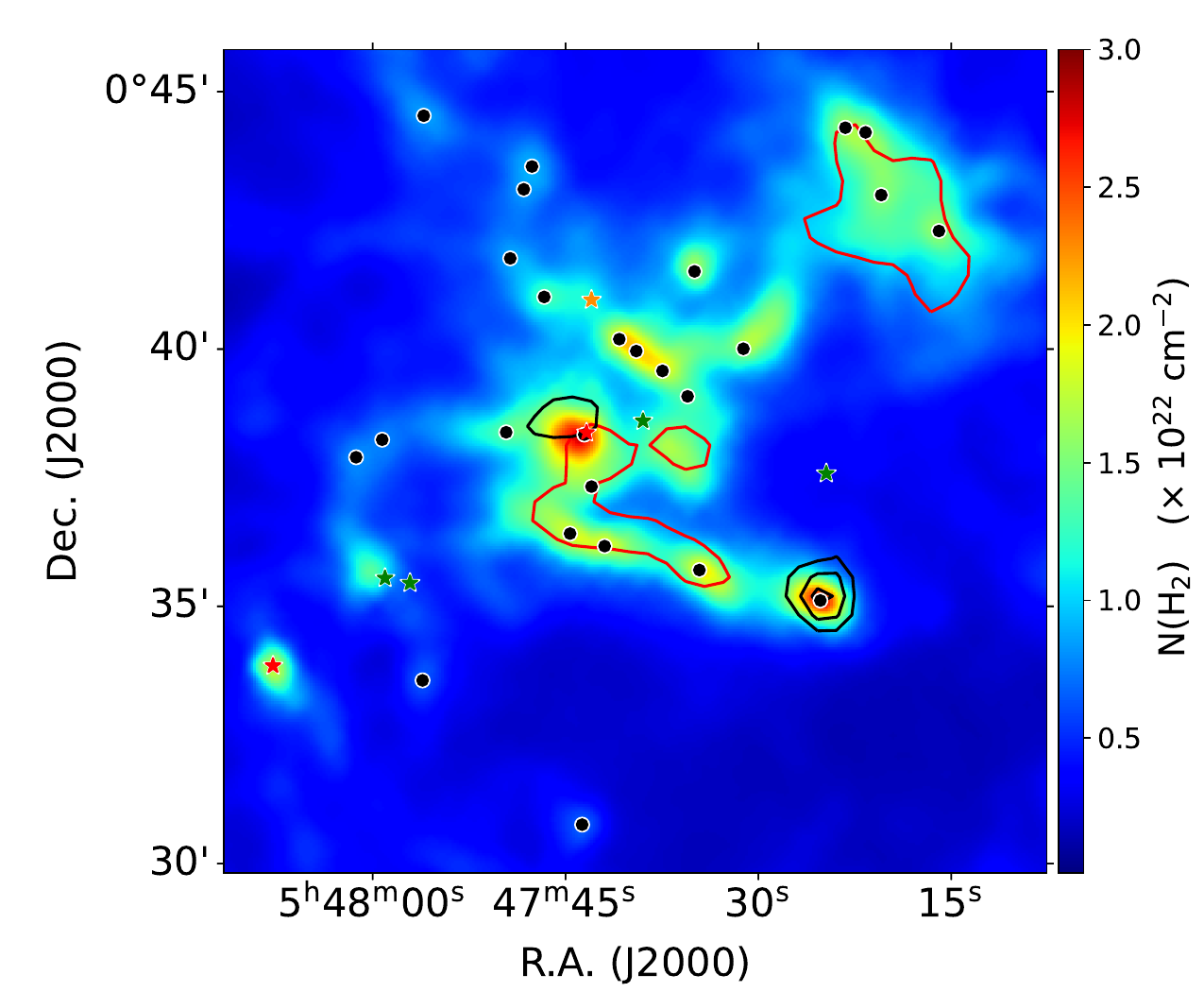}{0.5\textwidth}{(a) NGC2071-north}
          \fig{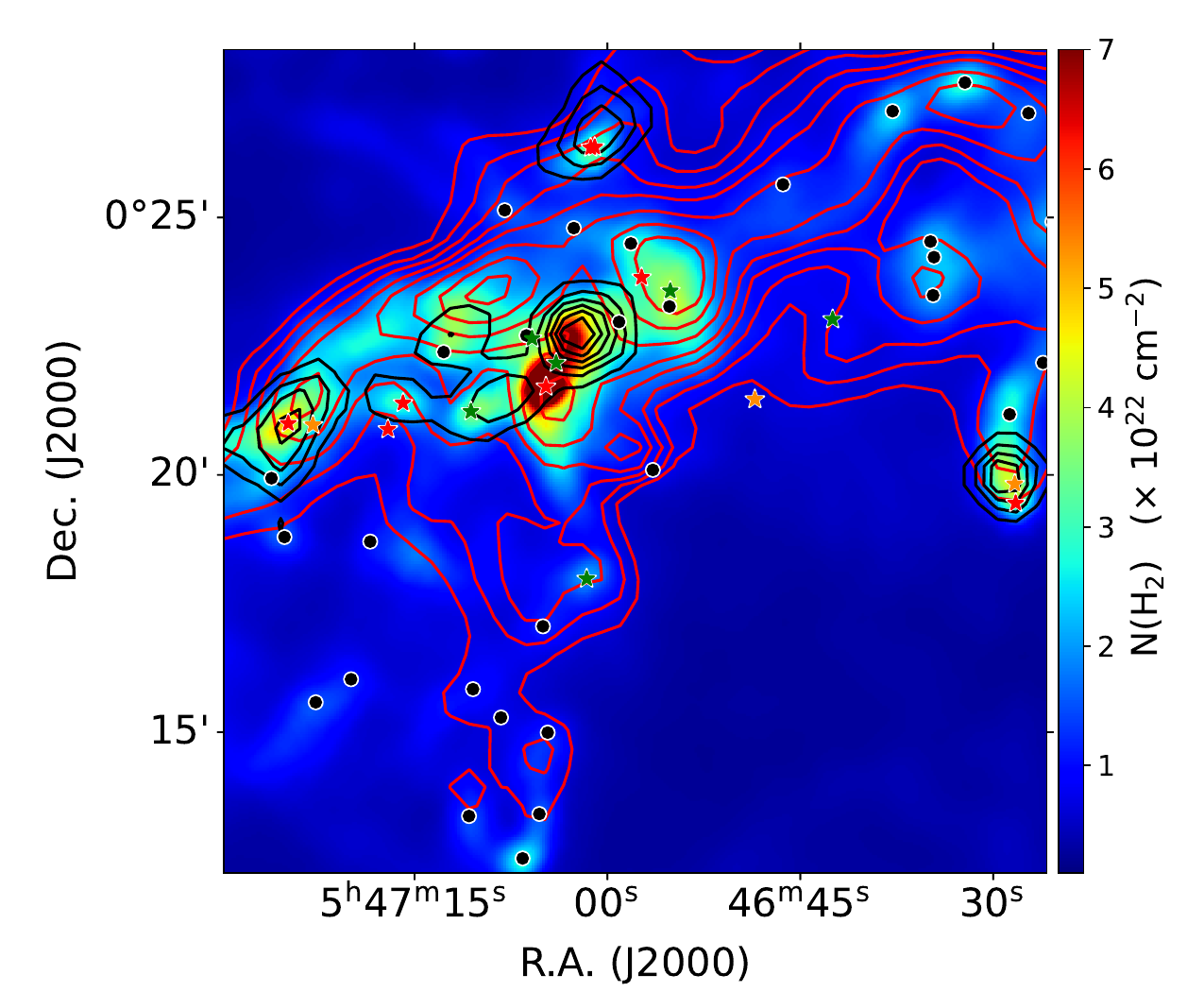}{0.5\textwidth}{(b) NGC2071}}
  \gridline{\fig{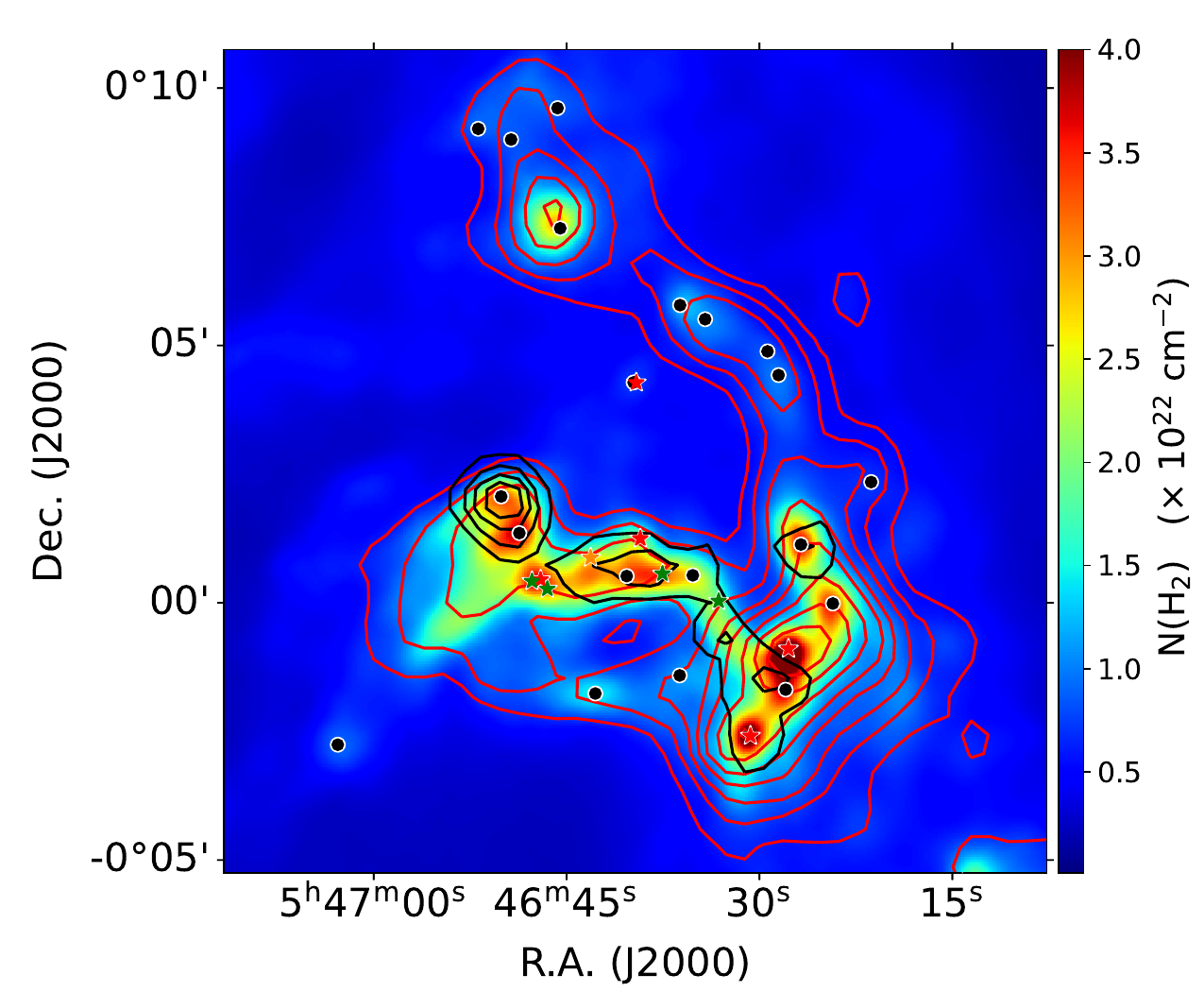}{0.5\textwidth}{(c) NGC2068}
          \fig{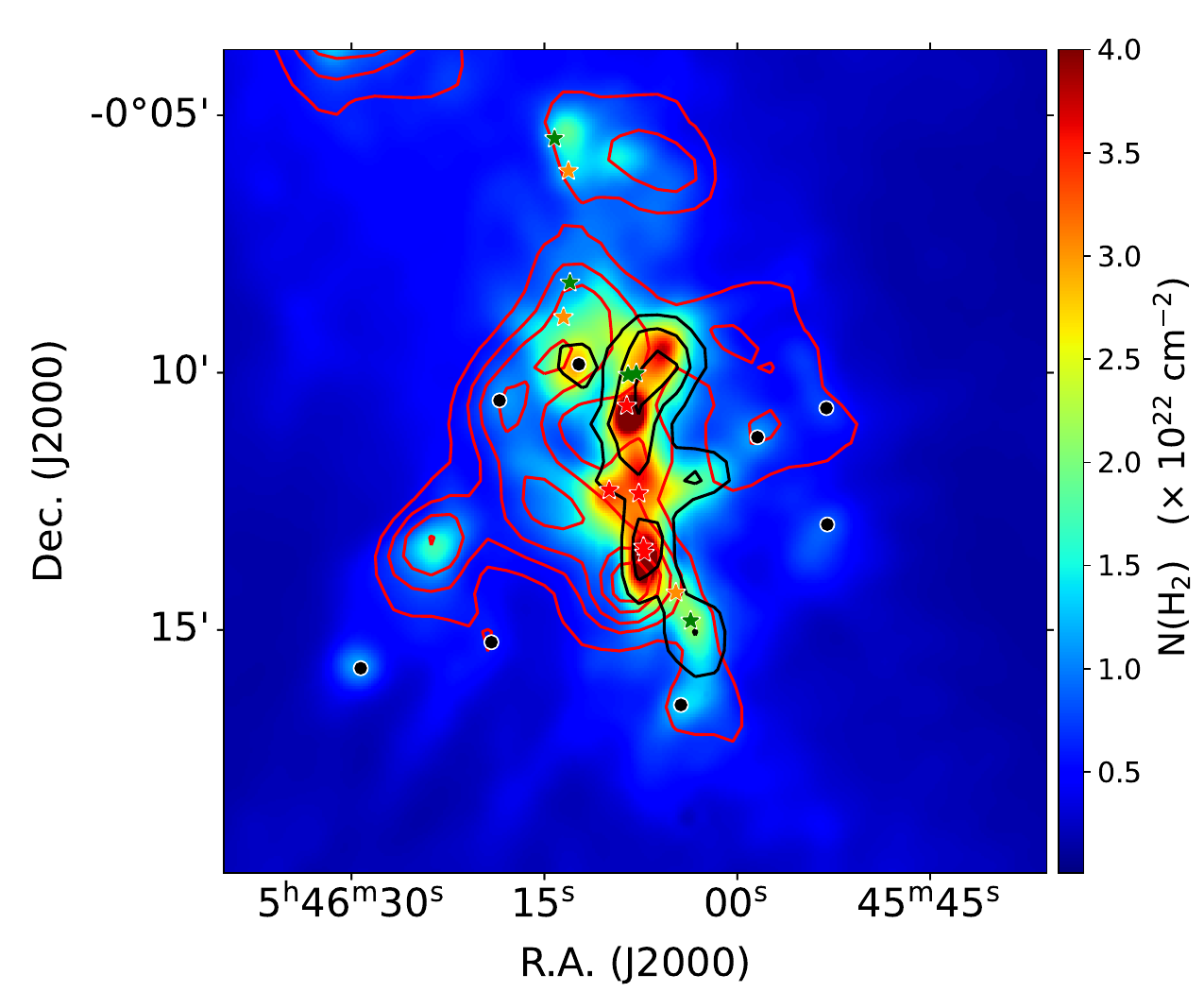}{0.5\textwidth}{(d) LBS23}}
\caption{Zoom-in distribution of SO and NH$_2$D in Orion B. The integrated intensities of \nhTdj\ and \soj\ lines are shown in black and red contours, respectively. The background color tones are for H$_2$ column density distribution \citep{arzoumanian19, konyves20}. The contour levels for \nhTd\ data are 0.031 K \kms~$\times$ [2, 5, 8, 11, 14, 17] and those for \so\ data are 0.036 K \kms~$\times$ [15, 22, 30, 40, 50, 65, 80, 95]. The H$_2$ column density level in background is indicated in color bar.  The YSOs from \citet{furlan16} are shown with asterisk symbols (red for Class 0 protostars, green for Class I protostars, and orange for flat-spectrum YSOs). Black circles are Herschel robust prestellar cores presented in  \citet{konyves20}.
\label{fig:h2_so_nh2d}}
\end{figure*}

\begin{figure*}[ht!]
\centering 
\includegraphics[width=1\textwidth]{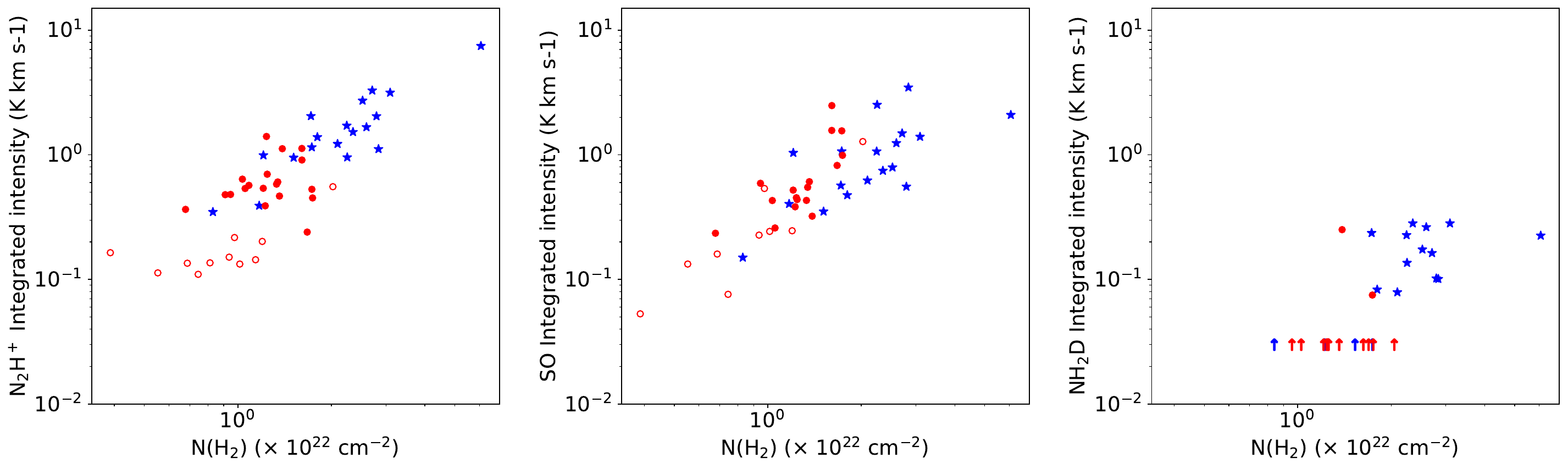}
\caption{Averaged integrated intensities of three molecular lines (\nThp, \so, and \nhTd) for dense cores as a function of H$_2$ column density measured with TRAO beam size at the \nThp\ emission peak position.
Blue stars, and filled red and open red cicles indicate gravitationally bound dense cores with YSOs, prestellar cores,  gravitationally unbound starless dense cores, respectively. 
Dense cores with (blue) and without (red) YSOs, where \nhTd\ emission is undetected in our sensitivity, are marked with arrows as upper limit (rms level at \nhTd\ integrated intensity map). 
\label{fig:chem_mom0}}
\end{figure*}

Figure~\ref{fig:so_nh2d} presents distributions of \soj\ and \nhTdj\ around \nThp\ dense cores of Orion~B. 
\citet{harju20} highlighted that the NH$_2$D~production is efficient inside the CO-depleted zone, where \nThp\ is bright. 
In fact, the \nhTdj\ emission is highly confined at the brightest part of \nThp~dense cores (see polygons and black contours in Figure~\ref{fig:so_nh2d}). 
In Figure~\ref{fig:h2_so_nh2d}, the \nhTd\ emission also shows strong correlation at the region with high H$_2$ column densities. 
On the other hand, the \so\ emission is widely distributed from the boundary to the central region of the H$_2$ column density distribution to its center.


In total, 15 dense cores were detected over 3$\sigma$ in \nhTdj. 
We call the smaller regions where \nhTdj\ was detected in the dense core as `\nhTd\ bright cores' in particular. 
Among these \nhTd, 13 \nhTd\ bright cores were found to coincide with YSOs
(Core IDs C1, C2, C3, C4, C5,   C7, C8, C9, C10, C12, C15, C16, and C19) and two \nhTd~bright cores were identified as starless cores (C14 and C22 in panels (a) and (c) of Figure~\ref{fig:so_nh2d}).
We note that the \nhTd\ bright cores are located in filaments F6, F17, F23, and F30, which are all the thermally supercritical and have YSOs (see Figure~\ref{fig:fil_vp} and Figure~\ref{fig:corenum}), indicating that the criticality of filament may have a good correlation with the chemical maturity of its embedded dense cores. 
In the virial analysis, C14 is gravitationally bound and C22 is marginally in virial equilibrium (see magenta and cyan circles in the right panel of Figure~\ref{fig:core_vp}).
According to the \nhTdj\ detection and the gravitational unstable status by the virial analysis, we suggest that C14 is the most evolved starless dense core in the Orion~B cloud and a good candidate site of star formation. 

The brightest \nThp\ core (C1) presents an anti-correlation between the \soj\ and \nhTdj\ emission 
(see the panel (b) in Figure~\ref{fig:so_nh2d}). 
Note that the region where the \nhTdj\ emission is strong tends to contain more YSOs inferring strong star formation activity. 
This could indicate that in these regions icy mantles, where NH$_2$D is expected to be mostly adsorbed during the prestellar stage of the core, have been recently sublimated, returning significant fractions of NH$_2$D into the gas phase \citep{caselli22}.
The distributions of both \soj\ and \nhTdj\ emission with multiple contour peaks in C1 show that there are several SO cores around multiple NH$_2$D cores at the central region of the \nThp\ core suggesting two possibilities. 
The first one is that there may be strong chemical differentiation in SO and NH$_2$D.
The second one is that active star formation unresolved with our current resolution may have been undergoing inside the region. 
Indeed, SO is copiously formed in shock regions (where icy mantles are sputtered) and it is used as tracer of accretion shocks \citep{sakai14}.

Different distribution between NH$_2$D and SO is also remarkably seen in the bulk cloudy region in the southernmost part of the cloud, where \nThp\ dense cores (C2, C4, and C5) are embedded. 
This region has been named as LBS23 in the CS molecular survey \citep{lada91} 
and is known to contain HH objects such as HH24, HH25, and HH26 \citep{herbig74, matthews83}. 
In Figure~\ref{fig:h2_so_nh2d}(d), the distribution of \nhTdj\ in the region shows a good correlation with H$_2$ column density, while \soj\ emission shows a distinct arc-like feature with an offset to the east side from the main distribution of \nhTdj.

We find bright SO emission in some regions where NH$_2$D and \nThp\ lines were not detected. 
For example, we find bright SO emission at the position of the fragment H1 where there is no detection of NH$_2$D and \nThp\ lines (see the panel (c) in Figure~\ref{fig:so_nh2d}) implying that it is chemically very young. 
Because of the infall signatures detected here as shown in Figure~\ref{fig:H1} and the chemical youth inferred from the bright \so\ emission, the fragment H1 seems to be still gathering material.

Moreover, in Figure~\ref{fig:h2_so_nh2d} we compare distributions of the Herschel H$_2$ column density, our \nhTd\ and \so\ emissions, YSOs \citep{furlan16}, and Herschel prestellar cores \citep{konyves20}.
Since the angular resolution of Herschel data is about three times higher, some column density enhancements can be explained by the local fragmentation, which can not be seen in our observing resolution. 

To demonstrate chemical differentiation, we present the averaged integrated intensities of dense cores in \nThpj\, and \soj, and \nhTdj\ lines as a function of H$_2$ column density in Figure~\ref{fig:chem_mom0}. The integrated intensities and H$_2$ column densities were averaged within the TRAO beam size ($\theta_{\rm B}$ $\sim$ 52\arcsec\ for \soj) at the emission peak positions.
Starless dense cores are marked with red circles. From the virial analysis, we divide them into gravitationally bound (red filled circle) and gravitationally unbound (red open circle). 
The dense cores having positional coincidence with YSOs are indicated with blue star symbols and they are all found to be gravitationally bound in virial analysis. 
The left and middle panels in Figure~\ref{fig:chem_mom0} show that \nThp\ emission and \so\ emission tends to be stronger at high H$_2$ column density regions as also seen in the emission distribution of \nThp\ and \so\ in Figure~\ref{fig:h2_so_nh2d}(a)--(c)). 

Understanding the chemical behavior in \nhTd\ is more complicated. 
Recently, \citet{caselli22} presented the first evidence that \nhTd\ is almost totally frozen within the central 2000~au of prestellar core and showed that \nhTdj\ line emission is not reproducible by the chemical models without \nhTd\ depletion at central region. 
In their study the frozen \nhTd\ preserves its chemical composition until being in protostellar stage and can be released back in the gas phase to the neighborhood of protostellar objects or shocked region.
That may explain the limited detection of \nhTd\ found in dense cores with YSOs (blue stars in Figure~\ref{fig:chem_mom0}), due to the poor angular resolution of the present observations. 
In prestellar cores (red circles and red arrows in Figure~\ref{fig:chem_mom0}), \nhTd\ intensity could indeed be too faint to be detected and the line profile could be too narrow to be resolved by our moderate spectral resolution, so that the line profile can be also affected by spectral dilution. 
As \nhTd, \so\ will be depleted onto dust grain surfaces and return back to the gas phase when the ice sublimates or to shocked regions (e.g. at the centrifugal radius, as explained by \citealt{sakai14}, or along outflow). 
However, unlike \nhTd, \so\ can also be present in the more extended and less dense material surrounding the dense cores, where sulfur is not yet heavily frozen onto dust grains (e.g. \citealt{spezzano17, laas19, lattanzi20}). 

Further detailed analysis and discussion on the chemical differentiation are out of our scope and will be given in a future study of chemical tracers at higher angular/spectral resolutions.

\subsection{Implications of the survey on the star formation}

The primary aim of TRAO-FUNS survey is to obtain the velocity distribution of low dense filaments and their dense cores for the study of their formation mechanism. Since the observed three-dimensional PPV data can trace the multiple line-of-sight components in velocity, it provides a much more comprehensive understanding on the velocity coherent structures in molecular cloud. 
As parts of the series, previous reports on L1478 in the California molecular cloud \citep{chung19} and IC~5146 \citep{chung21} successfully identified velocity coherent structures (filaments) and dense cores, and derived their physical properties.
Results on the relation between the filament's criticality and existence of dense cores for low-mass star forming molecular cloud L1478 and low- to intermediate-mass star forming cloud IC~5146 showed that the majority of dense cores are located in the supercritical filaments. 
The results shown in the high-mass star forming region Orion~B are also found to be consistent with these molecular clouds in previous studies. 
Therefore, the supercriticality of a filament is supposed to be a necessary condition for the formation of dense cores regardless of the physical environment of the entire clouds. 

The second goal of the survey is to understand the kinematic effect of the surrounding filaments on the dense core formation. Based on the nonthermal velocity dispersions obtained from \cEoj\ and \nThpj\ transitions in three series (\citealt{chung19, chung21} and this study), the observational evidences that the nonthermal velocity dispersions of dense cores are smaller than those of filaments may indicate the dissipation of turbulence during core formation, which supports the colliding scenario in the formation of dense cores in the cloud proposed by \citet{padoan01}. 

Lastly, the survey is designed to investigate the chemical properties of both filaments and dense cores, especially the chemical differentiation between them. Since the early series on L1478, IC~5146 and Orion~B focused on the identification of filaments and dense cores and the derivation of physical properties, the chemical understanding will be developed in the next series of papers. 

Table~\ref{tab:series} summarizes the representative physical properties of the three molecular clouds of TRAO-FUNS survey that have been analyzed so far. The filament widths of two sources (L1478 and Orion~B) at the distances of about 400~pc are smaller than the filament width of IC~5146 at the distance of $\sim$700 pc ($W \sim$ 0.2--0.7 pc). 
That is a similar result to what is recently reported on the effect of distance on the filament widths \citep{panopoulou22}.
They show a trend that a mean width of filaments depends on not only the distance of the molecular clouds but also the angular resolution. 
Note that the widths in TRAO-FUNS studies are measured from the Gaussian FWHM of mean radial profile and not corrected the beam convolution effect. The beam size is also larger than Herschel data used in \citet{panopoulou22}. 
Moreover, \citet{andre22} revisit the issue and argued that the typical width of 0.1~pc in filamentary structures of nearby molecular clouds is a true common scale. 
We will carefully discuss this in a future study, after analyzing more targets observed on TRAO-FUNS survey using the same measurement to estimate the filament width.

\begin{deluxetable}{ccccc}
\tablenum{2}
\tablecaption{A comparison of properties of molecular clouds observed by the TRAO-FUNS project. 
\label{tab:series}}
\tablewidth{0pt}
\tabletypesize{\small}
\tablehead{
\colhead{} & \colhead{Property} & \colhead{L1478} &
\colhead {IC~5146} & \colhead {Orion~B} 
}
\startdata
    (1) & Distance   &   450 &   600--800    &	423 \\
    (2) & Width & 0.08--0.19 & 0.18--0.67 & 0.05--0.26 \\
    (3) & Fil\# & 10 & 14 & 30 \\
    (4) & Core\# & 8 & 22 & 48 \\
    (5) & YSO\# & 3 & 40 & 51 \\
    (6) & $<\sigma_{tot}>$ & 0.18--0.31 & 0.28--0.43 & 0.31--0.54 \\
    (7) & $<$N(H$_2$)$>$ & 7--40 & 22--88 & 28--145 
\enddata
\tablecomments{
Refer to \citet{chung19, chung21} for the details on scientific results on L1478 and IC~5146, respectively. 
(1) Distance to the molecular cloud adopted in units of pc. 
(2) Measured widths of filaments in units of pc. 
(3) Number of identified \cEoj\ filaments. 
(4) Number of identified \nThpj\ dense cores. 
(5) Number of YSOs associated with the filaments. 
(6) Total velocity dispersion of filaments measured from \cEoj\ in units of \kms. 
(7) Average H$_2$ column densities of the identified filaments in units of 10$^{20}$~cm$^{-2}$.}
\end{deluxetable}

Table~\ref{tab:series} indicates that as the mean H$_2$ column density of filaments in the molecular cloud increases (in the order of L1478, IC~5146, and Orion B), the numbers of identified filaments, \nThpj\ dense cores, and YSOs associated with filaments tend to increase. 
Similar results are reported by \citet{hacar18} who showed that active high-mass star-forming molecular clouds are higher in H$_2$ number density and have more supercritical fibers than less active low-mass star-forming molecular clouds. 
Results in those works also indicate that the star-forming activity of a cloud is closely related to the number of filaments and dense cores and essentially to its H$_2$ density.
Moreover, the total velocity dispersion measured from decomposed \cEoj\ spectrum also increases as the mean H$_2$ column density increases being consistent to the relation between the total velocity dispersion and column density (refer to Figure 6 in \citealt{arzoumanian13}).

The forthcoming studies on molecular clouds with various physical environments observed by our TRAO-FUNS survey (e.g., Polaris flare, Taurus, Perseus, Serpens, Aquila) will be helpful to verify the tentative conclusion and to understand the environmental effect on the properties of filaments. Further studies for these clouds using our chemical tracers (e.g., \nhTdj, \soj, \nThpj) will expand our understanding on evolutionary stages of filaments and dense cores in various star forming regions. 

\section{Summary} \label{sec:sum}

This paper presents results of mapping observations of the NGC~2068 and NGC~2071 regions of the Orion~B molecular cloud in multiple molecular lines performed with the TRAO 14-m telescope to study the filamentary structures in the clouds and their roles in the formation of dense cores.  Our results are summarized as follows:

\begin{enumerate}

\item{Many of the observed \cEoj\ spectra in the clouds are shown to contain multiple velocity components and decomposed into an individual velocity component with a Gaussian fit. 
This allowed us to identify filaments with similar velocities, so called velocity-coherent filaments, using the  Friends-of-Friends algorithm. 
A total of 30 filaments were extracted, and their physical properties such as length, width, column density, and mass were measured. 
The longest filament is found to have its length of $\sim$6.5 pc and the median length for the entire sample of the filaments is 0.8 pc. 
The median value of the filament widths is estimated to be 0.1 pc with a range of 0.05 to 0.26 pc.
The H$_2$ column densities of the filaments range from 2.8 $\times$ 10$^{21}$ cm$^{-2}$ to 1.4 $\times$ 10$^{22}$ cm$^{-2}$. The filament mass is calculated to be in the range of 4 $M_{\odot}$ and 842 $M_{\odot}$ with a median of 38 $M_{\odot}$. }  

\item{We identified 48 \nThp\ dense cores using the $\texttt{FellWalker}$ algorithm, and estimated the core masses and virial masses to examine the dynamical state. 
From the virial analysis for these dense cores, all dense cores with YSOs and half of starless dense cores are found to be in gravitational bound status. }

\item{The distributions of the nonthermal velocity dispersions extracted from the \nThpj\ and \cEoj\ show that the dense cores are found to be mostly in transonic/subsonic motions while their natal filaments are in supersonic nonthermal velocity dispersions (excepting parts of the region in F17b). This may indicate that gas turbulent motions in the filaments around dense cores have been dissipated at the core scale to help to form the dense cores in their surrounding filaments. }

\item{The filaments with \nThp\ dense core are mostly found to be gravitationally bound, thermally supercritical, and thus under environments in which dense cores can be formed well. 
The filaments with larger ${M}_\mathrm{line}/{M}_\mathrm{line}^\mathrm{th,crit}$ tend to have more dense cores the majority of which tend to contain YSOs.  
Therefore, it is well understood that the criticality of the filament can be a key condition for its fragmentation and the formation of dense cores/YSOs in the filament. }

\item{From a case study for a filament F23, a filament might be forming dense cores in a process of the longitudinal fragmentation, in the sense that its mass is far higher than its thermal critical line mass, and its fragments are quasi-periodically located along the filament with an average of 0.39~pc which is comparable to its theoretical value (four times of the filament width: 0.52~pc) predicted in the gravitational collapse in an isothermal filament considering uncertain factors in its calculation such as inclination of the filament.}

\item{We identified infall signatures in at least nine \nThp\ dense cores using the observations of \hcopj\ and \hTcopj, finding that their parent filaments are all in the gravitationally critical status. This may imply that a gravitationally unstable environment in the filaments should help to enhance the evolution of the dense cores to finally form stars there.   
We find a fragment H1, which is not detected in \nThp\ but showing infall signatures over the fragment. It is believed to be in very early condensing process before the stage of a prestellar core, but not yet enough condensed to be detected with \nThpj\ line, again making us to believe that ``fertile'' environment in the filament is helping to produce a future prestellar core.}

\item{We examined how molecular tracers (\nhTd\ and \so) are differentiated in the \nThp\ dense cores and how this chemical differentiation is related to the physical status of their parent filaments.  
We find that the \nhTd\ emission is strongly correlated with the H$_2$ column density and is usually well detected in the central regions of the \nThp\ cores. 
We find that 13 out of 15 \nhTd\ bright cores are forming YSOs. 
These may indicate that \nhTd\ traces the most evolved status of the dense cores while \so\ does the least evolved status of the dense cores.  
It is found that these \nhTd\ cores are all located in the thermally supercritical filaments, indicating that the criticality of filaments may play a key role on the chemical maturity of dense cores. 
}

\end{enumerate}

\begin{acknowledgments}
The authors are grateful to the anonymous referee for the detailed and constructive comments, which helped to improve the paper. This work was supported by the Basic Science Research Program through the National Research Foundation of Korea (NRF) funded by the Ministry of Education, Science and Technology (NRF-2019R1A2C1010851), and by the Korea Astronomy and Space Science Institute grant funded by the Korea government  (MSIT) (Project No. 2023-1-84000). H. Y. is supported by Basic Science Research Program through the National Research Foundation of Korea (NRF) funded by the Ministry of Education (NRF-2021R1A6A3A01087238).  E.J.C. is supported by Basic Science Research Program through the National Research Foundation of Korea (NRF) funded by the Ministry of Education (NRF-2022R1I1A1A01053862). S.K. is supported by National Research Council of Science \& Technology (NST) -- Korea Astronomy and Space Science Institute (KASI) Postdoctoral Research Fellowship for Young Scientists at KASI in South Korea.
\end{acknowledgments}


\bibliography{ms_orionb}{}
\bibliographystyle{aasjournal}

\begin{deluxetable*}{cCCCDCCCCCCCCCC}
\tablecaption{Physical properties of the filaments in Orion~B. 
\label{tab:fil}}
\tablenum{3}
\tablewidth{500pt}
\tabletypesize{\scriptsize}
\tablehead{
\colhead{Fil. ID} &
\colhead{$l$} & 
\colhead{$W$} & 
\colhead{$AR$} & 
\multicolumn2c{$\langle V_\mathrm{LSR}\rangle_\mathrm{min}^\mathrm{max}$} &
\colhead{$\langle T_\mathrm{dust}\rangle$} &
\colhead{$\sigma_\mathrm{tot}$} & 
\colhead{$M$} &
\colhead{$M_\mathrm{line}$} & 
\colhead{$M_\mathrm{line}^\mathrm{tot,crit}$} & 
\colhead{$M_\mathrm{line}^\mathrm{th,crit}$} & 
\colhead{$N_\mathrm{H_2}^\mathrm{Herschel}$} & 
\colhead{$N_\mathrm{H_2}^\mathrm{C^{18}O}$} & 
\colhead{Core} &
\colhead{YSO}
} 
\decimalcolnumbers
\startdata
F1   &   1.8   & 0.09   & 19.3  & 8.9_{8.8}^{9.3}    &  14.7   & 0.35 \pm 0.03   &    45 \pm 6    &    26 \pm 4     &  57 \pm 10      & 25 \pm 11    & 4.2   & 9.6    & 0   & 0  \\
F2   &   0.8   & 0.18   & 4.8   & 8.7_{8.5}^{9.0}    &  14.3   & 0.38 \pm 0.04   &    27 \pm 5    &    40 \pm 6     &  67 \pm 14      & 23 \pm 14    & 4.0   & 9.1    & 0   & 0  \\
F3   &   2.0   & 0.15   & 13.5  & 9.5_{9.0}^{9.8}    &  13.9   & 0.38 \pm 0.03   &   106 \pm 13   &    66 \pm 7     &  67 \pm 11      & 23 \pm 10    & 6.2   & 14.1   & 3   & 1  \\
F4   &   0.9   & 0.26   & 3.4   & 8.5_{8.1}^{8.9}    &  13.9   & 0.39 \pm 0.03   &    79 \pm 9    &\phn95 \pm 11    &  71 \pm 11      & 23 \pm 10    & 6.1   & 13.8   & 1   & 0  \\
F5   &   0.6   & 0.07   & 7.8   & 8.7_{8.5}^{8.9}    &  14.4   & 0.31 \pm 0.03   &    10 \pm 2    &    15 \pm 3     &  45 \pm 9\phn   & 23 \pm 8\phn & 2.8   & 6.3    & 0   & 0  \\
F6   &   1.1   & 0.16   & 6.7   & 8.8_{8.4}^{9.2}    &  13.3   & 0.43 \pm 0.03   &    86 \pm 7    &    64 \pm 7     &  86 \pm 12      & 23 \pm 12    & 7.1   & 16.2   & 4   & 2  \\
F7   &   0.8   & 0.11   & 7.3   & 9.8_{9.6}^{10.1}   &  15.1   & 0.35 \pm 0.03   &    19 \pm 5    &    40 \pm 6     &  57 \pm 10      & 25 \pm 11    & 3.6   & 8.2    & 0   & 0  \\
F8   &   3.5   & 0.12   & 28.5  & 9.7_{9.0}^{10.6}   &  14.3   & 0.37 \pm 0.03   &   185 \pm 25   &    56 \pm 7     &  64 \pm 10      & 23 \pm 10    & 4.7   & 10.8   & 2   & 5  \\
F9   &   0.4   & 0.07   & 5.8   & 9.5_{9.3}^{9.6}    &  14.3   & 0.47 \pm 0.05   &    10 \pm 2    &    27 \pm 4     & 103 \pm 22\phn  & 23 \pm 20    & 4.9   & 11.2   & 1   & 0  \\
F10  &   0.3   & 0.07   & 4.9   & 9.1_{9.0}^{9.2}    &  14.2   & 0.33 \pm 0.03   & \phn6 \pm 1    &    14 \pm 2     &  51 \pm 9\phn   & 23 \pm 8\phn & 3.4   & 7.7    & 0   & 0  \\
F11  &   0.3   & 0.07   & 4.6   & 10.1_{9.9}^{10.3}  &  14.3   & 0.46 \pm 0.03   &    16 \pm 1    &    49 \pm 5     &  98 \pm 13      & 23 \pm 12    & 8.6   & 19.6   & 0   & 2  \\
F12  &   0.7   & 0.07   & 9.7   & 8.6_{8.4}^{8.9}    &  16.2   & 0.47 \pm 0.04   &    36 \pm 3    &    41 \pm 5     & 103 \pm 17\phn  & 27 \pm 18    & 7.5   & 17.0   & 0   & 2  \\
F13  &   0.8   & 0.10   & 8.4   & 9.6_{9.4}^{9.9}    &  20.1   & 0.54 \pm 0.03   &    62 \pm 5    &    97 \pm 7     & 136 \pm 15\phn  & 31 \pm 15    & 12.9  & 29.4   & 1   & 2  \\
F14  &   0.4   & 0.07   & 5.8   & 9.8_{9.6}^{9.9}    &  15.7   & 0.50 \pm 0.04   & \phn6 \pm 1    &    28 \pm 3     & 116 \pm 19\phn  & 25 \pm 19    & 6.6   & 15.1   & 0   & 1  \\
F15  &   0.3   & 0.06   & 4.3   & 8.9_{8.8}^{9.1}    &  12.8   & 0.39 \pm 0.02   & \phn9 \pm 1    &    41 \pm 4     &  71 \pm 7\phn   & 21 \pm 8\phn & 7.6   & 17.3   & 0   & 0  \\
F16  &   0.3   & 0.05   & 6.6   & 8.8_{8.7}^{8.9}    &  13.6   & 0.48 \pm 0.03   &    11 \pm 1    &    44 \pm 4     & 107 \pm 13\phn  & 23 \pm 14    & 8.2   & 18.6   & 0   & 0  \\
F17a &   1.4   & 0.21   & 6.8   & 9.5_{9.1}^{9.9}    &  14.8   & 0.34 \pm 0.07   &\phn84 \pm 13   &\phn77 \pm 10    &  54 \pm 22      & 25 \pm 24    & 4.8   & 11.0   & 0   & 0  \\
F17b &   4.6   & 0.14   & 32.6  & 9.3_{8.6}^{10.1}   &  14.6   & 0.48 \pm 0.06   &   565 \pm 42   &   134 \pm 11    & 107 \pm 27\phn  & 25 \pm 28    & 10.2  & 23.3   & 10  & 10  \\
F17c &   1.4   & 0.15   & 9.2   & 9.4_{8.9}^{10.1}   &  13.9   & 0.41 \pm 0.05   &   132 \pm 10   &   175 \pm 9\phn &  78 \pm 19      & 23 \pm 20    & 14.5  & 33.0   & 2   & 3  \\
F18  &   1.4   & 0.15   & 8.9   & 8.2_{7.8}^{8.6}    &  13.9   & 0.42 \pm 0.03   &\phn90 \pm 11   &   112 \pm 9\phn &  82 \pm 12      & 23 \pm 10    & 8.9   & 20.2   & 1   & 3  \\
F19  &   1.0   & 0.17   & 6.3   & 8.8_{8.5}^{9.0}    &  14.6   & 0.45 \pm 0.02   &    66 \pm 8    &   128 \pm 8\phn &  94 \pm 8\phn   & 25 \pm 11    & 11.2  & 25.5   & 0   & 1  \\
F20  &   1.4   & 0.10   & 13.7  & 9.6_{9.1}^{9.8}    &  14.7   & 0.36 \pm 0.03   &    49 \pm 6    &    43 \pm 5     &  60 \pm 10      & 25 \pm 9\phn & 5.7   & 13.0   & 0   & 0  \\
F21  &   0.5   & 0.08   & 6.9   & 10.6_{10.3}^{10.8} &  14.9   & 0.36 \pm 0.04   &    13 \pm 2    &    21 \pm 3     &  60 \pm 13      & 25 \pm 13    & 3.8   & 8.7    & 0   & 0  \\
F22  &   1.0   & 0.12   & 8.1   & 11.3_{11.0}^{11.6} &  14.6   & 0.37 \pm 0.03   &    43 \pm 4    &    46 \pm 5     &  64 \pm 10      & 25 \pm 11    & 5.9   & 13.4   & 2   & 0  \\
F23  &   4.7   & 0.13   & 37.0  & 10.6_{9.7}^{11.3}  &  18.2   & 0.48 \pm 0.03   &   352 \pm 34   &   104 \pm 8\phn & 107 \pm 13\phn  & 29 \pm 14    & 9.8   & 22.3   & 8   & 2  \\
F24  &   0.3   & 0.07   & 3.7   & 11.1_{11.1}^{11.2} &  22.3   & 0.34 \pm 0.02   & \phn3 \pm 1    &    16 \pm 2     &  54 \pm 6\phn   & 36 \pm 8\phn & 3.7   & 8.4    & 0   & 0  \\
F25  &   0.3   & 0.06   & 5.3   & 9.2_{8.9}^{9.5}    &  16.5   & 0.43 \pm 0.04   & \phn8 \pm 1    &    27 \pm 4     &  86 \pm 16      & 27 \pm 18    & 4.5   & 10.2   & 0   & 0  \\
F26  &   0.3   & 0.07   & 4.8   & 9.3_{9.1}^{9.5}    &  21.1   & 0.42 \pm 0.03   & \phn6 \pm 1    &    22 \pm 3     &  82 \pm 12      & 34 \pm 13    & 5.5   & 12.5   & 0   & 0  \\
F27  &   0.6   & 0.13   & 4.9   & 9.7_{9.2}^{10.3}   &  18.8   & 0.51 \pm 0.04   &    33 \pm 4    &    70 \pm 7     & 121 \pm 19\phn  & 31 \pm 19    & 8.0   & 18.3   & 0   & 1  \\
F28  &   0.2   & 0.06   & 3.1   & 10.8_{10.7}^{10.9} &  14.3   & 0.33 \pm 0.03   & \phn3 \pm 1    &    21 \pm 4     &  51 \pm 9\phn   & 23 \pm 10    & 3.1   & 7.1    & 0   & 0  \\
F29  &   0.5   & 0.10   & 53.8  & 11.0_{10.7}^{11.3} &  14.1   & 0.50 \pm 0.04   &    19 \pm 3    &    49 \pm 5     & 116 \pm 19\phn  & 23 \pm 18    & 6.8   & 15.6   & 0   & 0  \\
F30  &   6.5   & 0.24   & 26.7  & 10.3_{9.5}^{11.3}  &  15.5   & 0.42 \pm 0.03   &   700 \pm 67   &   130 \pm 12    &  82 \pm 12      & 25 \pm 11    & 7.6   & 17.2   & 8   & 23  \\
\enddata
\tablecomments{
Column 1: Fillament ID.
Column 2 and Column 3: Filament length and width in units of pc. 
Column 4: Aspect ratio (AR) defined as the ratio between the length and the width of the filament.  
Column 5:  The average LSR velocity of the filament with its maximum and minimum values in units of \kms. 
Column 6:  Averaged dust temperature of the filament in units of K.  
Column 7: The average total velocity dispersion (estimated with Equation~\ref{eq:sigtot}) of the filament  in units of \kms.  
Column 8: Mass of the filament derived from the \cEoj~emission in units of Solar mass. 
Column 9: Line mass of the filament (calculated by $l/W$) in units of Solar mass per pc.
Column 10: The total critical line mass of the filament in units of Solar mass per pc, which corresponds to the virial mass of filament per unit length \citep{fiege00}. 
Column 11: The thermal critical line mass of the filament in units of Solar mass per pc. 
Column 12 and Column 13: The average H$_2$ and \cEo\ column densities in units of 10$^{21}$~cm$^{-2}$ and 10$^{14}$~cm$^{-2}$, respectively.  
Column 14: The number of \nThp\ dense cores whose positions and LSR velocities coincide with those of the filaments. 
Column 15: The number of young stellar objects whose positions coincide with the filaments \citep{furlan16}.
}
\end{deluxetable*}

\newpage

\begin{deluxetable*}{ccccCcccccc}
\tablecaption{Physical properties of the dense cores in Orion~B. \label{tab:core}}
\tablewidth{500pt}
\tabletypesize{\scriptsize}
\tablenum{4}
\tablehead{
\colhead{Core ID} &
\colhead{R.A. (J2000)} &
\colhead{Dec. (J2000)}& 
\colhead{$R_\mathrm{core}$} & 
\colhead{$\langle V_\mathrm{LSR}\rangle_\mathrm{min}^\mathrm{max}$} &
\colhead{$\langle T_\mathrm{dust} \rangle$} & 
\colhead{$\sigma_\mathrm{tot}$} &
\colhead{$M_\mathrm{core}$} & 
\colhead{$M_\mathrm{vir}$} & 
\colhead{$\alpha_\mathrm{vir}^\mathrm{core}$} &
\colhead{Fil. ID}
}
\decimalcolnumbers
\startdata
C1   &  05:47:03.40 &   +00:21:59.98  &  0.13  &   9.4 _{ 8.7 }^{ 9.8 }   &  15.5  &  0.56$\pm$0.03   &     246.6$\pm$96.6  &     33.1$\pm$3.6  &  0.13$\pm$0.05  &  17 \\
C2   &  05:46:07.66 & $-$00:13:33.97  &  0.06  &  10.3 _{ 10.1 }^{ 10.4 } &  13.6  &  0.28$\pm$0.01   &  \phn91.0$\pm$35.7  &  \phn5.4$\pm$0.4  &  0.06$\pm$0.02  &  30 \\
C3   &  05:47:25.40 &   +00:20:53.98  &  0.08  &   9.3 _{ 9.0 }^{ 9.8 }   &  15.3  &  0.42$\pm$0.03   &     101.2$\pm$39.7  &     13.2$\pm$1.9  &  0.13$\pm$0.05  &  17 \\
C4   &  05:46:07.66 & $-$00:09:31.98  &  0.03  &  10.1 _{ 9.8 }^{ 10.6 }  &  14.0  &  0.38$\pm$0.02   &  \phn63.5$\pm$25.0  &  \phn3.6$\pm$0.4  &  0.06$\pm$0.02  &  30 \\
C5   &  05:46:07.66 & $-$00:12:05.97  &  0.04  &  10.0 _{ 9.7 }^{ 10.3 }  &  13.5  &  0.37$\pm$0.02   &  \phn81.4$\pm$31.9  &  \phn5.0$\pm$0.5  &  0.06$\pm$0.03  &  30 \\
C6   &  05:47:37.13 &   +00:20:09.97  &  0.03  &   9.2 _{ 9.0 }^{ 9.4 }   &  14.7  &  0.28$\pm$0.02   &  \phn54.2$\pm$21.3  &  \phn2.9$\pm$0.4  &  0.05$\pm$0.02  &  17 \\
C7   &  05:47:16.60 &   +00:21:37.98  &  0.09  &   9.3 _{ 9.1 }^{ 9.8 }   &  15.3  &  0.42$\pm$0.03   &  \phn34.8$\pm$13.6  &     14.5$\pm$2.1  &  0.42$\pm$0.17  &  17 \\
C8   &  05:46:38.46 &   +00:00:44.00  &  0.10  &  10.4 _{ 10.0 }^{ 10.8 } &  18.7  &  0.40$\pm$0.01   &  \phn31.3$\pm$12.3  &     15.3$\pm$0.8  &  0.49$\pm$0.19  &  30 \\
C9   &  05:46:48.73 &   +00:01:28.00  &  0.10  &  10.9 _{ 10.7 }^{ 11.1 } &  19.9  &  0.25$\pm$0.01   &  \phn37.0$\pm$14.6  &  \phn8.5$\pm$0.7  &  0.23$\pm$0.09  &  30 \\
C10  &  05:46:28.20 &   +00:20:09.99  &  0.08  &   9.5 _{ 9.4 }^{ 9.6 }   &  14.0  &  0.18$\pm$0.01   &      24.5$\pm$9.7   &  \phn4.5$\pm$0.5  &  0.18$\pm$0.08  &  17 \\
C11  &  05:47:13.66 &   +00:18:19.98  &  0.09  &  10.2 _{ 10.1 }^{ 10.4 } &  20.5  &  0.31$\pm$0.02   &      23.7$\pm$9.3   &  \phn9.8$\pm$1.3  &  0.41$\pm$0.17  &  23 \\
C12  &  05:47:44.46 &   +00:38:29.91  &  0.09  &   8.9 _{ 8.6 }^{ 9.7 }   &  13.4  &  0.37$\pm$0.02   &  \phn26.0$\pm$10.3  &     11.4$\pm$1.2  &  0.44$\pm$0.18  &  6  \\
C13  &  05:46:45.80 &   +00:07:19.99  &  0.06  &  10.6 _{ 10.6 }^{ 10.7 } &  22.4  &  0.21$\pm$0.01   &   \phn9.4$\pm$3.8   &  \phn5.0$\pm$0.5  &  0.53$\pm$0.22  &  23 \\
C14  &  05:47:25.40 &   +00:35:11.94  &  0.07  &   9.0 _{ 8.9 }^{ 9.1 }   &  13.8  &  0.17$\pm$0.01   &      13.8$\pm$5.5   &  \phn3.8$\pm$0.4  &  0.27$\pm$0.11  &  6  \\
C15  &  05:46:26.73 & $-$00:01:05.99  &  0.08  &  10.4 _{ 10.0 }^{ 10.7 } &  15.6  &  0.59$\pm$0.04   &      14.2$\pm$5.6   &     21.2$\pm$2.9  &  1.50$\pm$0.62  &  30 \\
C16  &  05:47:01.93 &   +00:26:23.98  &  0.05  &   9.8 _{ 9.6 }^{ 9.9 }   &  14.3  &  0.21$\pm$0.01   &  \phn26.6$\pm$10.7  &  \phn2.8$\pm$0.3  &  0.11$\pm$0.04  &  17 \\
C17  &  05:47:38.60 &   +00:39:57.91  &  0.08  &   9.5 _{ 9.3 }^{ 9.6 }   &  13.4  &  0.26$\pm$0.02   &      12.4$\pm$5.0   &  \phn5.8$\pm$0.9  &  0.47$\pm$0.20  &  -  \\
C18  &  05:47:01.93 &   +00:17:57.99  &  0.07  &   9.6 _{ 9.4 }^{ 9.7 }   &  21.1  &  0.29$\pm$0.03   &   \phn9.8$\pm$3.9   &  \phn7.7$\pm$1.6  &  0.79$\pm$0.36  &  13 \\
C19  &  05:46:31.13 & $-$00:02:33.99  &  0.07  &  10.3 _{ 10.2 }^{ 10.6 } &  14.9  &  0.28$\pm$0.02   &      12.2$\pm$4.8   &  \phn6.0$\pm$0.8  &  0.49$\pm$0.21  &  30 \\
C20  &  05:46:23.79 &   +00:00:00.00  &  0.06  &  10.3 _{ 9.5 }^{ 11.0 }  &  20.1  &  0.82$\pm$0.05   &   \phn4.8$\pm$1.9   &     31.5$\pm$3.8  &  6.58$\pm$2.77  &  -  \\
C21  &  05:46:37.00 &   +00:27:07.99  &  0.07  &   9.1 _{ 8.8 }^{ 9.2 }   &  12.9  &  0.28$\pm$0.04   &      12.7$\pm$5.1   &  \phn5.6$\pm$1.6  &  0.44$\pm$0.22  &  17 \\
C22  &  05:46:26.73 &   +00:01:06.00  &  0.06  &  11.1 _{ 11.1 }^{ 11.2 } &  17.5  &  0.24$\pm$0.02   &   \phn3.5$\pm$1.4   &  \phn4.6$\pm$0.8  &  1.33$\pm$0.59  &  23 \\
C23  &  05:46:32.60 &   +00:27:29.99  &  0.06  &   8.9 _{ 8.8 }^{ 9.0 }   &  13.0  &  0.31$\pm$0.03   &   \phn5.4$\pm$2.2   &  \phn5.5$\pm$1.1  &  1.02$\pm$0.46  &  17 \\
C24  &  05:48:00.59 &   +00:16:51.97  &  0.07  &   8.6 _{ 8.5 }^{ 8.8 }   &  13.6  &  0.30$\pm$0.03   &   \phn8.0$\pm$3.3   &  \phn6.5$\pm$1.3  &  0.81$\pm$0.37  &  -  \\
C25  &  05:47:47.39 &   +00:19:25.97  &  0.08  &   9.0 _{ 8.9 }^{ 9.2 }   &  13.6  &  0.28$\pm$0.03   &      11.7$\pm$4.6   &  \phn7.0$\pm$1.5  &  0.59$\pm$0.27  &  17 \\
C26  &  05:47:20.99 &   +00:05:29.99  &  0.06  &  10.9 _{ 10.8 }^{ 10.9 } &  15.0  &  0.21$\pm$0.02   &   \phn6.0$\pm$2.5   &  \phn4.1$\pm$0.8  &  0.68$\pm$0.31  &  23 \\
C27  &  05:47:06.33 &   +00:12:27.99  &  0.08  &  10.5 _{ 10.4 }^{ 10.6 } &  17.1  &  0.29$\pm$0.02   &   \phn8.7$\pm$3.6   &  \phn8.0$\pm$1.1  &  0.93$\pm$0.41  &  23 \\
C28  &  05:47:43.00 &   +00:36:17.92  &  0.07  &   9.0 _{ 8.8 }^{ 9.0 }   &  12.9  &  0.21$\pm$0.02   &   \phn7.7$\pm$3.1   &  \phn4.3$\pm$0.8  &  0.56$\pm$0.25  &  6  \\
C29  &  05:47:23.93 &   +00:15:23.98  &  0.06  &  10.0 _{ 10.0 }^{ 10.1 } &  17.3  &  0.19$\pm$0.02   &   \phn4.6$\pm$1.9   &  \phn4.0$\pm$0.8  &  0.87$\pm$0.41  &  23 \\
C30  &  05:47:45.93 &   +00:17:35.97  &  0.06  &   9.0 _{ 8.8 }^{ 9.2 }   &  14.0  &  0.30$\pm$0.03   &   \phn6.9$\pm$2.8   &  \phn5.6$\pm$1.1  &  0.82$\pm$0.37  &  17 \\
C31  &  05:46:13.53 & $-$00:05:29.99  &  0.07  &   9.6 _{ 9.5 }^{ 9.7 }   &  15.2  &  0.22$\pm$0.02   &   \phn3.9$\pm$1.7   &  \phn4.4$\pm$0.8  &  1.12$\pm$0.53  &  30 \\
C32  &  05:47:32.73 &   +00:05:29.99  &  0.07  &  11.4 _{ 11.3 }^{ 11.5 } &  14.3  &  0.15$\pm$0.01   &   \phn4.5$\pm$2.0   &  \phn3.2$\pm$0.4  &  0.70$\pm$0.32  &  22 \\
C33  &  05:48:00.60 &   +00:35:55.91  &  0.05  &   9.6 _{ 9.6 }^{ 9.7 }   &  14.0  &  0.20$\pm$0.02   &   \phn2.7$\pm$1.1   &  \phn3.1$\pm$0.6  &  1.15$\pm$0.53  &  8  \\
C34  &  05:47:15.13 &   +00:42:09.92  &  0.06  &   8.6 _{ 8.5 }^{ 8.7 }   &  13.7  &  0.17$\pm$0.01   &   \phn4.9$\pm$2.1   &  \phn3.2$\pm$0.4  &  0.65$\pm$0.29  &  4 \\
C35  &  05:46:15.00 &   +00:23:27.99  &  0.07  &   9.6 _{ 9.4 }^{ 11.0 }  &  14.0  &  0.16$\pm$0.01   &   \phn5.2$\pm$2.3   &  \phn3.5$\pm$0.4  &  0.68$\pm$0.31  &  17 \\
C36  &  05:47:34.20 &   +00:35:55.93  &  0.06  &   9.0 _{ 8.9 }^{ 9.0 }   &  13.2  &  0.18$\pm$0.01   &   \phn3.7$\pm$1.6   &  \phn3.4$\pm$0.4  &  0.91$\pm$0.39  &  6 \\
C37  &  05:47:41.53 &   +00:27:29.95  &  0.06  &   9.2 _{ 9.1 }^{ 9.2 }   &  13.8  &  0.10$\pm$0.01   &   \phn3.6$\pm$1.6   &  \phn3.2$\pm$0.6  &  0.88$\pm$0.43  &  9 \\
C38  &  05:46:42.86 & $-$00:01:49.99  &  0.05  &  11.2 _{ 11.2 }^{ 11.3 } &  15.6  &  0.43$\pm$0.01   &   \phn1.4$\pm$0.7   &  \phn8.2$\pm$0.4  &  5.90$\pm$2.89  &  30 \\
C39  &  05:47:29.80 &   +00:40:41.91  &  0.05  &   9.2 _{ 9.2 }^{ 9.2 }   &  13.4  &  0.19$\pm$0.02   &   \phn1.2$\pm$0.6   &  \phn2.7$\pm$0.6  &  2.20$\pm$1.09  &  3 \\
C40  &  05:46:35.53 &   +00:24:33.99  &  0.05  &   8.0 _{ 8.0 }^{ 8.0 }   &  13.4  &  0.13$\pm$0.01   &   \phn1.8$\pm$0.8   &  \phn2.1$\pm$0.3  &  1.18$\pm$0.55  &  18 \\
C41  &  05:47:34.20 &   +00:41:47.90  &  0.04  &   9.3 _{ 9.3 }^{ 9.3 }   &  13.6  &  0.17$\pm$0.01   &   \phn0.6$\pm$0.4   &  \phn2.2$\pm$0.3  &  3.64$\pm$2.29  &  3 \\
C42  &  05:47:31.26 &   +00:16:29.98  &  0.05  &   9.4 _{ 9.3 }^{ 9.4 }   &  16.5  &  0.26$\pm$0.02   &   \phn0.5$\pm$0.3   &  \phn4.1$\pm$0.6  &  7.75$\pm$4.19  &  17 \\
C43  &  05:47:23.93 &   +00:10:37.99  &  0.04  &  10.6 _{ 10.6 }^{ 10.7 } &  16.1  &  0.11$\pm$0.01   &   \phn0.5$\pm$0.2   &  \phn1.6$\pm$0.3  &  3.54$\pm$1.95  &  23 \\
C44  &  05:47:43.00 &   +00:30:47.94  &  0.04  &   9.2 _{ 9.2 }^{ 9.3 }   &  15.1  &  0.18$\pm$0.02   &   \phn1.1$\pm$0.5   &  \phn2.5$\pm$0.6  &  2.34$\pm$1.17  &  - \\
C45  &  05:47:57.66 &   +00:22:43.95  &  0.04  &  10.0 _{ 10.0 }^{ 10.0 } &  14.5  &  0.10$\pm$0.01   &   \phn0.7$\pm$0.4   &  \phn1.6$\pm$0.3  &  2.31$\pm$1.28  &  8 \\
C46  &  05:47:47.40 &   +00:43:37.88  &  0.05  &   9.8 _{ 9.8 }^{ 9.8 }   &  14.0  &  0.10$\pm$0.01   &   \phn0.8$\pm$0.6   &  \phn1.9$\pm$0.4  &  2.33$\pm$1.75  &  3 \\
C47  &  05:47:20.99 &   +00:07:41.99  &  0.05  &  10.6 _{ 10.6 }^{ 10.6 } &  15.3  &  0.10$\pm$0.01   &   \phn0.7$\pm$0.4   &  \phn2.0$\pm$0.4  &  2.67$\pm$1.66  &  23 \\
C48  &  05:47:40.06 &   +00:08:03.99  &  0.07  &  11.2 _{ 11.2 }^{ 11.3 } &  13.6  &  0.19$\pm$0.03   &   \phn0.4$\pm$0.2   &  \phn3.1$\pm$1.0  &  8.35$\pm$4.53  &  22 \\
\enddata
\tablecomments{
Column 1: Core name given in order of its peak intensity, from the highest to the lowest. 
Column 2 and Column 3:  Coordinates at peak intensity positions of cores. 
Column 4: The deconvolved core size as the geometric mean of major and minor axis size assuming its elliptical shape in units of pc. 
Column 5: The LSR velocity obtained from hyperfine fitting of the \nThpj~spectrum in units of \kms. 
Column 6: The average dust temperature of dense core in units of K. 
Column 7: The average total velocity dispersion of the dense core from the \nThpj~line profiles. 
Column 8 and Column 9: The core mass and its virial mass in units of Solar mass. 
Column 10: The core virial parameter, the ratio of the core mass to the virial mass. 
Column 11: The name of the filament surrounding the core. 
}
\end{deluxetable*}

\newpage

\appendix

\restartappendixnumbering
\section{All identified \cEo\ filaments in NGC~2068 and NGC~2071 regions of Orion~B and analysis of their physical quantities} \label{A1}

This figure set presents the physical properties of each velocity-coherent filament and some schematic methods used deriving some of them.
Figure Set~\ref{fig:figA} show the atlas of filaments identified using the decomposed \cEoj\ data. 
The atlas figure of each filament contains nine panels as follows:

\begin{itemize}

\item[] (a) Distribution of the integrated intensity of a filament. 

\item[] (b) Distribution of the systemic velocity of a filament.

\item[] (c) Distribution of the velocity dispersion of a filament. 

\item[] (d) Determination of the  spine of a filament and radial cuts for deriving radial profiles.
Red dots indicate path of spine and the cyan curve is the smoothed path that is used for deriving the length and width of the filament. The black lines display radial cuts that are perpendicular to the smoothed spine and used for drawing radial profiles in panel (e) and (f).

\item[] (e) Radial profiles of the intensity along the black lines drawn in panel (d). Black and red thick lines show the averaged profile and its Gaussian fit result, respectively.


\item[] (f) The distribution of the YSOs and the dense cores over the filament. The filament is depicted with its integrated intensity distribution in color. The gray contours indicate the \nThpj~integrated intensity distribution with contour levels of [0.08, 0.3, 0.8, 2.0, 4.0, 6.0, 8.0, 10.0] K \kms. The \nThp~dense cores were identified from the $\texttt{FellWalker}$ algorithm. The cyan and green polygon shapes are to indicate the dense cores to be and not to be associated with their surrounding filament, respectively. The asterisks are the young stellar objects classified in \citet{furlan16}. Red : Class 0 protostars. Green : Class I protostars. Yellow : Flat-spectrum YSOs. Note that the coordinate scales given in this panel are the same as the cases for the panels (a), (b), (c), and (d). 
The red arrow shows the direction measured along the spine of the filament for the variations of integrated intensity, velocity centroid, and nonthermal velocity dispersion given in the panel (g) to (i).

\item[] (g) The variation of the integrated intensity along the spine of filament.

\item[] (h) The variation of the velocity centroid along the spine of filament. 

\item[] (i) The variation of the nonthermal velocity dispersion divided by sound speed along the spine of filament.
The orange and blue dots are estimated from \cEoj\ (for filament)
and \nThpj\ (for dense cores), respectively. The inset depicts the number distribution of the nonthermal velocity dispersion with histograms of the same color codes for the filament and cores as those for dots. 
The vertical blue lines indicate a representative position of the dense cores. 

\end{itemize}

\newpage

\figsetstart
\figsetnum{A}
\label{fig:figA}
\figsettitle{The information of velocity-coherent filament and analysis of its physical quantities. The caption details can be found in Appendix~\ref{A1}.}

\figsetgrpstart
\figsetgrpnum{A.1}
\figsetgrptitle{Filament F1}
\figsetplot{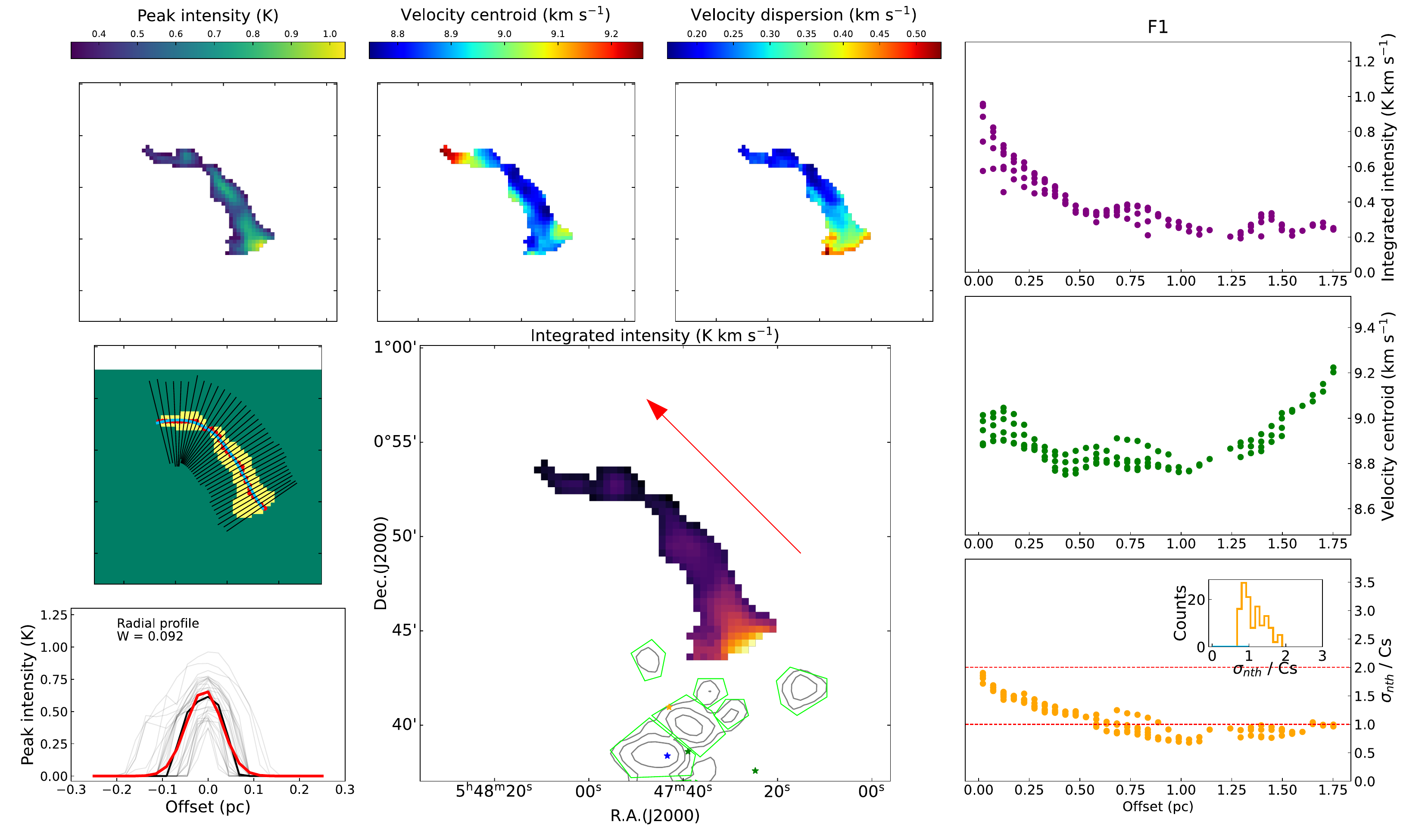}
\figsetgrpnote{Filament quantities}
\figsetgrpend

\figsetgrpstart
\figsetgrpnum{A.2}
\figsetgrptitle{Filament F2}
\figsetplot{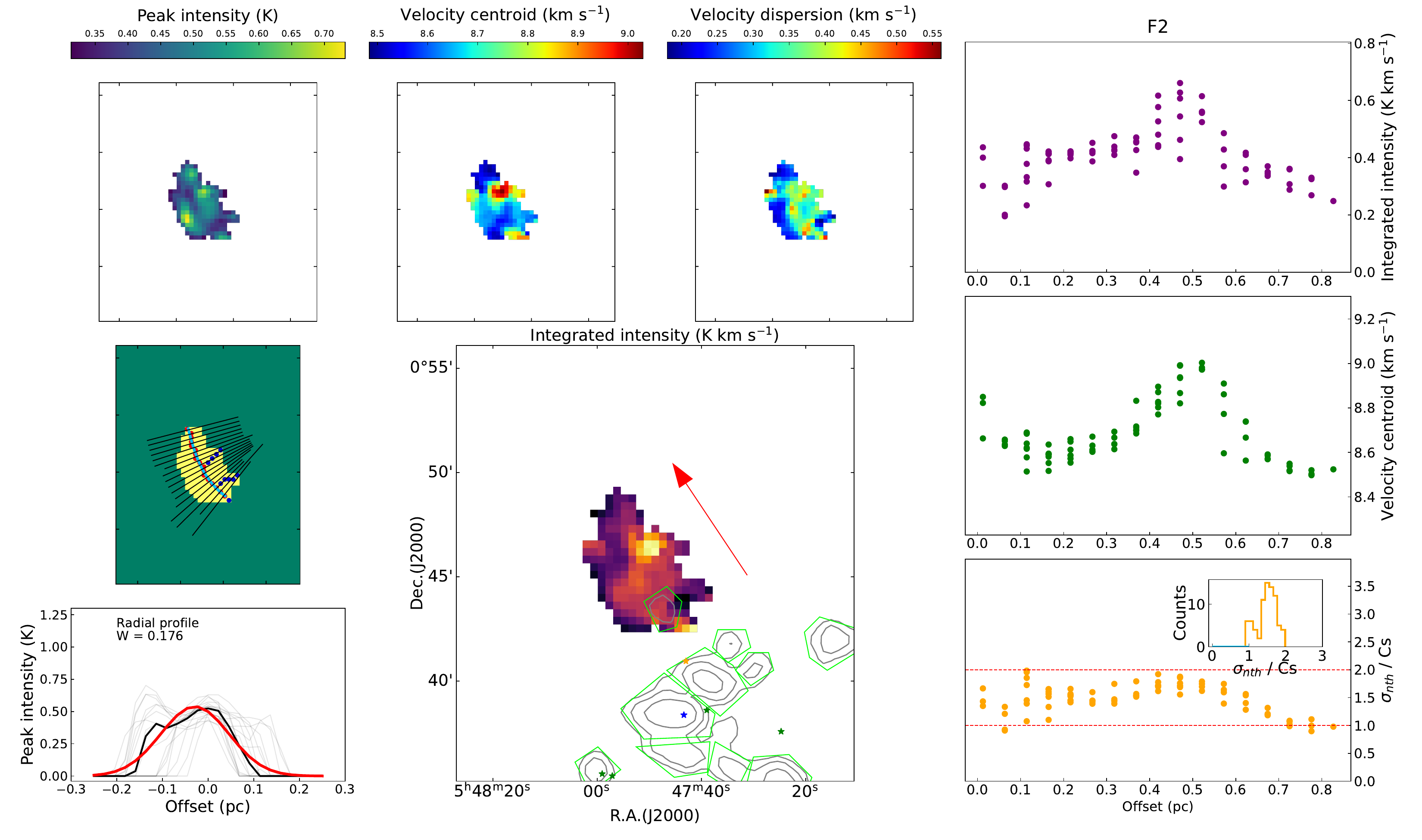}
\figsetgrpnote{Filament quantities}
\figsetgrpend

\figsetgrpstart
\figsetgrpnum{A.3}
\figsetgrptitle{Filament F3}
\figsetplot{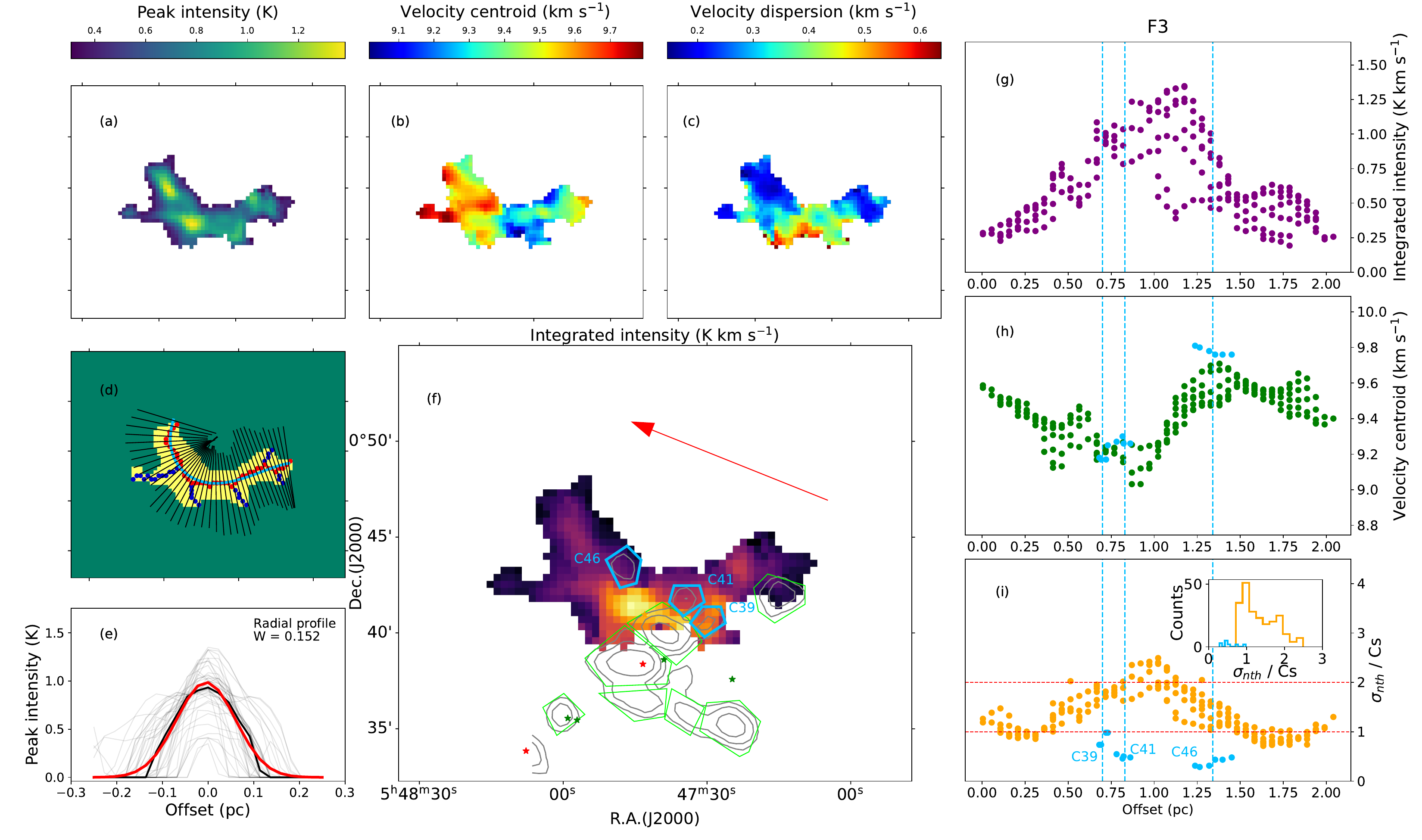}
\figsetgrpnote{Filament quantities}
\figsetgrpend

\figsetgrpstart
\figsetgrpnum{A.4}
\figsetgrptitle{Filament F4}
\figsetplot{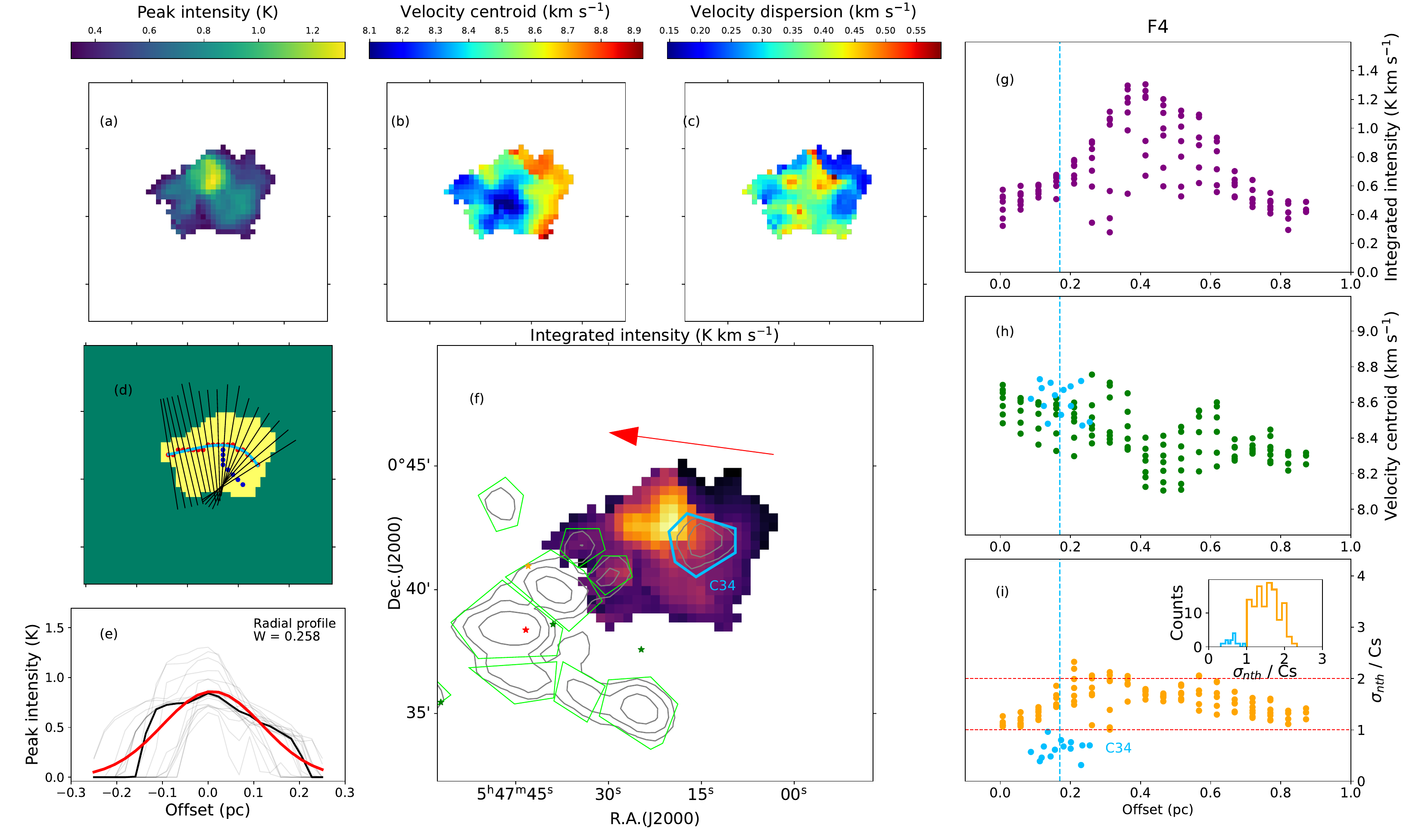}
\figsetgrpnote{Filament quantities}
\figsetgrpend

\figsetgrpstart
\figsetgrpnum{A.5}
\figsetgrptitle{Filament F5}
\figsetplot{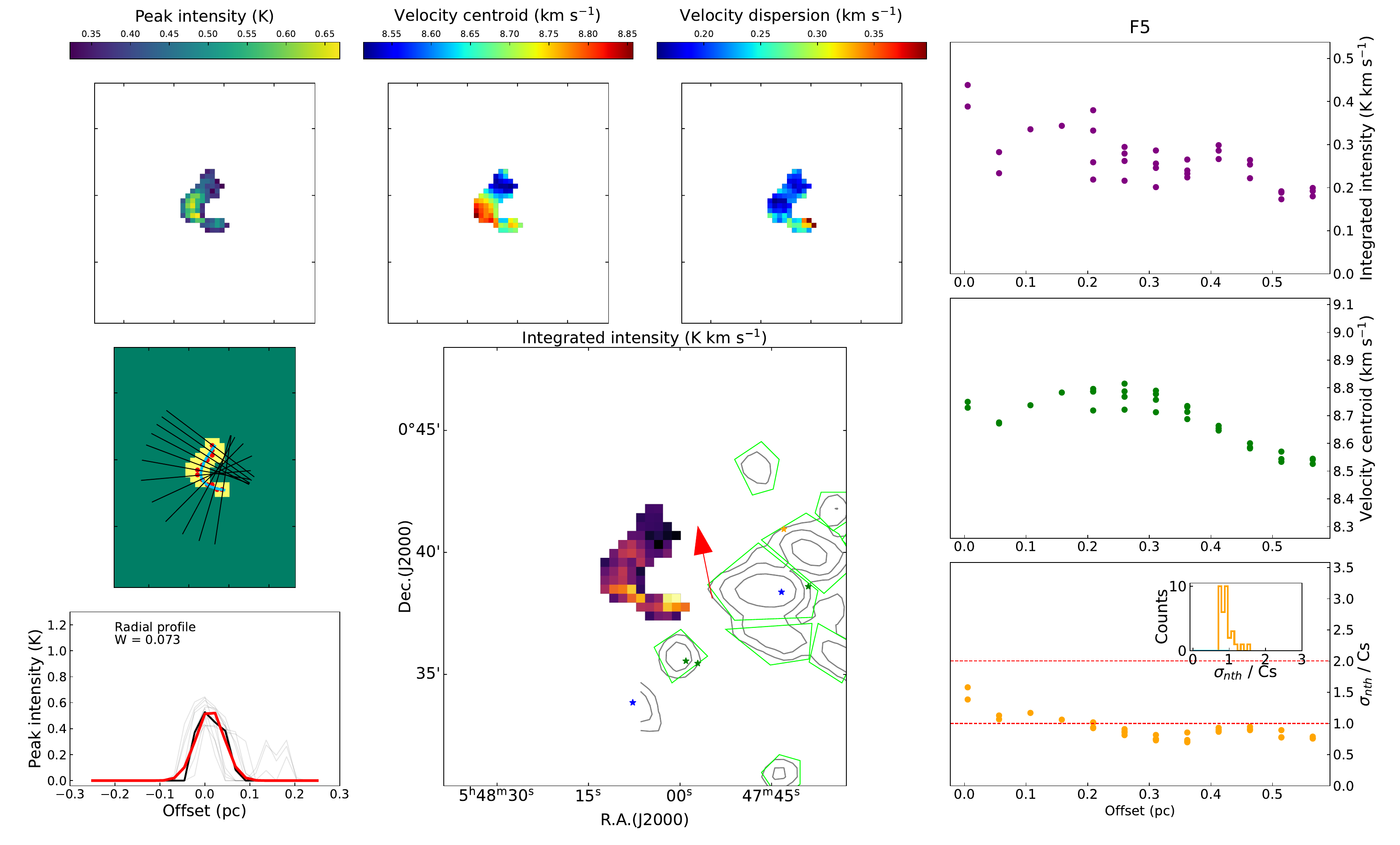}
\figsetgrpnote{Filament quantities}
\figsetgrpend

\figsetgrpstart
\figsetgrpnum{A.6}
\figsetgrptitle{Filament F6}
\figsetplot{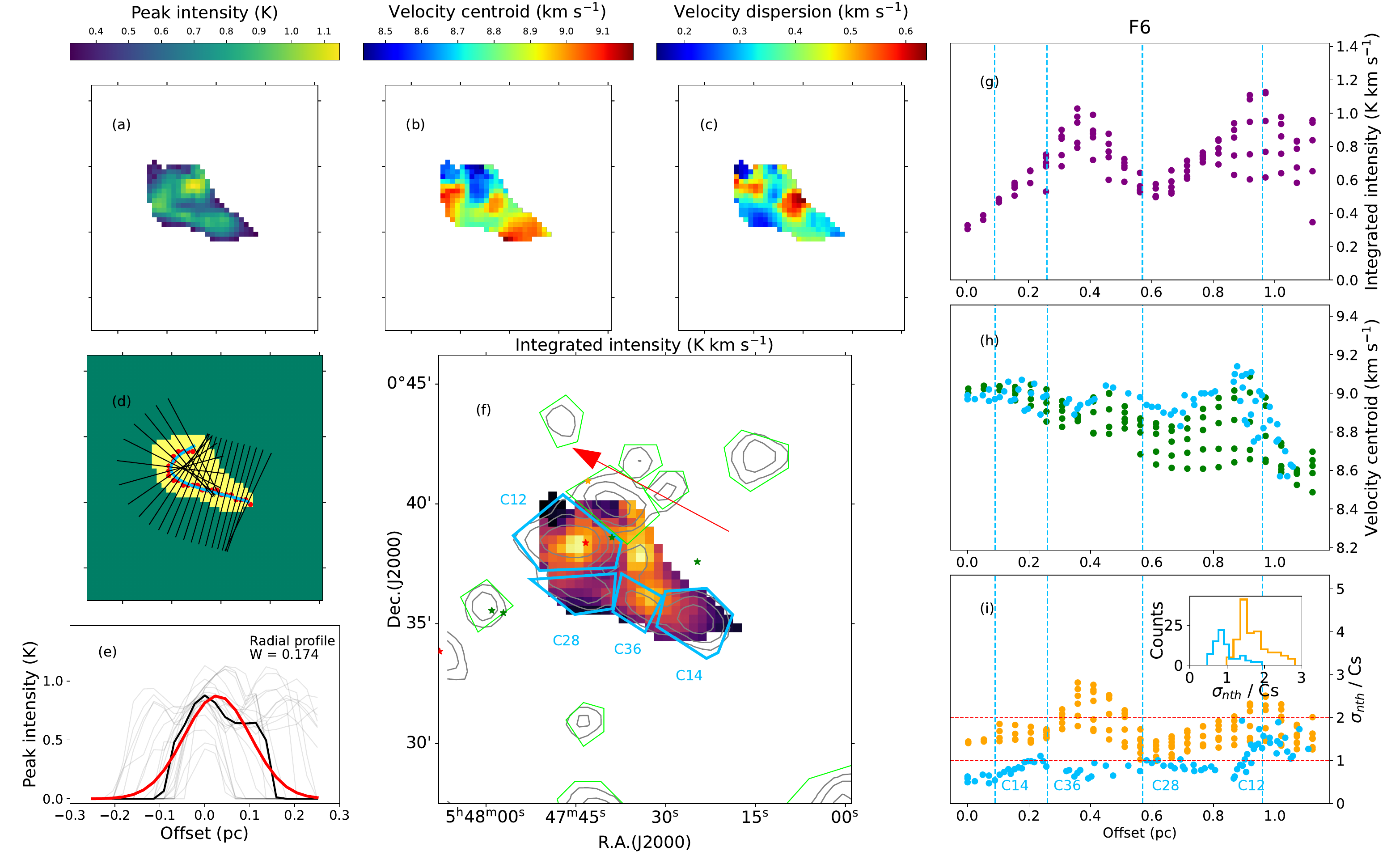}
\figsetgrpnote{Filament quantities}
\figsetgrpend

\figsetgrpstart
\figsetgrpnum{A.7}
\figsetgrptitle{Filament F7}
\figsetplot{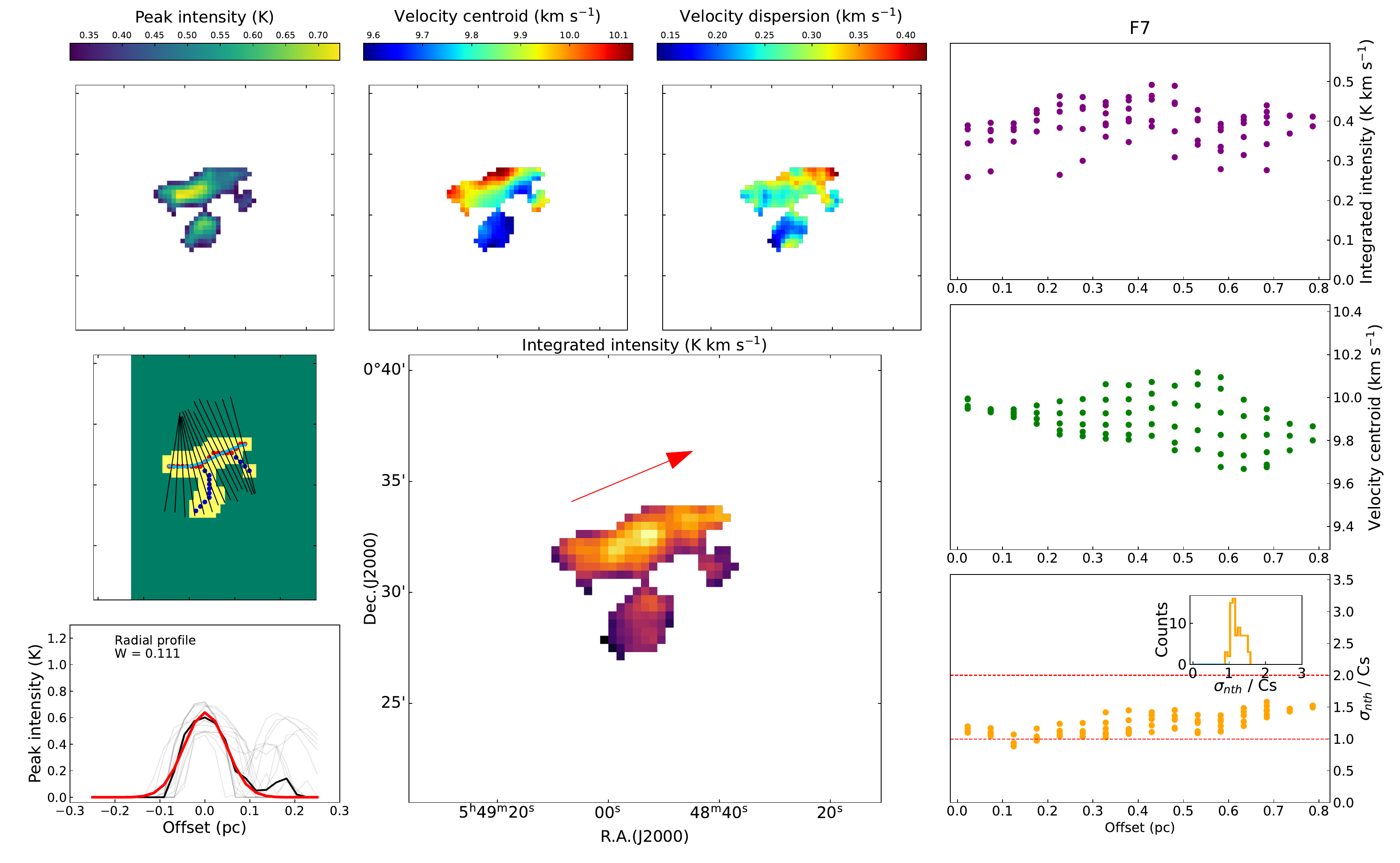}
\figsetgrpnote{Filament quantities}
\figsetgrpend

\figsetgrpstart
\figsetgrpnum{A.8}
\figsetgrptitle{Filament F8}
\figsetplot{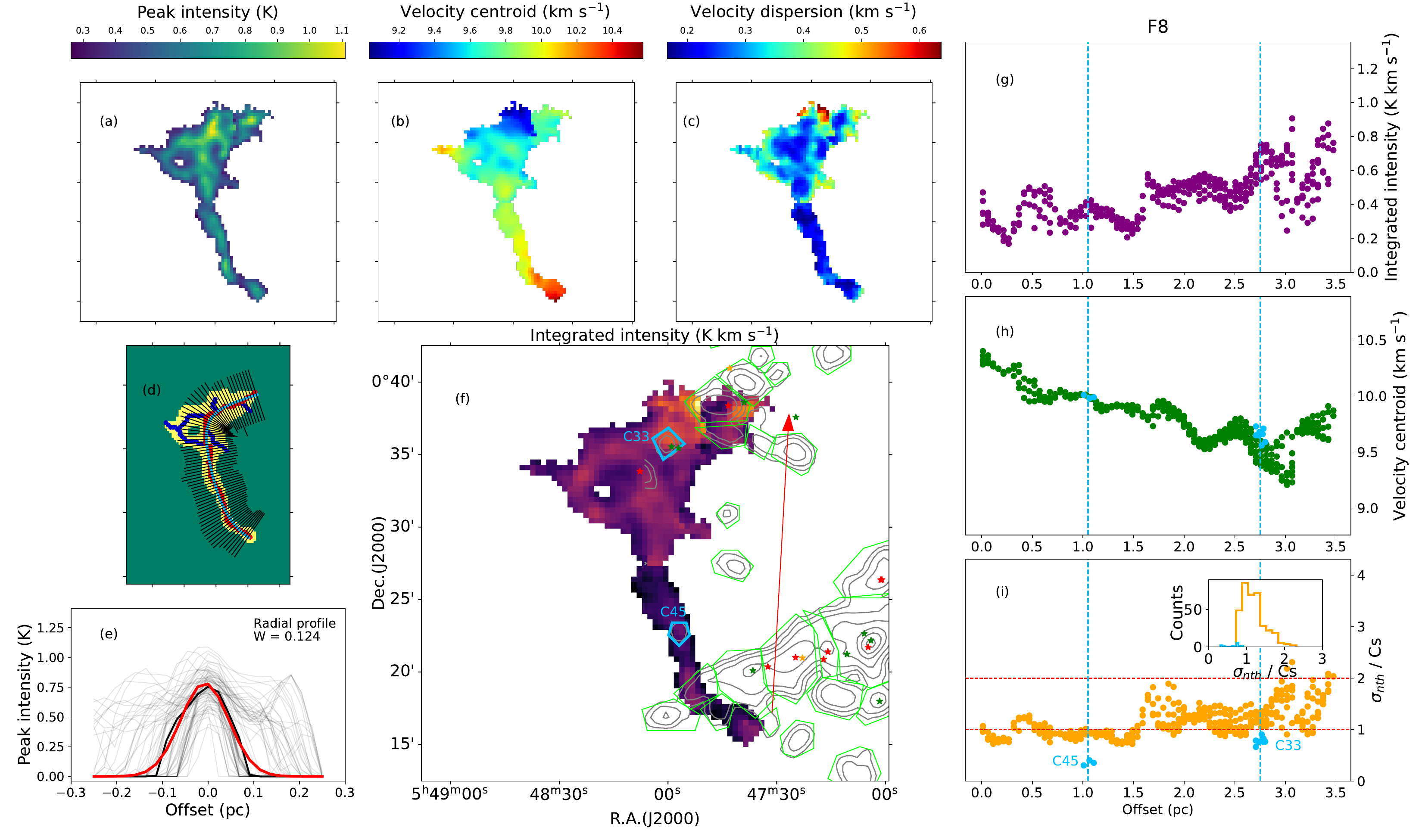}
\figsetgrpnote{Filament quantities}
\figsetgrpend

\figsetgrpstart
\figsetgrpnum{A.9}
\figsetgrptitle{Filament F9}
\figsetplot{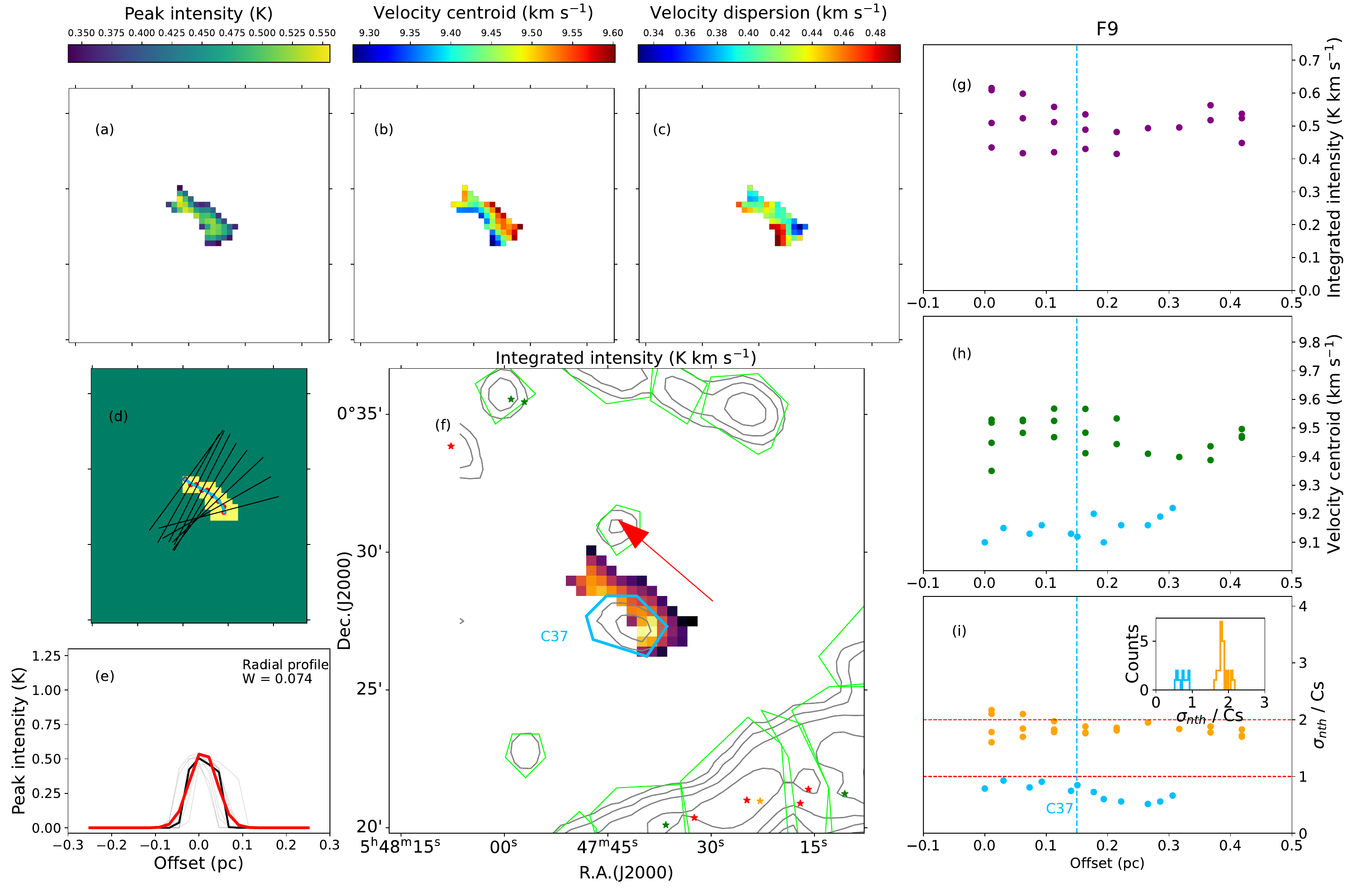}
\figsetgrpnote{Filament quantities}
\figsetgrpend

\figsetgrpstart
\figsetgrpnum{A.10}
\figsetgrptitle{Filament F10}
\figsetplot{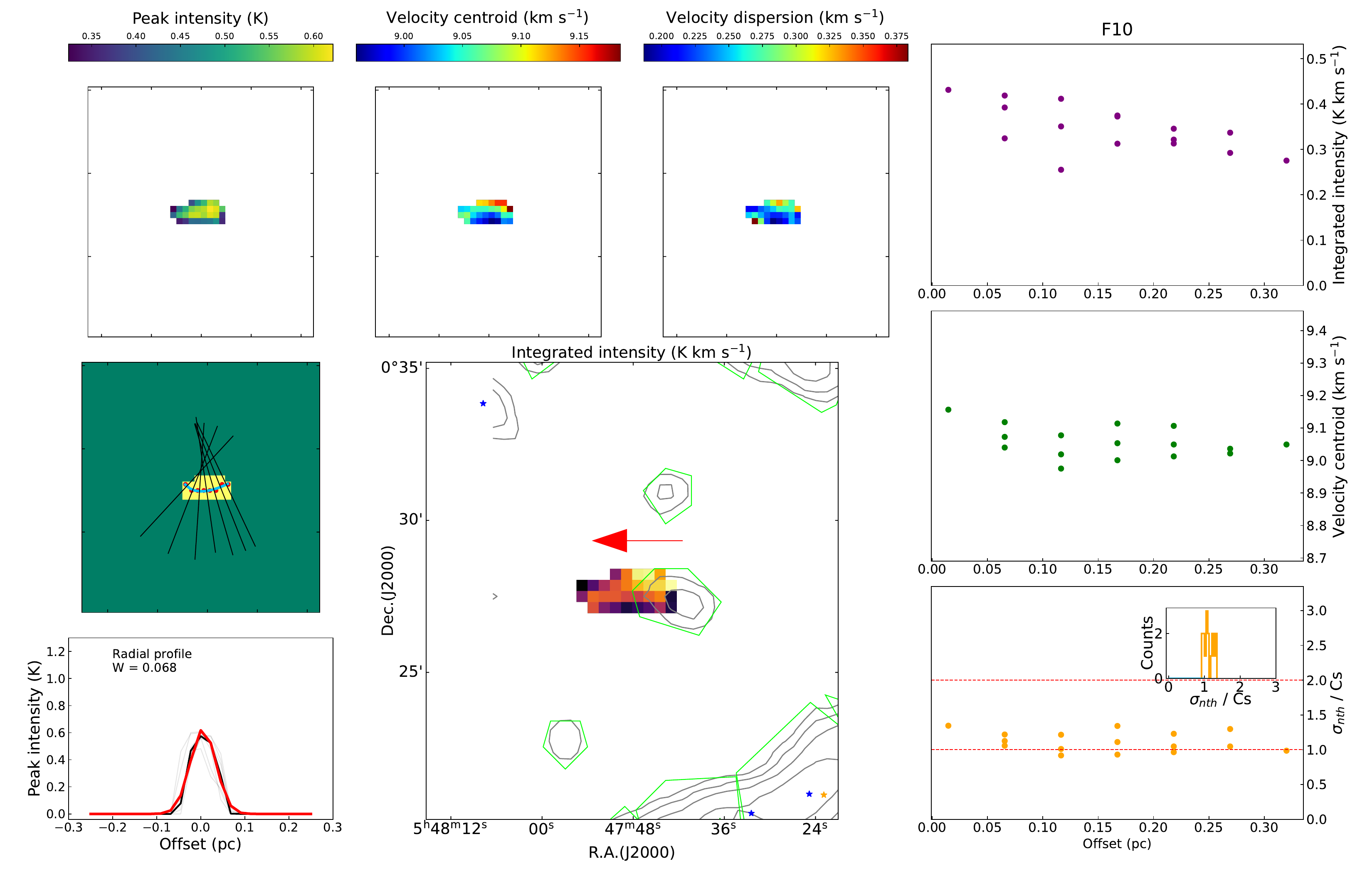}
\figsetgrpnote{Filament quantities}
\figsetgrpend

\figsetgrpstart
\figsetgrpnum{A.11}
\figsetgrptitle{Filament F11}
\figsetplot{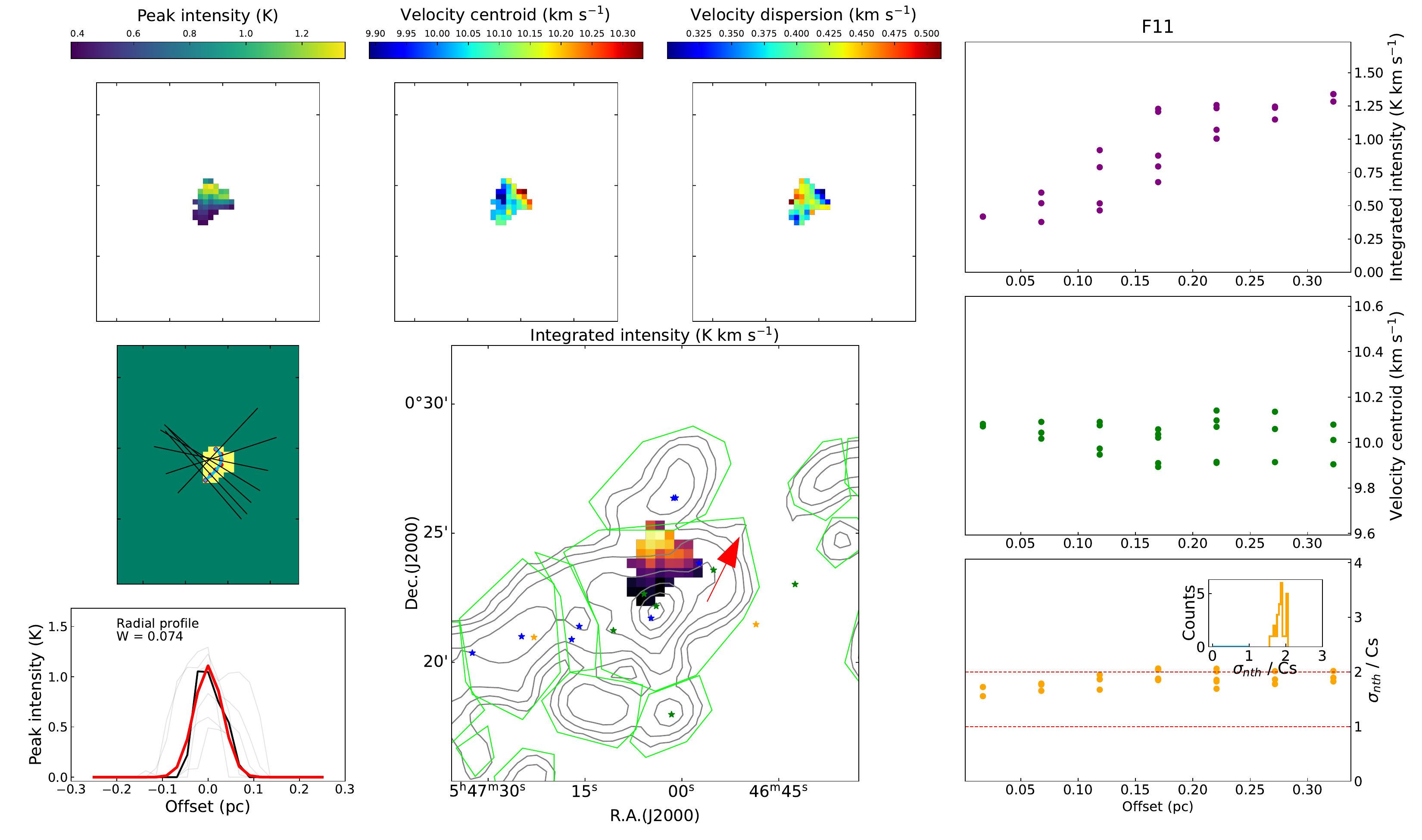}
\figsetgrpnote{Filament quantities}
\figsetgrpend

\figsetgrpstart
\figsetgrpnum{A.12}
\figsetgrptitle{Filament F12}
\figsetplot{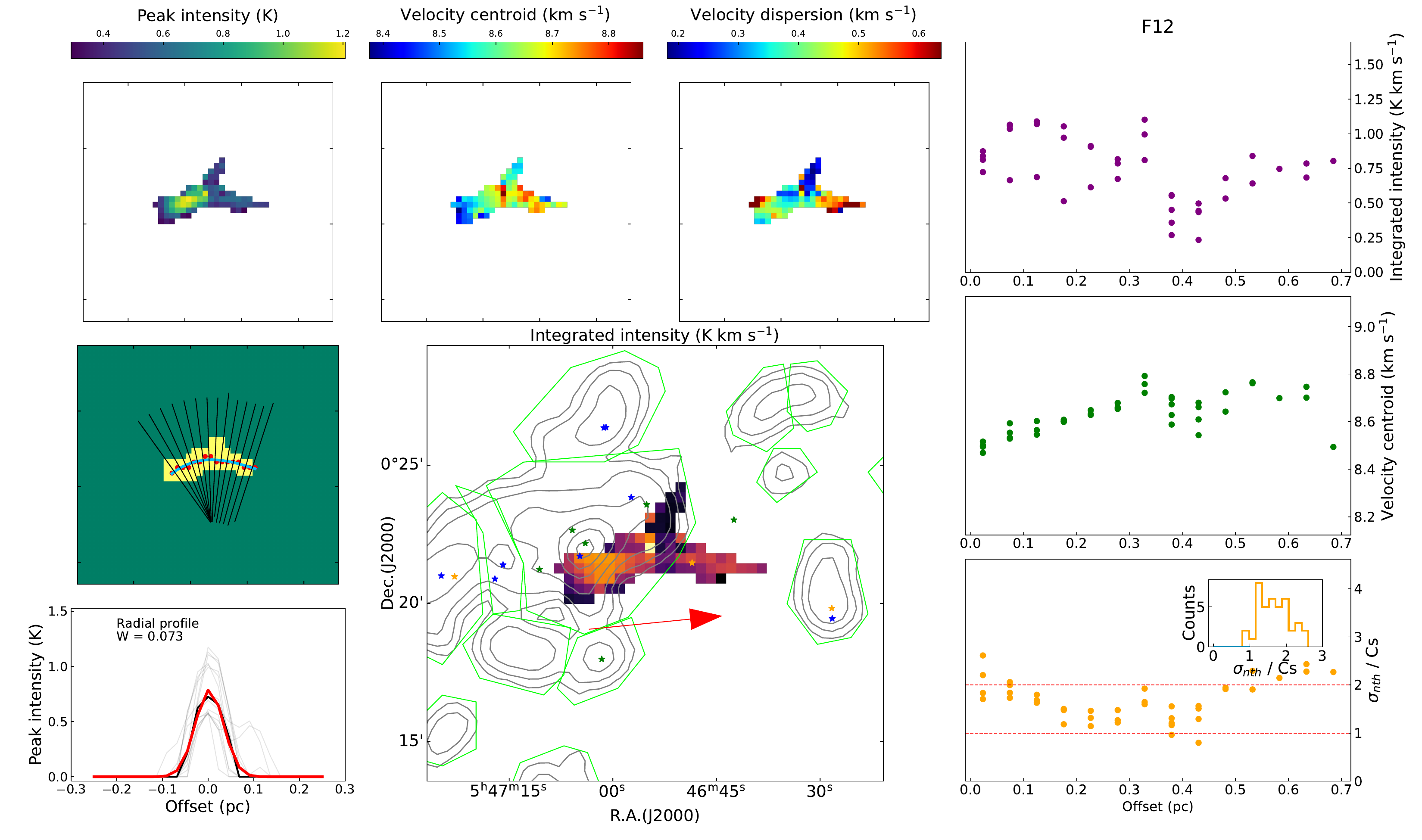}
\figsetgrpnote{Filament quantities}
\figsetgrpend

\figsetgrpstart
\figsetgrpnum{A.13}
\figsetgrptitle{Filament F13}
\figsetplot{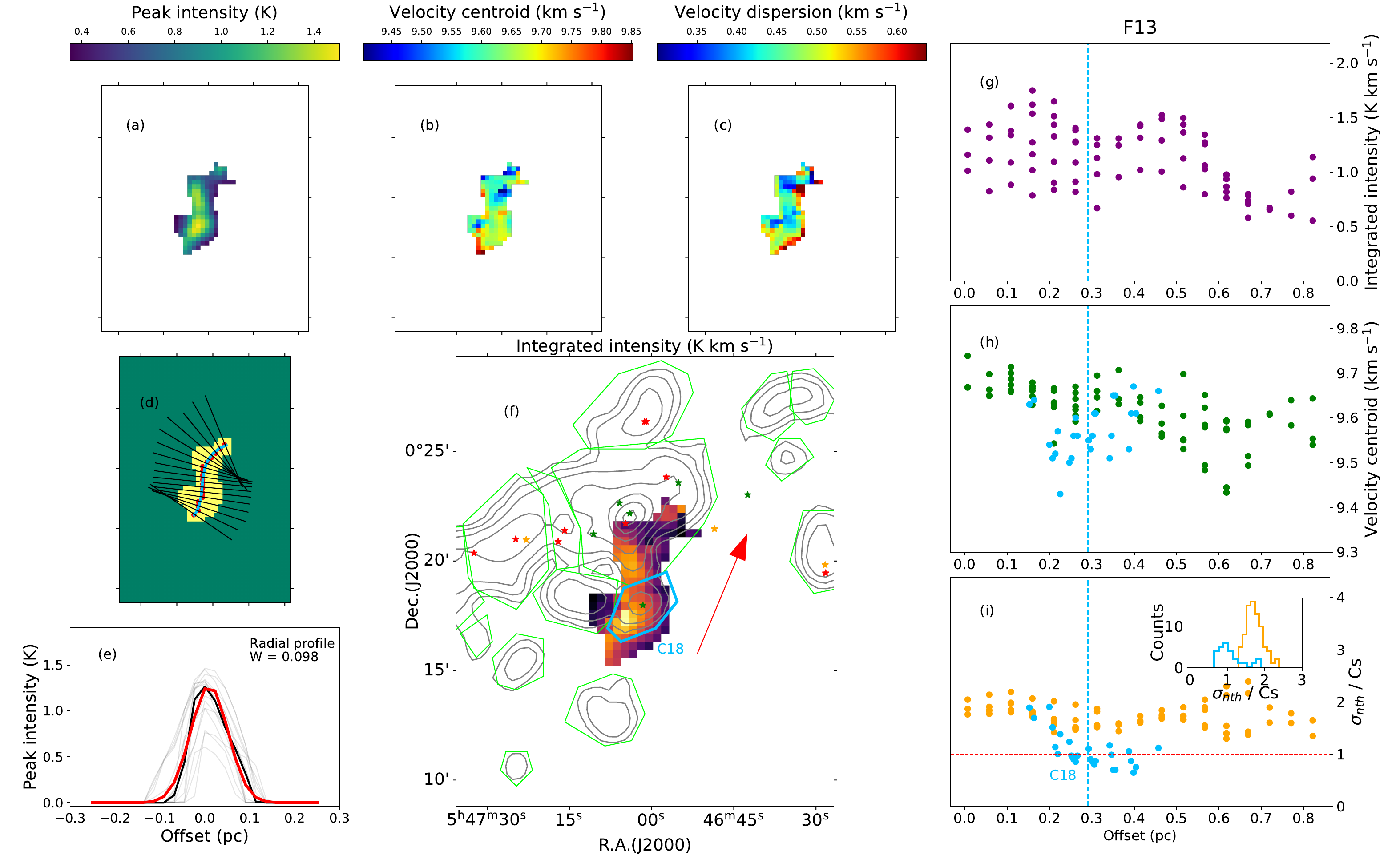}
\figsetgrpnote{Filament quantities}
\figsetgrpend

\figsetgrpstart
\figsetgrpnum{A.14}
\figsetgrptitle{Filament F14}
\figsetplot{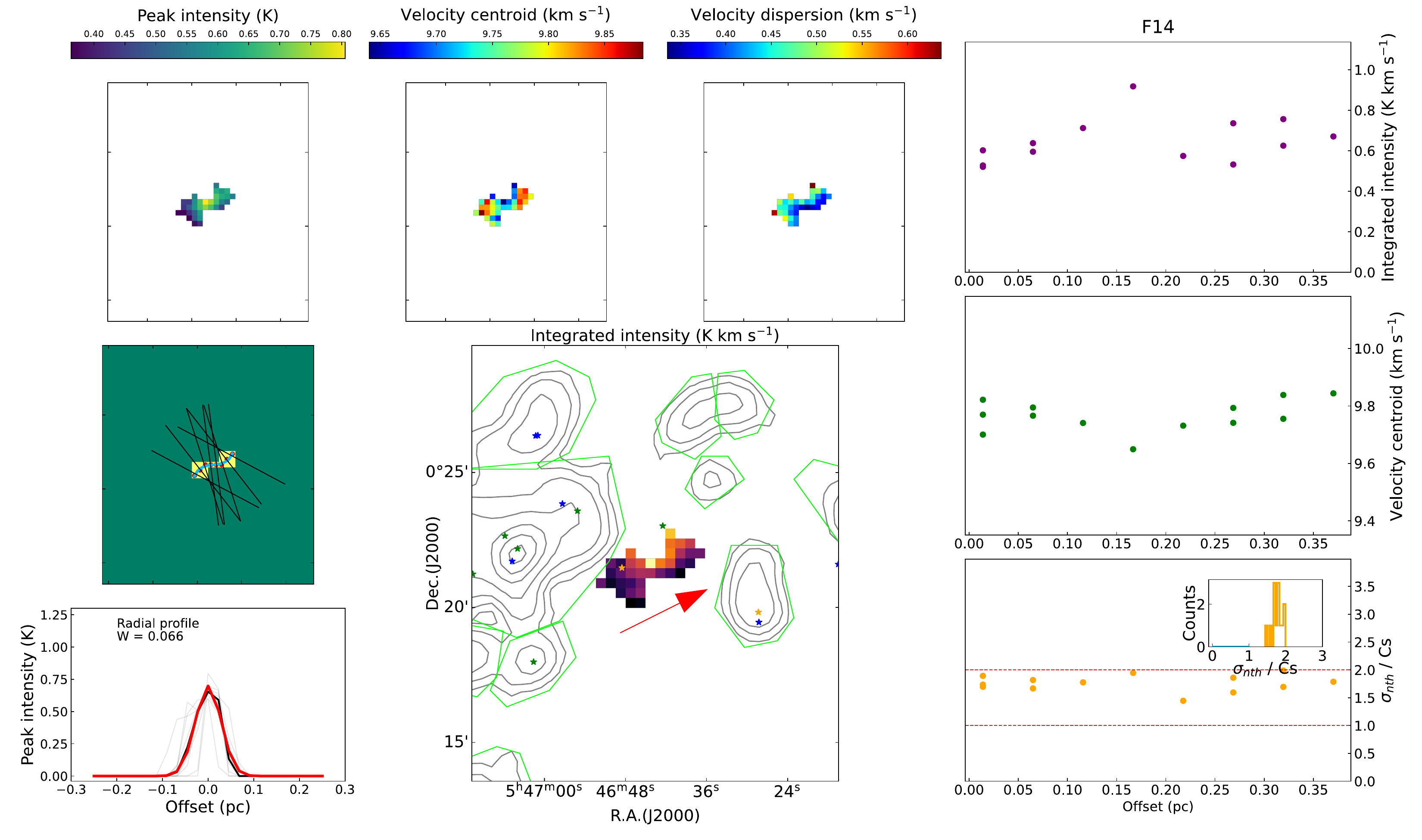}
\figsetgrpnote{Filament quantities}
\figsetgrpend

\figsetgrpstart
\figsetgrpnum{A.15}
\figsetgrptitle{Filament F15}
\figsetplot{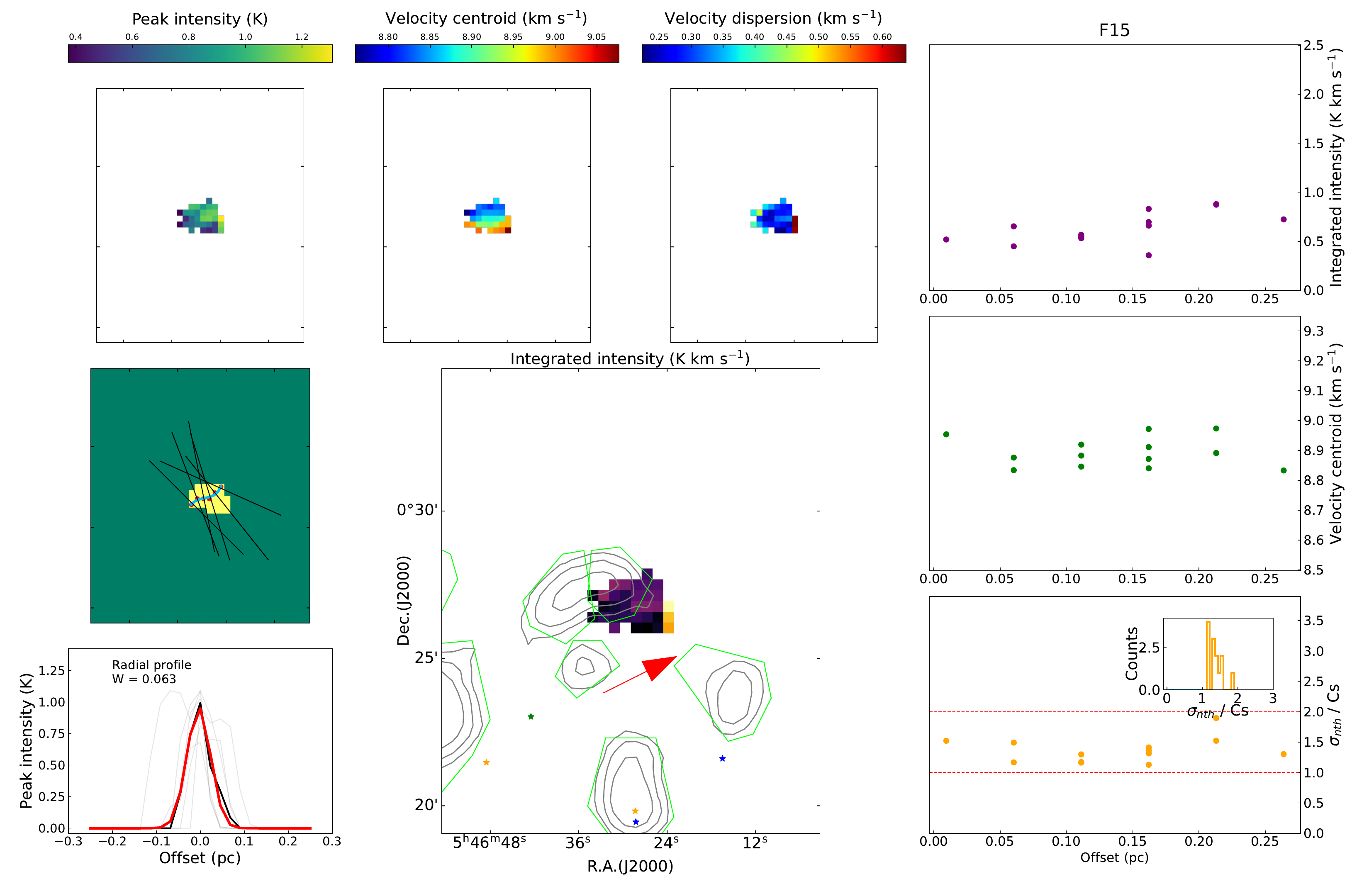}
\figsetgrpnote{Filament quantities}
\figsetgrpend

\figsetgrpstart
\figsetgrpnum{A.16}
\figsetgrptitle{Filament F16}
\figsetplot{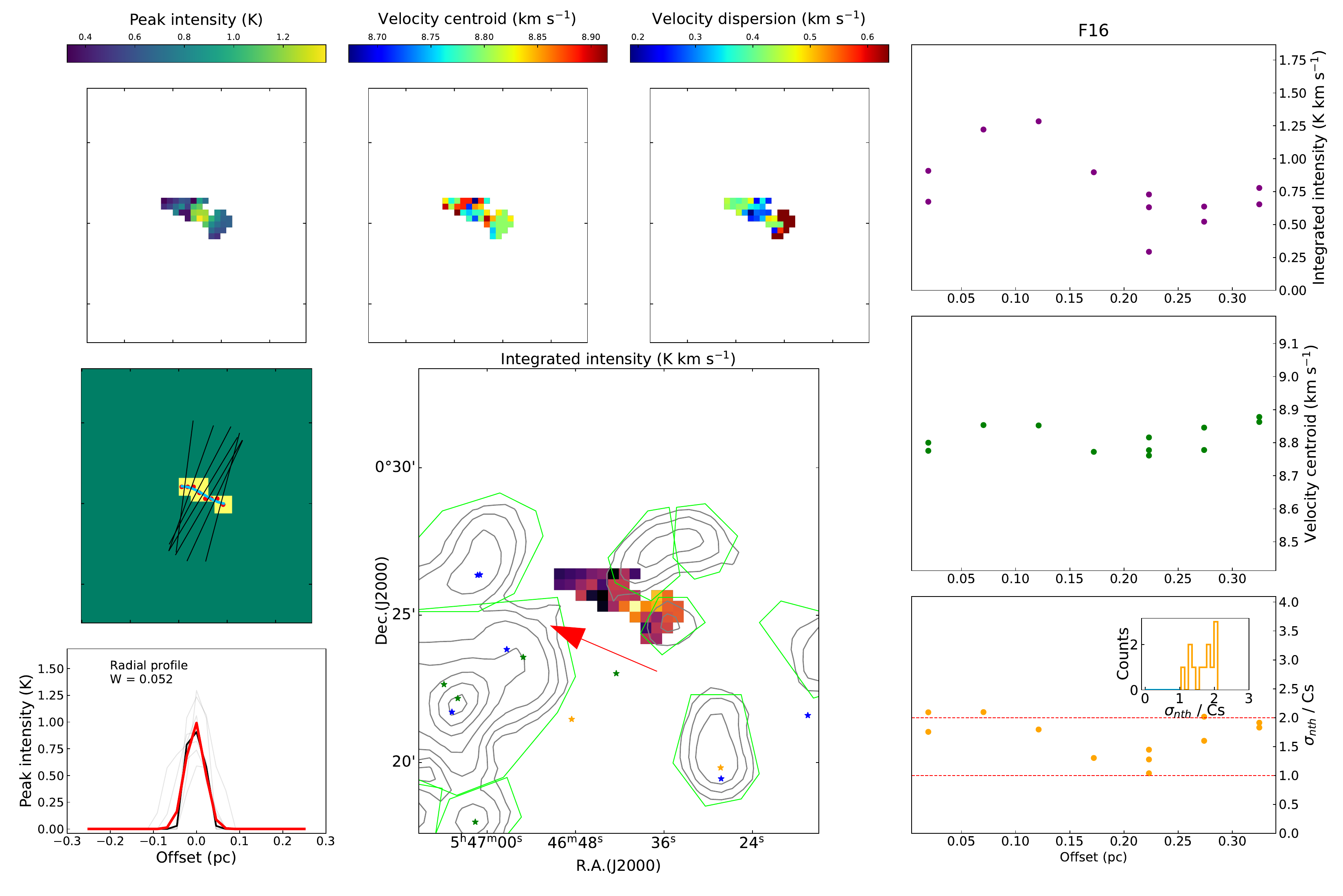}
\figsetgrpnote{Filament quantities}
\figsetgrpend

\figsetgrpstart
\figsetgrpnum{A.17}
\figsetgrptitle{Filament F17a}
\figsetplot{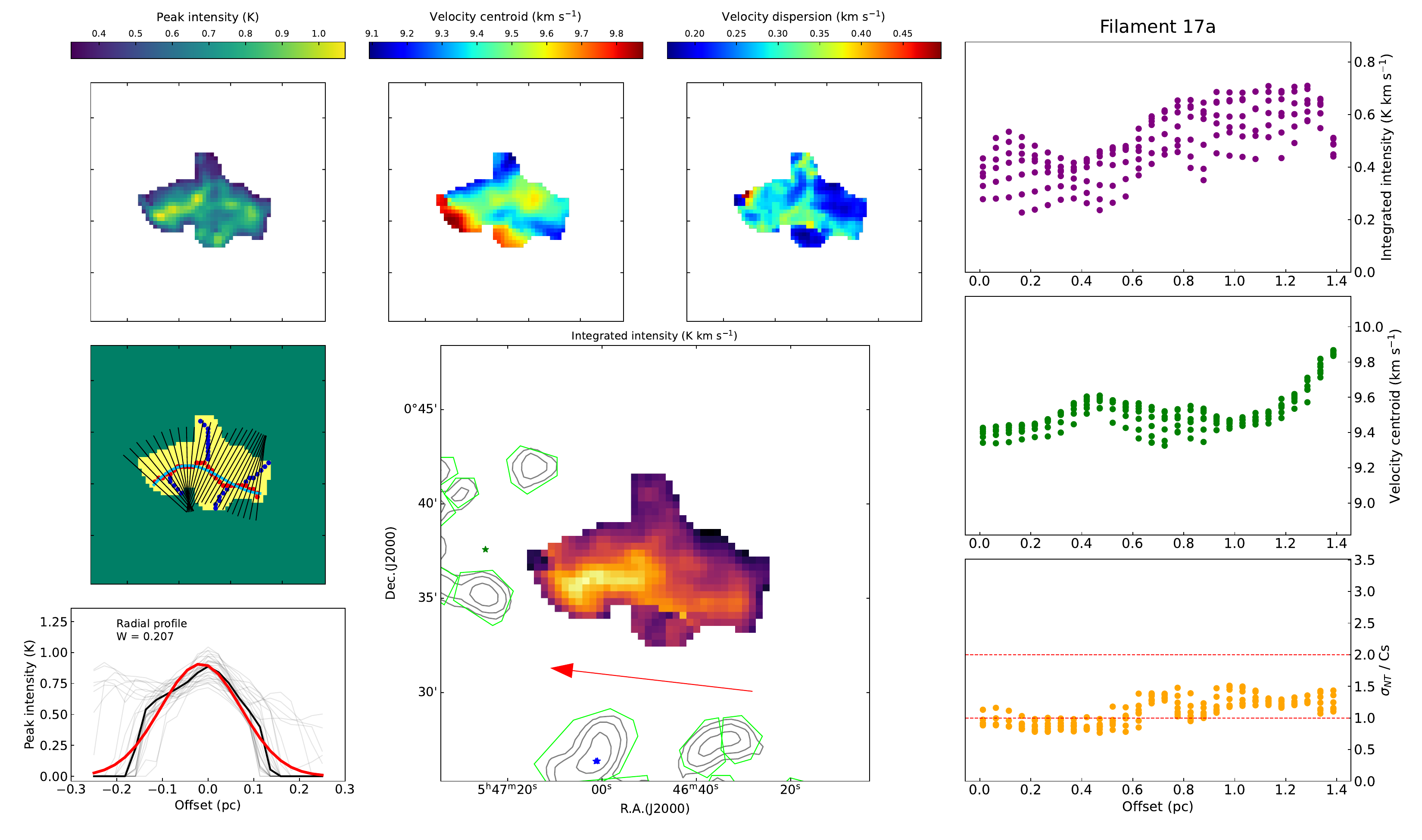}
\figsetgrpnote{Filament quantities}
\figsetgrpend

\figsetgrpstart
\figsetgrpnum{A.18}
\figsetgrptitle{Filament F17b}
\figsetplot{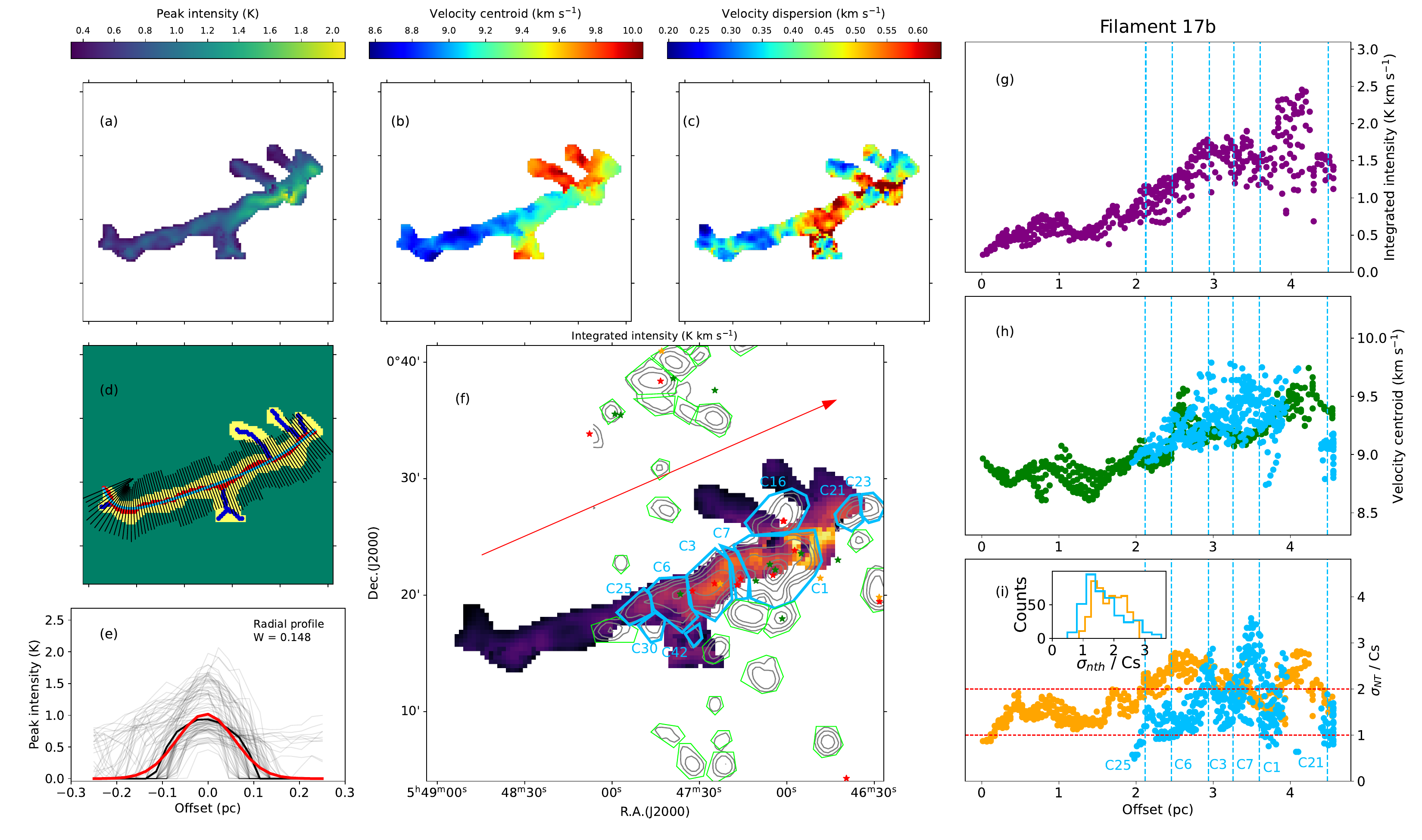}
\figsetgrpnote{Filament quantities}
\figsetgrpend

\figsetgrpstart
\figsetgrpnum{A.19}
\figsetgrptitle{Filament F17c}
\figsetplot{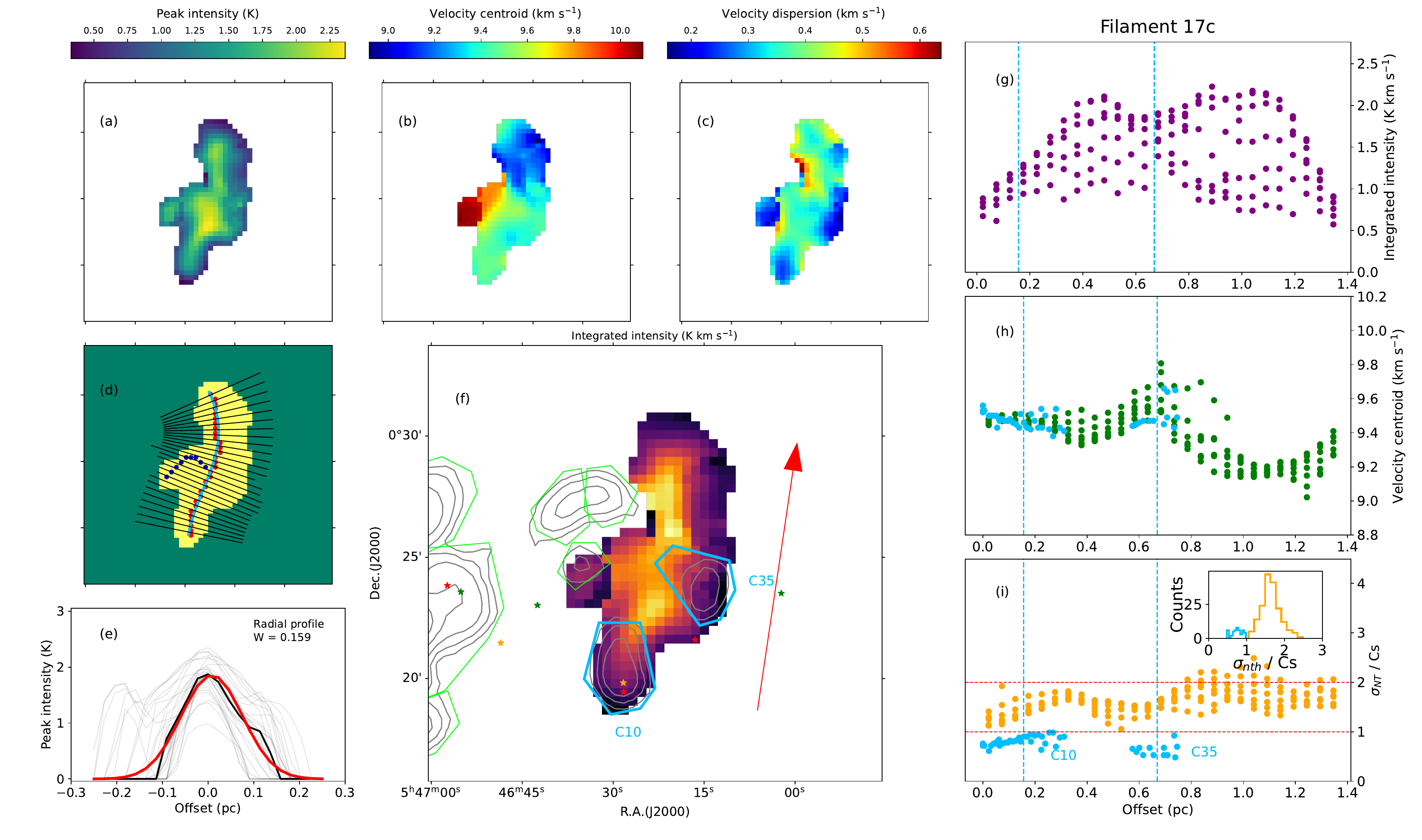}
\figsetgrpnote{Filament quantities}
\figsetgrpend

\figsetgrpstart
\figsetgrpnum{A.20}
\figsetgrptitle{Filament F18}
\figsetplot{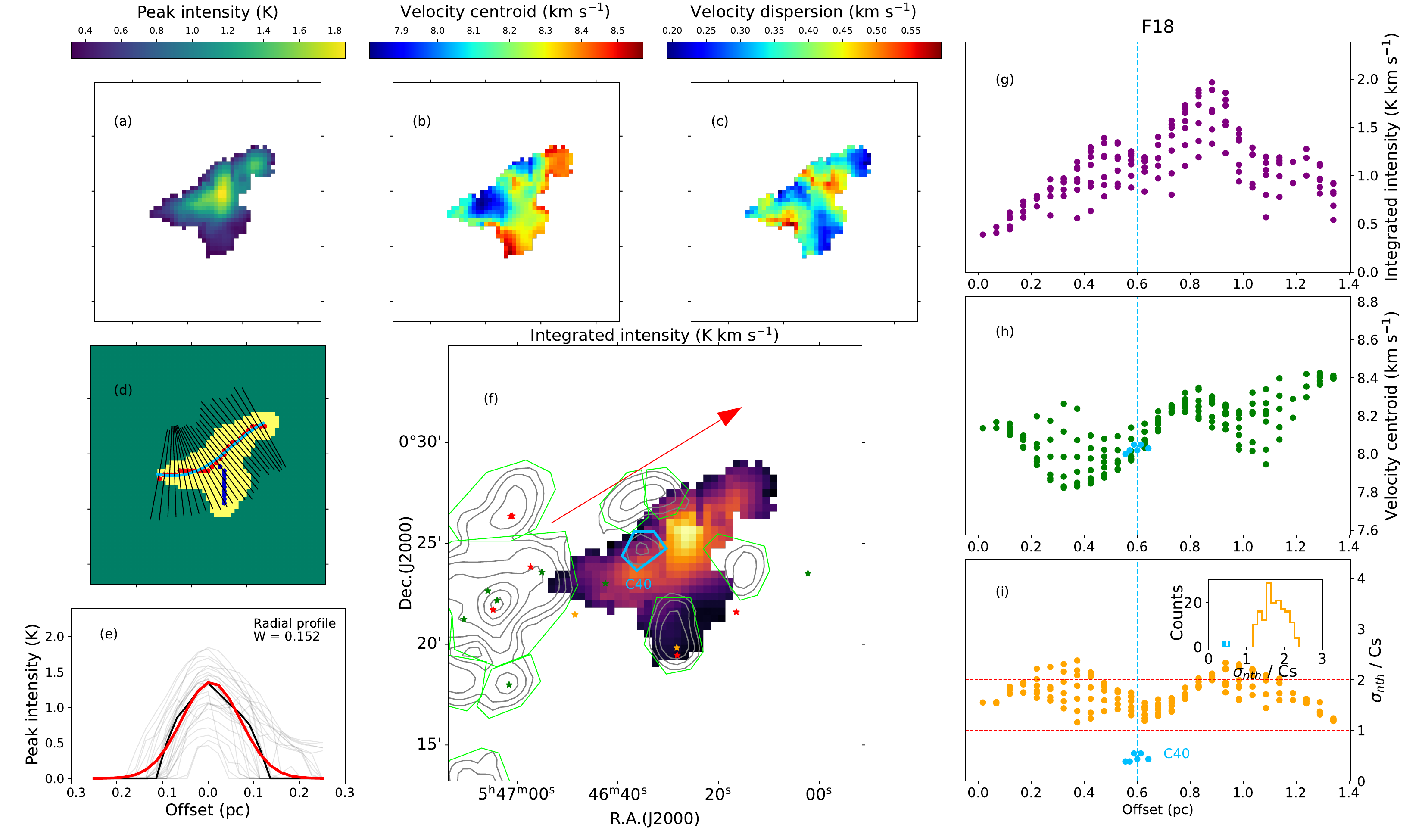}
\figsetgrpnote{Filament quantities}
\figsetgrpend

\figsetgrpstart
\figsetgrpnum{A.21}
\figsetgrptitle{Filament F19}
\figsetplot{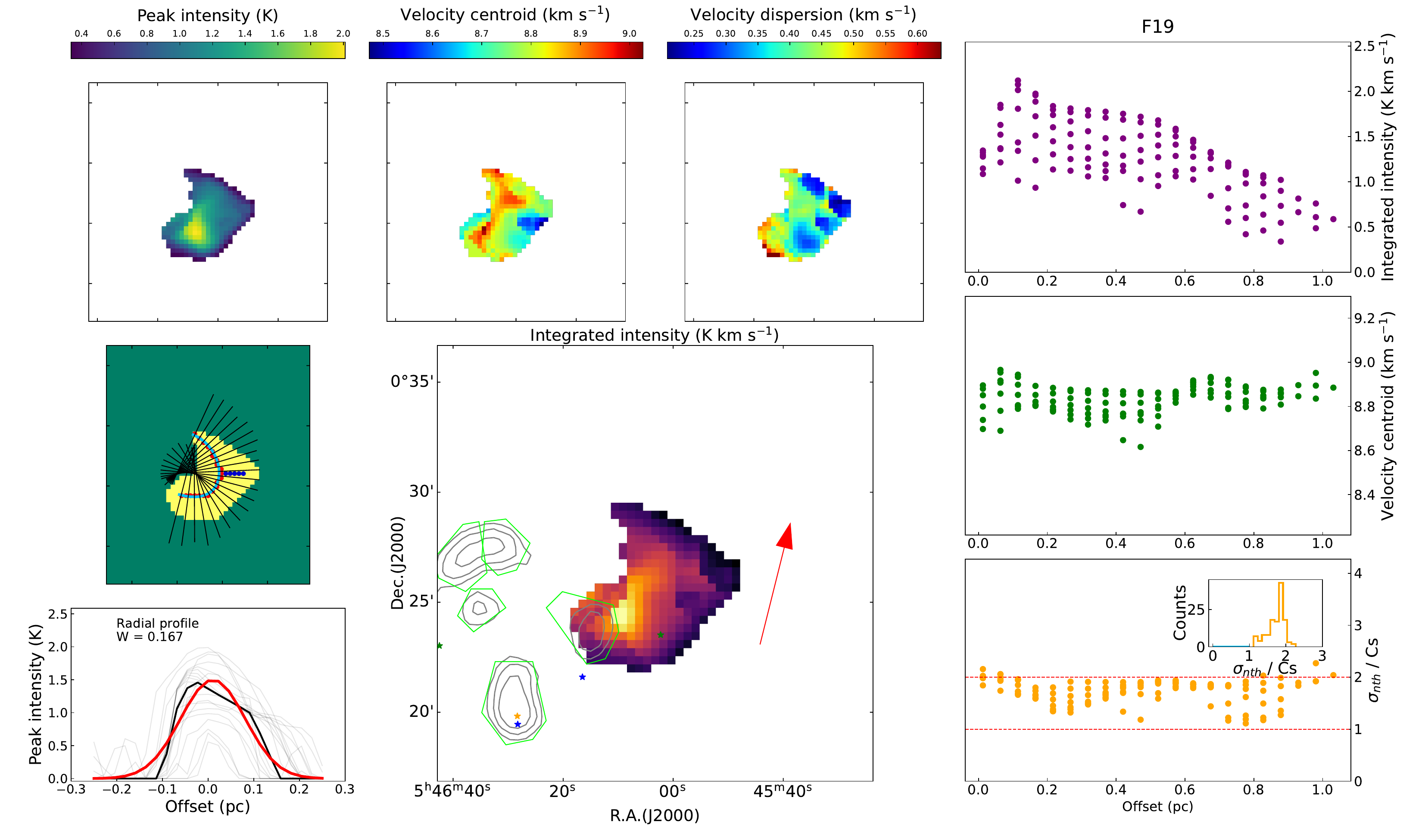}
\figsetgrpnote{Filament quantities}
\figsetgrpend

\figsetgrpstart
\figsetgrpnum{A.22}
\figsetgrptitle{Filament F20}
\figsetplot{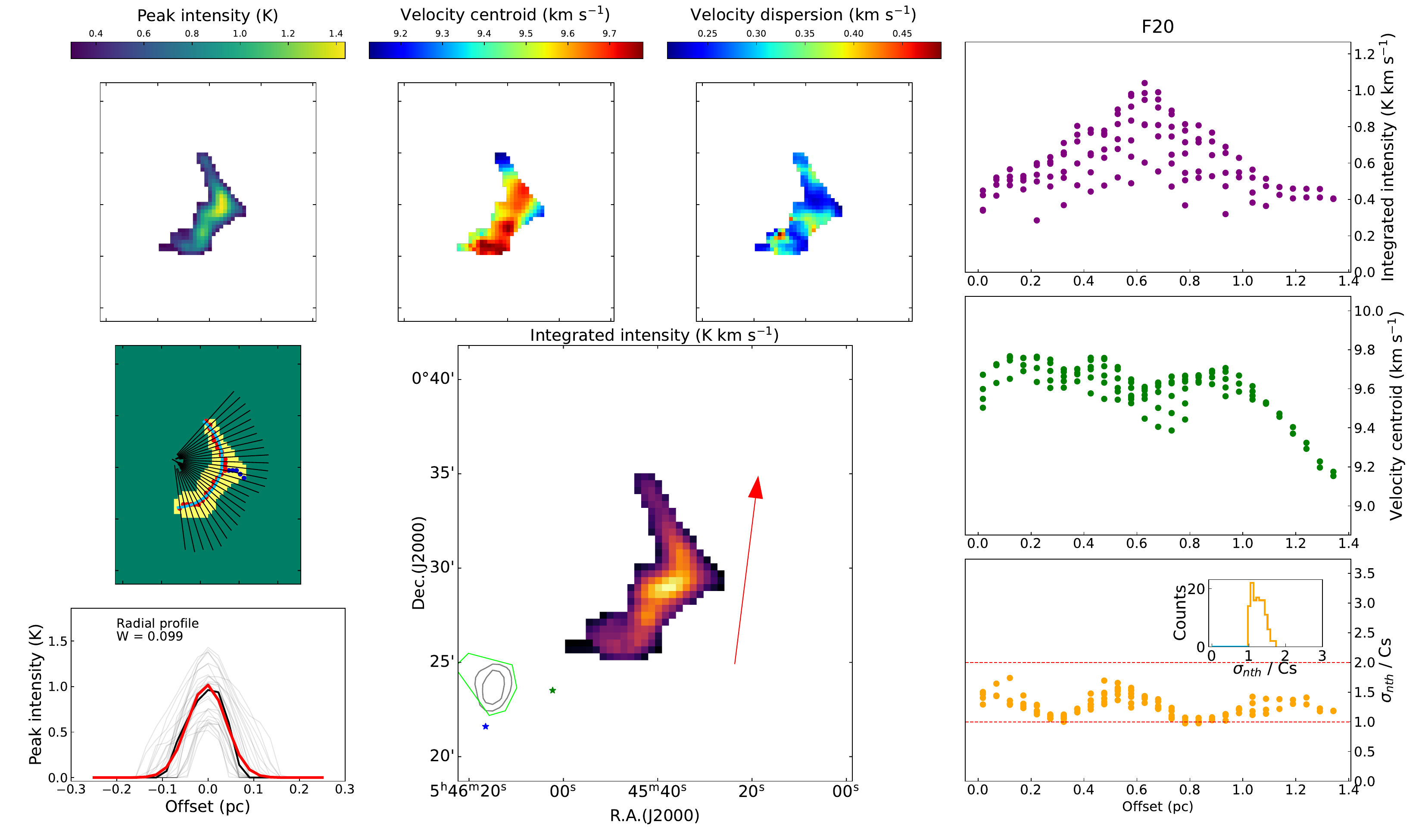}
\figsetgrpnote{Filament quantities}
\figsetgrpend

\figsetgrpstart
\figsetgrpnum{A.23}
\figsetgrptitle{Filament F21}
\figsetplot{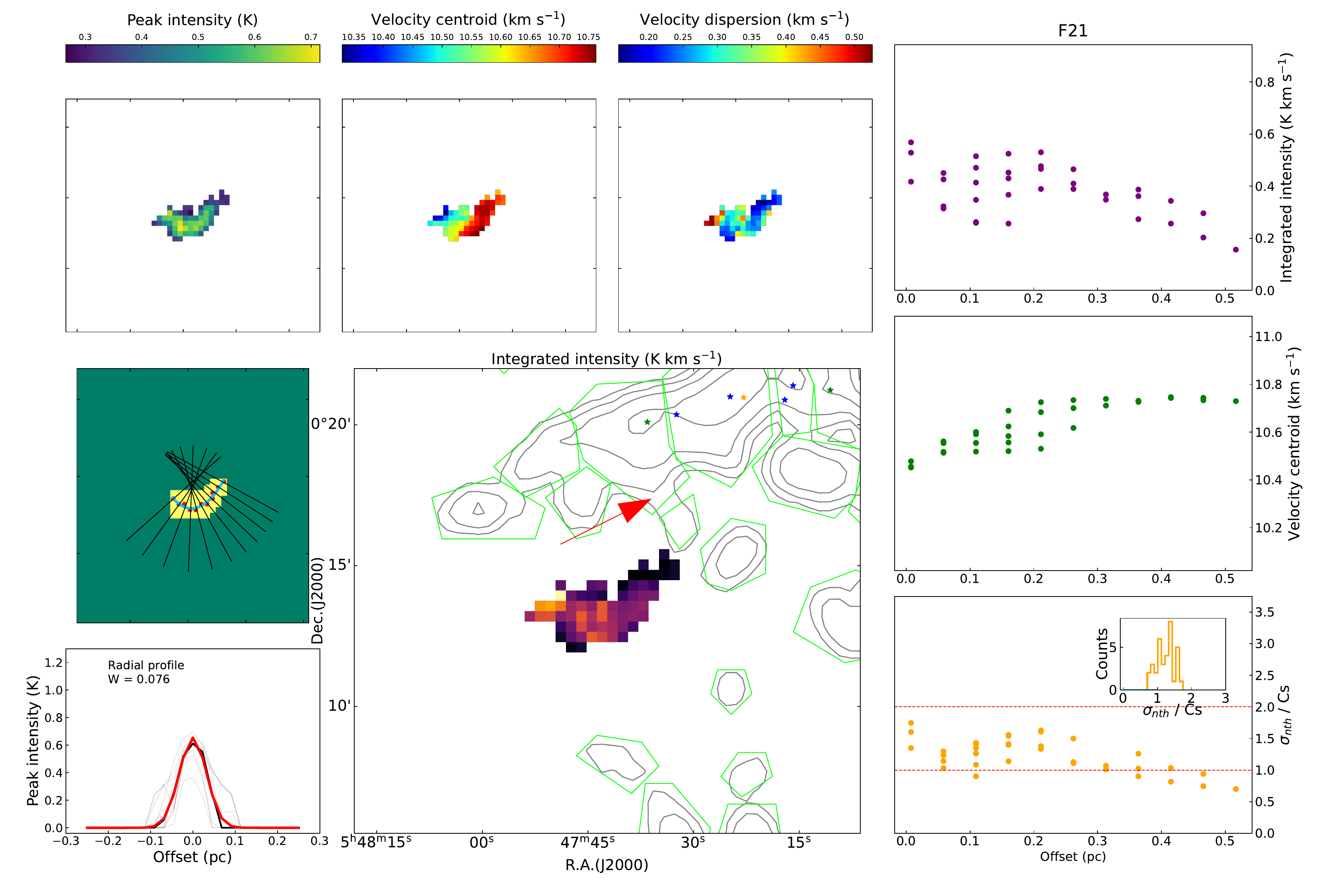}
\figsetgrpnote{Filament quantities}
\figsetgrpend

\figsetgrpstart
\figsetgrpnum{A.24}
\figsetgrptitle{Filament F22}
\figsetplot{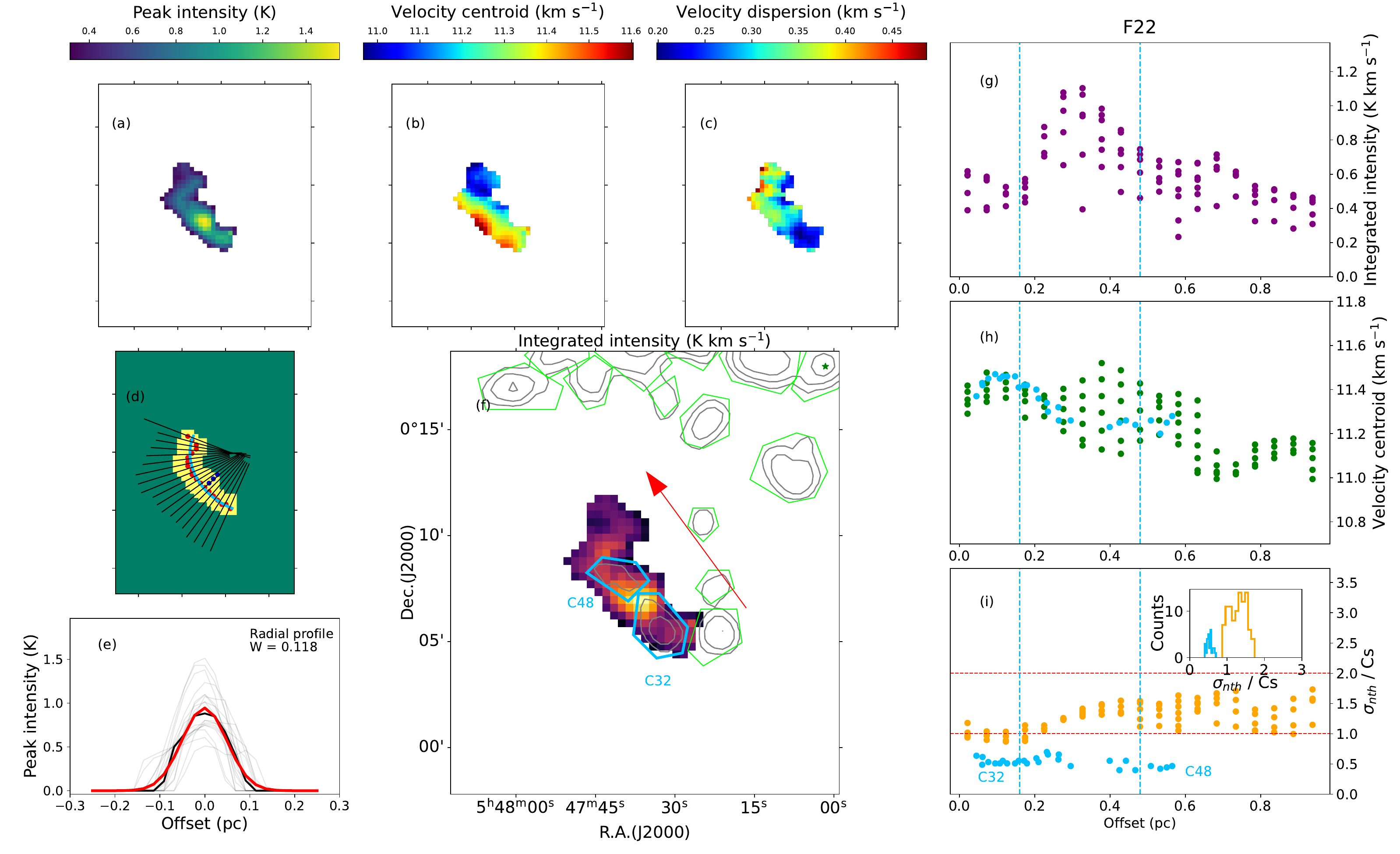}
\figsetgrpnote{Filament quantities}
\figsetgrpend

\figsetgrpstart
\figsetgrpnum{A.25}
\figsetgrptitle{Filament F23}
\figsetplot{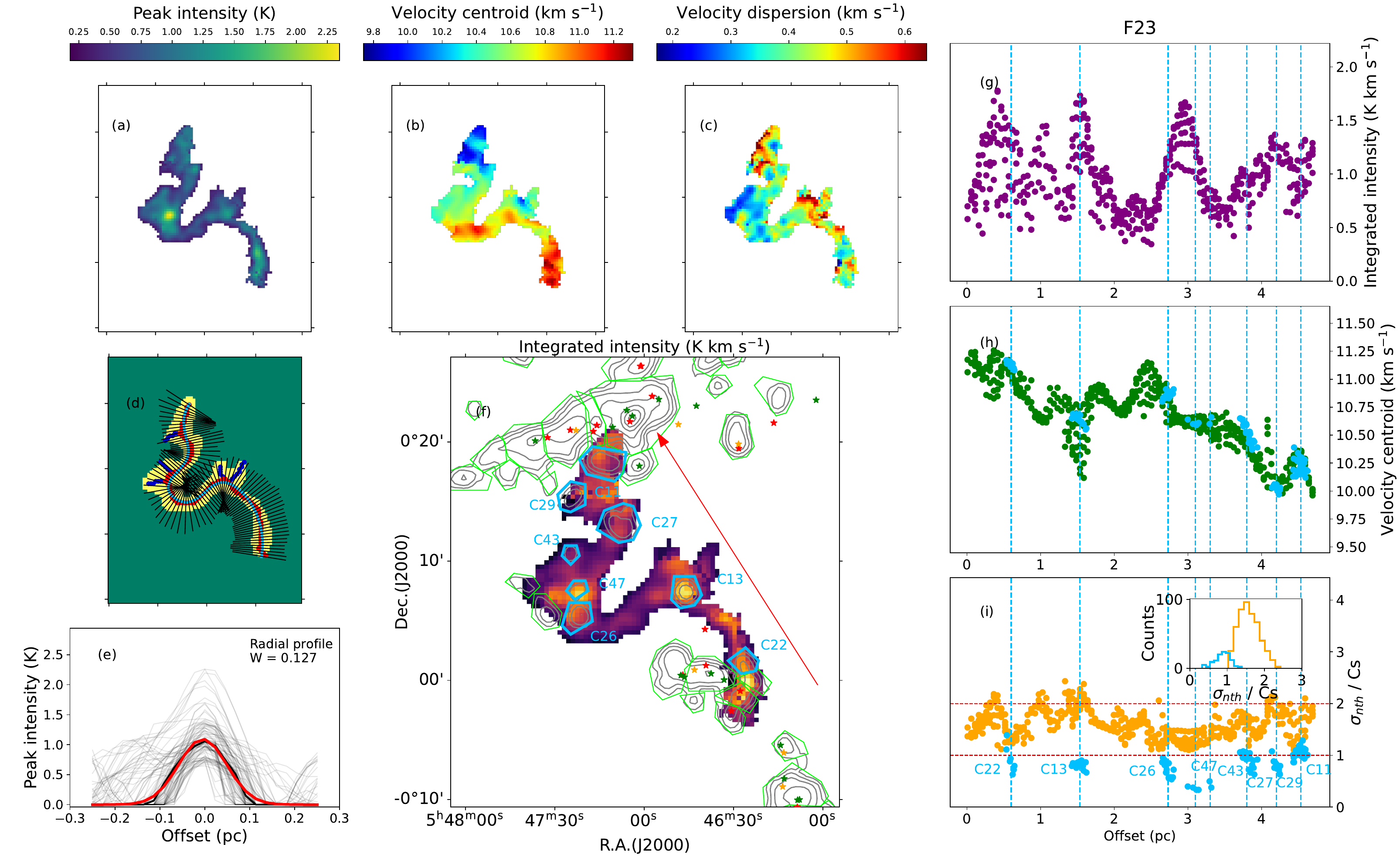}
\figsetgrpnote{Filament quantities}
\figsetgrpend

\figsetgrpstart
\figsetgrpnum{A.26}
\figsetgrptitle{Filament F24}
\figsetplot{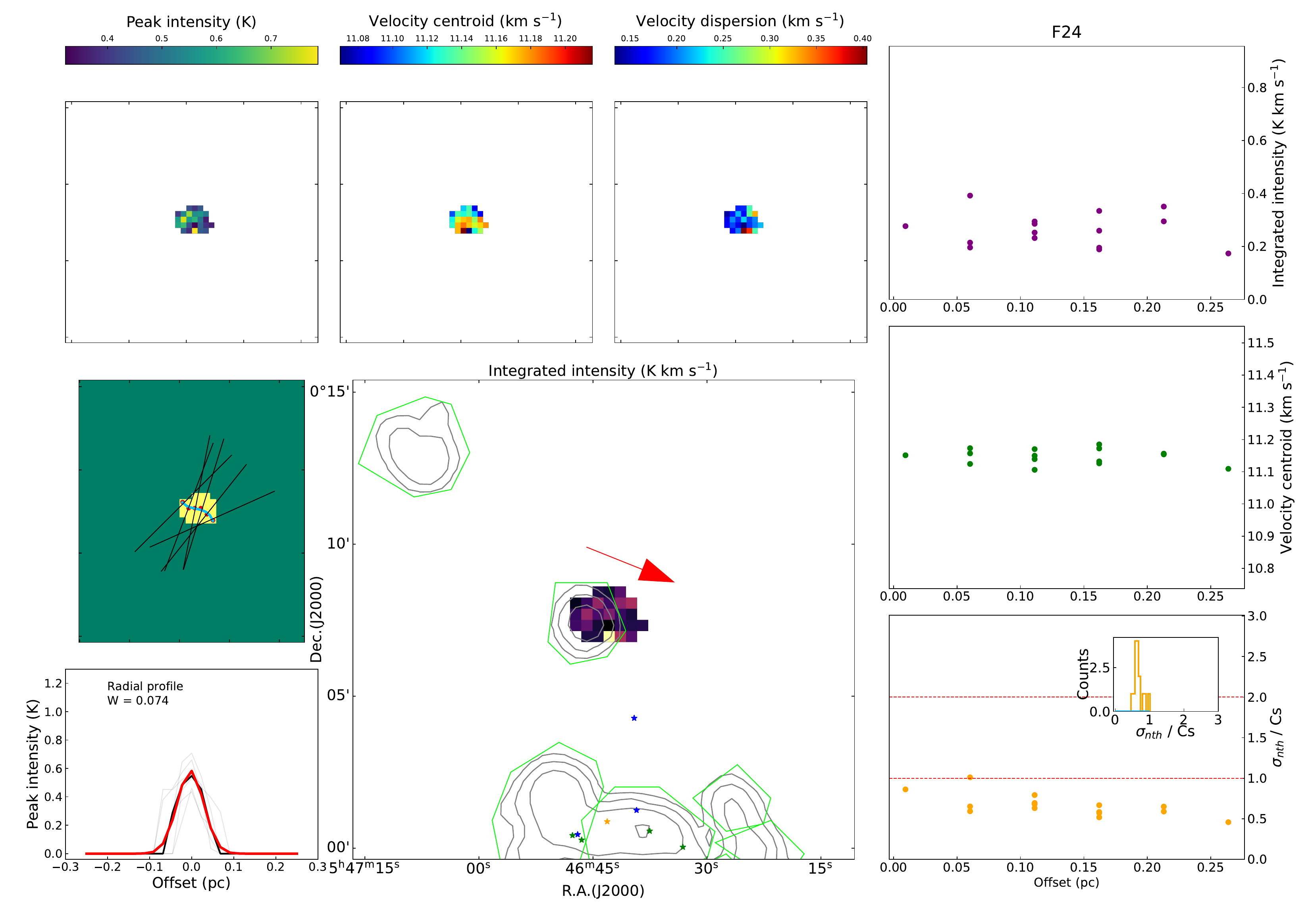}
\figsetgrpnote{Filament quantities}
\figsetgrpend

\figsetgrpstart
\figsetgrpnum{A.27}
\figsetgrptitle{Filament F25}
\figsetplot{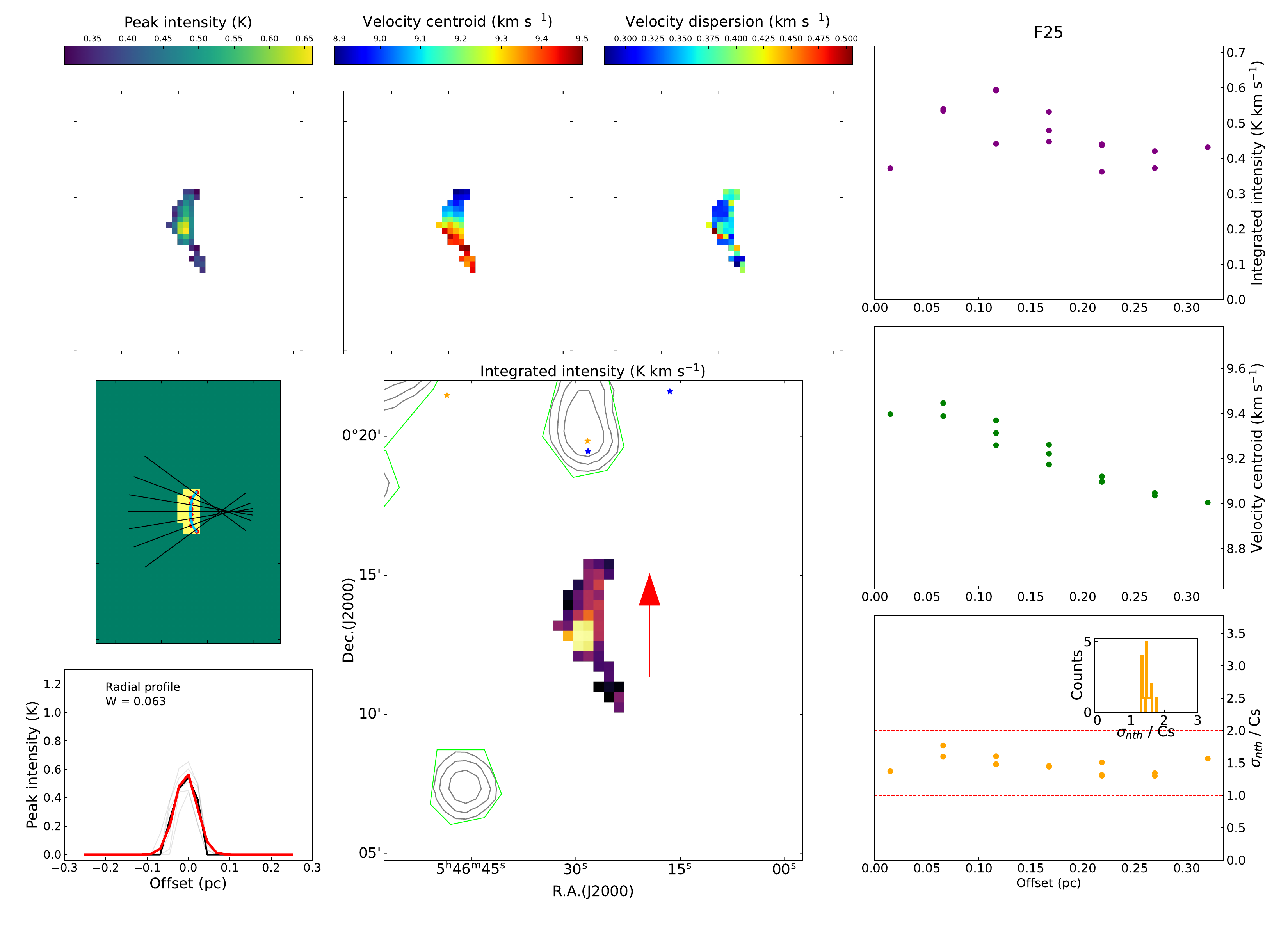}
\figsetgrpnote{Filament quantities}
\figsetgrpend

\figsetgrpstart
\figsetgrpnum{A.28}
\figsetgrptitle{Filament F26}
\figsetplot{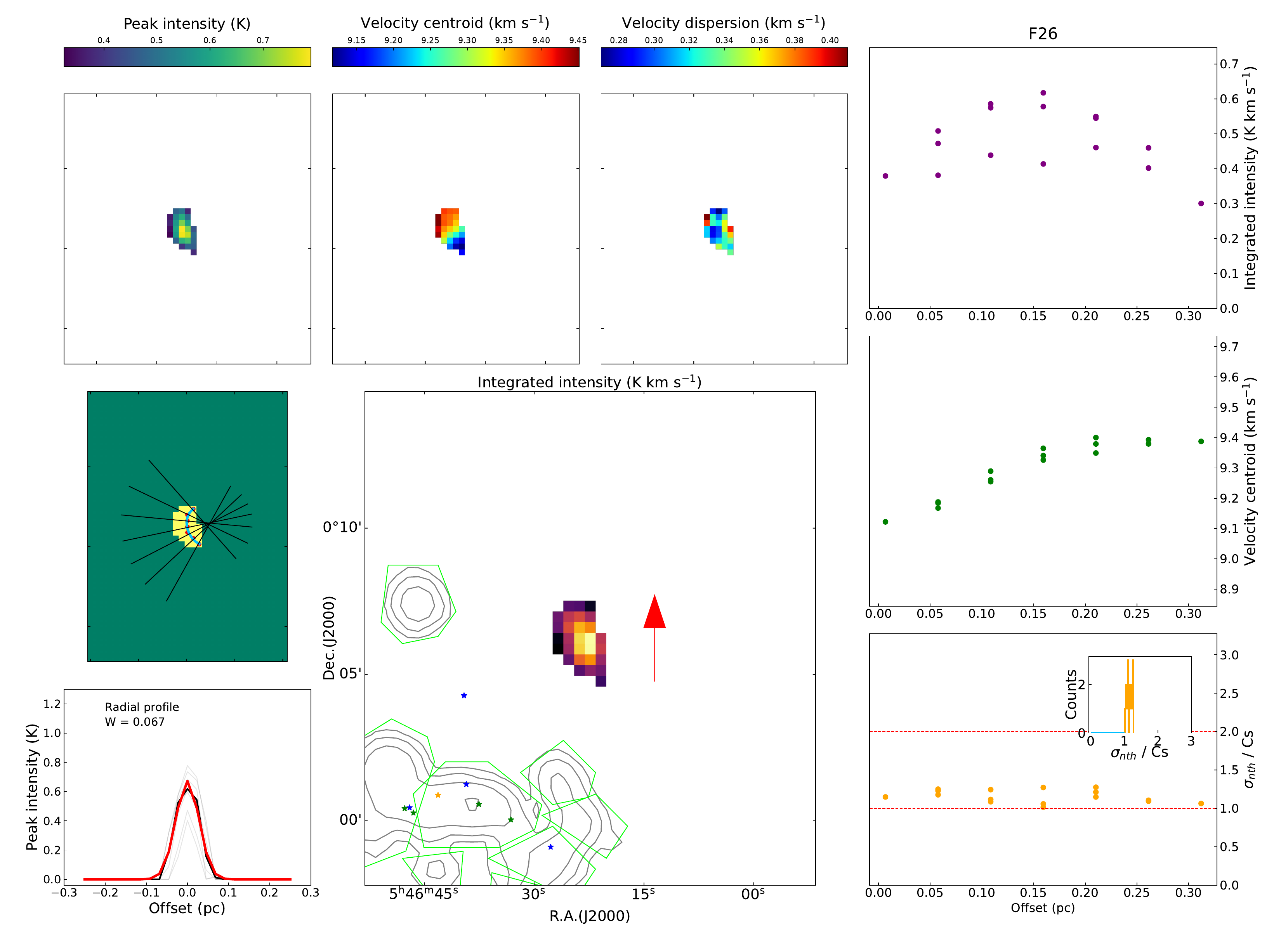}
\figsetgrpnote{Filament quantities}
\figsetgrpend

\figsetgrpstart
\figsetgrpnum{A.29}
\figsetgrptitle{Filament F27}
\figsetplot{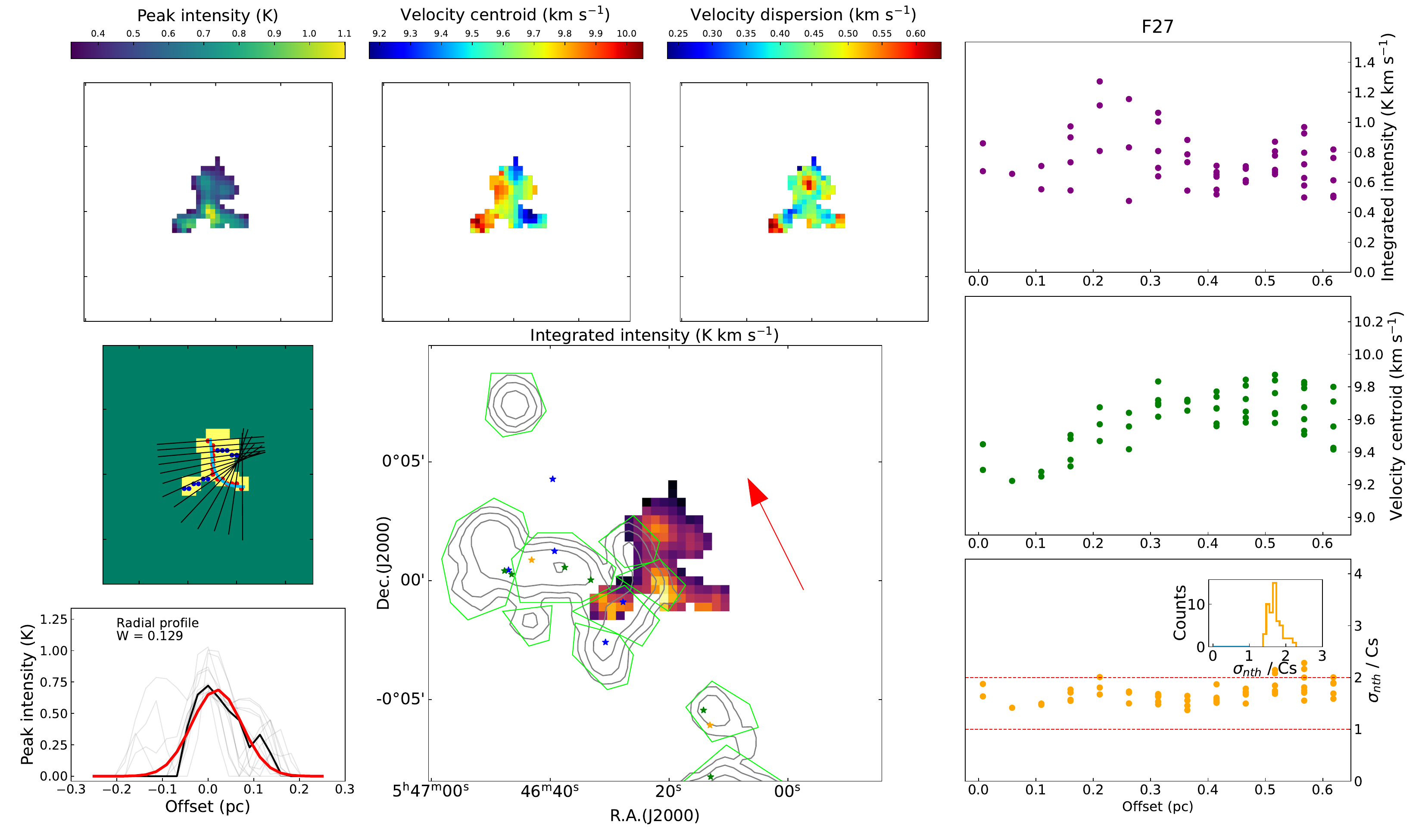}
\figsetgrpnote{Filament quantities}
\figsetgrpend

\figsetgrpstart
\figsetgrpnum{A.30}
\figsetgrptitle{Filament F28}
\figsetplot{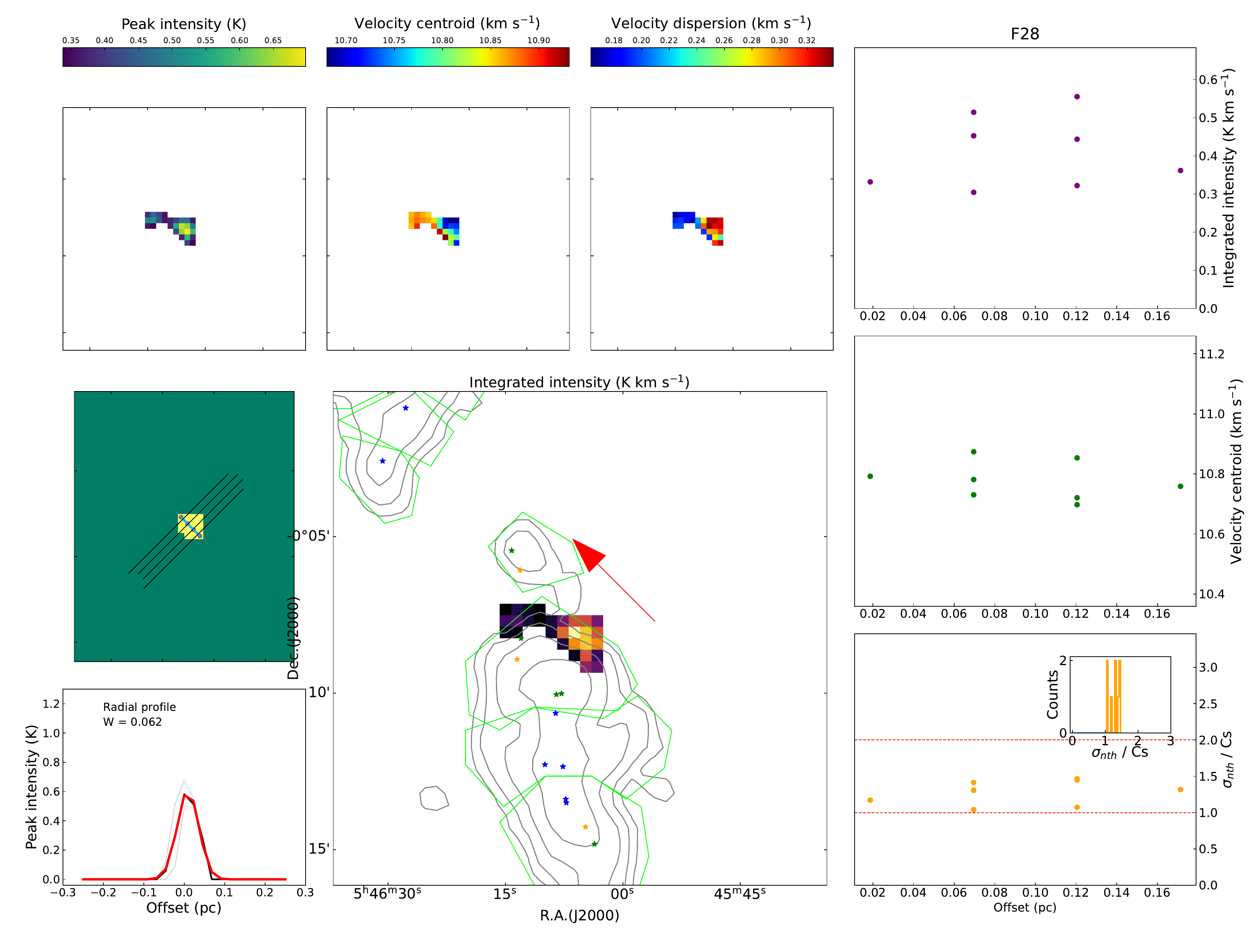}
\figsetgrpnote{Filament quantities}
\figsetgrpend

\figsetgrpstart
\figsetgrpnum{A.31}
\figsetgrptitle{Filament F29}
\figsetplot{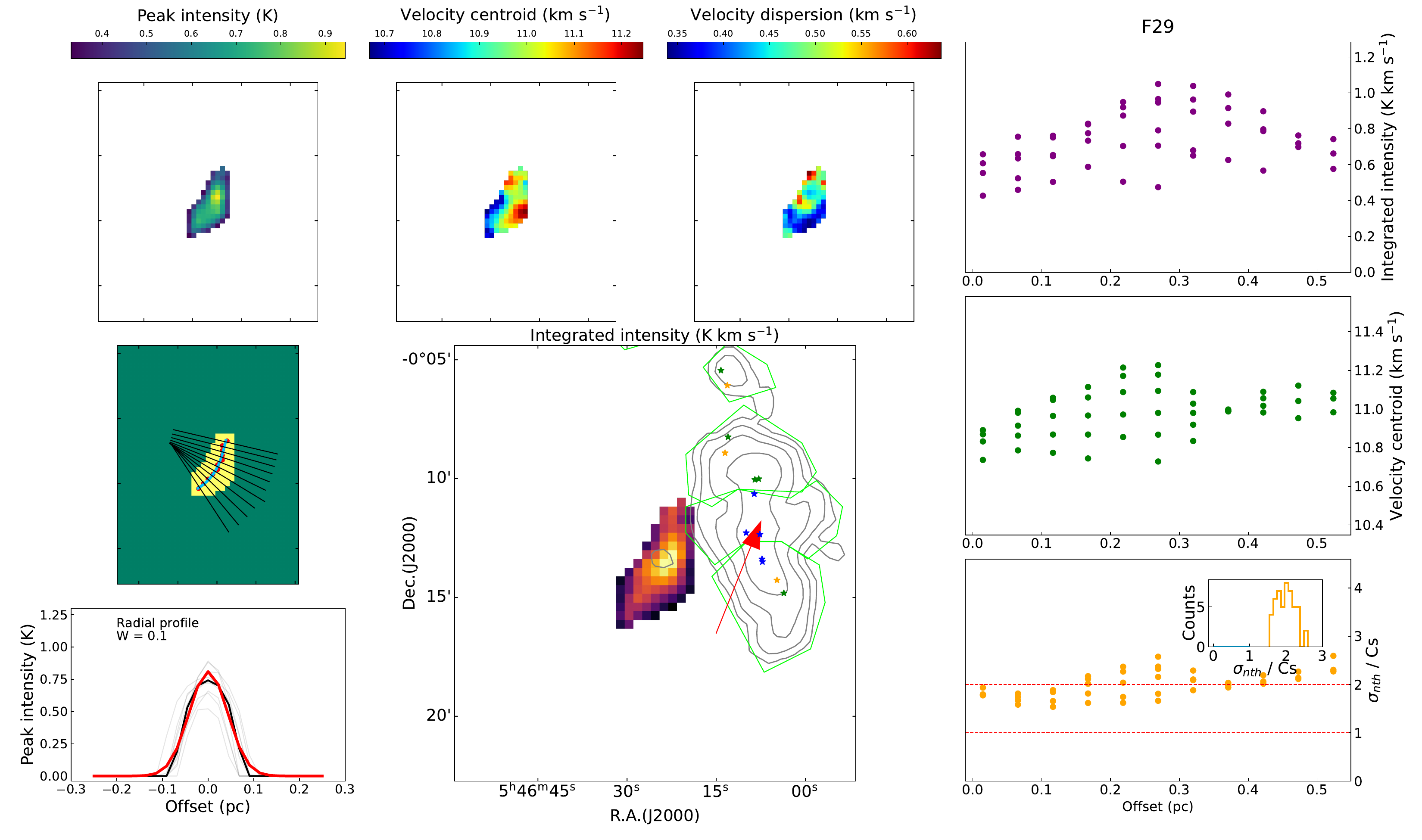}
\figsetgrpnote{Filament quantities}
\figsetgrpend

\figsetgrpstart
\figsetgrpnum{A.32}
\figsetgrptitle{Filament F30}
\figsetplot{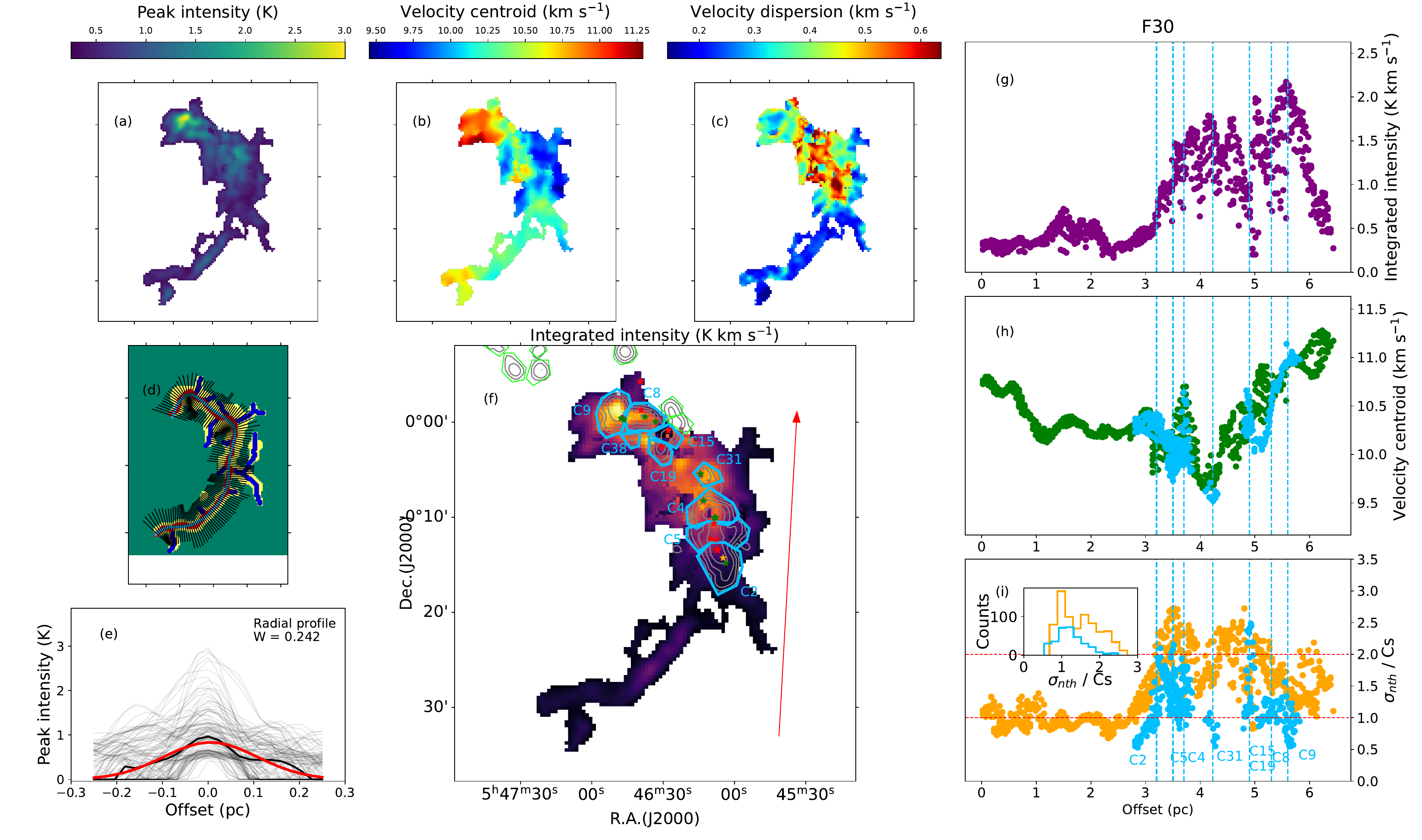}
\figsetgrpnote{Filament quantities}
\figsetgrpend

\figsetend

\begin{figure}
\figurenum{A.18}
\epsscale{1.12}
\plotone{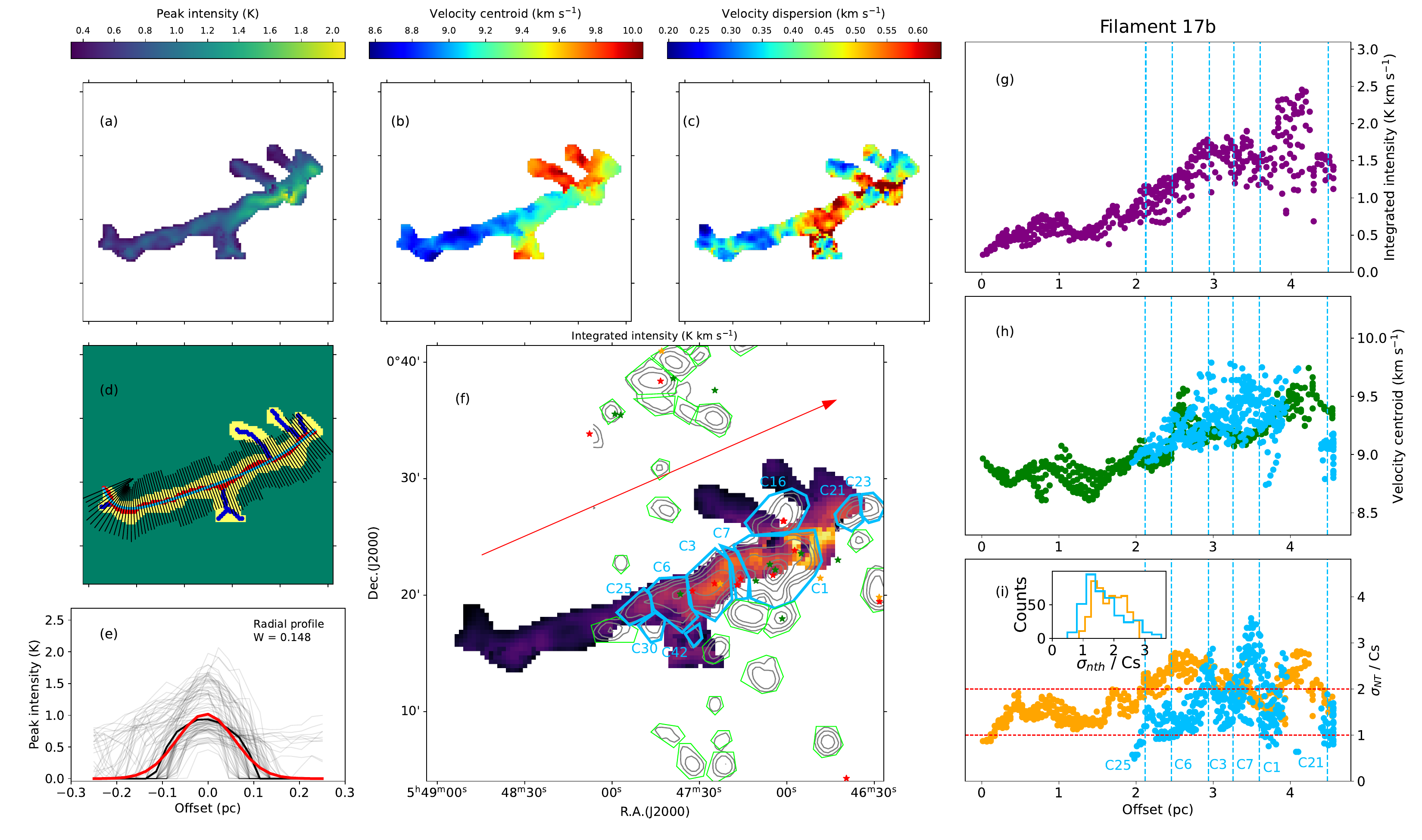}
\label{fig:fA18}
\caption{Physical quantities of filament F17b. 
The complete figure set (32 images) is available in the online journal.}
\end{figure}

\figsetend

\restartappendixnumbering
\section{Distribution of \nThp\ dense cores with respect to \cEo\ filaments} \label{A2}
\begin{figure*}
 \begin{center}
  \includegraphics[width=0.9\textwidth]{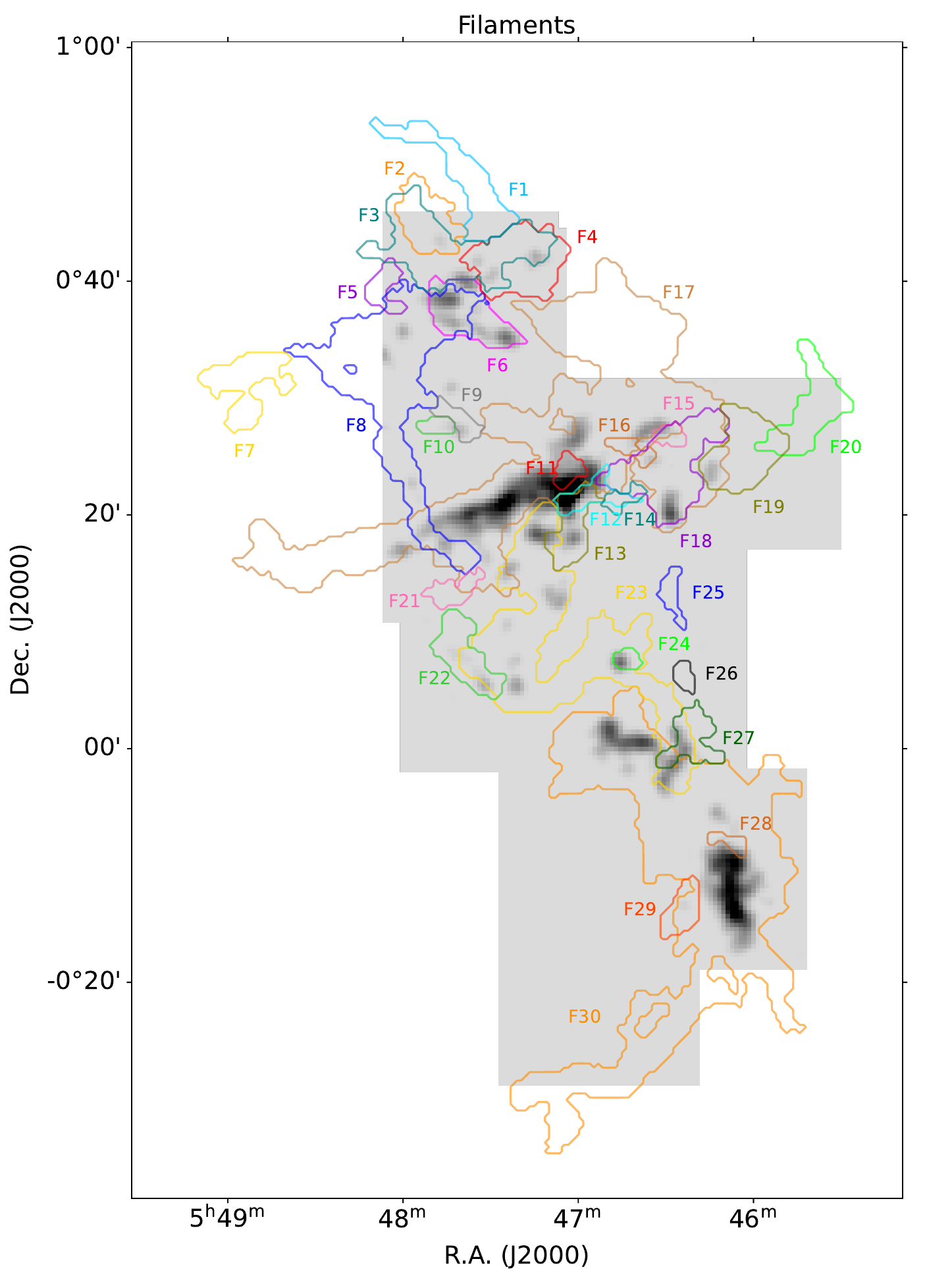}
 \end{center}
\caption{Distribution of filaments and the N$_2$H$^+$ cores. The contours in various colors are the same as filaments presented in Figure~\ref{fig:fil}(a) and the shade box and gray tones indicate the \nThp~observed region and \nThpj~integrated intensity. }
\label{fig:dist_fil_core}
\end{figure*}

Here we present supplementary information on the filaments and the 48 \nThp\ dense cores. 
Figure~\ref{fig:dist_fil_core} shows the distribution of filaments identified from \cEoj\ and the N$_2$H$^+$ cores described with \nThpj\ integrated intensity image.

\figsetstart
\figsetnum{B2}
\label{fig:figB2}
\figsettitle{Comparison between the \cEo\ spectrum and \nThp\ spectrum at the position of each dense core to explore the relation between LSR velocities of filamentary structure and that of dense cores along the line-of-sight.}

\figsetgrpstart
\figsetgrpnum{B2.1}
\figsetgrptitle{fB1}
\figsetplot{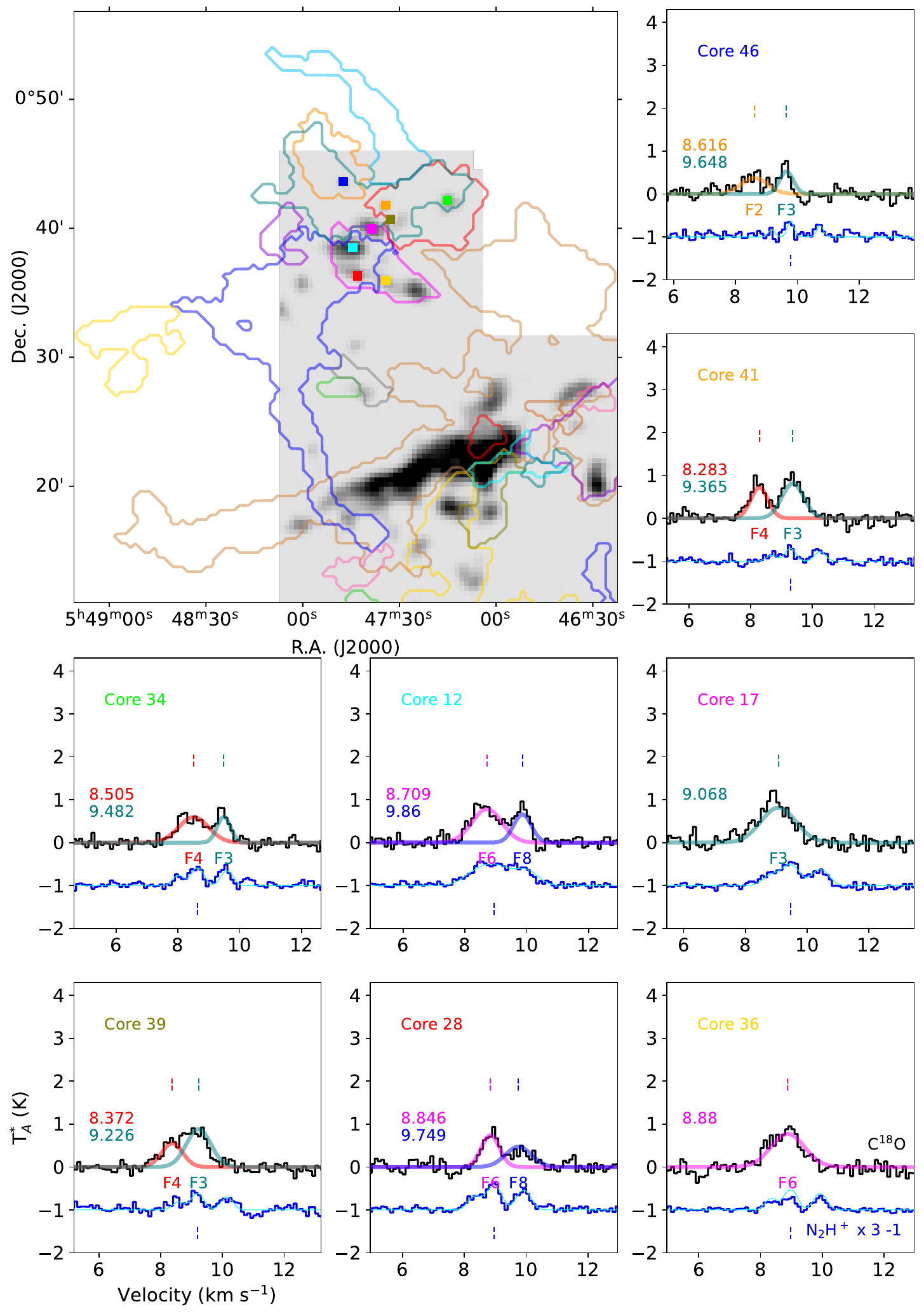}
\figsetgrpnote{The spectra for the filaments and dense cores}
\figsetgrpend

\figsetgrpstart
\figsetgrpnum{B2.2}
\figsetgrptitle{fB2}
\figsetplot{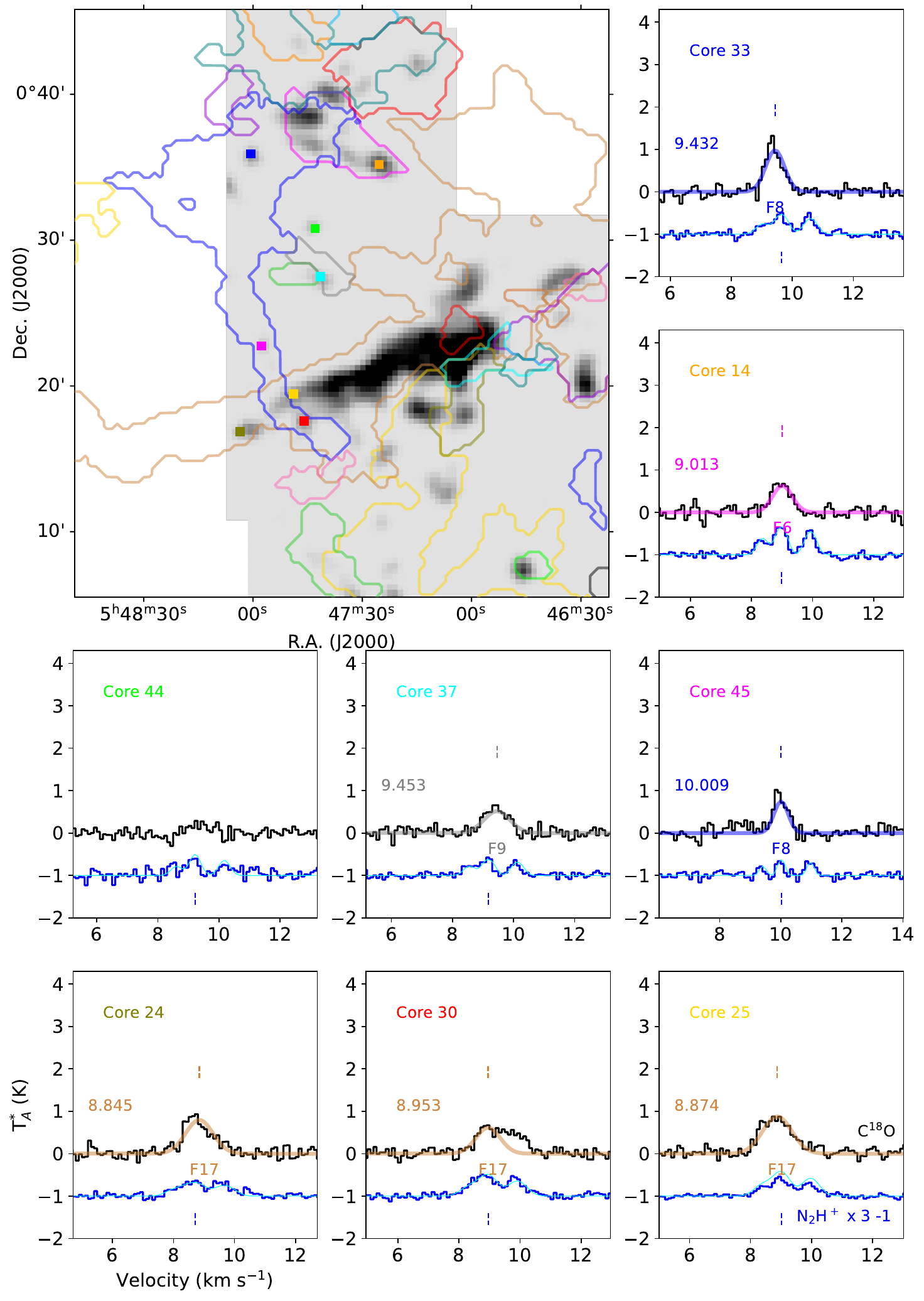}
\figsetgrpnote{The spectra for the filaments and dense cores}
\figsetgrpend

\figsetgrpstart
\figsetgrpnum{B2.3}
\figsetgrptitle{fB3}
\figsetplot{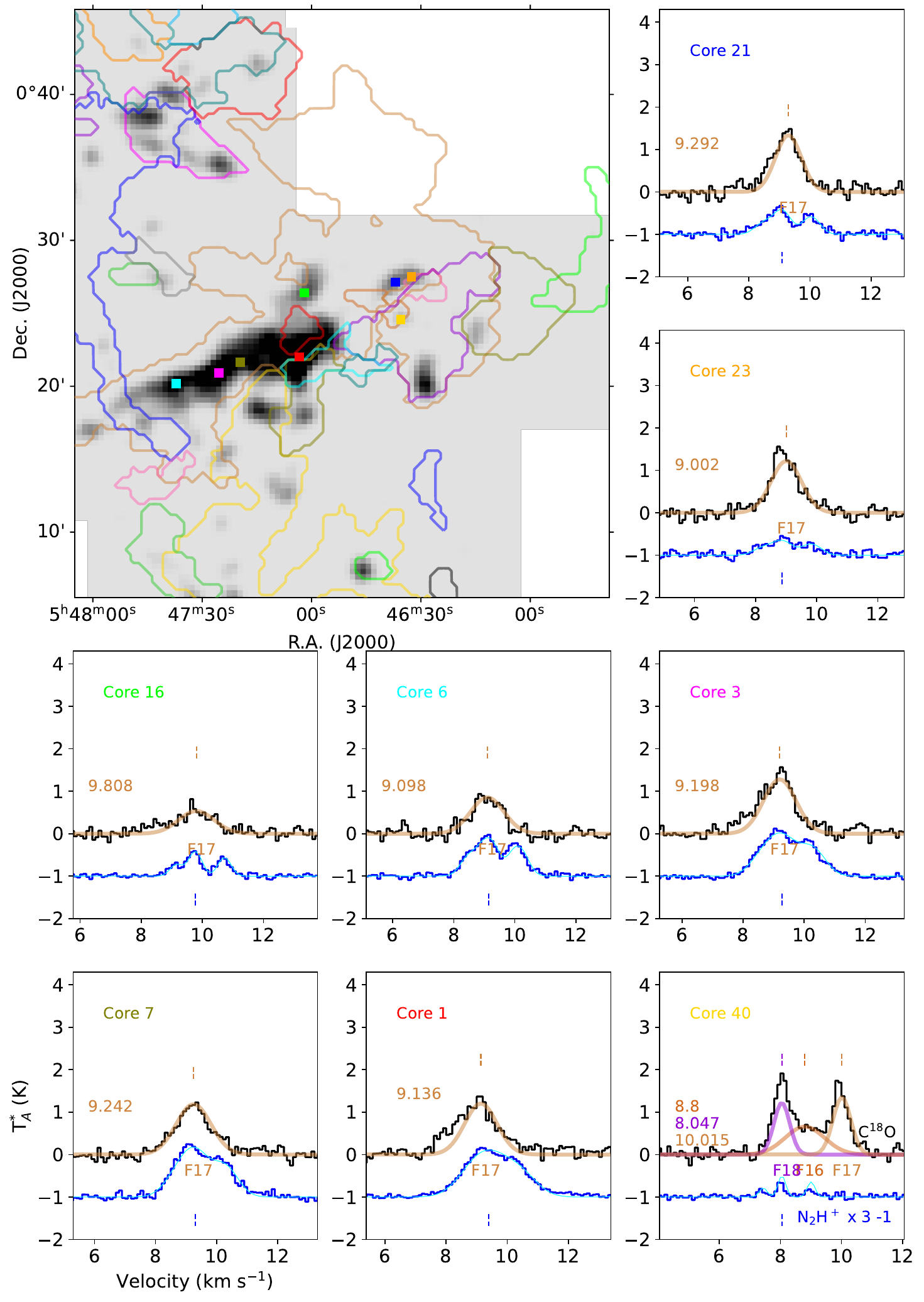}
\figsetgrpnote{The spectra for the filaments and dense cores}
\figsetgrpend

\figsetgrpstart
\figsetgrpnum{B2.4}
\figsetgrptitle{fB4}
\figsetplot{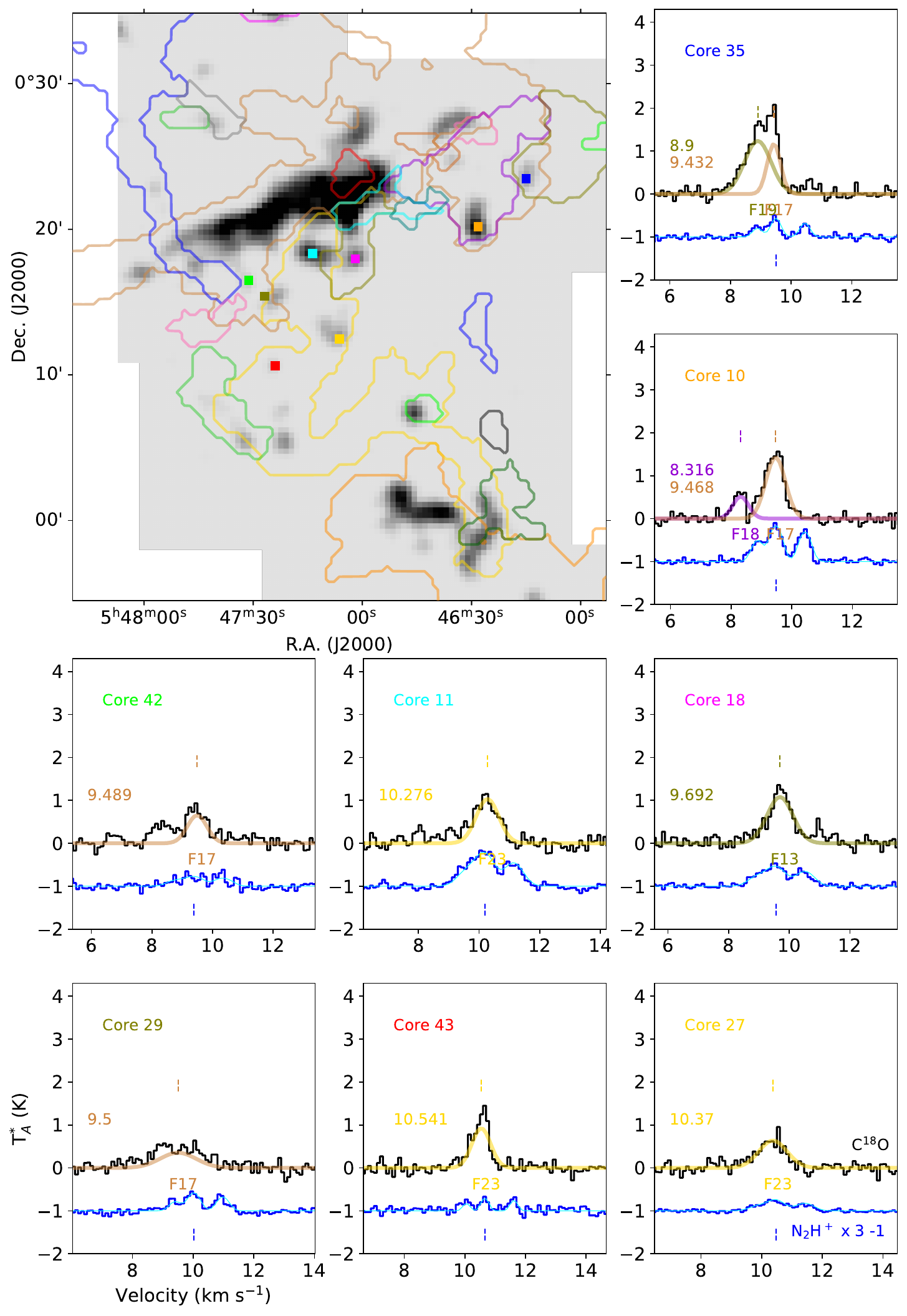}
\figsetgrpnote{The spectra for the filaments and dense cores}
\figsetgrpend

\figsetgrpstart
\figsetgrpnum{B2.5}
\figsetgrptitle{fB5}
\figsetplot{FigureB5.pdf}
\figsetgrpnote{The spectra for the filaments and dense cores}
\figsetgrpend

\figsetgrpstart
\figsetgrpnum{B2.6}
\figsetgrptitle{fB6}
\figsetplot{FigureB6.pdf}
\figsetgrpnote{The spectra for the filaments and dense cores}
\figsetgrpend

\figsetend

\begin{figure*}
\centering
\figurenum{B2.1}
\includegraphics[width=0.75\textwidth]{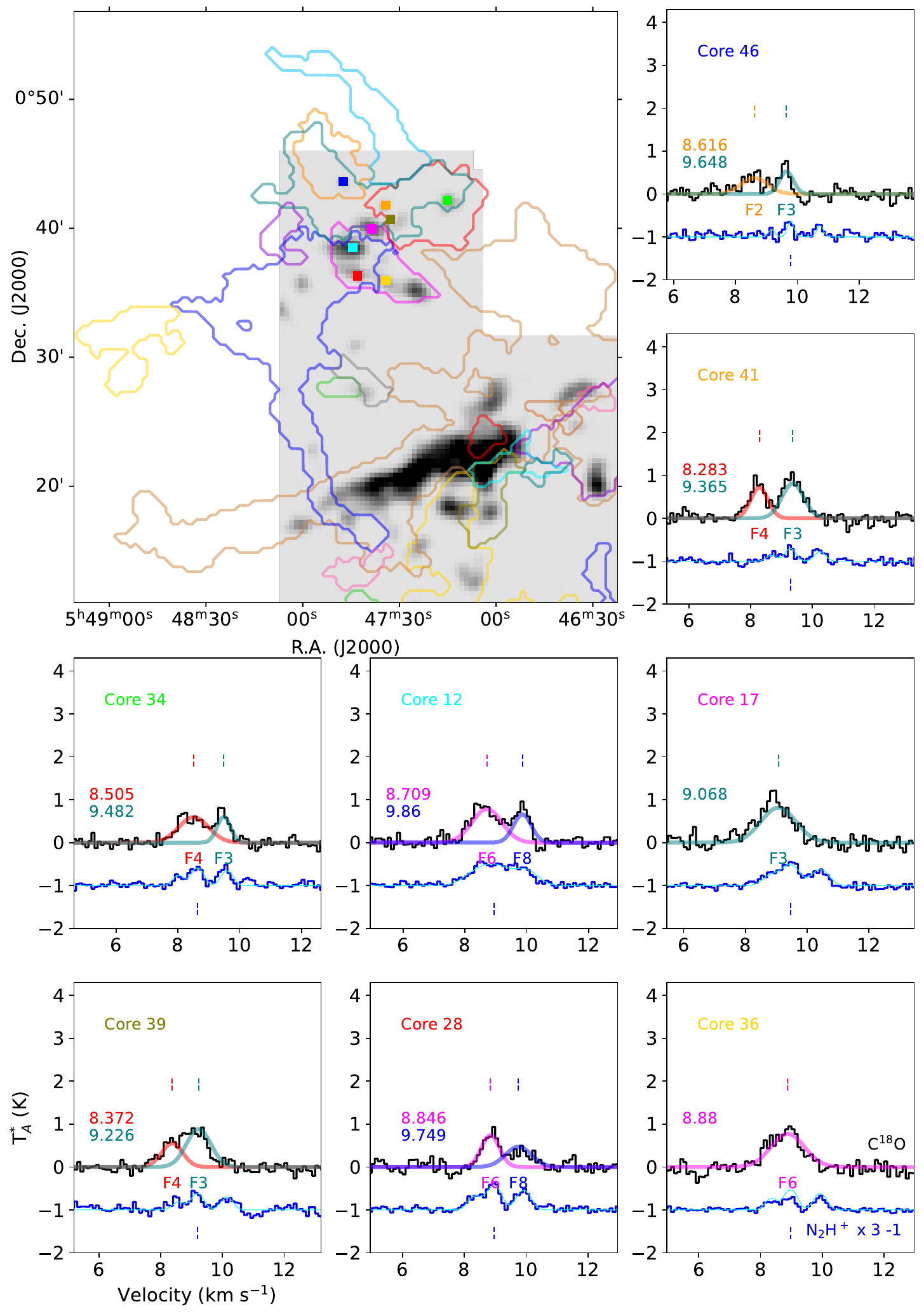}
\caption{The spectra for the filaments and dense cores. The top-left panel displays the distribution of the filaments (colored contour lines) and the positions of the dense cores (colored square symbols) on the \nThpj~integrated intensity map (gray scale). Eight small panels show the \cEoj~spectra (upper black line) at the position of the peak intensity of the dense cores and averaged \nThp~profiles (lower blue line) for the dense cores. We multiplied 3 for the \nThp~spectrum with \textminus1 K offset for visualization. The color codes of filaments in the top-left panel correspond to those for the decomposed \cEo~line profile and the filament name in the small panels. The positions of velocity centroids of each decomposed profile are marked with the small vertical dashed lines and numbers above the profile. The filament IDs are shown below the profile. The hyperfine structure fit result of \nThpj\ is drawn on its profile with cyan line. 
The complete figure set (6 images) is available in the online journal.}

\end{figure*}

\restartappendixnumbering
\section{\cs, \hcop, \cEo, \hTcop, and \nThp\ line spectra for 48 dense cores} \label{A3}
We present the distributions of \hcopj, \hTcopj, and \csj\ in Figure~\ref{fig:hcop} and the averaged spectra of \csj, \hcopj, \cEoj, \hTcopj, and \nThpj~lines for dense cores in Appendix~\ref{A3}. 
As for \cs~line, we could not find any significant infall signature toward dense cores in this study, although \cs~line was known to one of the best infall tracers \citep{lee99, lee01}. 

Figure~\ref{fig:line_spec1} to Figure~\ref{fig:line_spec3} show \csj, \hcopj, \cEoj, \hTcopj~and \nThpj~ molecular line profiles for 48 dense cores. All the profiles shown here are the average one for the spectra over the dense core. We do not present spectra of several dense cores if the transition is not observed or the observed emission has S/N $<$ 3. The core names are labeled on the top of each panel.

\begin{figure*}
  \includegraphics[width=0.98\textwidth, angle=0]{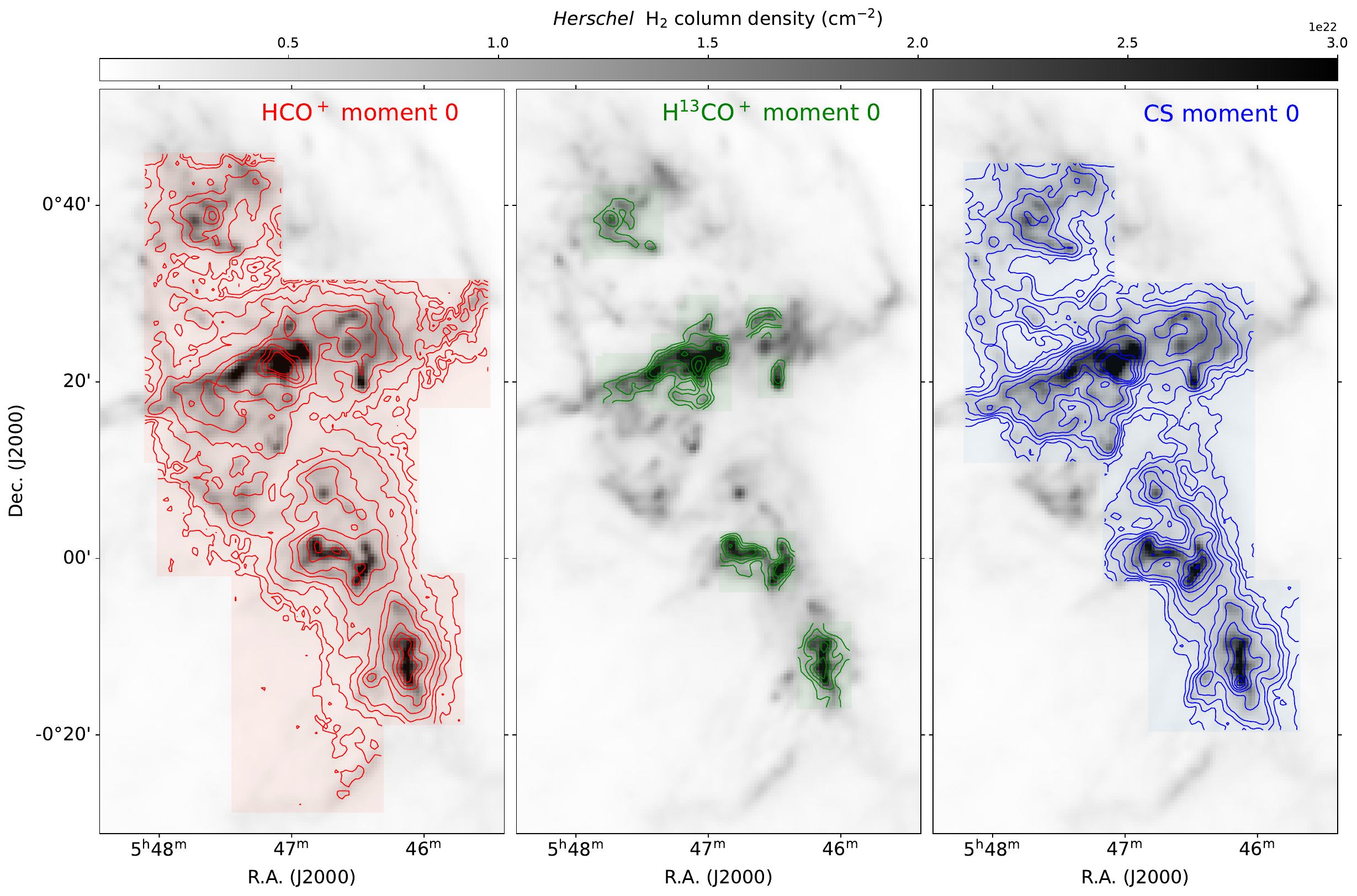}
  \caption{The integrated intensity maps of \hcopj~(red), \hTcopj~(green), and \csj~(blue) lines. The contour levels for \hcopj\ are 0.041 K \kms~$\times$ [5, 10, 15, 20, 30, 40, 70, 100, 130], those for \hTcopj\ are 0.026 K \kms~$\times$ [5, 8, 13, 18, 25, 30, 40], and those for  \csj\ are 0.037 K \kms~$\times$ [5, 10, 15, 20, 30, 45, 60, 80, 100]. The gray scale and color shaded regions in the background indicate Herschel H$_2$ column density distribution and  observing regions for each line, respectively. 
\label{fig:hcop}}
\end{figure*}

\begin{figure*}
\begin{tabbing}
  \includegraphics[width=0.25\textwidth]{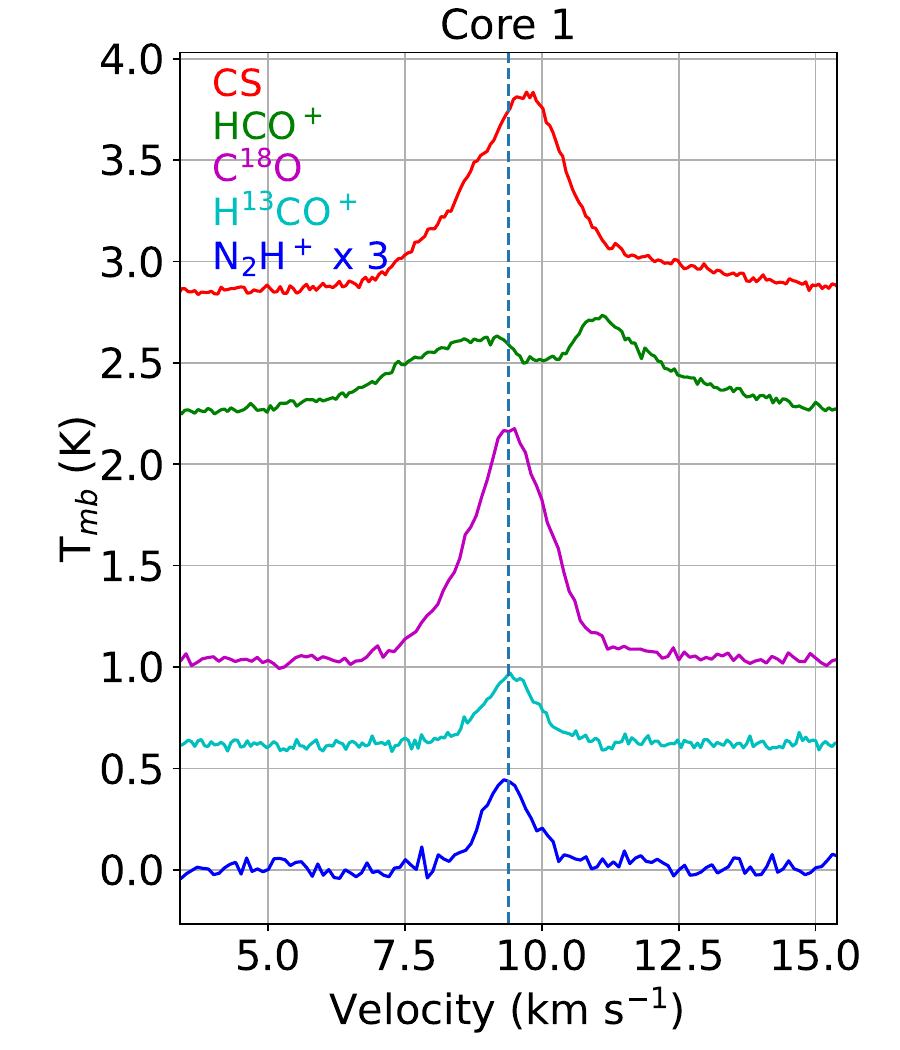}
  \includegraphics[width=0.25\textwidth]{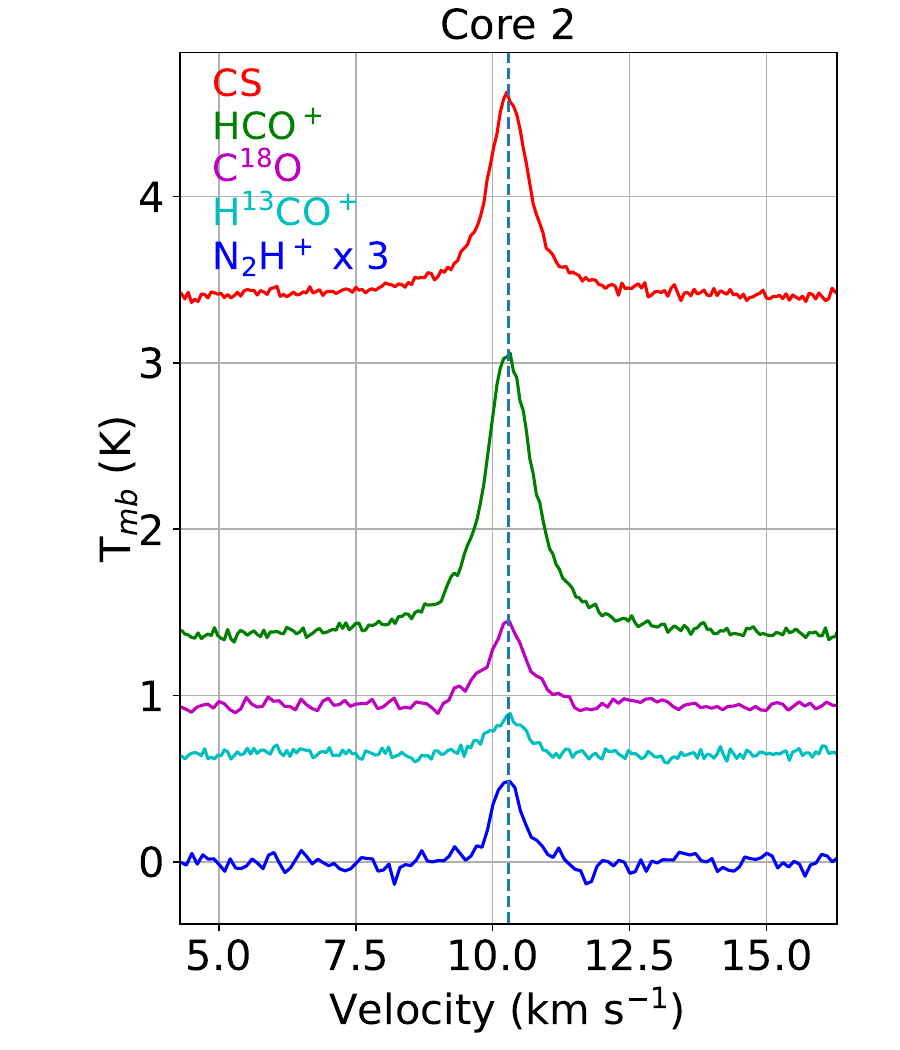} 
  \includegraphics[width=0.25\textwidth]{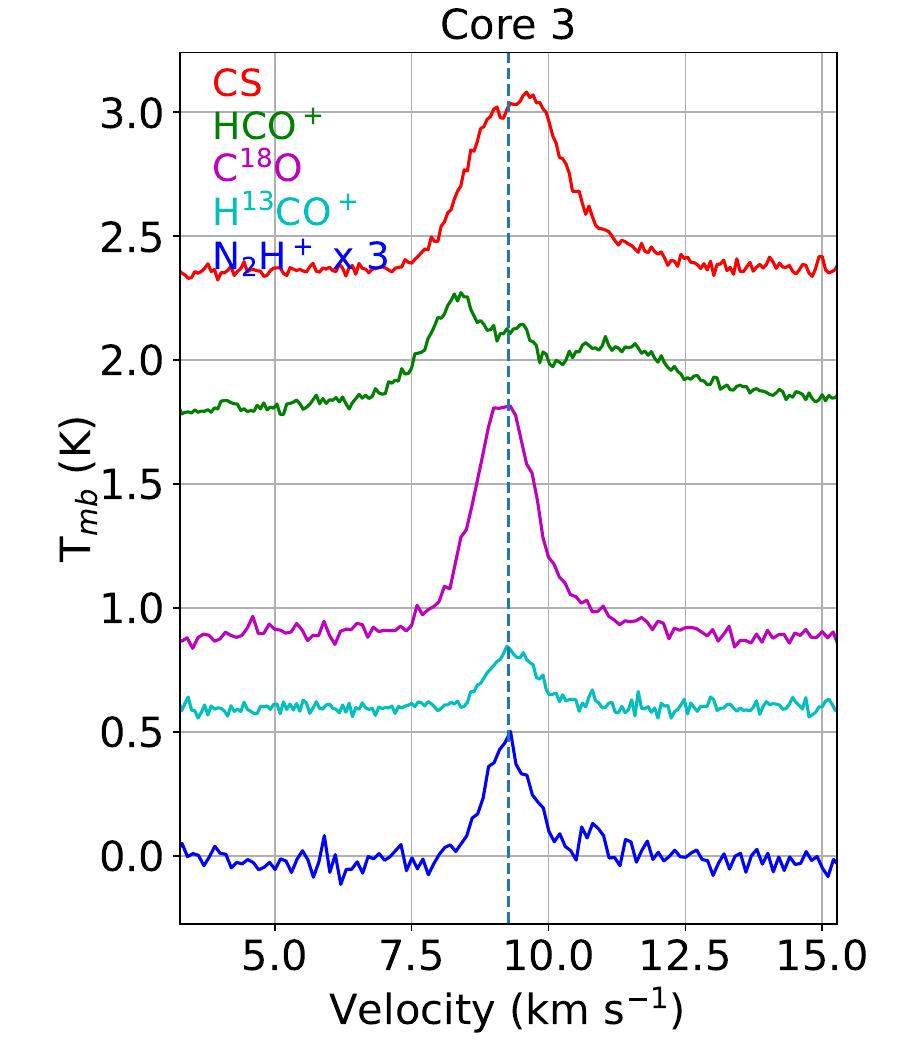}  
  \includegraphics[width=0.25\textwidth]{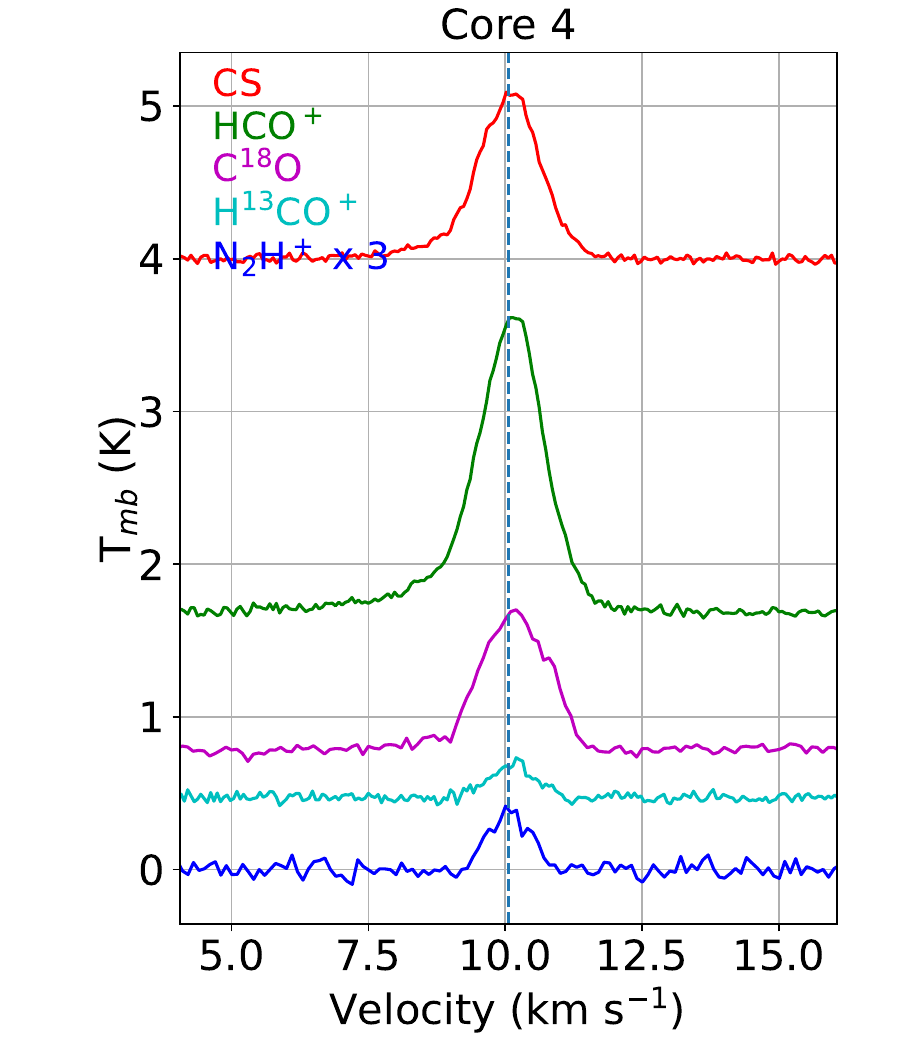} \\
  \includegraphics[width=0.25\textwidth]{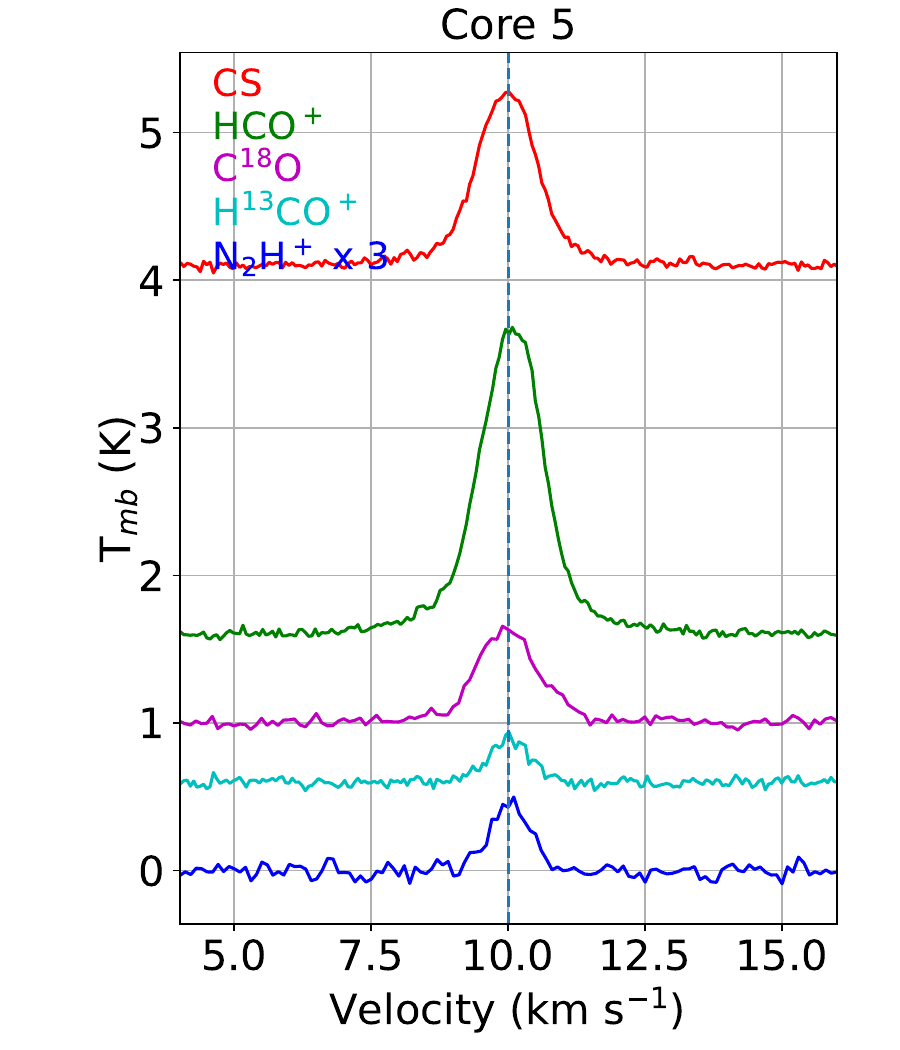} 
  \includegraphics[width=0.25\textwidth]{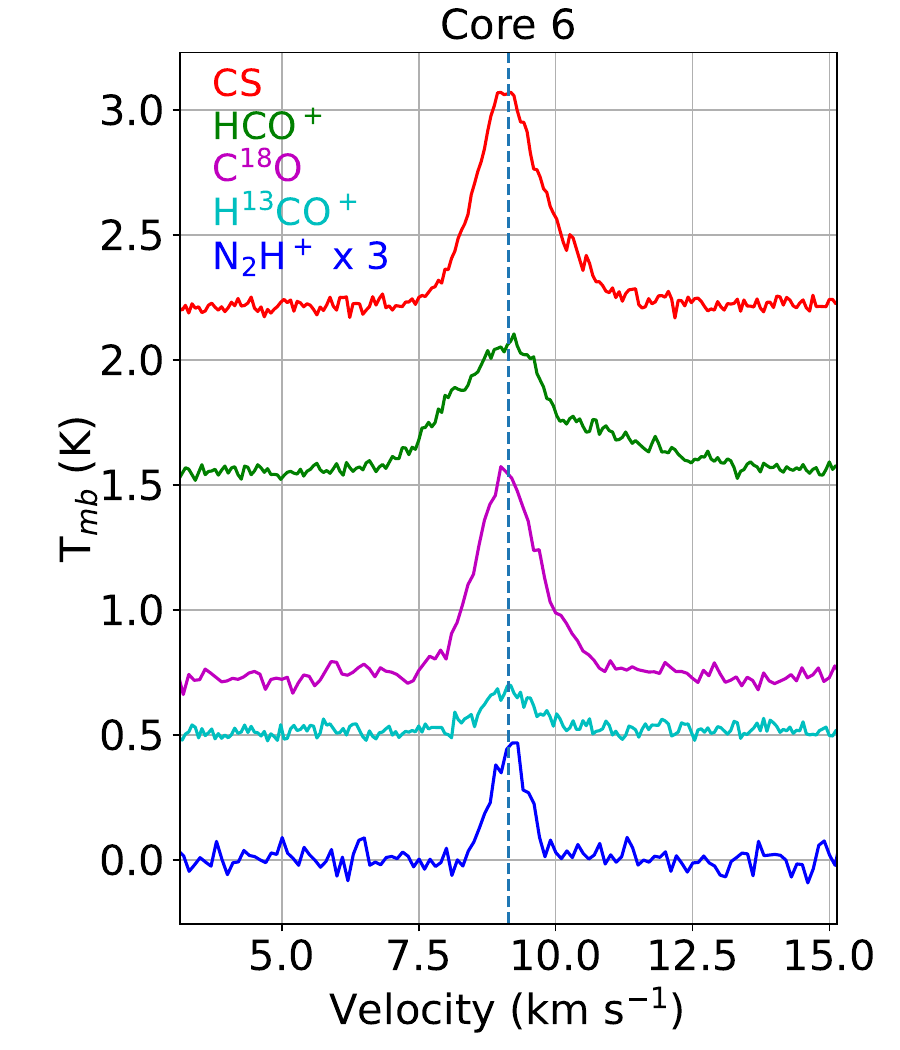}
  \includegraphics[width=0.25\textwidth]{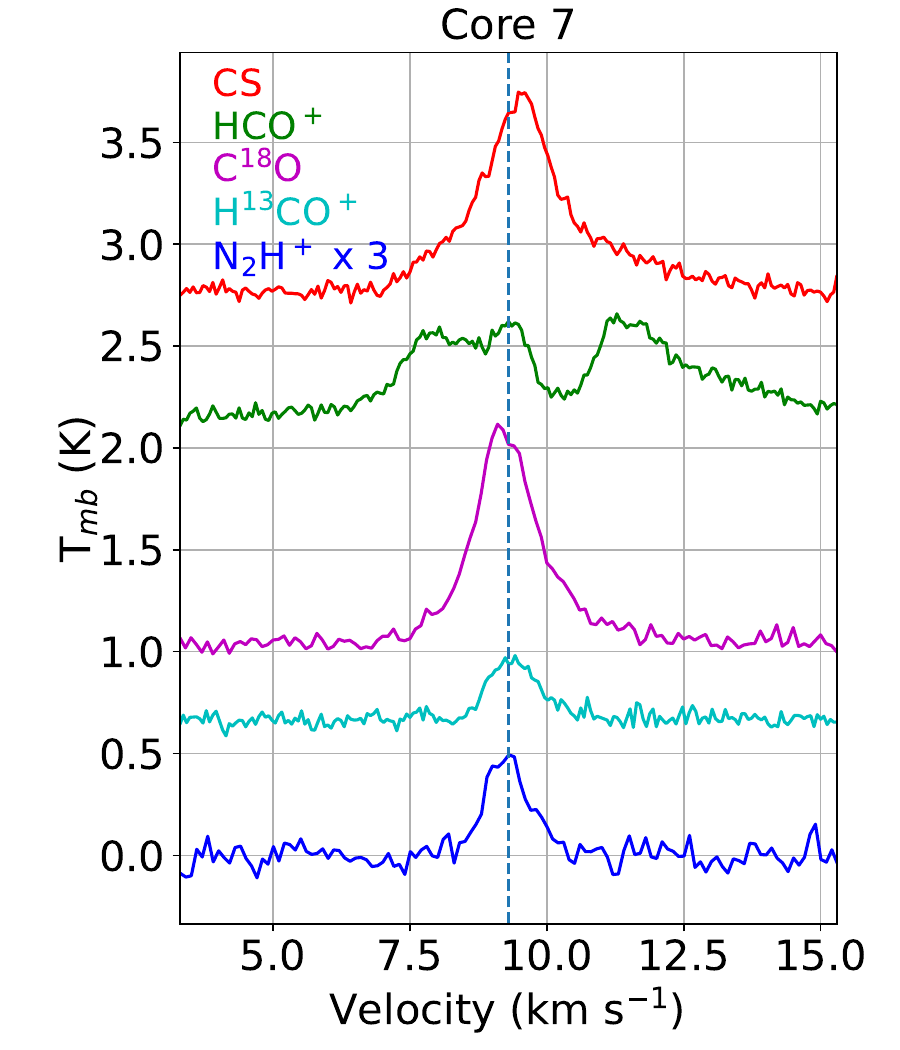}
  \includegraphics[width=0.25\textwidth]{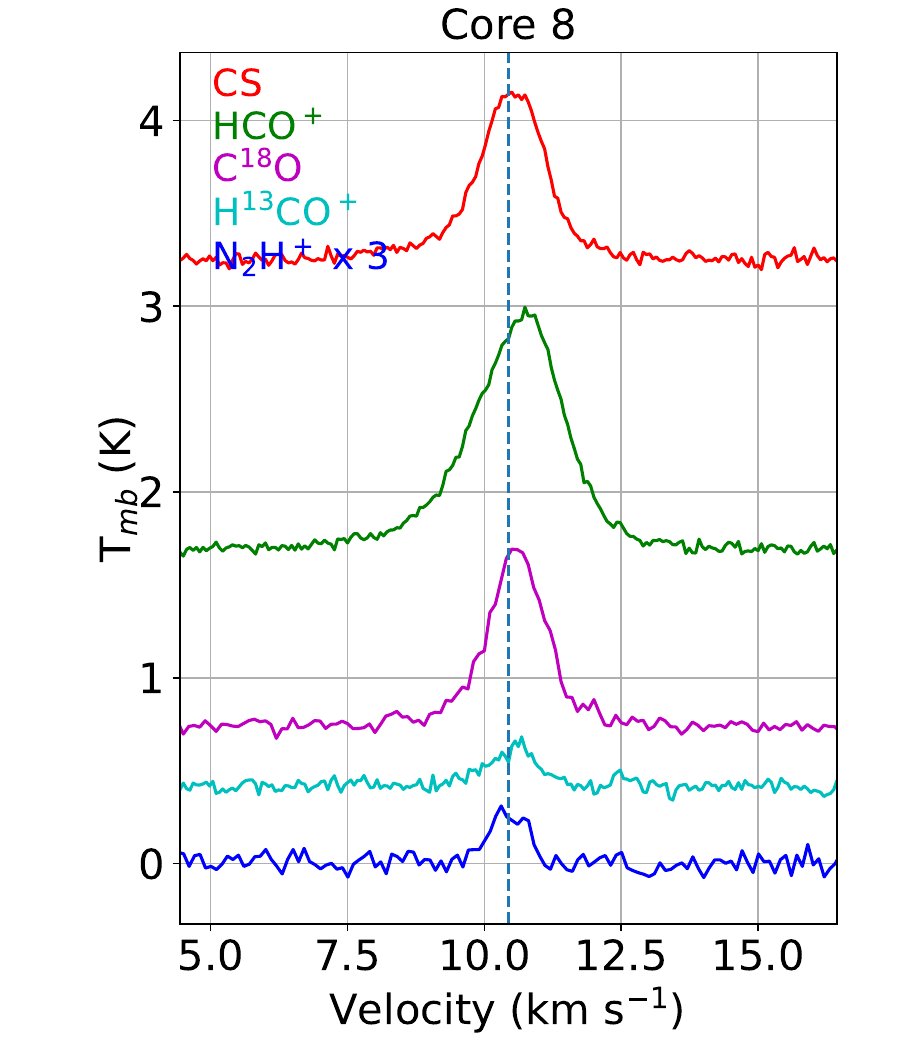} \\
  \includegraphics[width=0.25\textwidth]{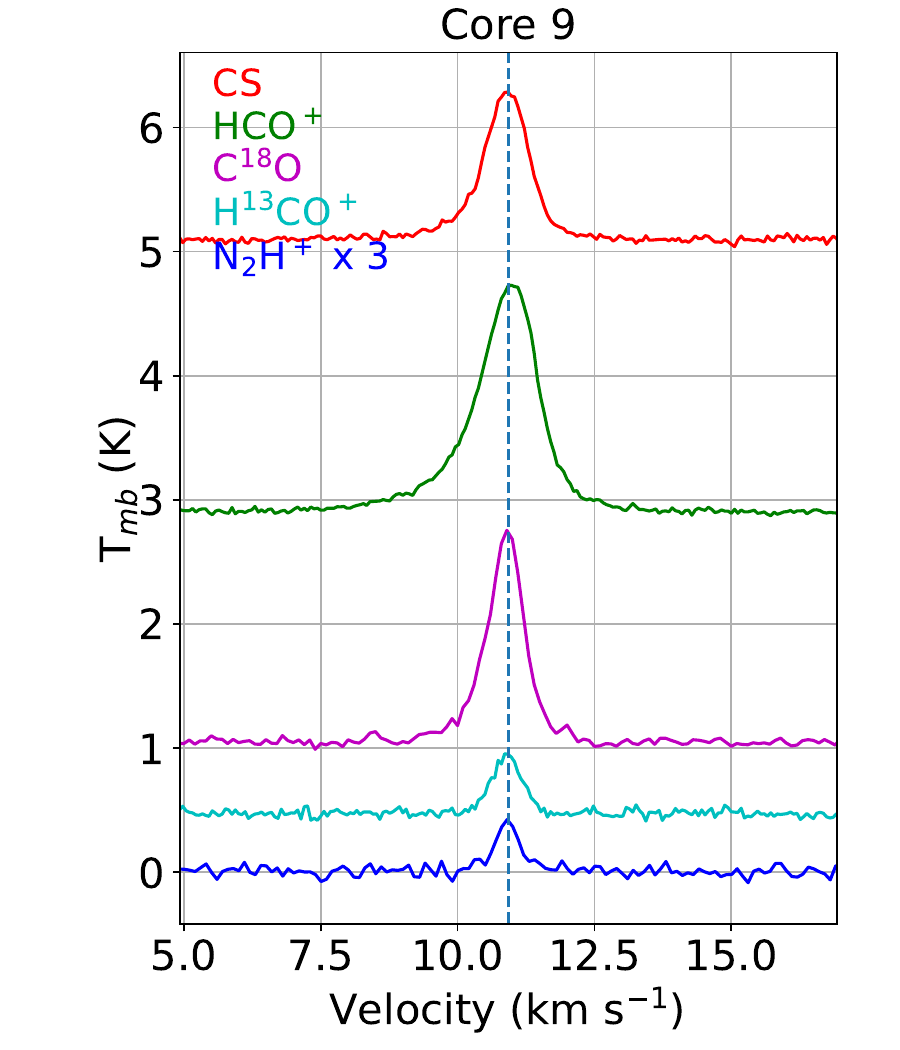}
  \includegraphics[width=0.25\textwidth]{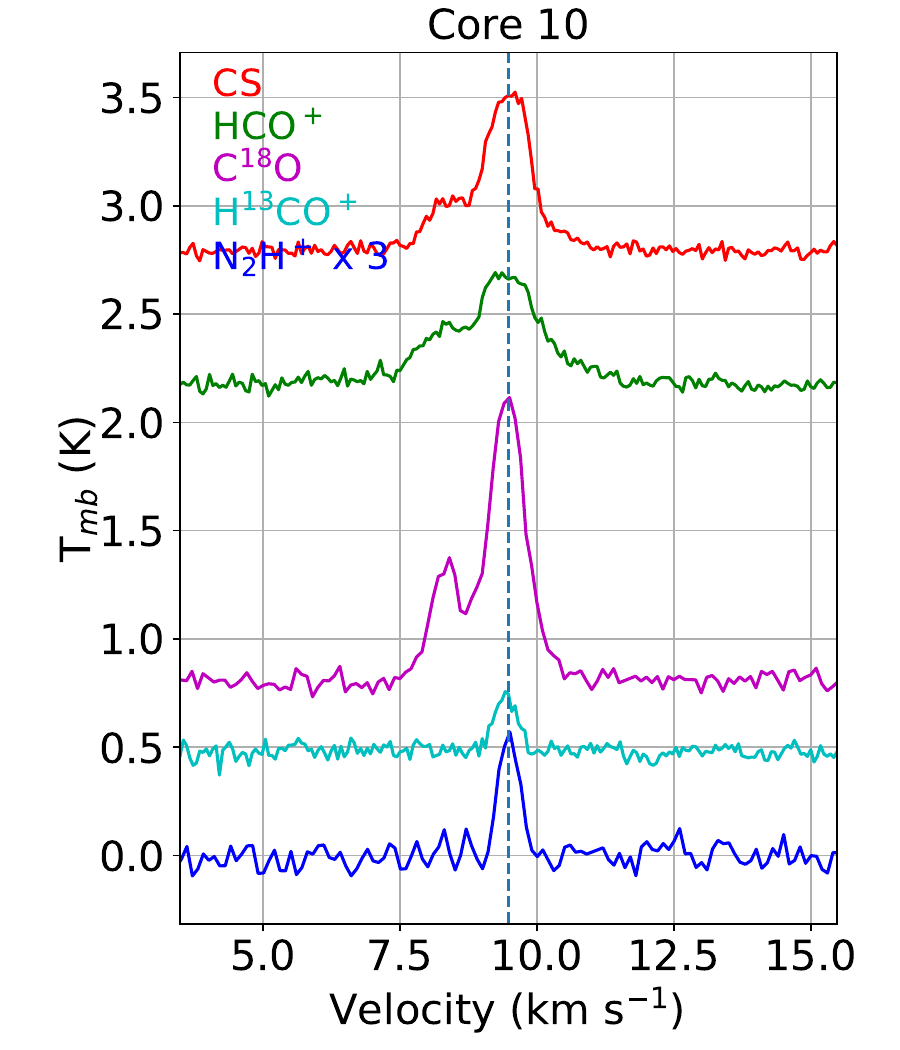}
  \includegraphics[width=0.25\textwidth]{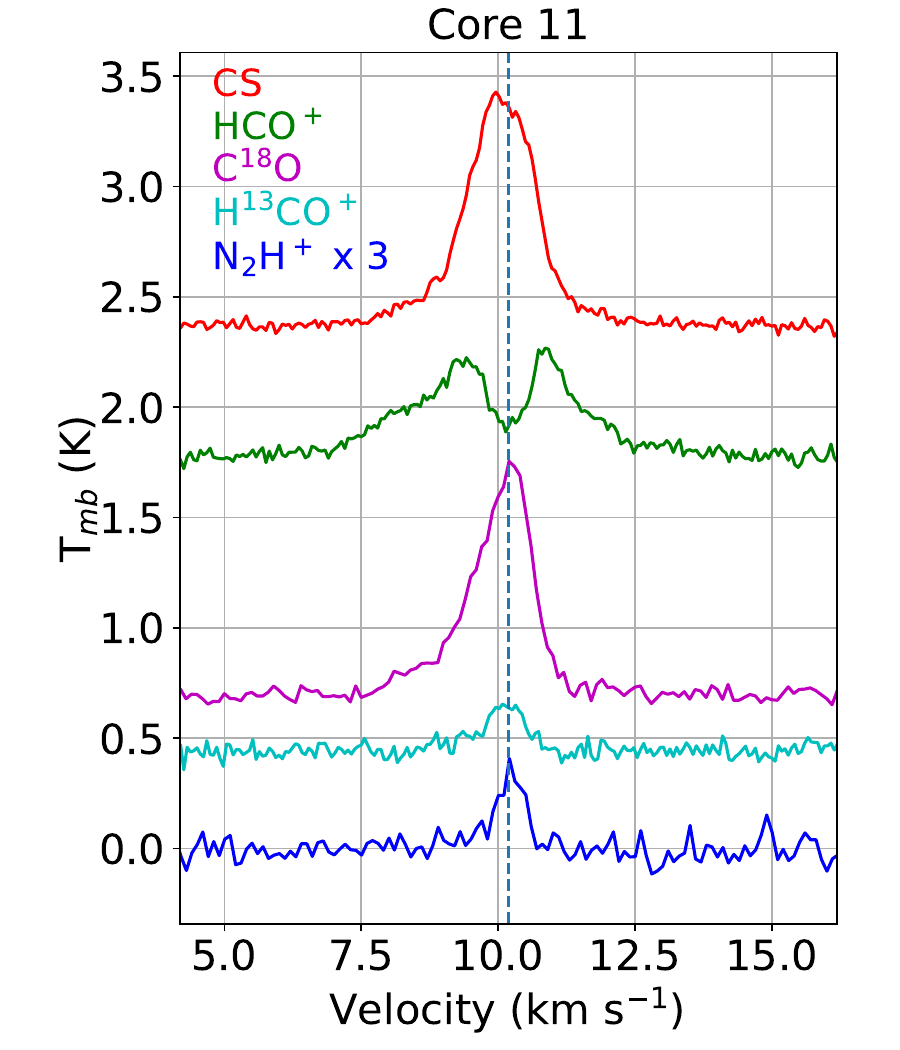}
  \includegraphics[width=0.25\textwidth]{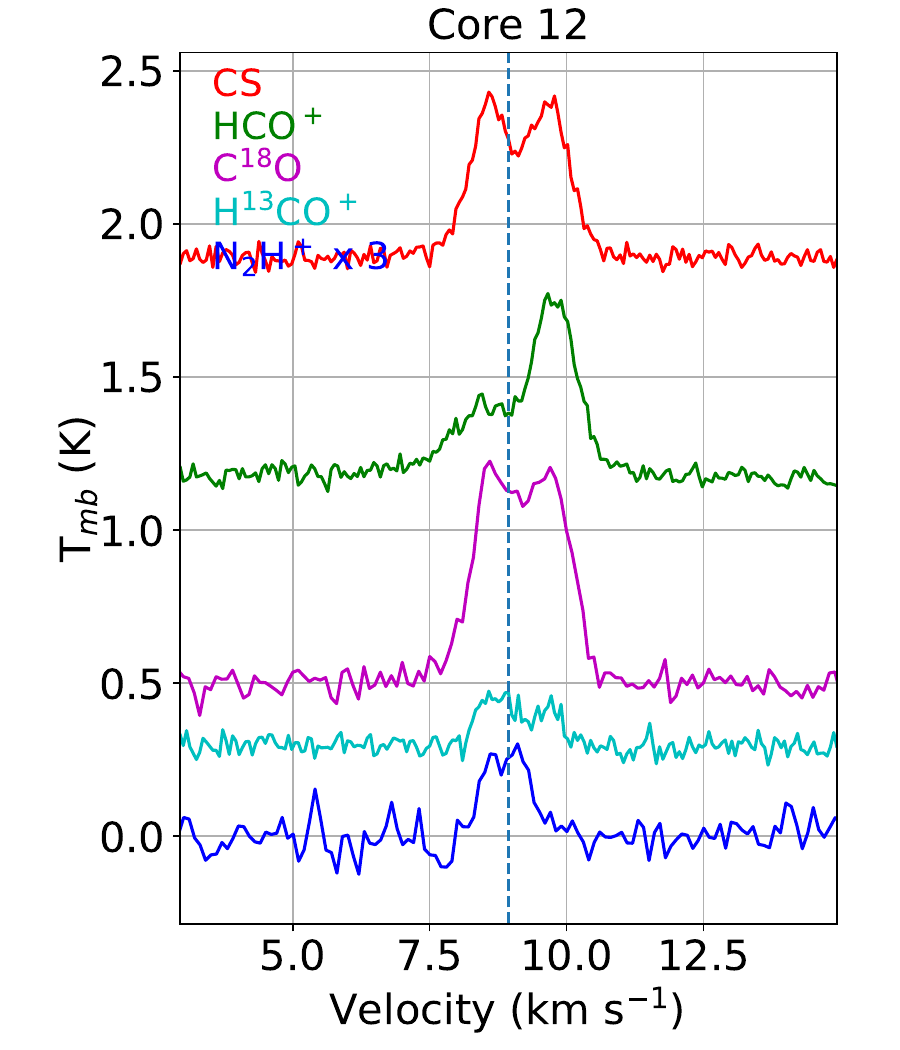}  \\
  \includegraphics[width=0.25\textwidth]{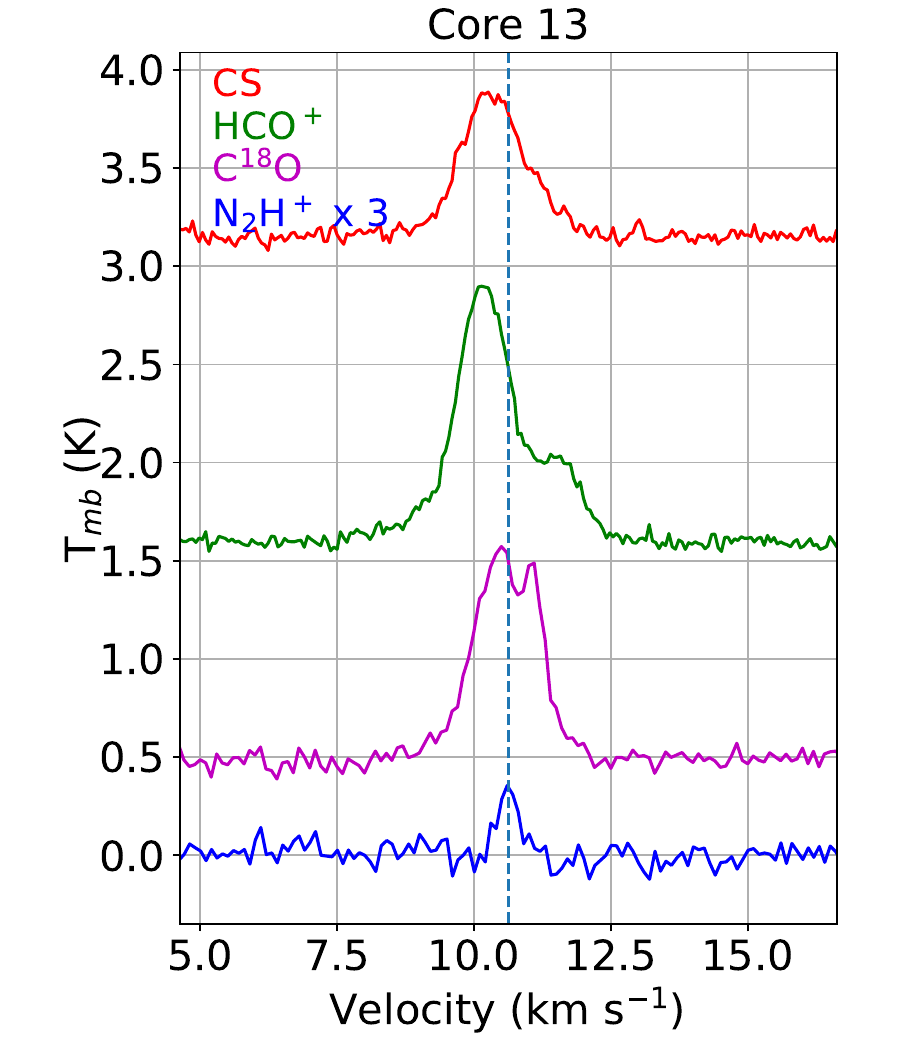}
  \includegraphics[width=0.25\textwidth]{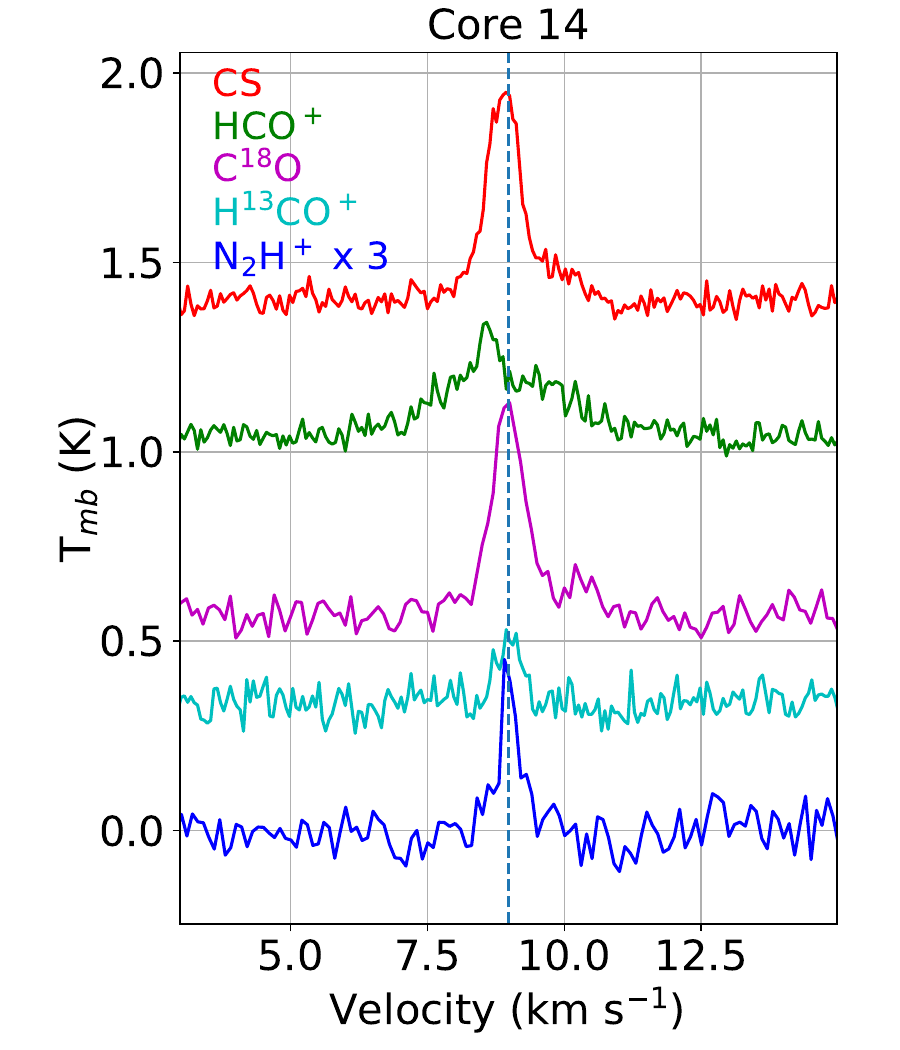} 
  \includegraphics[width=0.25\textwidth]{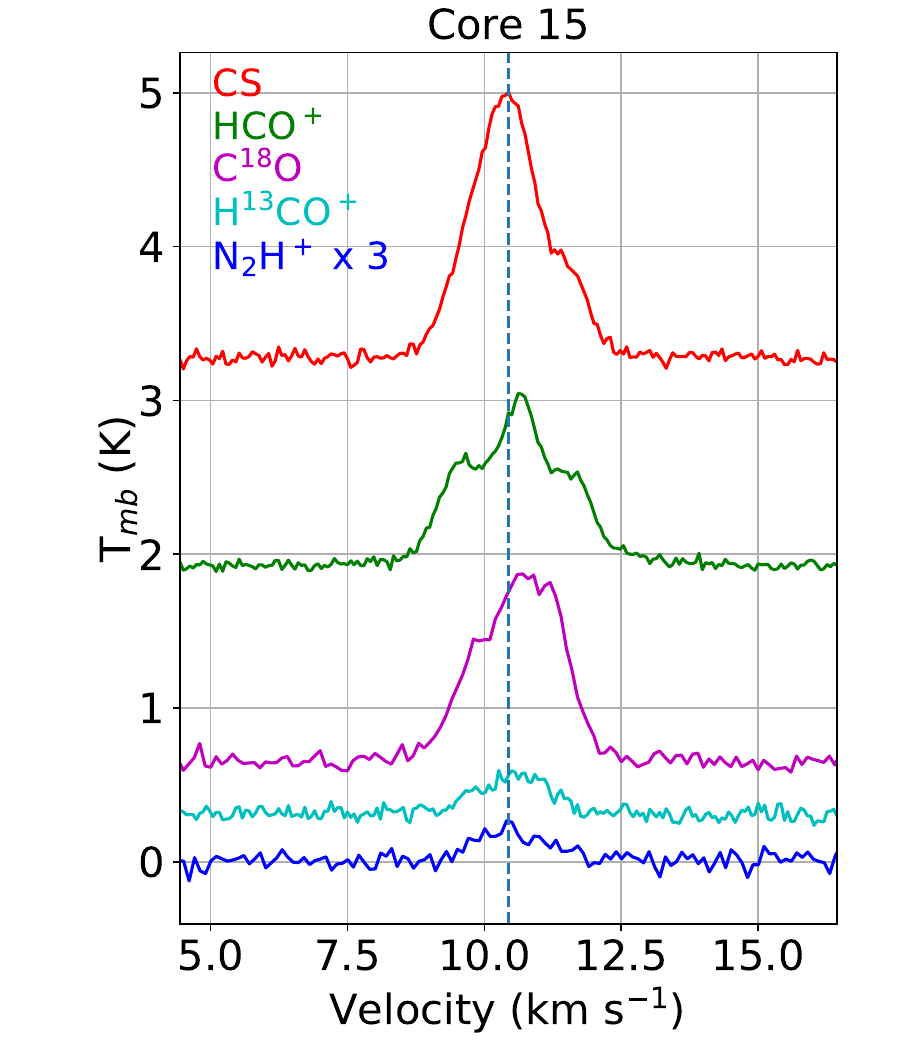}  
  \includegraphics[width=0.25\textwidth]{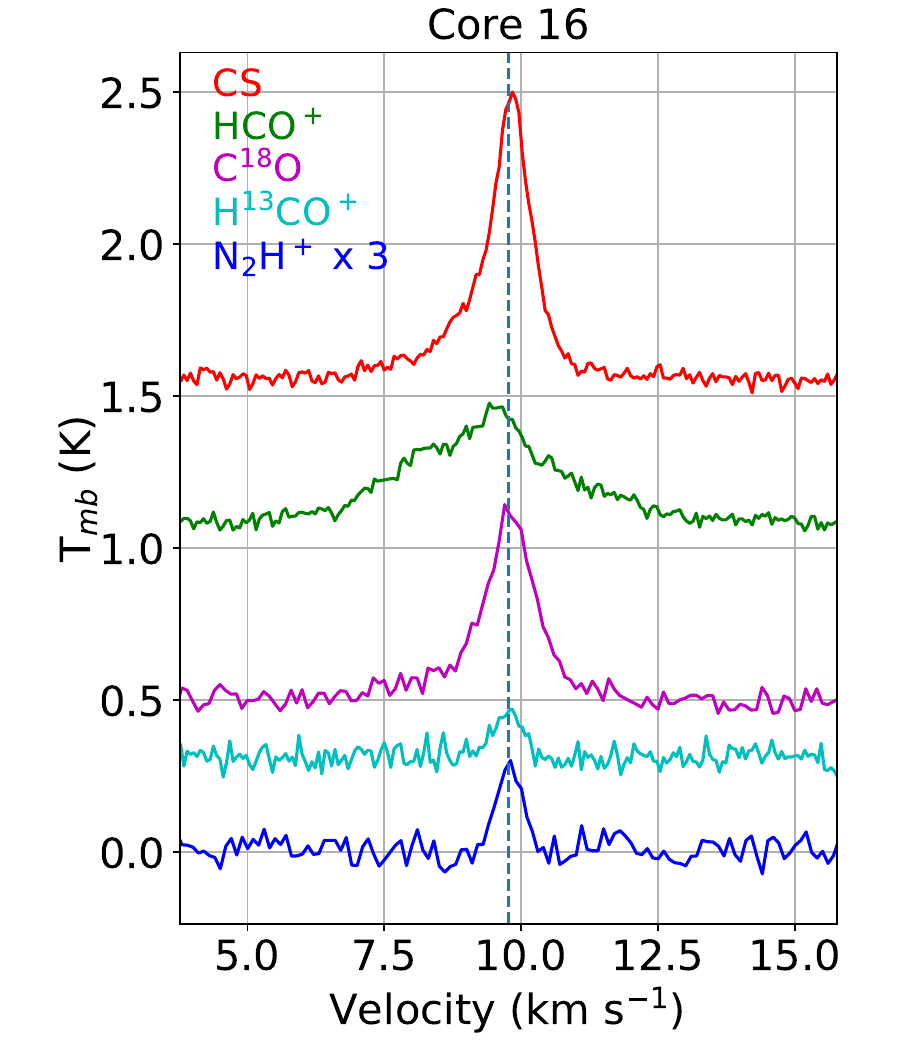} 

\end{tabbing}
 \caption{\csj, \hcopj, \cEoj, \hTcopj\ and the isolated hyperfine component of \nThpj\ line profiles of dense cores. All the profiles shown here are the average one for the spectra over the dense core. The vertical dashed line is the systemic velocity obtained from \nThp~hyperfine fitting. }
\label{fig:line_spec1}
\end{figure*}

\begin{figure*}
\begin{tabbing}
  \includegraphics[width=0.25\textwidth]{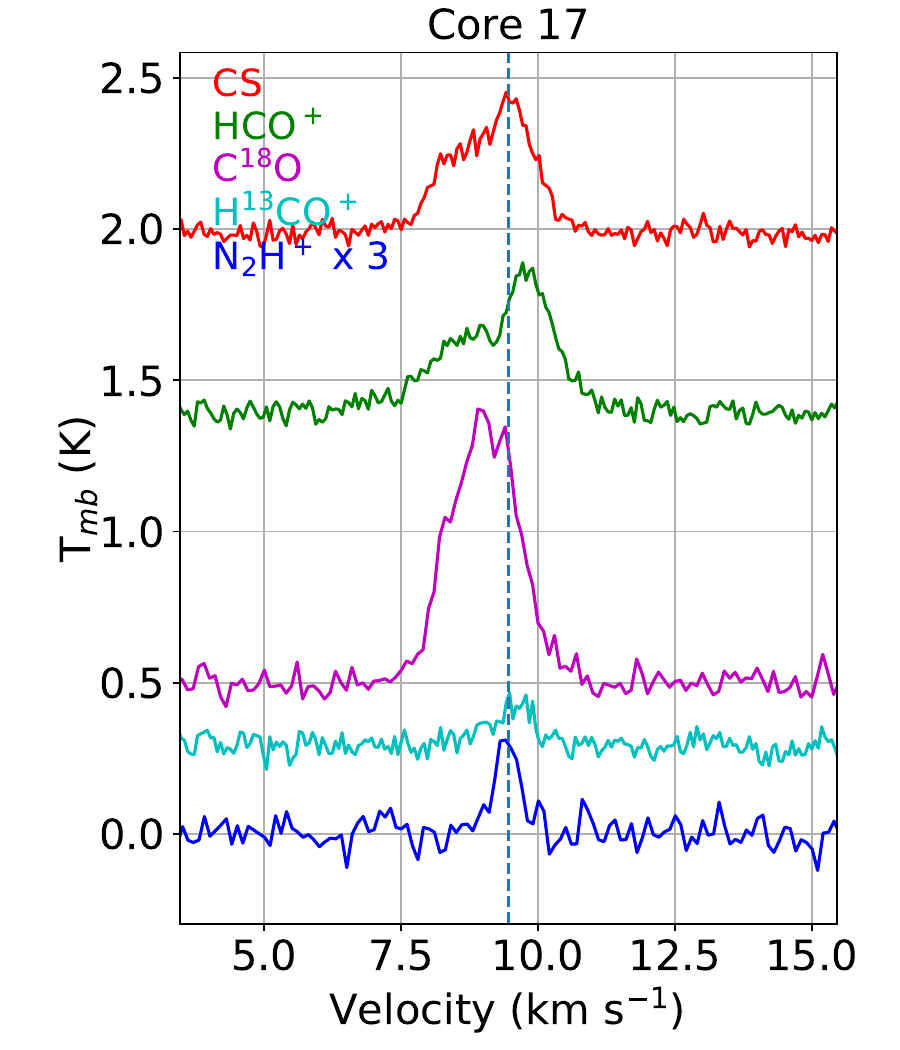} 
  \includegraphics[width=0.25\textwidth]{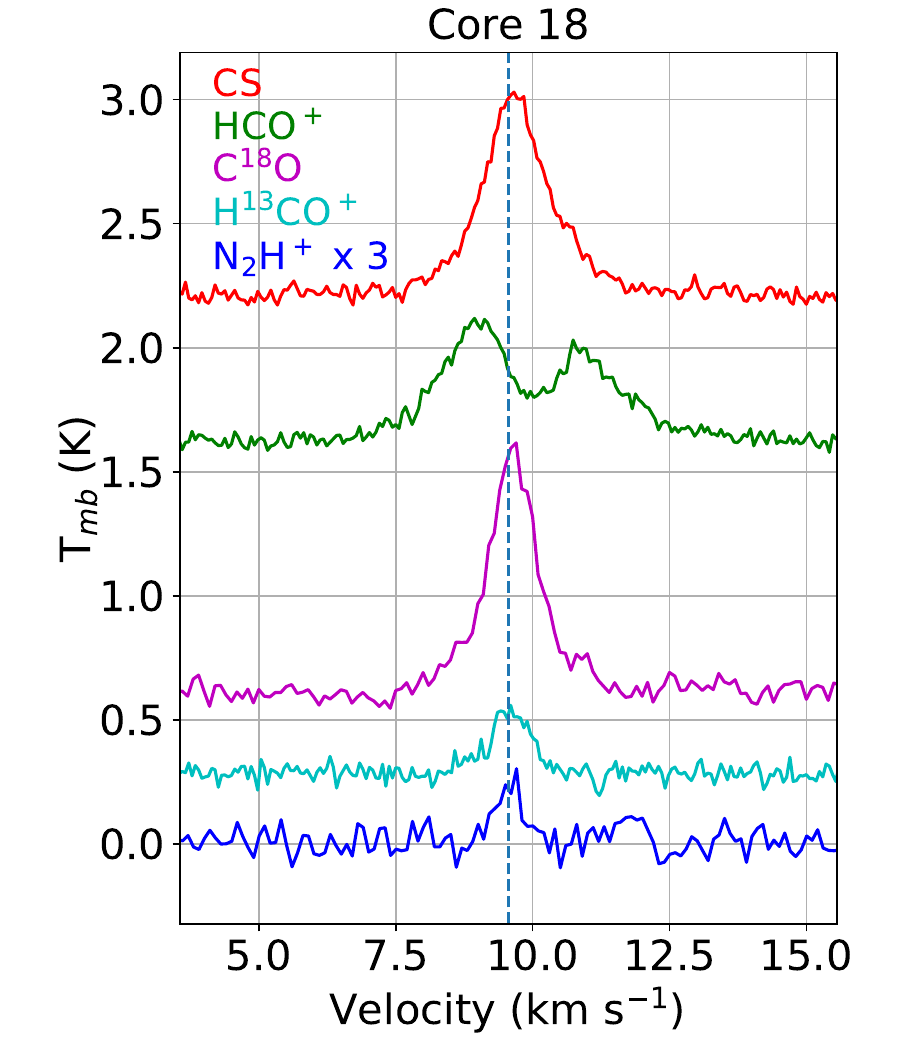}
  \includegraphics[width=0.25\textwidth]{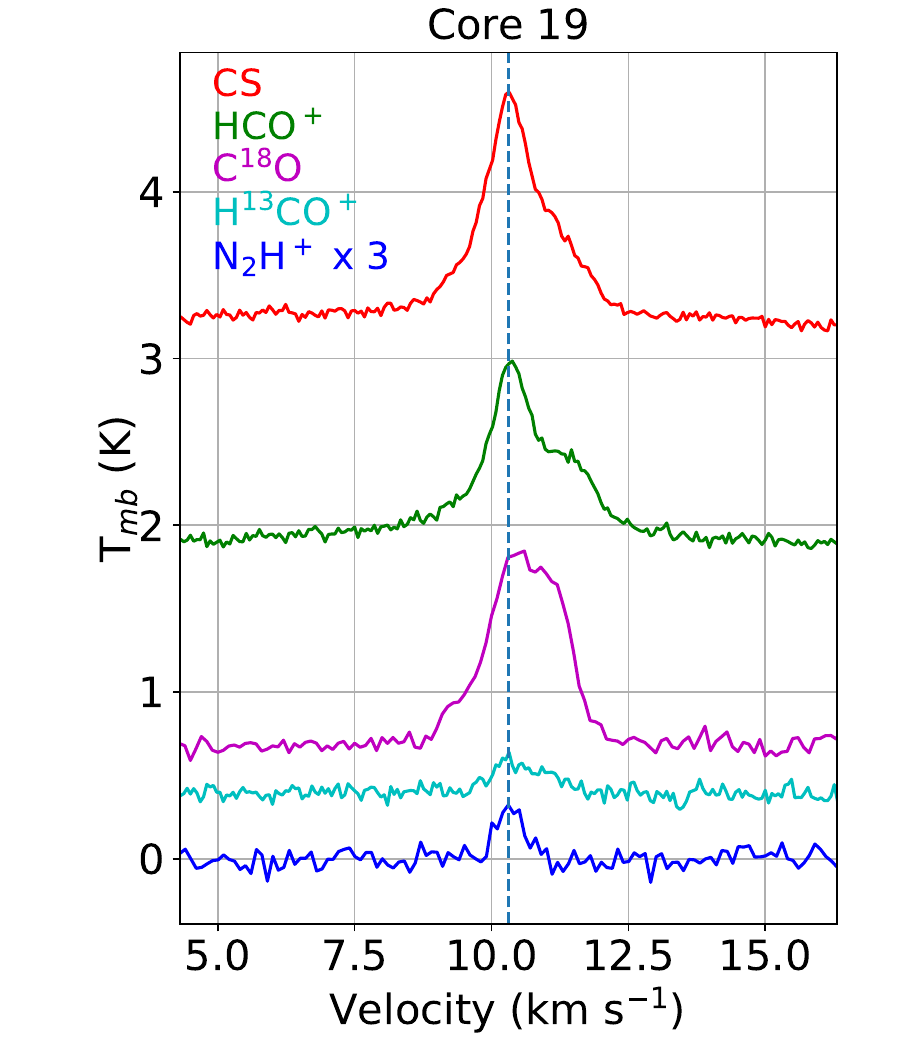}
  \includegraphics[width=0.25\textwidth]{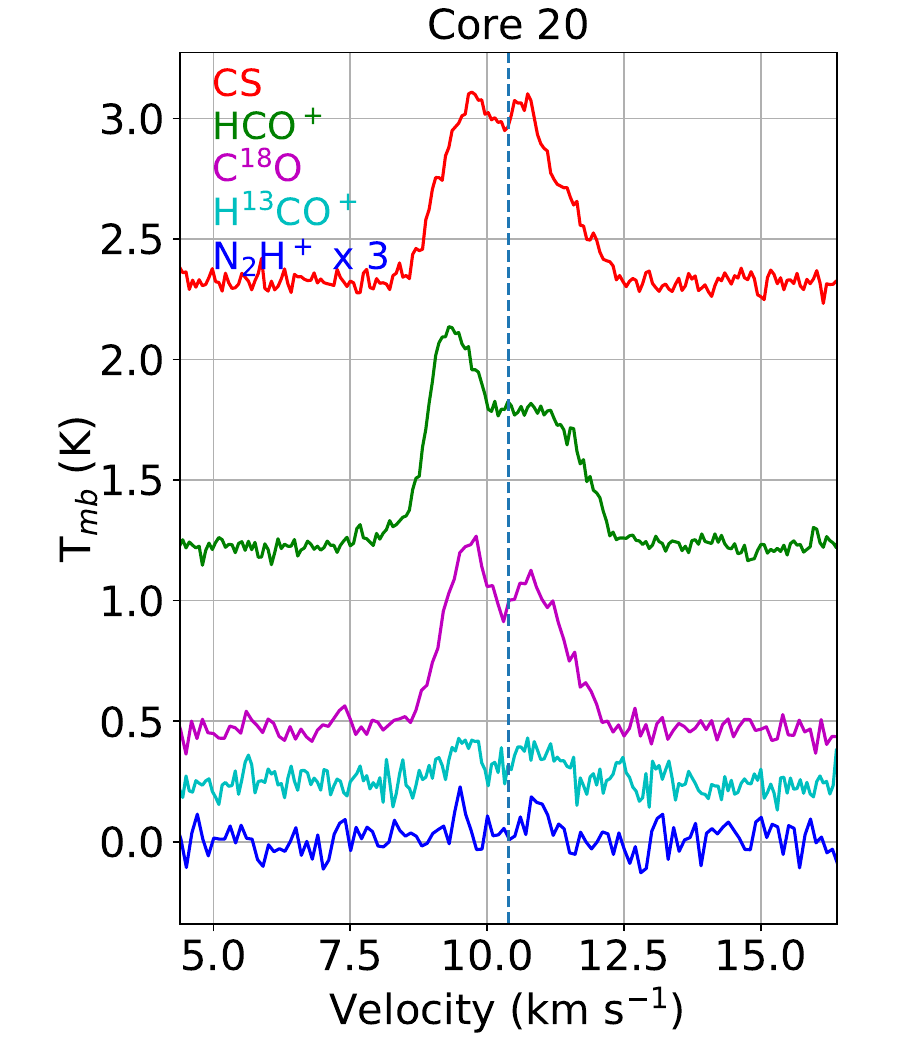} \\
  \includegraphics[width=0.25\textwidth]{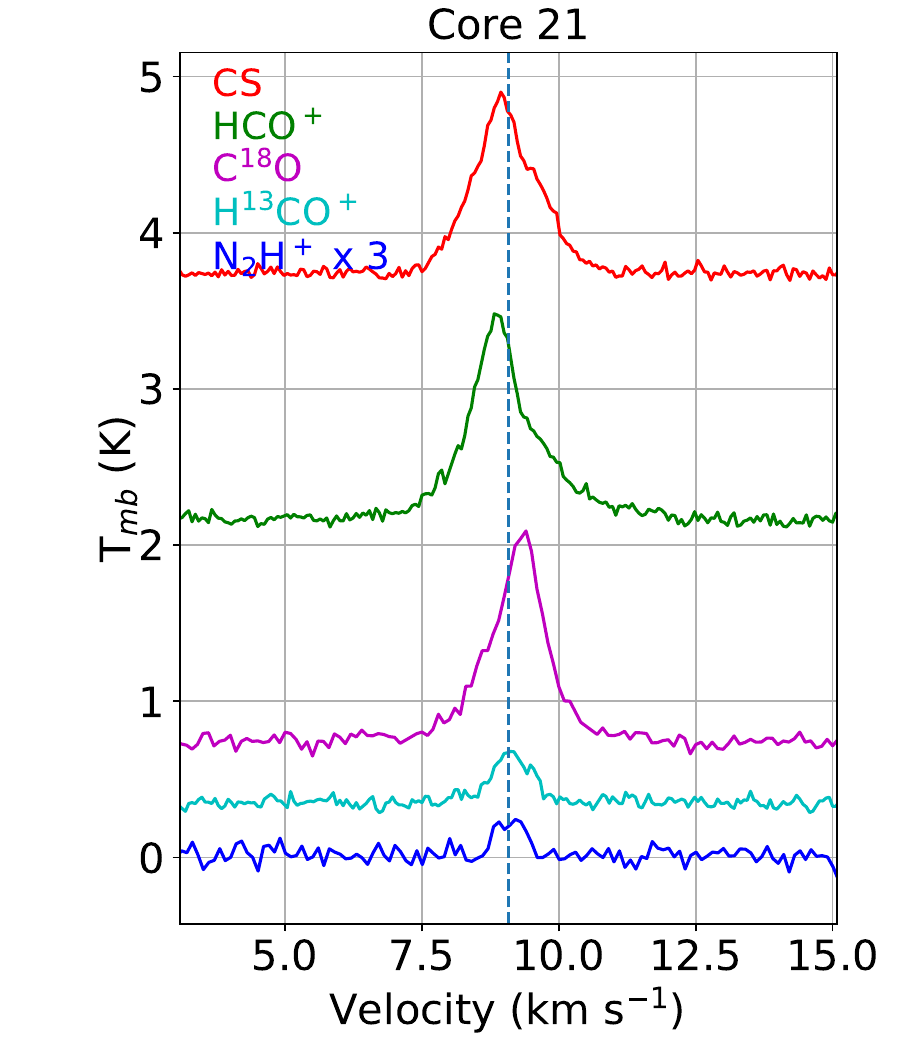}
  \includegraphics[width=0.25\textwidth]{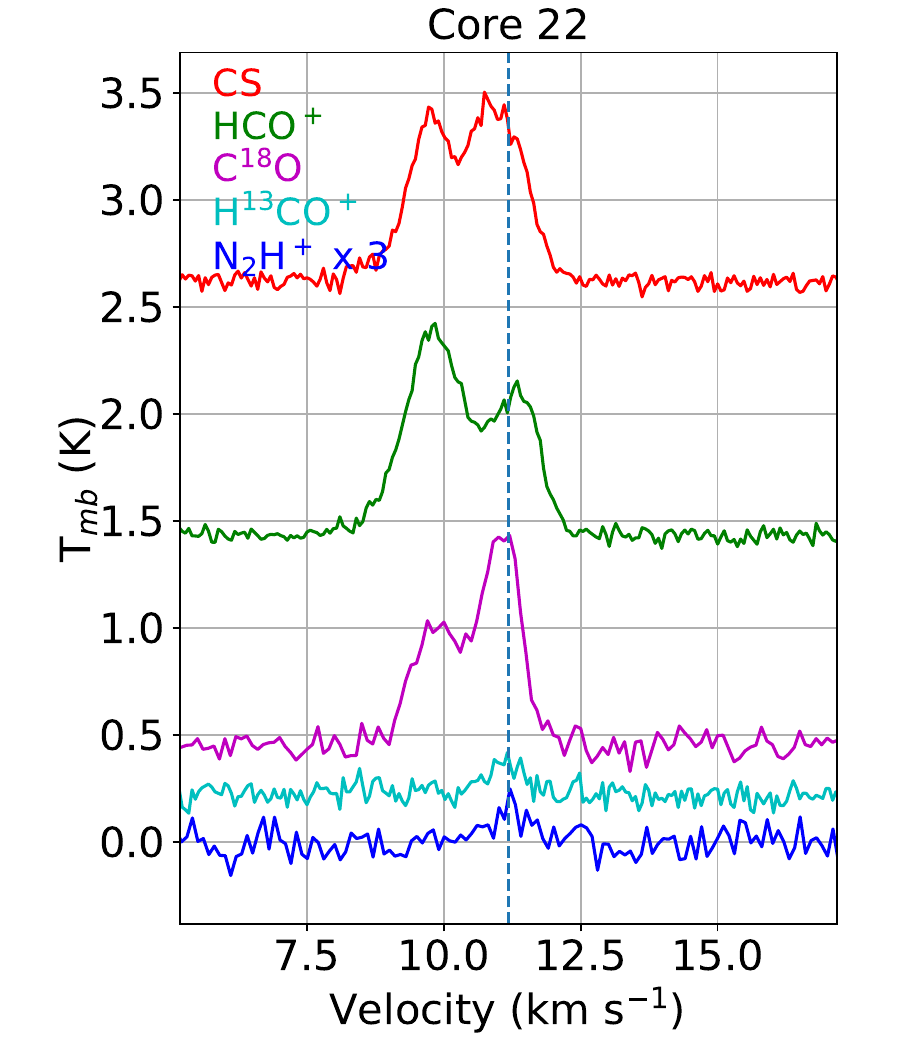}
  \includegraphics[width=0.25\textwidth]{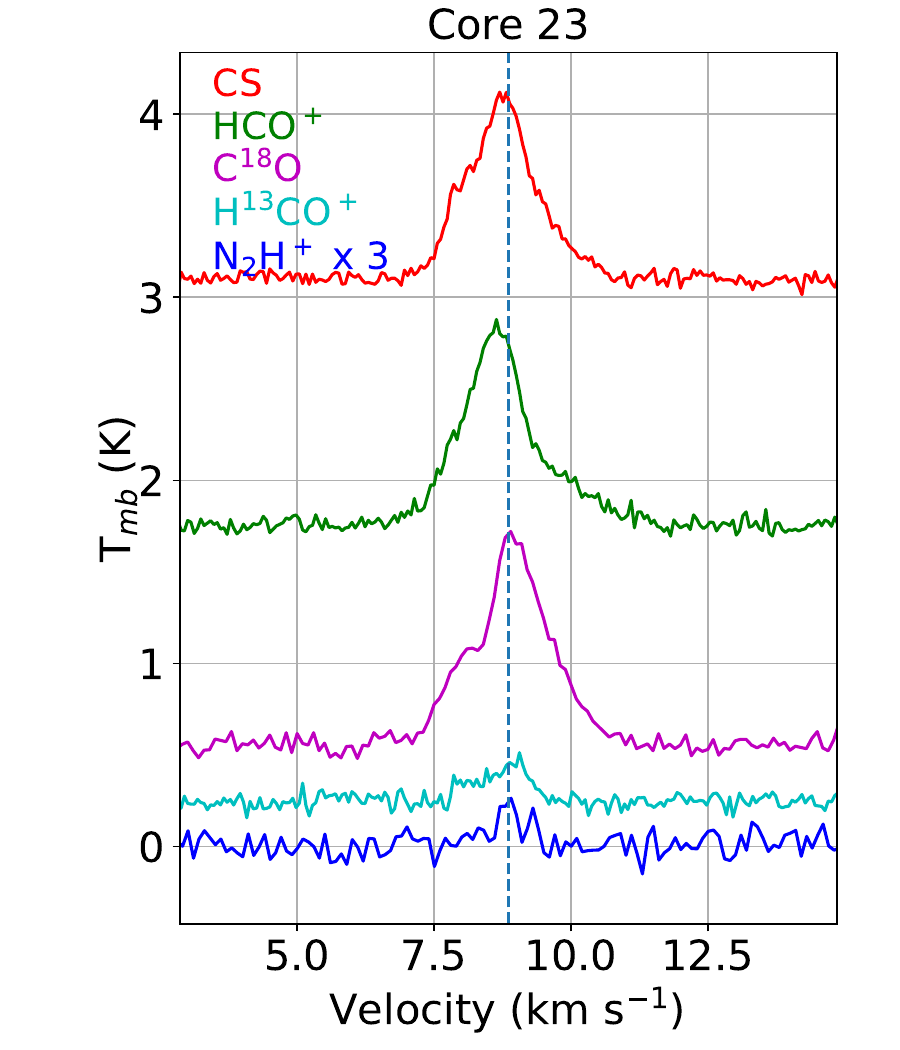}
  \includegraphics[width=0.25\textwidth]{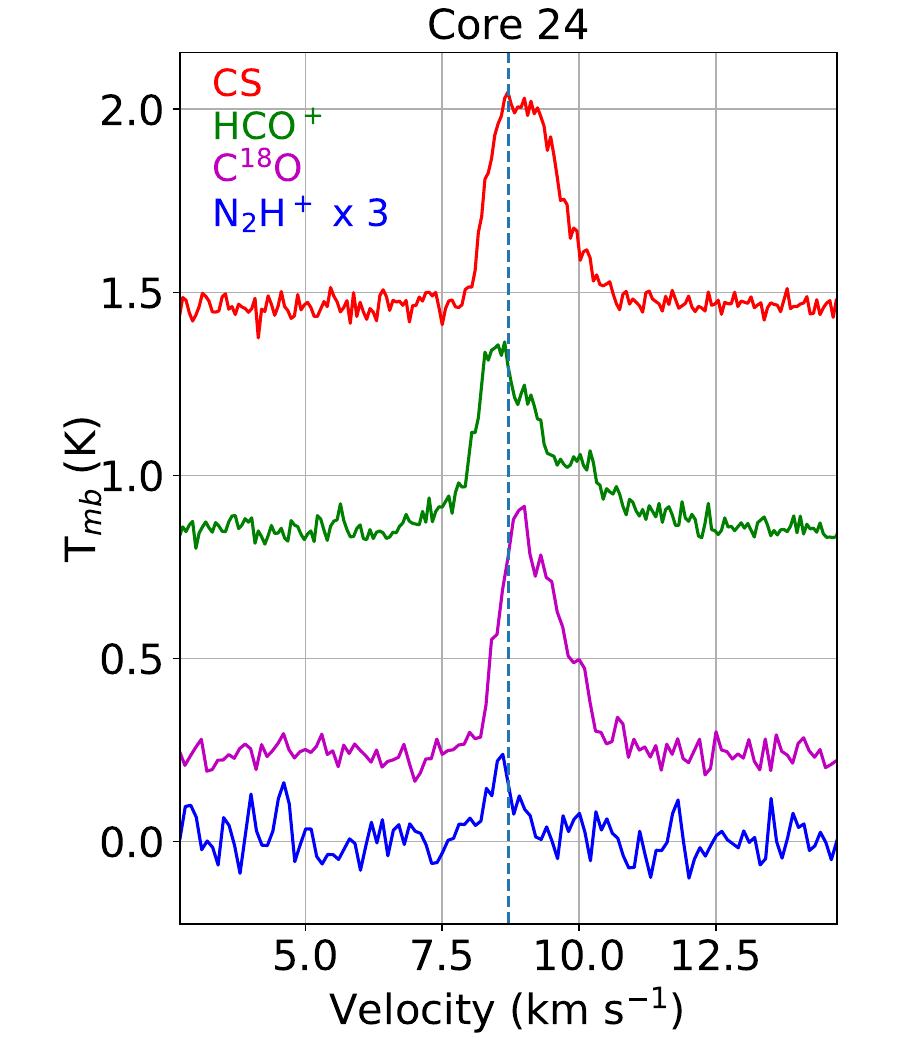} \\
  \includegraphics[width=0.25\textwidth]{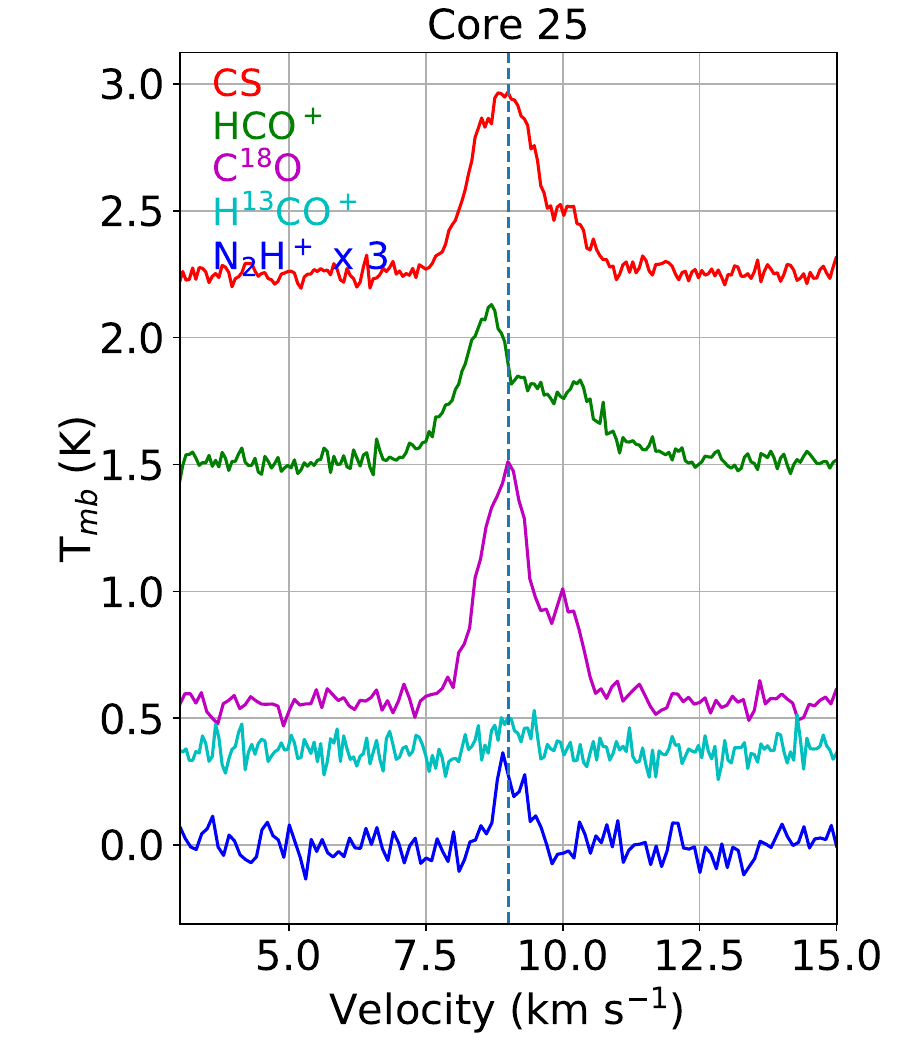}
  \includegraphics[width=0.25\textwidth]{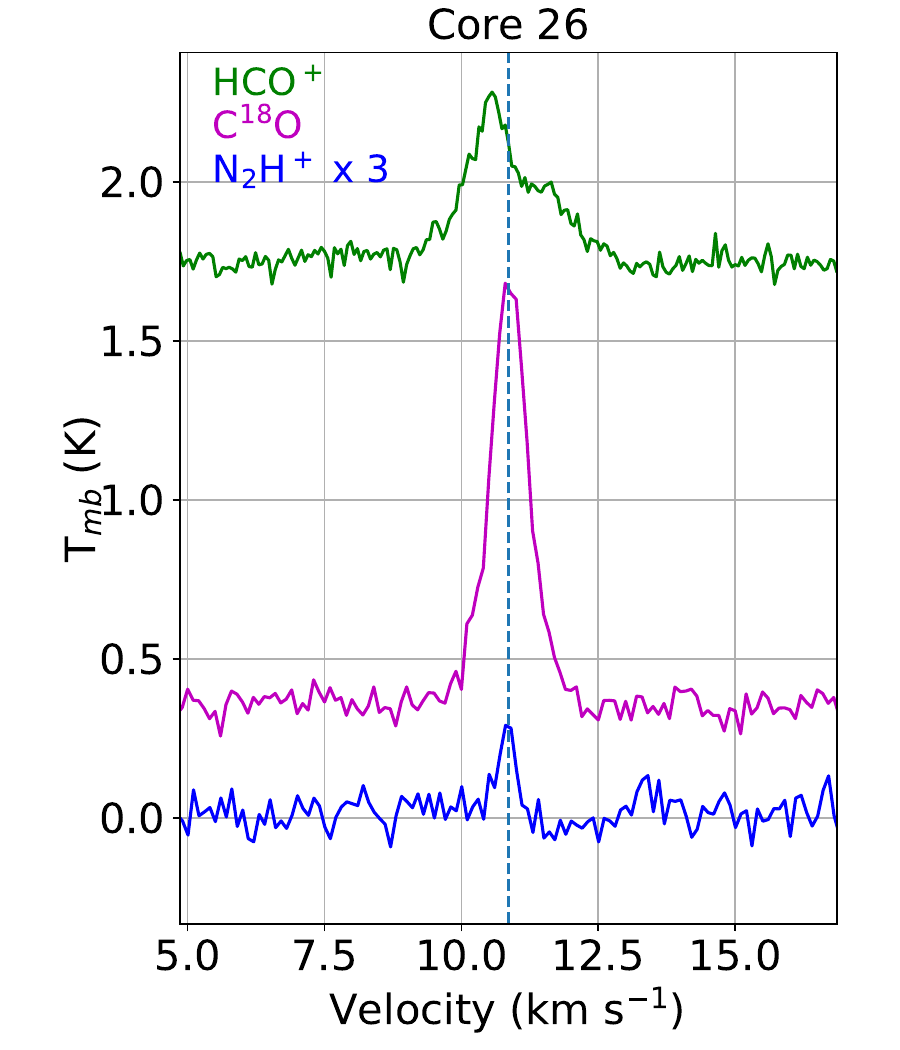} 
  \includegraphics[width=0.25\textwidth]{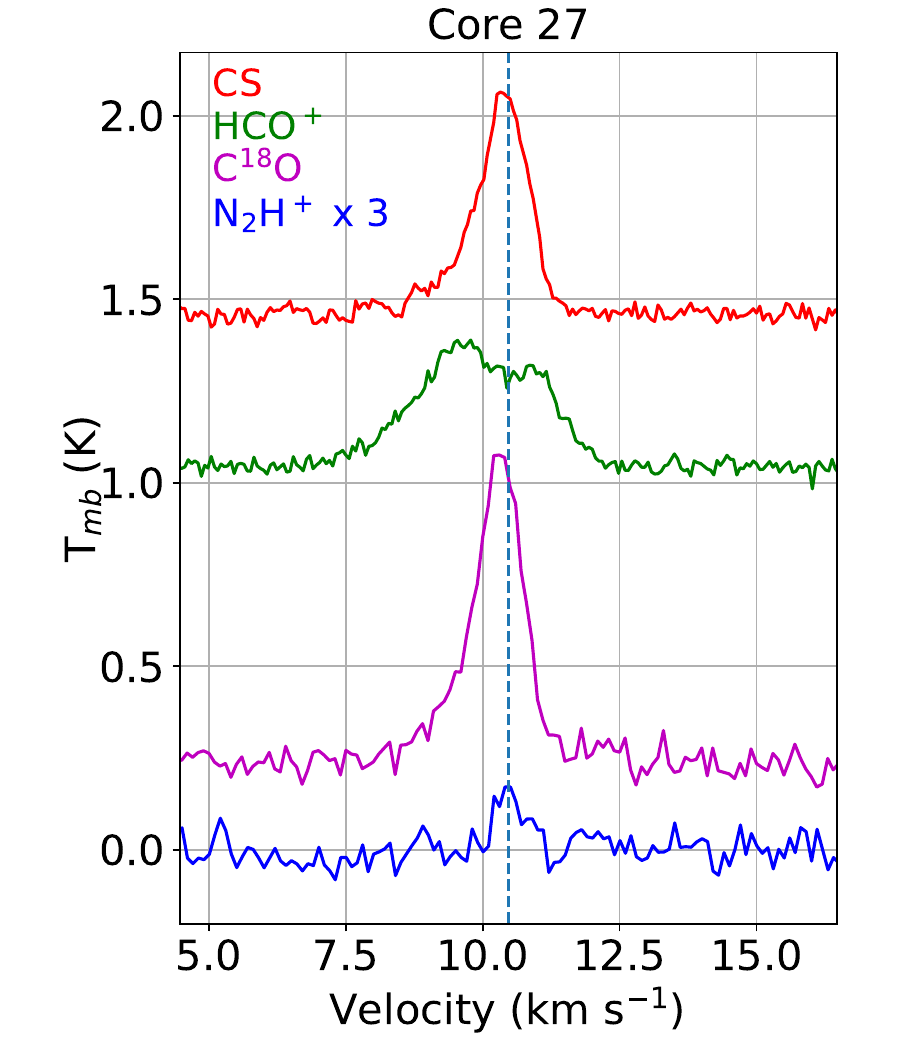}  
  \includegraphics[width=0.25\textwidth]{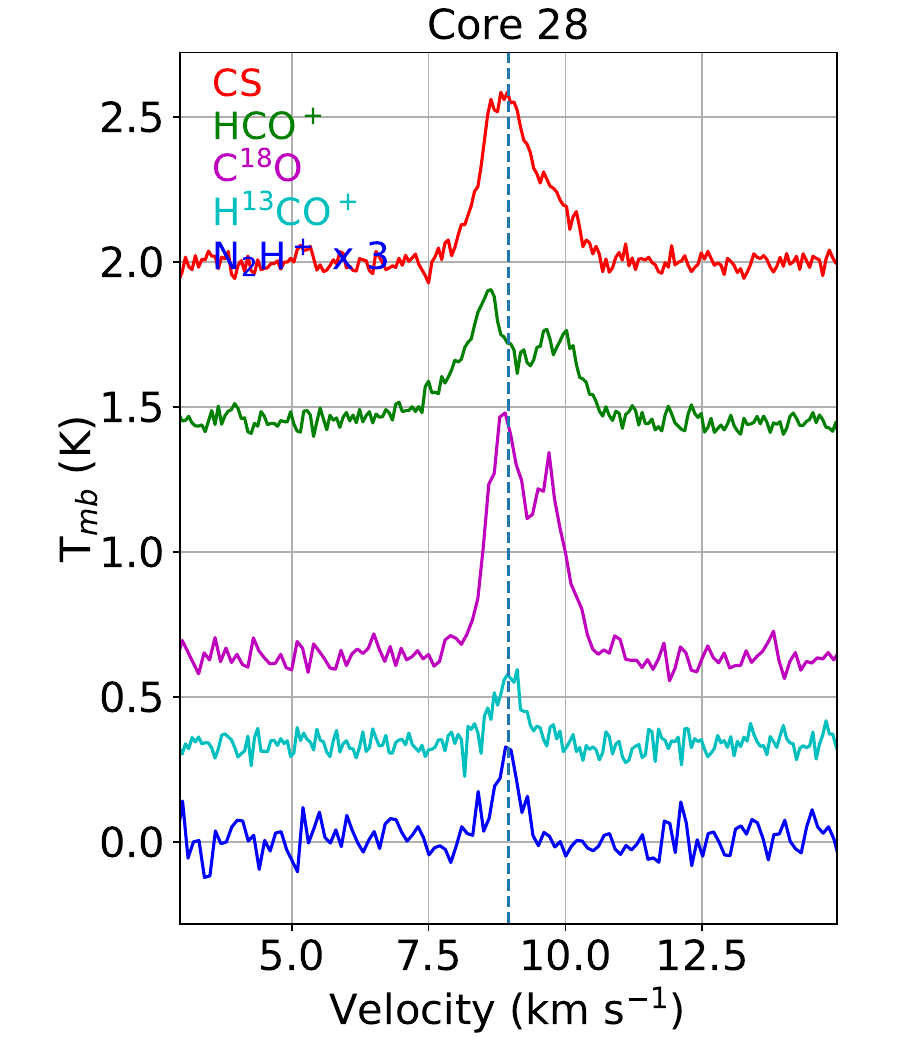} \\
  \includegraphics[width=0.25\textwidth]{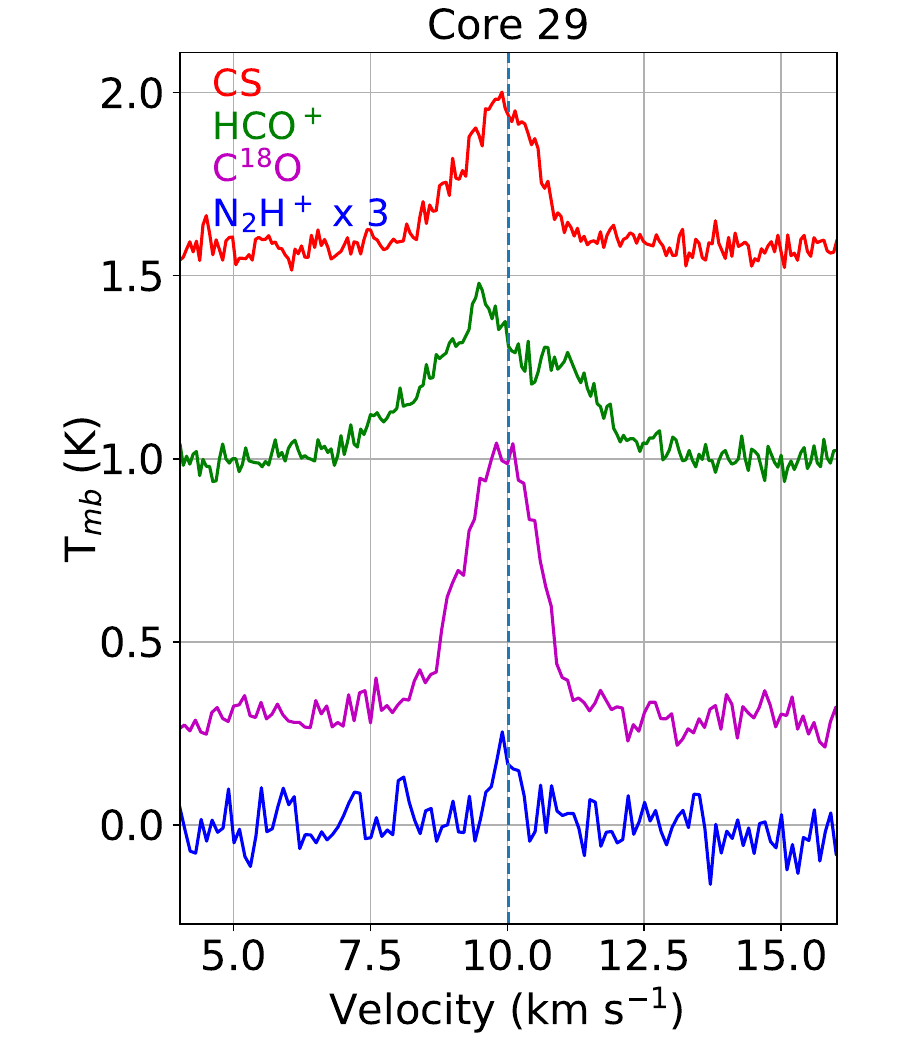} 
  \includegraphics[width=0.25\textwidth]{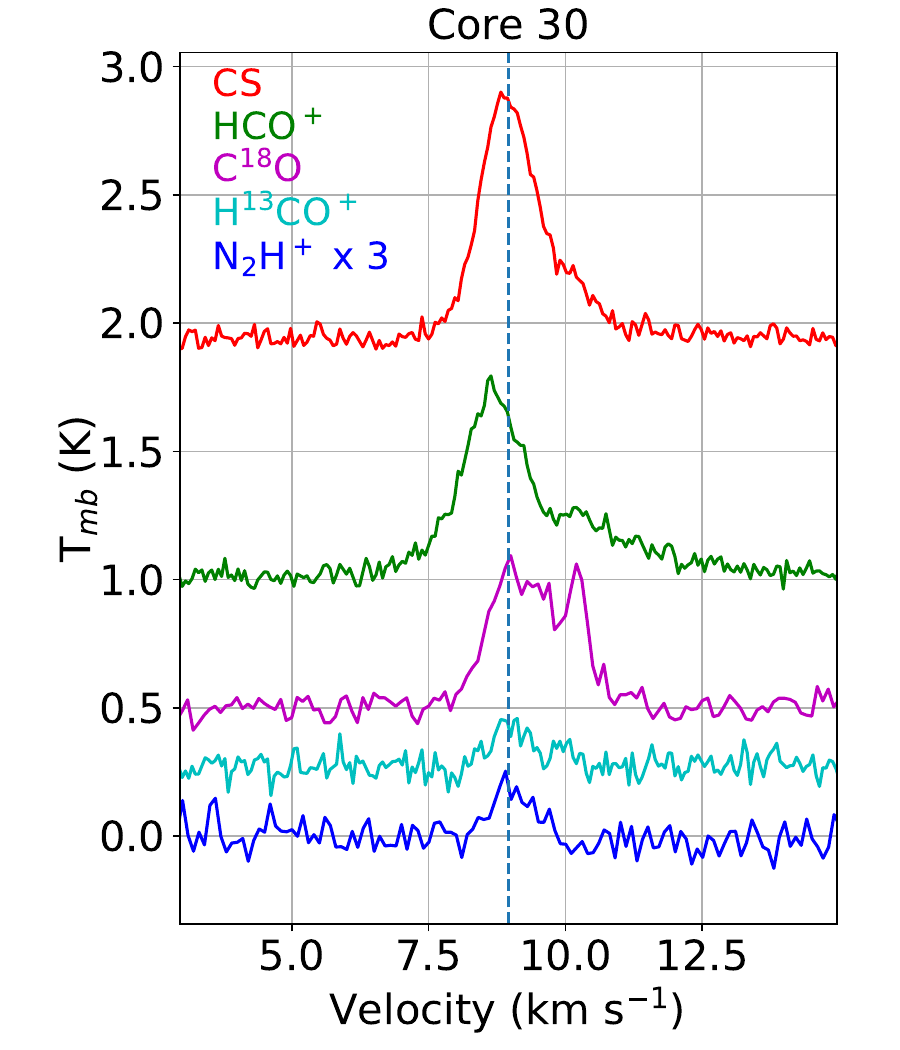}
  \includegraphics[width=0.25\textwidth]{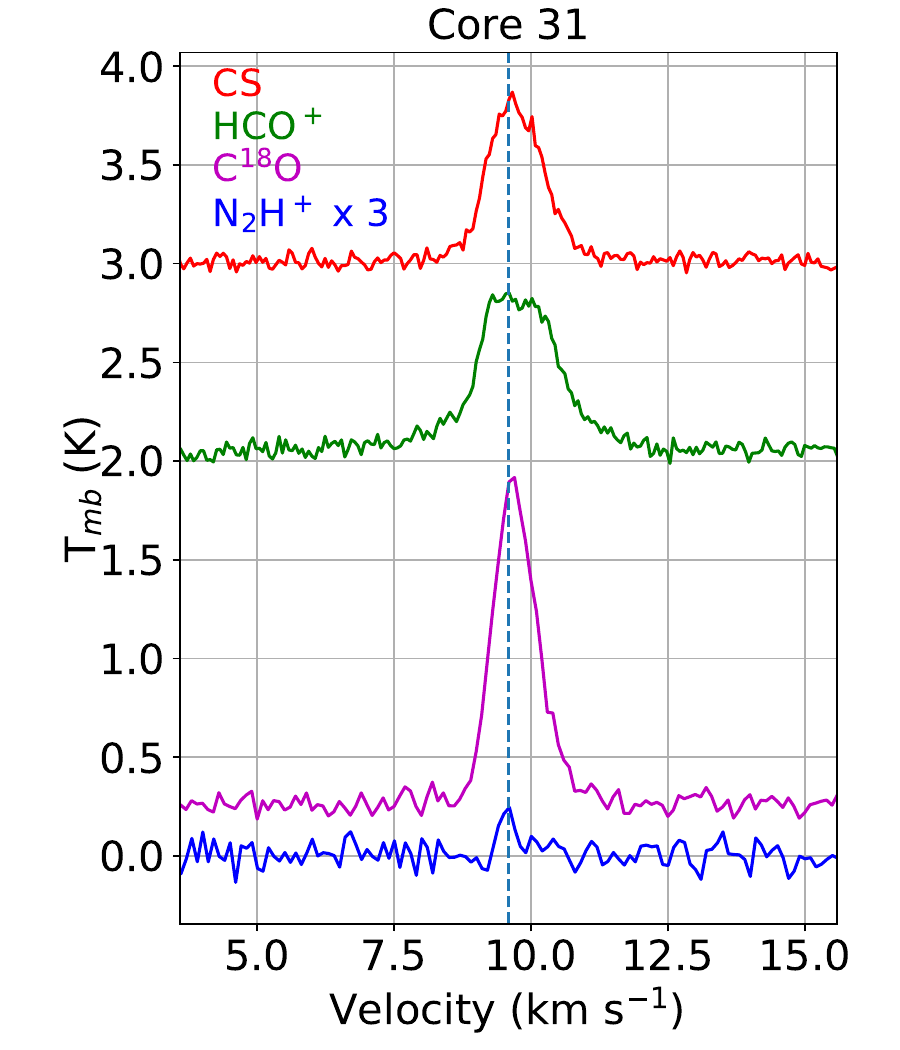}
  \includegraphics[width=0.25\textwidth]{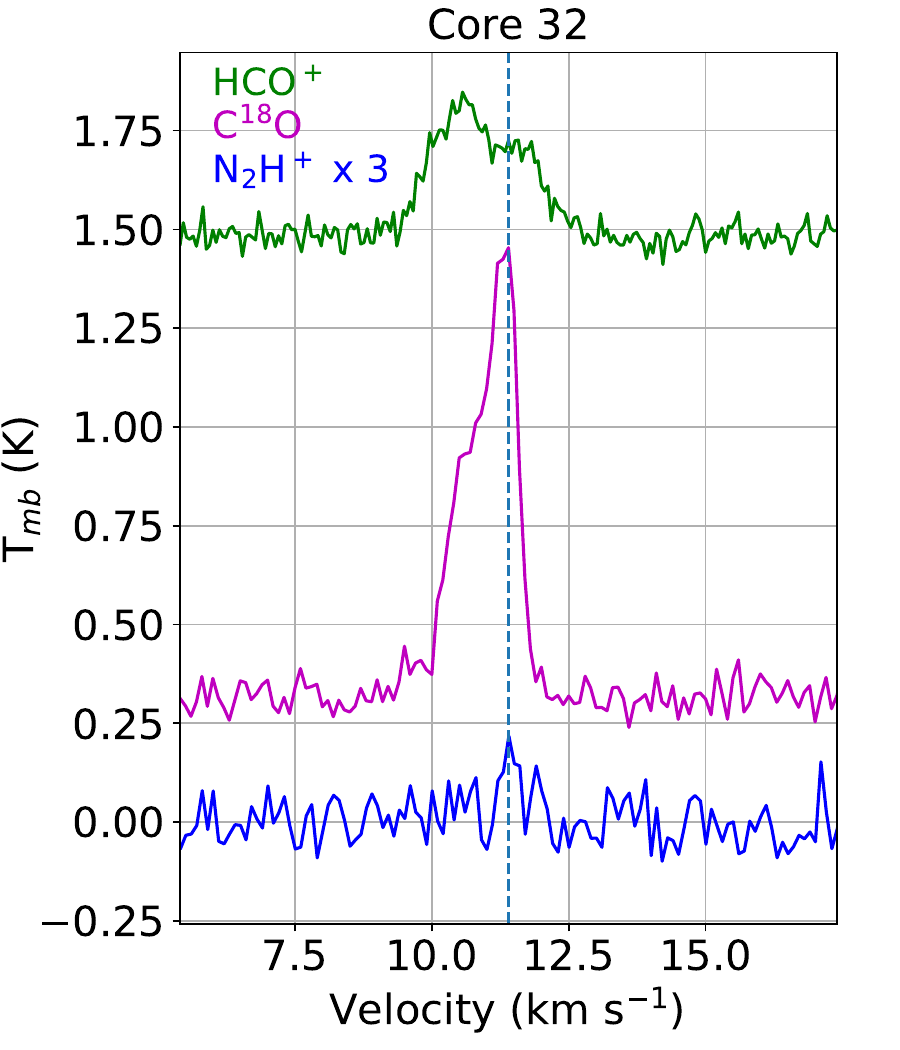}
\end{tabbing}
 \caption{Continued.}
\label{fig:line_spec2}
\end{figure*}

\begin{figure*}
\begin{tabbing}
  \includegraphics[width=0.25\textwidth]{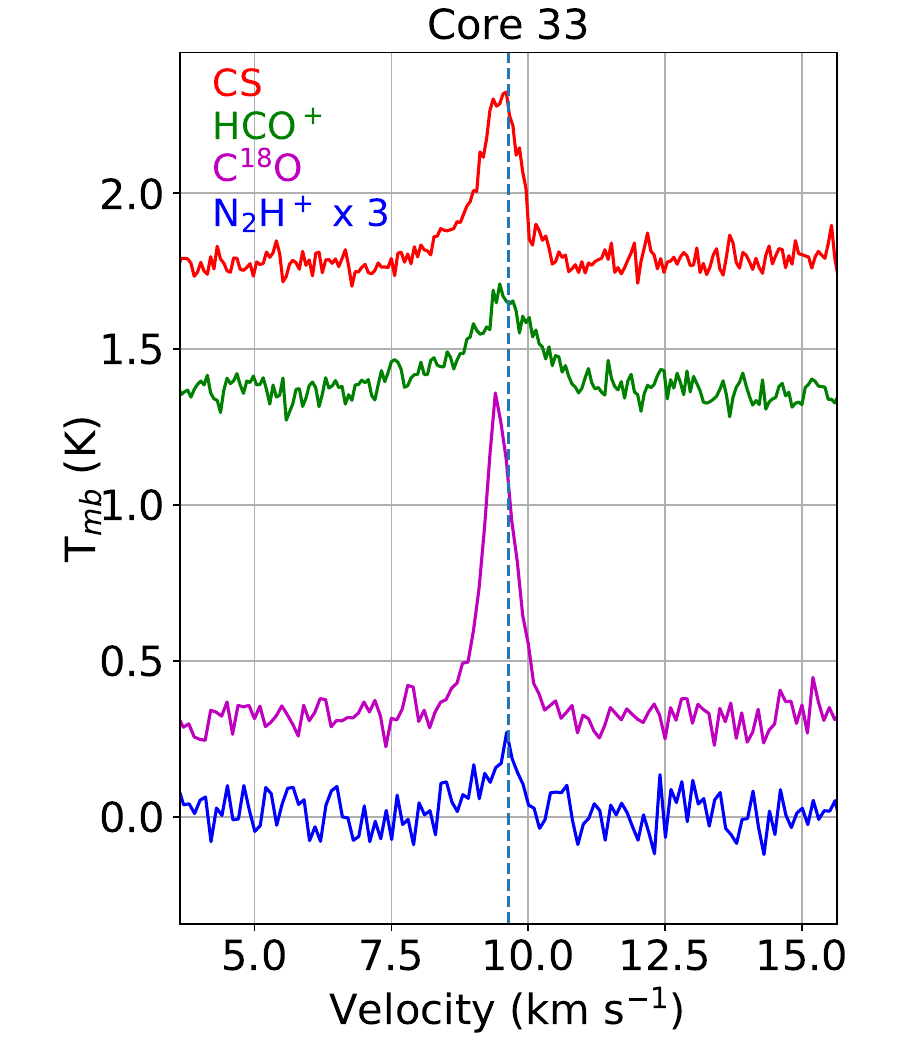}
  \includegraphics[width=0.25\textwidth]{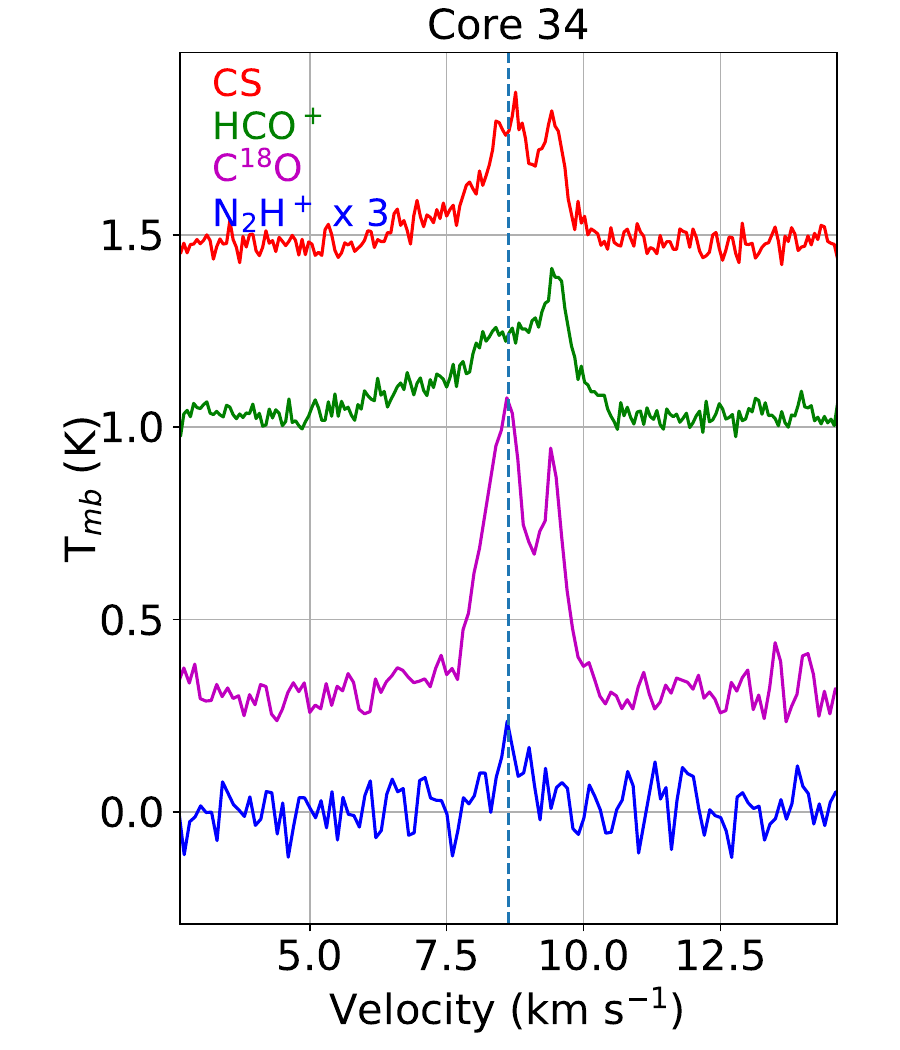}
  \includegraphics[width=0.25\textwidth]{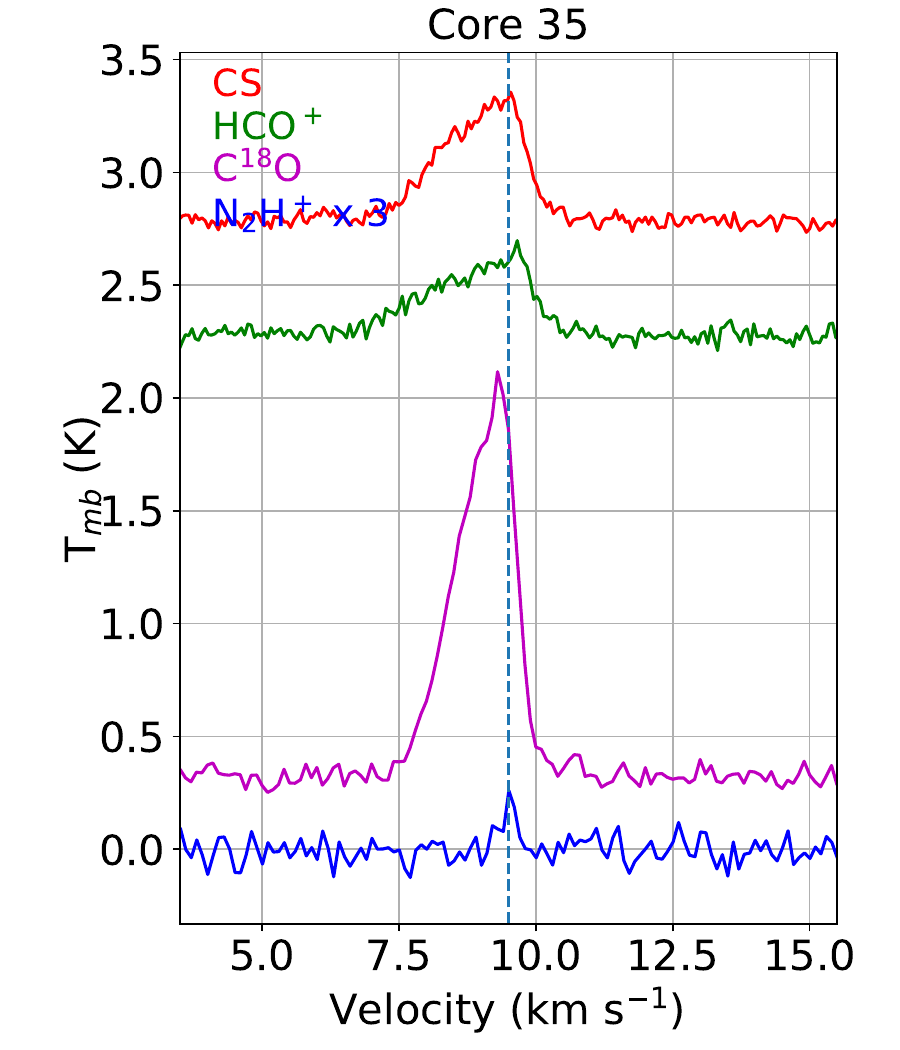}
  \includegraphics[width=0.25\textwidth]{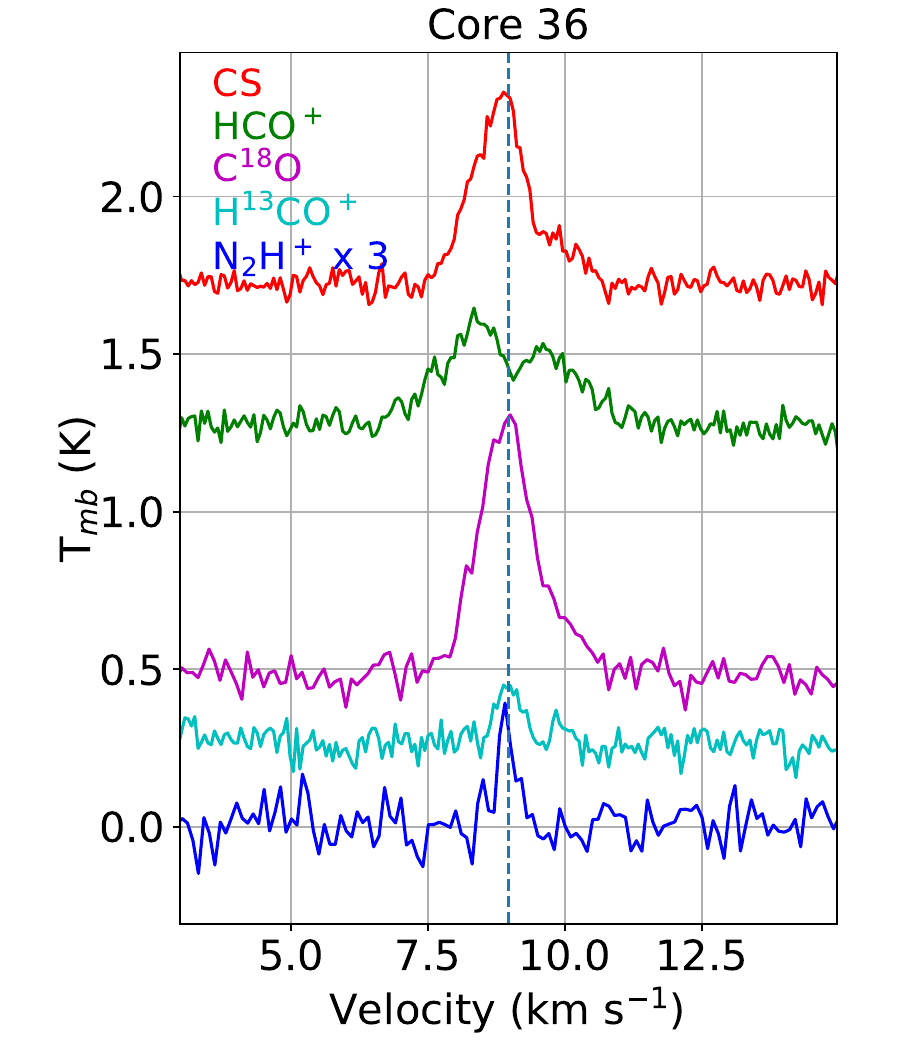} \\
  \includegraphics[width=0.25\textwidth]{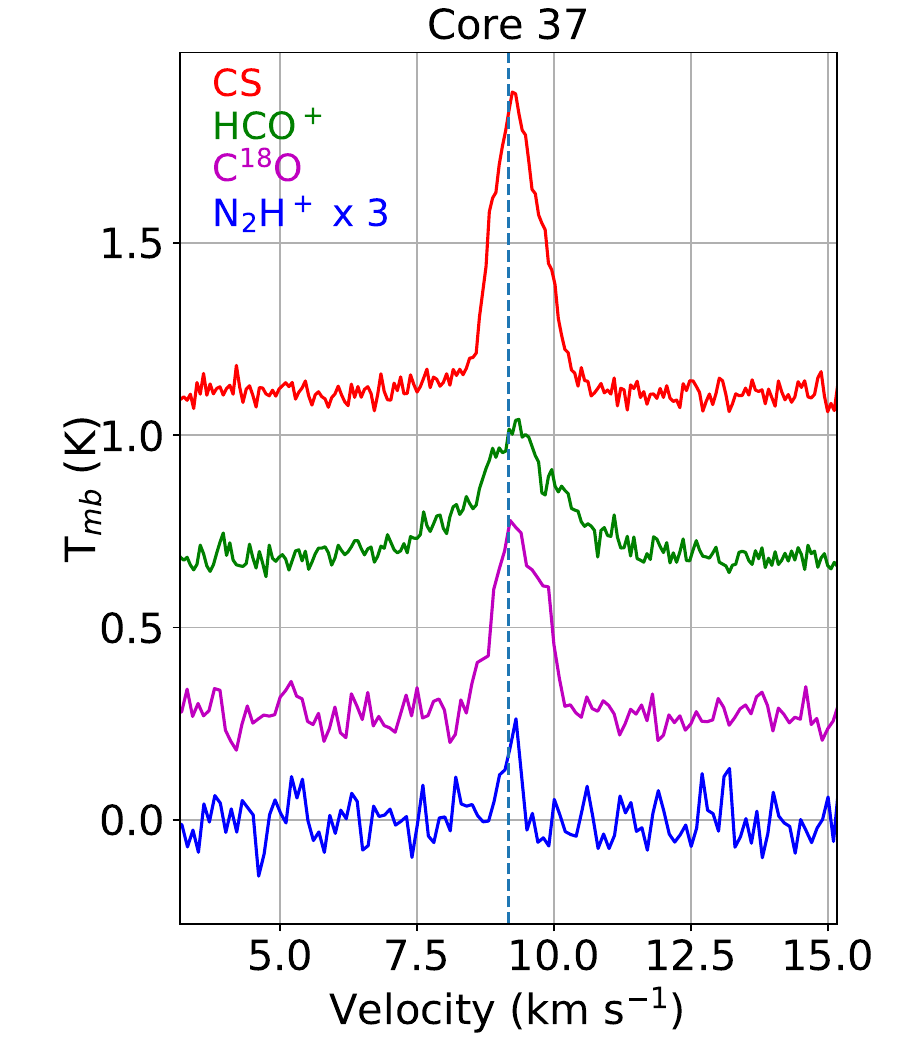}
  \includegraphics[width=0.25\textwidth]{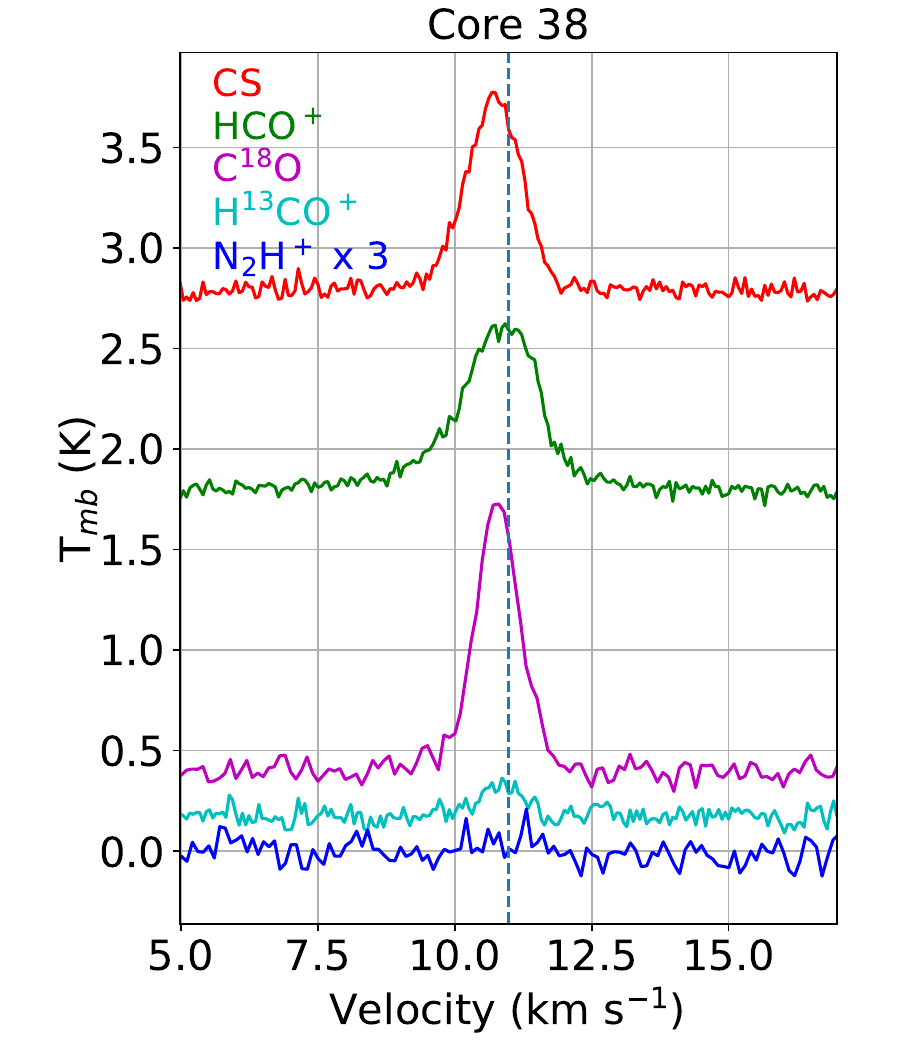} 
  \includegraphics[width=0.25\textwidth]{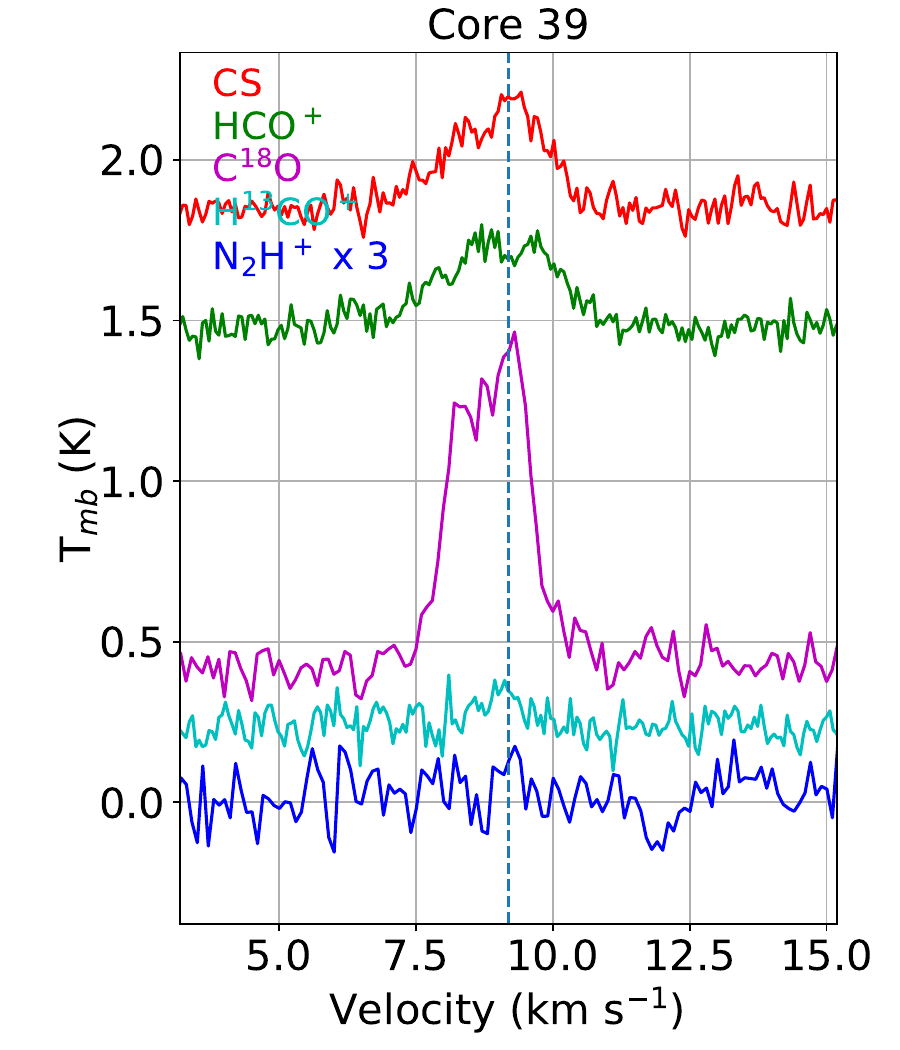}  
  \includegraphics[width=0.25\textwidth]{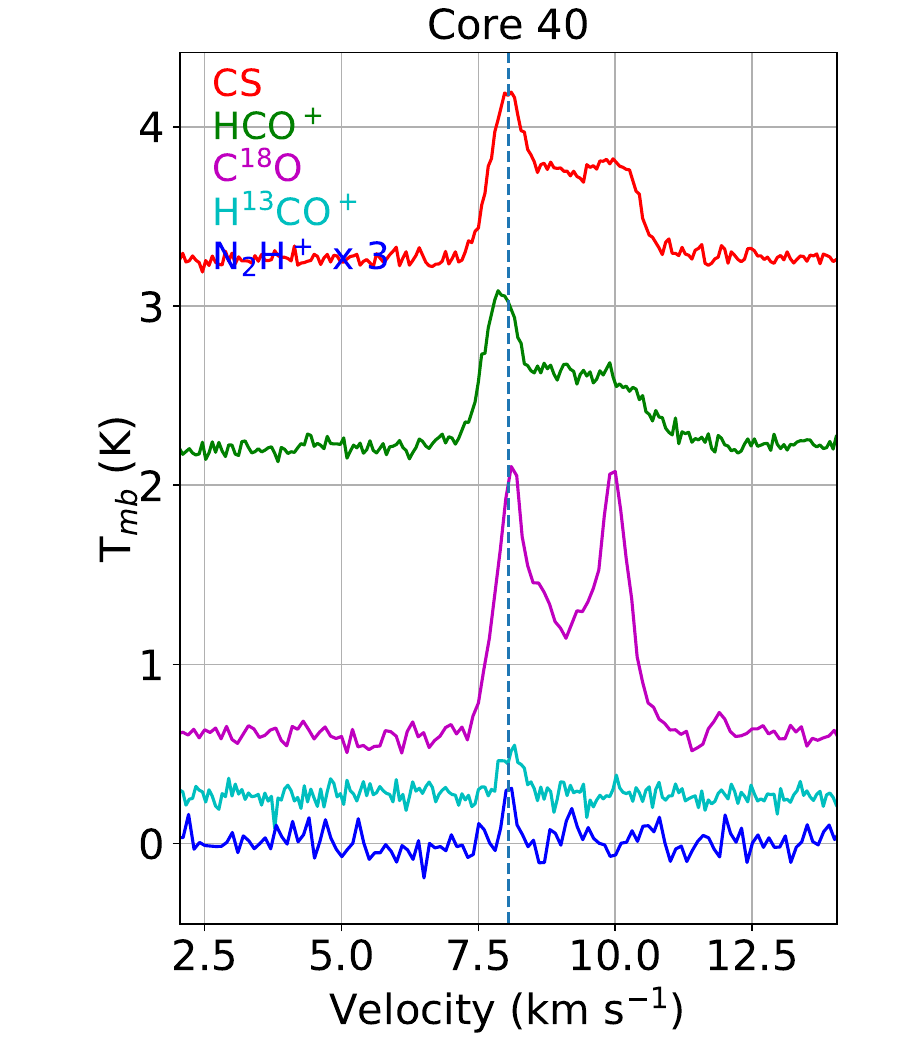} \\
  \includegraphics[width=0.25\textwidth]{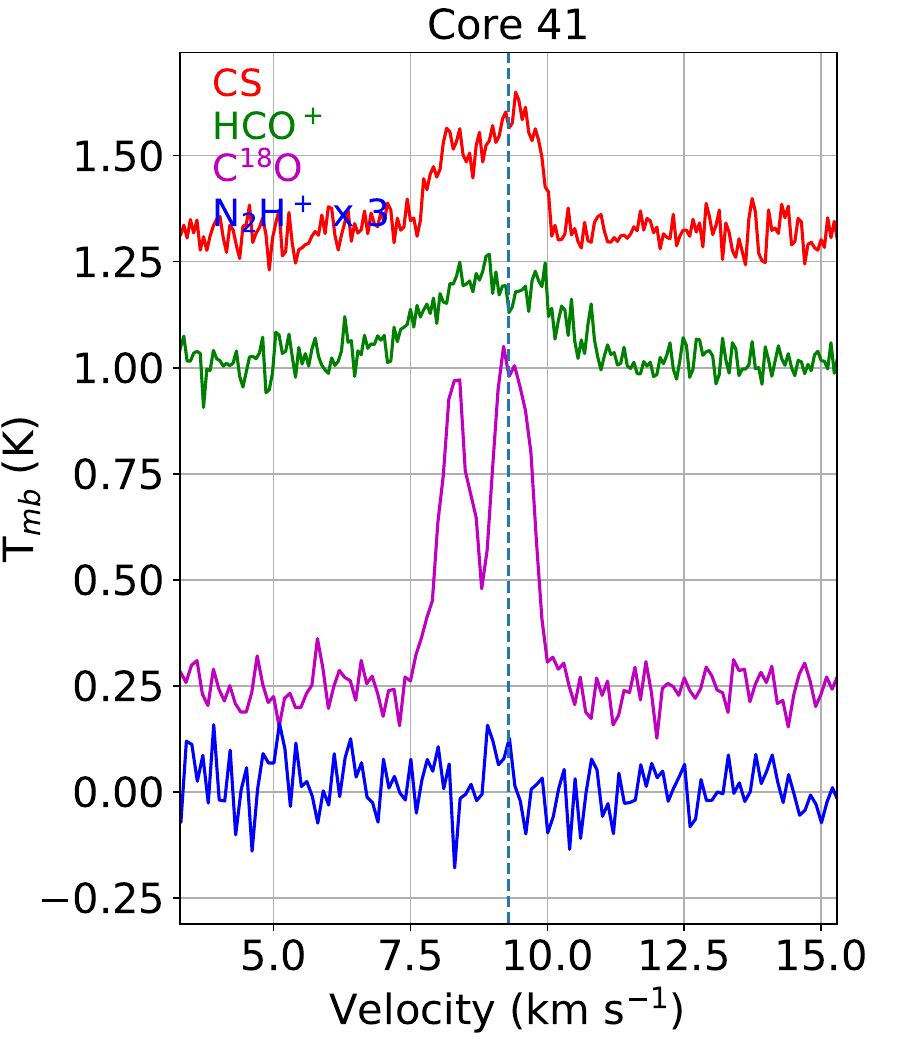} 
  \includegraphics[width=0.25\textwidth]{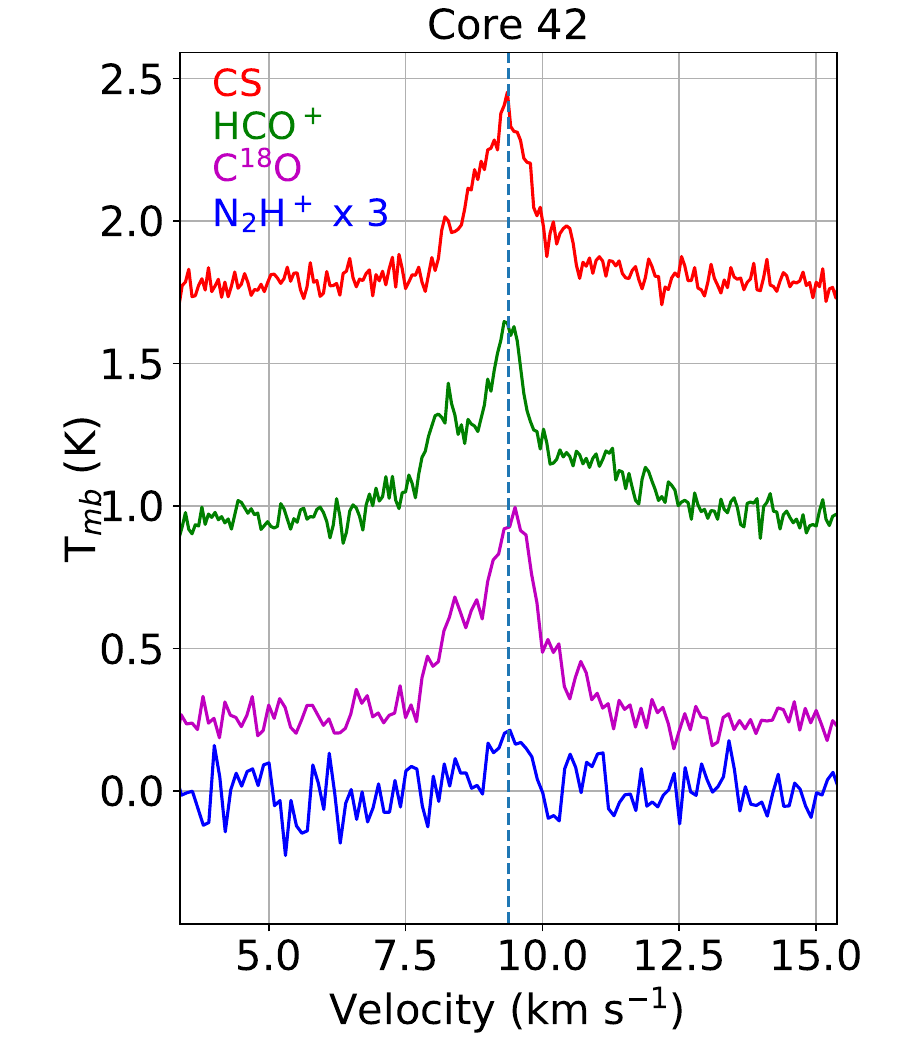}
  \includegraphics[width=0.25\textwidth]{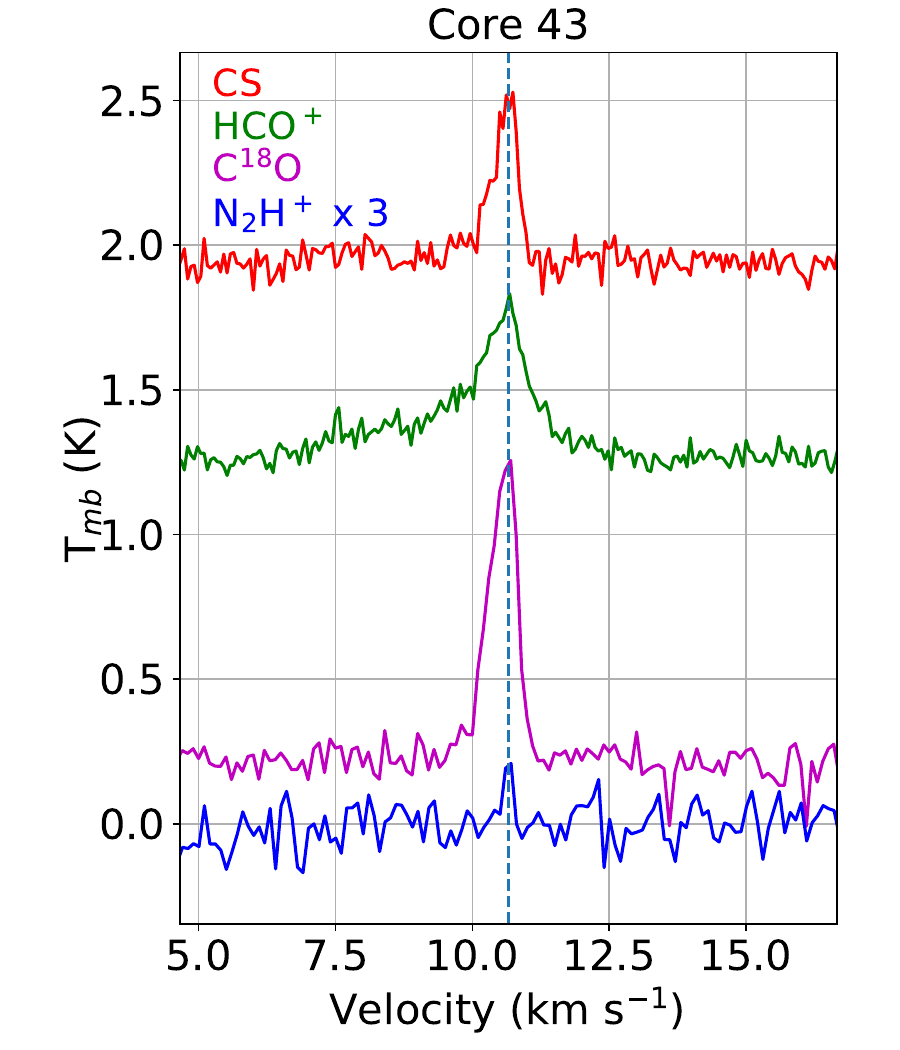}
  \includegraphics[width=0.25\textwidth]{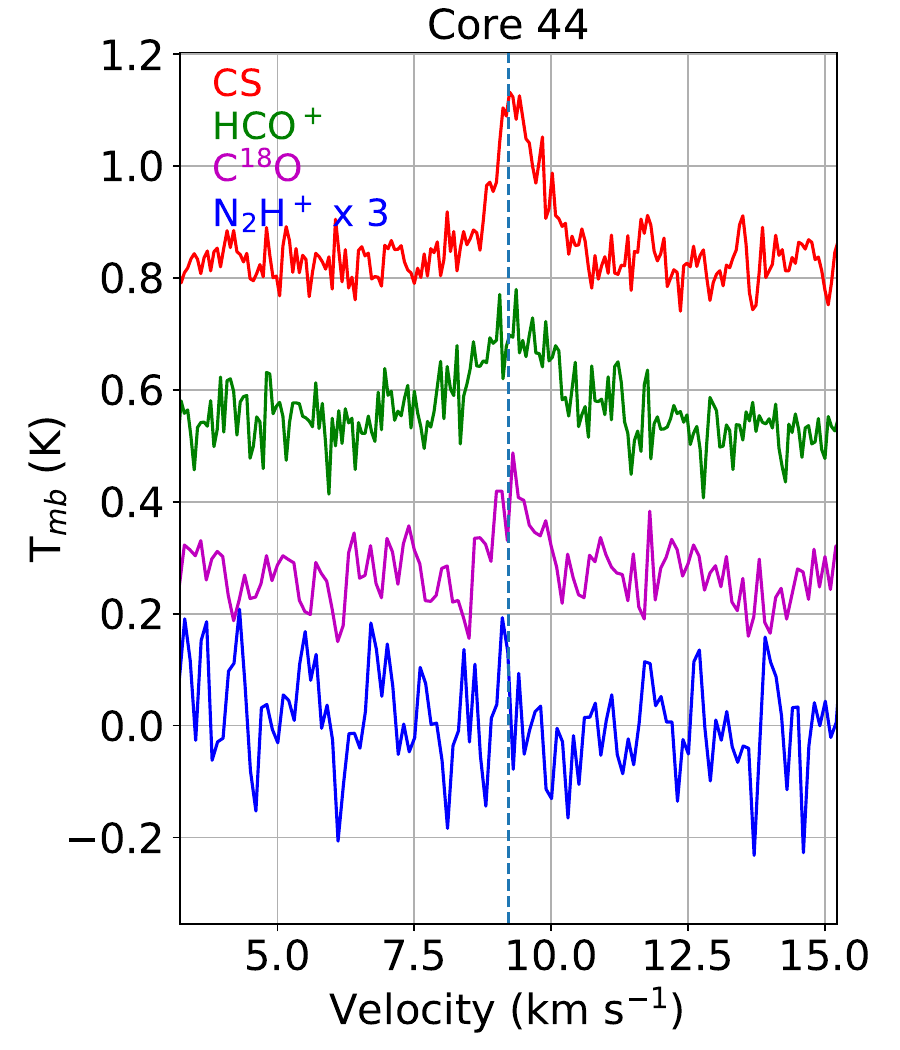} \\
  \includegraphics[width=0.25\textwidth]{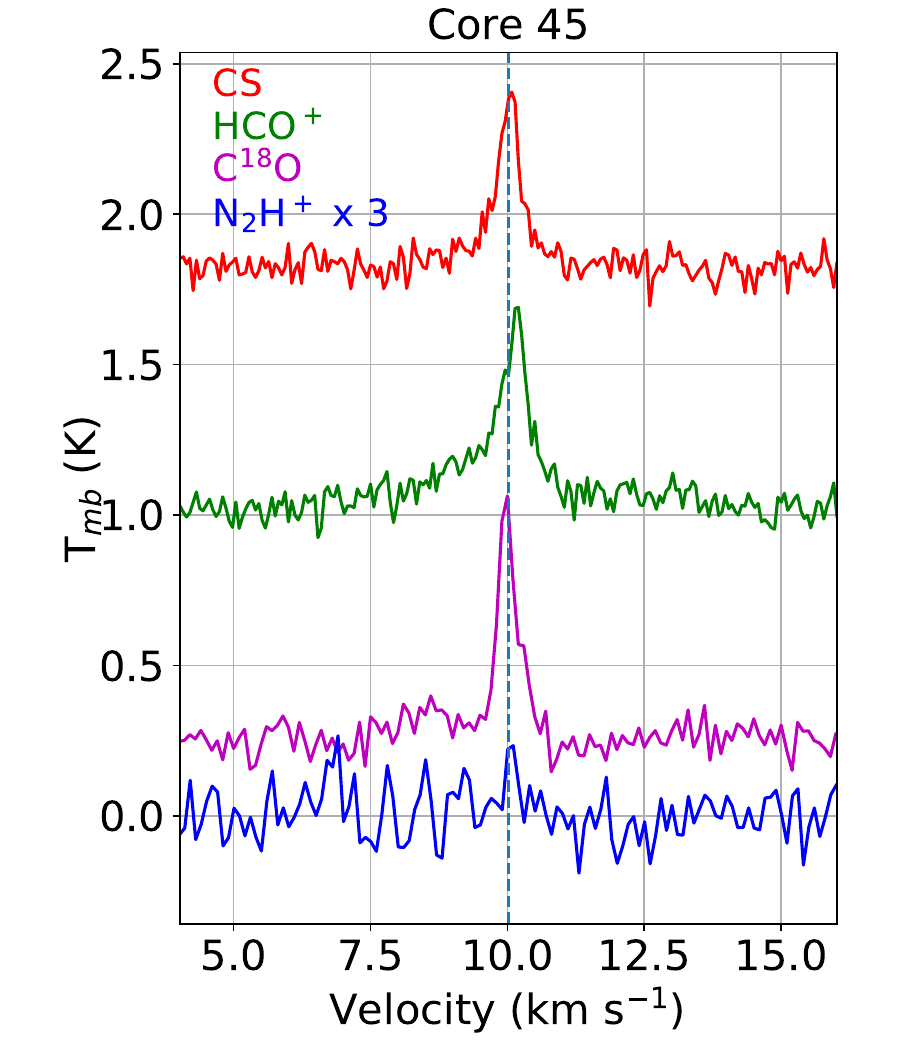}
  \includegraphics[width=0.25\textwidth]{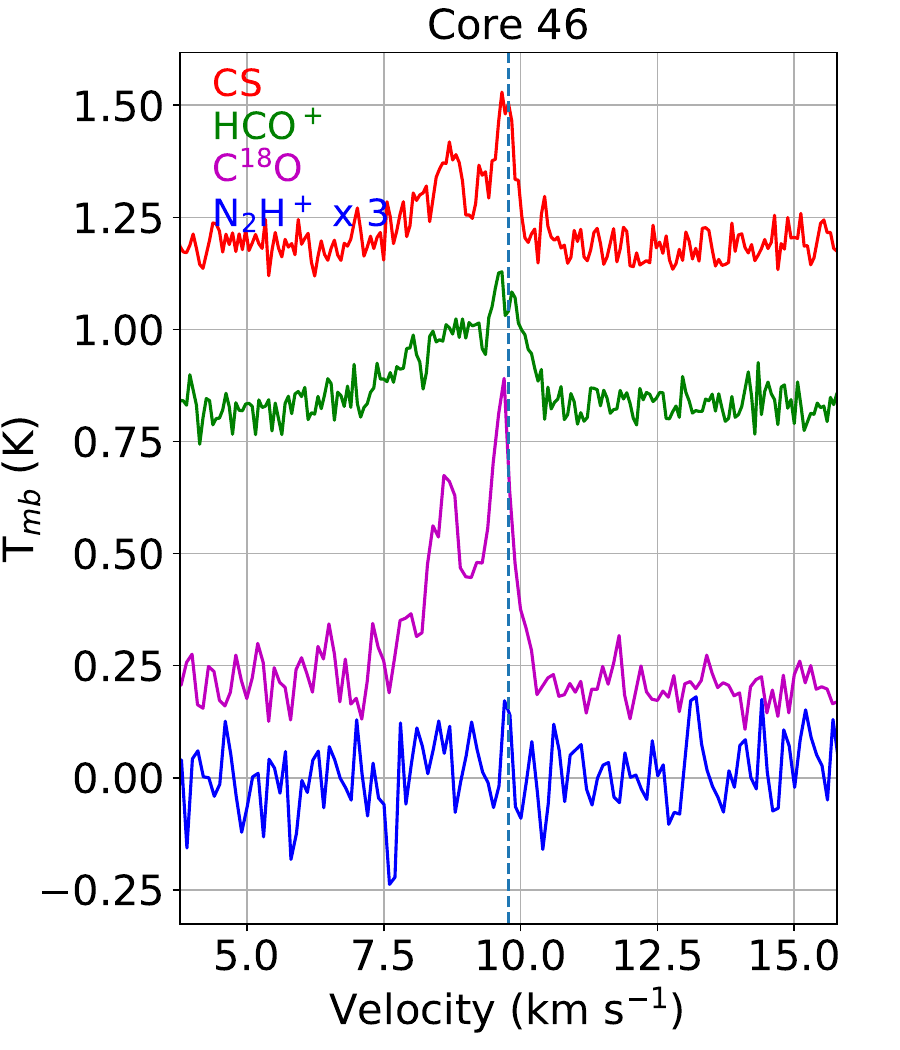}
  \includegraphics[width=0.25\textwidth]{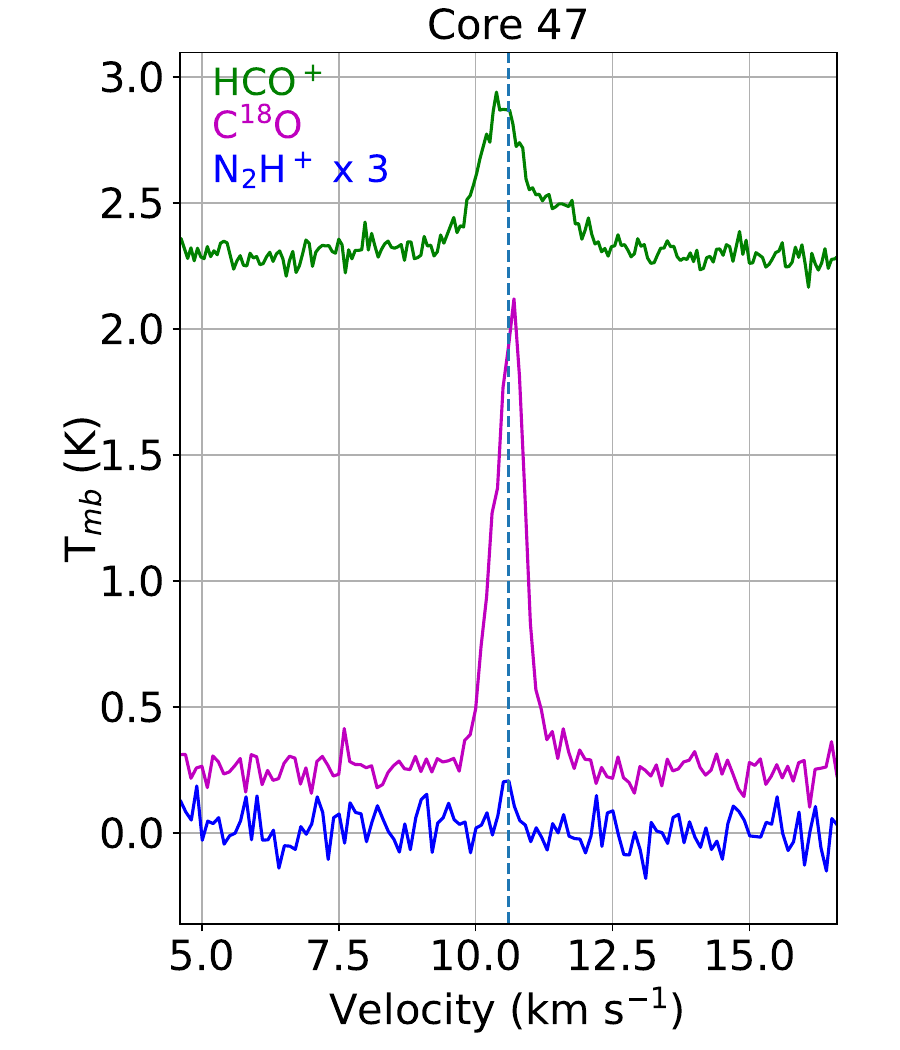}
  \includegraphics[width=0.25\textwidth]{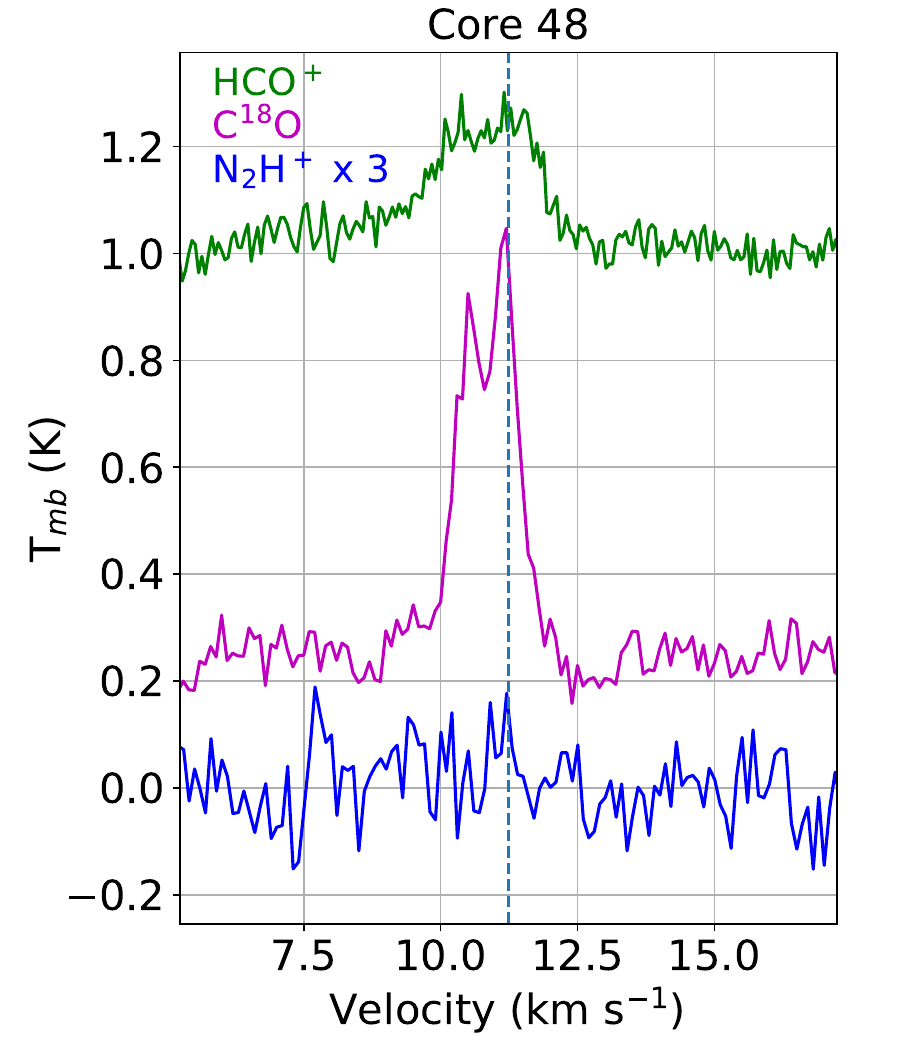} 
\end{tabbing}
 \caption{Continued.}
\label{fig:line_spec3}
\end{figure*}



\end{document}